\documentclass{emulateapj}
\slugcomment{{\sc Accepted to ApJ:} April 20, 2012} 

\usepackage{hyperref}

\usepackage[usenames]{color}
\definecolor{red}{rgb}{0.5,0.,0.}
\definecolor{green}{rgb}{0.,0.5,0.}

\newcommand{\um}{\ensuremath{ \mu\mbox{m}}}
\newcommand{\Deg}{\ensuremath{^\circ}}
\newcommand{\ngal}{45}
\newcommand{\ngalminor}{27}
\newcommand{\nbarred}{29}
\newcommand{\noval}{6}

\shorttitle{Kinematics across bulge types}
\shortauthors{Fabricius et al.}

\begin{document}

\title{Kinematic Signatures of Bulges Correlate with Bulge Morphologies \\and
\\S\'ersic Index$^*$}

\author{
Maximilian H. Fabricius\altaffilmark{1,2}
and Roberto P. Saglia \altaffilmark{1} 
and David B. Fisher\altaffilmark{3}  
and Niv Drory\altaffilmark{4} 
and Ralf Bender\altaffilmark{1,2}
and Ulrich Hopp\altaffilmark{1,2}}

\altaffiltext{1}{Max Planck Institute for Extraterrestrial Physics,
Giessenbachstrasse, 85748 Garching, Germany} 
\altaffiltext{2}{University Observatory Munich, Schienerstrasse 1, 81679
Munich, Germany}
\altaffiltext{3}{Laboratory of Millimeter Astronomy, University of Maryland,
College Park, MD 29742} 
\altaffiltext{4}{Instituto de Astronomia, Universidad Nacional Autonoma de
Mexico (UNAM), A.P. 70-264, 04510 Mexico, D.F.}

\begin{abstract}
We use the Marcario Low Resolution Spectrograph (LRS) at the
Hobby-Eberly-Telescope (HET) to study the kinematics of pseudobulges
and classical bulges in the nearby universe. We present major-axis rotational
velocities, velocity dispersions, and $h_3$ and $h_4$ moments derived from
high-resolution ($\sigma_{inst} \approx 39$~kms$^{-1}$) spectra for
\ngal~S0~to~Sc galaxies; for \ngalminor~of the galaxies we also present
minor axis data.  We combine our kinematics with bulge-to-disk decompositions.
We demonstrate for the first time that purely kinematic diagnostics of the
bulge dichotomy agree systematically with those based on S\'ersic index. Low
S\'ersic index bulges have both increased rotational support (higher $v/\sigma$
values) and on average lower central velocity dispersions. Furthermore, we
confirm that the same correlation also holds when visual morphologies are used
to diagnose bulge type.
The previously noted trend of photometrically flattened bulges to have shallower velocity
dispersion profiles turns to be significant and systematic if the S\'ersic
index is used to distinguish between pseudobulges and classical bulges.
The correlation between $h_3$ and $v/\sigma$ observed in elliptical
galaxies is also observed in intermediate type galaxies, irrespective
of bulge type.  Finally, we present evidence for formerly undetected
counter rotation in the two systems NGC\,3945 and NGC\,4736.\\
{\scriptsize $^*$Based on observations obtained with the Hobby-Eberly
Telescope, which is a joint project of the University of Texas at Austin, the
Pennsylvania State University, Stanford University,
Ludwig-Maximilians-Universit\"at München, and Georg-August-Universit\"ot
G\"ottingen.}
\end{abstract} 

\keywords{galaxies: bulges --- galaxies: evolution --- galaxies: formation ---
galaxies: structure --- galaxies: dynamics}

\section{Introduction}
There is ample observational evidence that bulges in early type spiral galaxies
come in different varieties. They are not all just like small elliptical
galaxies which happen to live in the centre of a spiral disk \citep{Kormendy93,
KK04}.

While classical bulges seem to lie on 
photometric projections of the 
the fundamental plane of elliptical
galaxies \citep{Fisher2010} pseudobulges resemble disks more than little
ellipticals. They are still photometrically distinct from the outer disk as
they appear as a central brightening above the inwards extrapolation of the
outer exponential disk profile. 
However, as opposed to classical bulges,
their S\'ersic indices fall close-to or below two \citep{Fisher2008}.
Other groups have shown that a large fraction of galaxies with boxy or peanut
shaped bulges do show signs of inner disks \citep{BF99,CB04, Kormendy10a}. 
The phenomenon of inner disks is however not limited to bulges that
morphologically resemble disks as a whole
\citep{Scorza1995,Emsellem2004,Falcon-Barroso2003,Falcon-Barroso2004,
Sarzi2006,Falcon-Barroso2006,Comeron2010}, although it seems ubiquitous in this
class of objects.

Internal secular evolution is commonly seen as an important channel for the
formation of central disk-like structures \citep{KK04, Athanassoula2005}.  In
this picture, asymmetries such as spiral structure and bars support the
angular momentum transfer of disk material and thereby the transport of gas
into the inner bulge regions. In their recent study of bulges within the local
11~Mpc volume \cite{Fisher2011} show that a majority of bulges in the local
universe are pseudobulges. Their existence in large quantities in our local (low
density) environment may seem to pose a problem for the understanding of the
baryonic physics of galaxy formation \citep{Kormendy10b} as, at first sight, it
is not clear how central disks would survive the large quantity of low redshift
($z <$ $\approx$1) mergers \citep{Stewart2008} typical of $\Lambda$CDM
simulations \citep{White78,White91}. But disk structures do not have to be
destroyed in all mergers. \citet{Hopkins2008} show that the heating of the
disks in a minor merging event is a non-linear function of progenitor mass
ratio once the satellite rigidity and the orbits are modelled properly. In
addition to the impact parameters and the mass fraction, the baryonic gas
content within the progenitors is
an important parameter to the final result of a merging process
\citep{Hopkins2009,Stewart2009, Governato2009} --- {\it wet mergers} are more
likely to produce disks. Minor mergers may also create inner disk structures
\citep{Eliche-Moral2011} while increasing the S\'ersic index only moderately
\citep{Eliche-Moral2006}.  Based on semi-analytical models for hierarchical
growth which include prescriptions for the survival of disks,
\citet{Fontanot2011} show that the existence of the majority of the galaxies
with no significant bulge component in the local volume
can be understood in the context of $\Lambda$CDM.

\cite{KK04} identify a number of criteria to differentiate between classical
and pseudobulges such as the bulge vs.~the disk ellipticity, their location in
the $v_{max}/\sigma$ diagram and the bulge morphology.
\citet{Fisher2008,Fisher2010} show that the bulge S\'ersic Index is successful
in differentiating bulge types --- pseudobulges have S\'ersic indices that fall
near or below $n = 2$, unlike classical bulges and elliptical galaxies which
have higher S\'ersic indices.

The identification of the bulge morphology as well as the accurate derivation
of photometric structural parameters heavily relies on high spatial resolution
imaging. 
Recently, such data has become available for a large number of bulges. Adding
sufficiently high-resolution spectroscopic data allows us to ask the question
whether all morphologically disk-like bulges also show kinematic disk-like
behaviour, such as high $v/\sigma$ values and/or flattening or drops in the
$\sigma$ profile? Also, whether differences seen in structural parameters such
as S\'ersic index are reflected in the kinematic structure as well?

In \S\ref{sec:sample} we describe the sample selection and characteristics, in
\S\ref{sec:obs} we describe the long-slit observations, in
\S\ref{sec:reduction} we give account on
the derivation of $H$-band surface brightness profiles and their decomposition,
as well as 
the details of the data reduction, especially the removal of emission features, 
and finally 
the kinematic extraction.  Our
results are presented in \S\ref{sec:results}, in \S\ref{sec:discussion} we
discuss correlations between kinematic parameters and morphological parameters
and indications for an increased rotational support of pseudobulges.  We
finally discuss and summarize our findings in \S\ref{sec:summary}.

\section{Sample}
\label{sec:sample}
\begin{figure*}
        \begin{center}
	\includegraphics[width=0.8\textwidth]{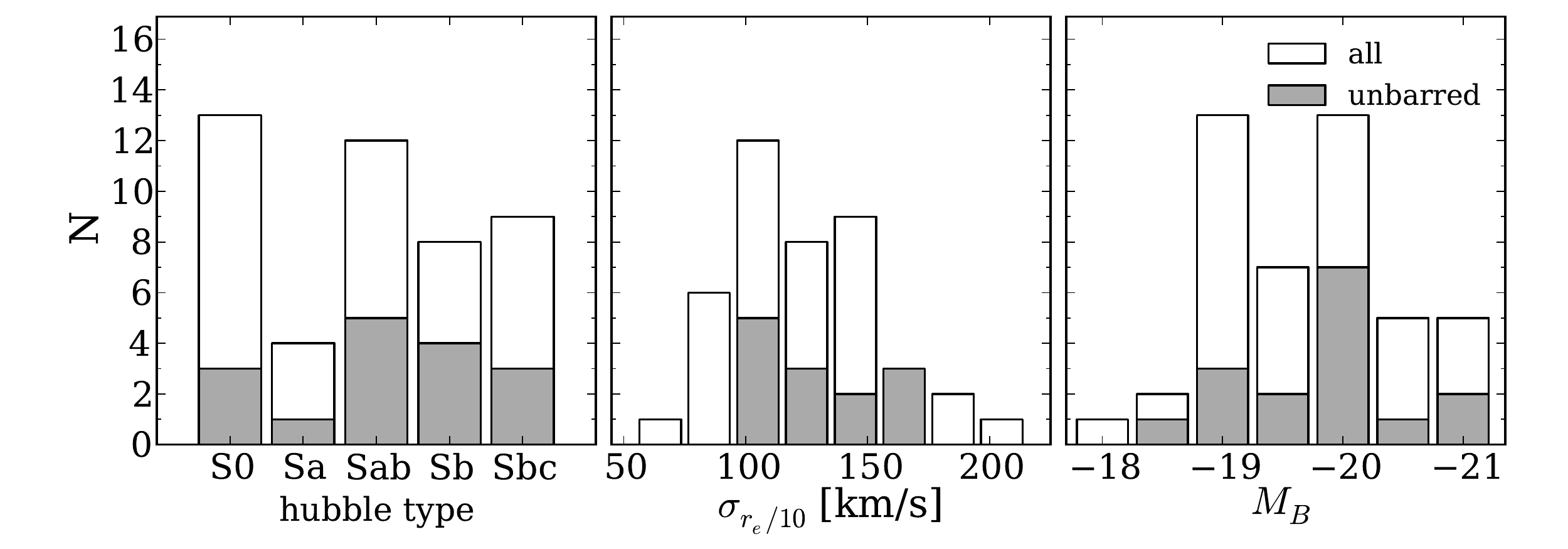}
        \end{center}
	\caption{\label{fig:sample}Distribution of Hubble types, 
	central velocity dispersions and total magnitudes
	in the sample.} 
\end{figure*}

As we aim to study the kinematics of bulges, our sample consists of
\ngal~galaxies spanning the full range of Hubble types that do contain
bulges: S0 to Sbc.  Further, roughly two-thirds of our galaxies are
barred, a similar fraction to the total fraction of bars observed in
the local universe (see Fig.~\ref{fig:sample}).
Table~\ref{tab:sample} lists the objects in our sample. 
For signal-to-noise ($S/N$) reasons we are
biased towards high-luminosity objects.  Absolute B-Band magnitudes span the
range from $M_B =-17.3$ to $M_B = -21.3$. Central velocity dispersions lie
between 60~kms$^{-1}$ and 220~kms$^{-1}$. We select our targets to be located
close enough in distance to properly resolve the bulge regions in typical
seeing conditions.  
With the exception of NGC\,2964 and NGC\,4030, all galaxies have bulge
radii larger than 5~arcseconds and are typically located at distances
closer than 25~Mpc.  Only NGC\,4030, NGC\,4260, and NGC\,4772 are located
at significantly larger distances of 29.3~Mpc, 48.4~Mpc and 40.9~Mpc,
respectively.  The bulge radii of 7.3~arcseconds and 23.5~arcseconds
of the latter two leave us confident that we are able to nevertheless
sufficiently resolve their bulges. NGC\,2964 and NGC\,4030 have bulge
radii of 3.1~arcseconds and 3.0~arcseconds, and are excluded from all
structural analysis concerning the bulges, we restrict ourselves to
presenting their kinematic data.

%
%
In order to break the known degeneracy between the bulge effective radius and
S\'ersic index in 1D surface brightness decompositions \citep{Graham97} we
require all our targets to have HST imaging in F160W, I, or R band.
Most of the objects are found in \cite{Fisher2008} and/or \cite{Fisher2010} and
have extensive HST and ground-based multi-wavelength coverage.
%
To allow for a visual inspection and morphological classification of the bulge
region we select objects which have close-to $V$-band HST images of their bulge
region (see \S\ref{sec:pseudoid}) available from the archive and we do not
observe edge-on or close-to edge-on objects ($i > 70\Deg$).
Four objects in our sample do have a larger inclination.  NGC\,1023 and
NGC\,4371 are S0-types and contain very little dust and allow an undisturbed
view into the bulge region.  The situation is different for NGC\,3593 and
NGC\,7331, where the inclination and --- in the case of NGC\,3593 --- the
absence of an optical HST image inhibits a morphological classification. We
present the kinematic data for those objects refrain from classifying them as
classical or pseudobulges.

%
\begin{deluxetable}{clrrrrlr}
\tablecolumns{6}
\tabletypesize{\scriptsize}
\tablewidth{0pc}
\tablecaption{Galaxy sample. \label{tab:sample}}
\tablehead{
\colhead{Galaxy} &\colhead{$htype$} &\colhead{D} &\colhead{$src_D$} &\colhead{$M_B$} & \colhead{\it i} \\
\colhead{} &\colhead{} &\colhead{[Mpc]} &\colhead{} &\colhead{mag} &\colhead{[\Deg]} 
\\
\colhead{(1)} &\colhead{(2)} &\colhead{(3)} &\colhead{(4)} &\colhead{(5)} &\colhead{(6)} }
\startdata
NGC\,1023 & .LBT-.. & 11.5 & 2 & -20.0 & 77 \\
NGC\,2460 & .SAS1.. & 23.6 & 1 & -19.1 & 44 \\
NGC\,2681 & PSXT0.. & 17.2 & 2 & -20.1 & 0 \\
NGC\,2775 & .SAR2.. & 14.4 & 1 & -19.8 & 41 \\
NGC\,2841 & .SAR3*. & 9.0 & 1 & -19.7 & 68 \\
NGC\,2859 & RLBR+.. & 25.4 & 1 & -20.2 & 33 \\
NGC\,2880 & .LB.-.. & 21.9 & 2 & -19.2 & 68 \\
NGC\,2964 & .SXR4*. & 19.9 & 1 & -19.1 & 58 \\
NGC\,3031 & .SAS2.. & 3.9 & 2 & -20.1 & 59 \\
NGC\,3166 & .SXT0.. & 22.0 & 1 & -20.4 & 56 \\
NGC\,3245 & .LAR0*. & 20.9 & 2 & -19.9 & 67 \\
NGC\,3351 & .SBR3.. & 8.6 & 1 & -19.1 & 42 \\
NGC\,3368 & .SXT2.. & 8.6 & 1 & -19.6 & 55 \\
NGC\,3384 & .LBS-*. & 8.6 & 1 & -18.8 & 62$^a$ \\
NGC\,3521 & .SXT4.. & 8.1 & 1 & -19.7 & 42 \\
NGC\,3593 & .SAS0*. & 8.8 & 1 & -17.9 & 75 \\
NGC\,3627 & .SXS3.. & 12.6 & 4 & -20.9 & 57 \\
NGC\,3675 & .SAS3.. & 10.7 & 1 & -19.1 & 60 \\
NGC\,3898 & .SAS2.. & 21.9 & 1 & -20.1 & 57 \\
NGC\,3945 & RLBT+.. & 19.0 & 1 & -19.6 & 63 \\
NGC\,3953 & .SBR4.. & 13.2 & 1 & -19.8 & 62 \\
NGC\,3992 & .SBT4.. & 22.9 & 5 & -21.2 & 47 \\
NGC\,4030 & .SAS4.. & 29.3 & 6 & -21.1 & 40 \\
NGC\,4203 & .LX.-*. & 15.1 & 2 & -19.1 & 27 \\
NGC\,4260 & .SBS1.. & 48.4 & 7 & -20.7 & 70 \\
NGC\,4274 & RSBR2.. & 12.5 & 1 & -19.1 & 66 \\
NGC\,4314 & .SBT1.. & 12.5 & 1 & -19.1 & 16 \\
NGC\,4371 & .LBR+.. & 14.3 & 1 & -19.0 & 79 \\
NGC\,4379 & .L..-P* & 15.9 & 8 & -18.4 & 42 \\
NGC\,4394 & RSBR3.. & 14.3 & 1 & -19.0 & 20 \\
NGC\,4448 & .SBR2.. & 12.5 & 1 & -18.5 & 52 \\
NGC\,4501 & .SAT3.. & 14.3 & 1 & -20.4 & 61 \\
NGC\,4536 & .SXT4.. & 12.2 & 9 & -19.3 & 59 \\
NGC\,4569 & .SXT2.. & 14.3 & 1 & -20.5 & 66 \\
NGC\,4698 & .SAS2.. & 14.3 & 1 & -19.3 & 51 \\
NGC\,4736 & RSAR2.. & 4.2 & 1 & -19.1 & 35 \\
NGC\,4772 & .SAS1.. & 40.9 & 5 & -21.1 & 68 \\
NGC\,4826 & RSAT2.. & 7.5 & 2 & -20.0 & 60 \\
NGC\,5055 & .SAT4.. & 7.8 & 1 & -20.2 & 56 \\
NGC\,5248 & .SXT4.. & 14.8 & 1 & -19.9 & 56 \\
NGC\,5566 & .SBR2.. & 20.1 & 5 & -20.1 & 61 \\
NGC\,7177 & .SXR3.. & 19.8 & 1 & -19.5 & 42 \\
NGC\,7217 & RSAR2.. & 16.6 & 1 & -19.8 & 36 \\
NGC\,7331 & .SAS3.. & 15.5 & 1 & -20.8 & 75 \\
NGC\,7743 & RLBS+.. & 19.2 & 10 & -19.0 & 40 \\
\enddata
\tablenotetext{Notes:}{
1) Galaxy name. 
2) Hubble type (RC3). 
3) Distance. 
4) Source for distance: 
1= \citet{Tully1988}
2= \citet{Tonry2001}
3= \citet{deVaucouleurs1991}
4= \citet{Saha2006}
5= \citet{Tully2009}
6= \citet{Springob2009}
7= \citet{Ekholm2000}
8= \citet{Blakeslee2009}
9= \citet{Riess2009}
10= \citet{Jensen2003}
5) Total $B$-band magnitude (Hyperleda).
6) Inclination (Hyperleda). 
{\it Comments:} a) No value in Hyperleda, from Peter Erwin (private communication).
}
\end{deluxetable}
\section{Observations}
\label{sec:obs}
We obtain major axis spectra for all and minor axis spectra for about half of
the galaxies in our sample.  In some cases the observed position angle is not
identical to the one published in Hyperleda\footnote{\tt http://leda.univ-lyon1.fr}
\citep{Paturel2003}. Also, in a few cases the {\it minor axis}
position angle is not orthogonal to the major axis position angle. We label
observations accordingly in Tab.~\ref{tab:observations}.

Observations were carried out in service mode during the period from April 2005
to April 2010  (see Tab.~\ref{tab:observations}) at the Hobby-Eberly Telescope
(HET; \citealp{RAMSEY98}) at McDonald Observatory. We use the Marcario Low
Resolution Spectrograph (LRS; \citealp{HILL98}) with a one arcsecond wide and
3.5~arcminute long-slit, the E2 phase volume holographic GRISM, and a Ford
Aerospace 3072$\times$1024 15~\um~pixel (usable range 2750$\times$900 pixel)
CCD detector yielding a spatial scale of 0.235 arcseconds per pixel. The
spectra cover the wavelength range from 4790~\AA~to 5850~\AA~with 0.38~\AA~per
pixel and a median instrumental resolution of $\sigma_{inst}$~=~39.3~kms$^{-1}$
(as measured on the 5577~\AA~night-sky line). The seeing varies from 1.2 to
4~arcseconds with a median value of 2.2~arcseconds. Integration times vary from
1.800~s to 3.800~s and on-object exposures are typically split into two for
cosmic rejection. For large galaxies where the DSS image of the galaxy exceeds the
slit length, we obtain separate exposures of empty sky with an exposure time of
420~s at the end of the science observation. In order to avoid an azimuth move
of the telescope, the sky exposures are typically obtained one
hour in RA after the object but at similar DEC.

\begin{center}
\begin{deluxetable*}{cllrlrl}
\tabletypesize{\scriptsize}
\tablecaption{List of observations. \label{tab:observations}}
\tablewidth{0pt}
\tablehead{\colhead{Galaxy} &\colhead{axis} &\colhead{date} &\colhead{seeing}
&\colhead{exp.-time} &\colhead{angle} &\colhead{sky}\\ 
\colhead{} &\colhead{} &\colhead{} &\colhead{["]} &\colhead{[s]} &\colhead{[\Deg]}
&\colhead{} \\
\colhead{(1)} &\colhead{(2)} &\colhead{(3)} &\colhead{(4)} &\colhead{(5)} &\colhead{(6)}
&\colhead{(7)}
}
\startdata
NGC\,1023 & MJ$^{}$ & 2009-10-24 & 1.9 & 2,400 & 87 & yes \\
NGC\,1023 & MN$^{}$ & 2009-10-23 & 2.2 & 2,400 & 177 & yes \\
NGC\,2460 & MJ$^{}$ & 2005-11-08 & 2.0 & 1,800 & 30 & no \\
NGC\,2460 & MN$^{}$ & 2005-11-08 & 2.0 & 1,800 & 120 & no \\
NGC\,2681 & MJ$^{}$ & 2007-10-21 & 2.0 & 2,700 & 114 & yes \\
NGC\,2775 & MJ$^{}$ & 2008-03-05 & 2.6 & 2,700 & 156 & no \\
NGC\,2775 & MN$^{}$ & 2008-12-24 & 3.4 & 2,400 & 66 & no \\
NGC\,2841 & MJ$^{}$ & 2007-11-08 & 1.6 & 1,800 & 152 & no \\
NGC\,2841 & MN$^{c}$ & 2008-12-23 & 3.2 & 2,280 & 58 & no \\
NGC\,2859 & MJ$^{}$ & 2005-11-09 & 1.6 & 3,000 & 80 & no \\
NGC\,2859 & MN$^{}$ & 2006-05-25 & 2.5 & 1,800 & 170 & no \\
NGC\,2880 & MJ$^{}$ & 2009-11-16 & 3.4 & 2,200 & 142 & no \\
NGC\,2880 & MN$^{}$ & 2009-12-18 & 2.2 & 2,400 & 52 & no \\
NGC\,2964 & MJ$^{}$ & 2010-02-18 & 1.9 & 2,400 & 96 & yes \\
NGC\,2964 & MN$^{}$ & 2010-03-21 & 2.2 & 2,400 & 7 & yes \\
NGC\,3031 & MJ$^{b}$ & 2007-02-22 & 2.2 & 2,700 & 137 & yes \\
NGC\,3031 & MN$^{c}$ & 2005-12-28 & 3.1 & 1,800 & 67 & no \\
NGC\,3166 & MJ$^{}$ & 2008-02-06 & 2.0 & 2,454 & 85 & yes \\
NGC\,3166 & MN$^{}$ & 2008-12-25 & 2.4 & 2,400 & 175 & no \\
NGC\,3245 & MJ$^{}$ & 2008-02-06 & 1.7 & 2,700 & 174 & yes \\
NGC\,3245 & MN$^{}$ & 2008-12-25 & 2.5 & 2,400 & 84 & no \\
NGC\,3351 & MJ$^{b}$ & 2008-02-09 & 1.5 & 2,550 & 165 & yes \\
NGC\,3351 & MN$^{}$ & 2008-12-27 & 5.0 & 2,400 & 75 & yes \\
NGC\,3368 & MJ$^{b}$ & 2007-02-26 & 3.1 & 2,420 & 153 & yes \\
NGC\,3368 & MN$^{c}$ & 2008-12-09 & 2.6 & 2,400 & 87 & yes \\
NGC\,3384 & MJ$^{}$ & 2009-12-13 & 1.7 & 2,400 & 53 & yes \\
NGC\,3384 & MN$^{}$ & 2010-02-19 & 1.8 & 2,400 & 143 & yes \\
NGC\,3521 & MJ$^{}$ & 2007-04-18 & 1.6 & 2,700 & 161 & yes \\
NGC\,3521 & MN$^{}$ & 2009-01-03 & 2.3 & 2,528 & 74 & yes \\
NGC\,3593 & MJ$^{}$ & 2010-02-17 & 1.2 & 2,400 & 84 & yes \\
NGC\,3627 & MJ$^{b}$ & 2006-12-27 & 2.3 & 1,800 & 10 & no \\
NGC\,3627 & MN$^{}$ & 2007-02-23 & 2.2 & 1,800 & 100 & no \\
NGC\,3675 & MJ$^{}$ & 2008-03-05 & 2.6 & 2,700 & 178 & yes \\
NGC\,3898 & MJ$^{}$ & 2007-04-19 & 1.6 & 2,700 & 108 & no \\
NGC\,3945 & MJ$^{b}$ & 2009-12-17 & 2.1 & 2,400 & 154 & yes \\
NGC\,3945 & MN$^{}$ & 2010-04-12 & 1.6 & 4,200 & 64 & yes \\
NGC\,3953 & MJ$^{b}$ & 2008-02-06 & 2.0 & 2,700 & 32 & yes \\
NGC\,3992 & MJ$^{b}$ & 2008-12-28 & 2.7 & 2,700 & 66 & yes \\
NGC\,4030 & MJ$^{}$ & 2005-04-05 & 2.3 & 1,800 & 27 & no \\
NGC\,4203 & MJ$^{}$ & 2007-04-12 & 1.3 & 2,520 & 7 & yes \\
NGC\,4260 & MJ$^{}$ & 2008-12-29 & 2.7 & 2,700 & 62 & no \\
NGC\,4274 & MJ$^{}$ & 2007-04-19 & 1.7 & 2,623 & 99 & yes \\
NGC\,4314 & MJ$^{}$ & 2007-02-20 & 2.3 & 2,700 & 127 & no \\
NGC\,4371 & MJ$^{b}$ & 2006-12-27 & 2.4 & 1,800 & 85 & no \\
NGC\,4371 & MN$^{}$ & 2006-06-19 & 1.8 & 1,800 & 175 & no \\
NGC\,4379 & MJ$^{}$ & 2007-02-21 & 2.7 & 1,800 & 97 & yes \\
NGC\,4394 & MJ$^{}$ & 2007-05-11 & 1.5 & 2,556 & 123 & yes \\
NGC\,4448 & MJ$^{b}$ & 2007-04-16 & 2.3 & 2,700 & 85 & yes \\
\enddata
\end{deluxetable*}
\end{center}
\setcounter{table}{1}
\begin{center}
\begin{deluxetable*}{cllrlrl}
\tabletypesize{\scriptsize}
\tablecaption{--- Continued}
\tablewidth{0pt}
\tablehead{\colhead{Galaxy} &\colhead{axis} &\colhead{date} &\colhead{seeing}
&\colhead{exp.-time} &\colhead{angle} &\colhead{sky}\\ 
\colhead{} &\colhead{} &\colhead{} &\colhead{["]} &\colhead{[s]} &\colhead{[\Deg]}
&\colhead{} \\
\colhead{(1)} &\colhead{(2)} &\colhead{(3)} &\colhead{(4)} &\colhead{(5)} &\colhead{(6)}
&\colhead{(7)}
}
\startdata
NGC\,4501 & MJ$^{}$ & 2010-04-06 & 2.0 & 3,340 & 140 & yes \\
NGC\,4501 & MN$^{}$ & 2010-04-08 & 2.1 & 2,505 & 50 & yes \\
NGC\,4536 & MJ$^{}$ & 2010-04-08 & 3.6 & 2,385 & 120 & yes \\
NGC\,4536 & MN$^{}$ & 2010-04-10 & 2.1 & 2,500 & 30 & yes \\
NGC\,4569 & MJ$^{}$ & 2007-06-15 & 2.1 & 2,700 & 14 & yes \\
NGC\,4569 & MN$^{}$ & 2010-04-12 & -$^a$ & 2,880 & 115 & yes \\
NGC\,4698 & MJ$^{}$ & 2008-12-28 & 2.4 & 2,700 & 166 & no \\
NGC\,4736 & MJ$^{}$ & 2009-12-12 & 1.9 & 2,400 & 105 & yes \\
NGC\,4736 & MN$^{c}$ & 2008-04-01 & 2.2 & 2,700 & 30 & yes \\
NGC\,4772 & MJ$^{b}$ & 2008-12-29 & 2.5 & 2,700 & 145 & no \\
NGC\,4826 & MJ$^{b}$ & 2008-01-09 & 2.0 & 2,187 & 96 & yes \\
NGC\,4826 & MN$^{c}$ & 2009-06-25 & 1.5 & 2,122 & 25 & yes \\
NGC\,5055 & MJ$^{}$ & 2008-03-05 & 4.0 & 2,700 & 103 & yes \\
NGC\,5055 & MN$^{}$ & 2009-06-26 & 2.2 & 2,400 & 13 & yes \\
NGC\,5248 & MJ$^{b}$ & 2007-04-18 & 1.6 & 2,700 & 109 & yes \\
NGC\,5566 & MJ$^{}$ & 2005-07-07 & 2.5 & 1,800 & 30 & no \\
NGC\,5566 & MN$^{}$ & 2006-05-24 & 2.0 & 1,750 & 120 & no \\
NGC\,7177 & MJ$^{b}$ & 2007-08-11 & 2.3 & 2,700 & 60 & yes \\
NGC\,7177 & MN$^{c}$ & 2009-11-12 & 1.9 & 2,600 & 173 & yes \\
NGC\,7217 & MJ$^{}$ & 2007-08-12 & 1.7 & 2,700 & 81 & yes \\
NGC\,7217 & MN$^{c}$ & 2008-12-29 & 1.4 & 2,400 & 178 & no \\
NGC\,7331 & MJ$^{}$ & 2007-08-11 & 1.7 & 2,700 & 171 & yes \\
NGC\,7743 & MJ$^{b}$ & 2008-12-28 & 2.3 & 2,400 & 100 & no \\
NGC\,7743 & MN$^{c}$ & 2009-10-17 & 2.3 & 2,400 & 167 & no 
\enddata
\tablenotetext{Notes:}{ 
1) Galaxy name.
2) MJ=major axis, MN=minor axis.
3) Date of observation. 
4) Seeing FWHM. 
5) Total exposure time. 
6) Slit position angle east of north.
7) Dedicated sky frame was taken.\nl
{\it Comments:} 
a) No seeing information available. b) The position angle 
differs by more than 10\Deg\ from the Hyperleda published value. 
c) MN axis PA not orthogonal to MJ axis PA.
}
\end{deluxetable*}
\end{center}
Furthermore we observe a collection of kinematic template stars (G and K
giants, see Tab.~\ref{tab:templates}, metallicity: [Fe/H]~=~-0.35~--~0.46) at
the beginning and spectroscopic standards throughout the duration of this
campaign.  The stars are wiggled and trailed along the slit such that a
spectrum is recorded at each position where the star crosses the slit.  This is
used to map out the anamorphic distortion of the spectrograph.   
%
%
\tabletypesize{\scriptsize}
\begin{center}
\begin{deluxetable}{llrl}
\tablecaption{Observed kinematic templates.\label{tab:templates}}
\tablewidth{0pt}
\tablehead{
\colhead{Identifier} & \colhead{type} & \colhead{[Fe/H]} & \colhead{date of obs.} \\ 
\colhead{(1)} & \colhead{(2)} & \colhead{(3)} & \colhead{(4)}    
} 
\startdata
HR\,2600   & K2III   & -0.35  &  04/03/2005 \\ 
HR\,3369   & G9III   &  0.17  &  04/02/2005 \\ 
HR\,3418   & K2III   &  0.09  &  04/03/2005 \\ 
HR\,3427   & K0III   &  0.16  &  04/03/2005 \\ 
HR\,3428   & K0III   &  0.24  &  04/03/2005 \\
HR\,3905   & K2IIIb  &  0.46  &  04/02/2005 \\ 
HR\,6018   & K0III-IV&  0.01  &  04/02/2005 \\ 
HR\,6159   & K7III   & -0.13  &  04/02/2005 \\ 
HR\,6770   & G8III   &  -0.05 &  04/03/2005 \\ 
HR\,6817   & K1III   & -0.06  &  04/02/2005 \\ 
HR\,7576   & K3III   &  0.42  &  04/03/2005 \\ 
\enddata
\tablenotetext{Notes:}{
1)~Identifier. 
2)~Stellar classification$^a$. 
3)~Metallicity$^a$.  
4)~Date of observation.\\
{\it Comments:} a) From \citet{Worthey1994}.
}
\end{deluxetable}
\end{center}
%
\section{Data Reduction}
\label{sec:reduction}
We reduce the long-slit spectra following standard procedures of bias
subtraction, cosmic ray rejection and flat fielding under MIDAS described in
\citet{Mehlert00} with additional steps needed to correct for spectral
alignment and anamorphism. We correct a two degree tilt between the spectra
and the CCD rows by appropriate sub-pixel shifting of the CCD columns.  Two bad
columns at positions corresponding to $\lambda$~=~4850~\AA~are corrected
through interpolation. 
We perform the wavelength calibration on neon and cadmium arc frames
with typically ten lines. Where the line signal is low we bin over a
few rows along the spatial direction but never over more than five
pixels corresponding to 1.2~arcseconds. After the original line
identification we first fit a 4th-order polynomial to the line
positions along the spatial direction in order to remove noise-induced
row to row jitter, and then fit a 3rd-order polynomial along the
spectral direction. The remaining RMS scatter in line position is
below one pixel.
We then rebin the spectra in log-wavelength and correct for anamorphic
distortion.  The distortion of LRS is measured using stars which are
trailed along the slit in order to generate several spectra or {\it
  traces} along the whole length of the slit.
We centroid the traces by calculating the first moment of the 
photon count in a 10~pixel wide window around the trace. We then first fit a
3rd-order polynomial to describe the trace position as function of wavelength
and then a firther 3rd-order polynomial to the trace positions in each
column to model the distortion. We find a distortion of up to ten pixel in the
corners of the CCD with respect to the centre of the detector (see also Fig~2. 
in \citealt{Saglia2010}).
We correct for the distortion by means of sub-pixel shifting. Counts of
individual pixel are distributed into pixel of the target frame according the
their overlapping surface area.
We measure the distortion on several stellar spectra taken in
similar manner at different nights. We find that the residual distortion at
the edges of the chip --- after correcting one stellar spectra with the
distortion information of a different night --- is never larger than
1.5~arcseconds. This is below the typical FWHM of the PSF of our observations
and, more importantly, well below the typical spatial bin sizes that we
use at the ends of the slit.
To correct for flexure of the instrument during the night we measure the
wavelength position of the  5577~\AA~skyline at the slit ends and correct the
wavelength calibration to zeroth order by adding a constant offset. The median absolute
offset of all observations is 17~kms$^{-1}$.  
Where dedicated sky spectra are available, we collapse them along the spatial
direction in order to obtain a single maximum signal-to-noise sky spectrum.
This spectrum is then scaled according to the exposure time of the object and
subtracted from the full frame. In cases where no sky frame is available, the
sky signal is determined from the slit ends. One advantage of long-slit
spectroscopy is that often the slit ends do contain sufficient non-object
contaminated sky.  However, the differential slit illumination is subject to
change with time because the HET prime focus assembly moves across the
telescope pupil during the duration of an observation. We test the effect of
this differential illumination on 44 blank sky spectra obtained over
the course of this survey.  We use the slit ends to determine the sky signal
in the same way as we do for the galaxy spectra. We then determine the
differences between those and the sky signal that we measure from the slit
centre.  We find that the residuals amount to no more than 5\% of the sky
signal in all cases. We then derive kinematics using 5\% higher and 5\%
lower sky values. The resulting errors are significantly smaller than the reported
error bars in all cases.
In the case of the major axis observation of NGC\,3368, NGC\,4569, and
the minor axis observation of NGC\,4569, the use of the dedicated sky
frame results in an over-subtraction of the sky (i.e.\ clearly
negative residuals) possibly because of stray light or an increased
level of sky background at the time the sky frame was taken.  In these
cases we use the sky from the slit ends instead.  In the cases of
large galaxies such as NGC\,3031, we test for object contamination by
using different window sizes at the slit ends for the sky extraction.
We find the effect of object contamination to be negligible in all
cases.
Finally, all frames go through an extensive visual inspection. Artifacts like
residuals of comic ray removal are corrected though interpolation of the
neighbouring pixel.
%
\subsection{Derivation of the Kinematics and Template Library} 
\label{sec:kin_derive}
We derive stellar kinematics using the Fourier Correlation Quotient (FCQ)
method of \cite{Bender1990, Bender1994}.  The log-wavelength calibrated and
sky-subtracted spectra are spatially binned to reach $S/N$-values of at least
30 per pixel.  An 8th-order polynomial is then fitted to the continuum and the
first and last three channels in Fourier space are filtered out to remove low
and high frequency variations in the continuum level.

FCQ measures the full line of sight velocity distribution (LOSVD). By means of
deconvolving the autocorrelation function, the FCQ method is more robust
against template mismatches than other Fourier or pixel-space based methods.
Nevertheless, nebular emission lines can significantly affect the derived
higher moments of the LOSVDs and therefore need to be taken into account for
the derivation of the kinematics.  This introduces the necessity of a very
accurate model spectrum because otherwise residual mismatches between the
observed galaxy spectrum and the broadened model spectrum will mimic emission
signatures which are then incorrectly removed.  We therefore form a pool of
template spectra by combining actual observed stellar spectra (see
Tab.~\ref{tab:templates}) with synthetic simple stellar population templates
(SSP) from \citet{Vazdekis99}.  Those include varying metallicities and ages.
We use a sub-sample of the published spectra with Salpeter IMF
\citep{Salpeter55}, and all combinations of ages of 1,~2,~5,~10, and 17.78~Gyr and
metallicities of
$\mbox{[Fe/H] = -1.68,~-1.28,~-0.68,~-0.38,~+0.00,~+0.20}$.  The published SEDs have a
nominal resolution of 1.8~\AA~(FWHM) which corresponds to $\sigma^*$~=~45~kms$^{-1}$
and therefore slightly lower than the spectral resolution of
$\sigma_{inst}$~=~39.3~kms$^{-1}$. 
\begin{figure}
        \begin{center}
        \begin{tabular}{c}
                \includegraphics[width=0.45\textwidth]{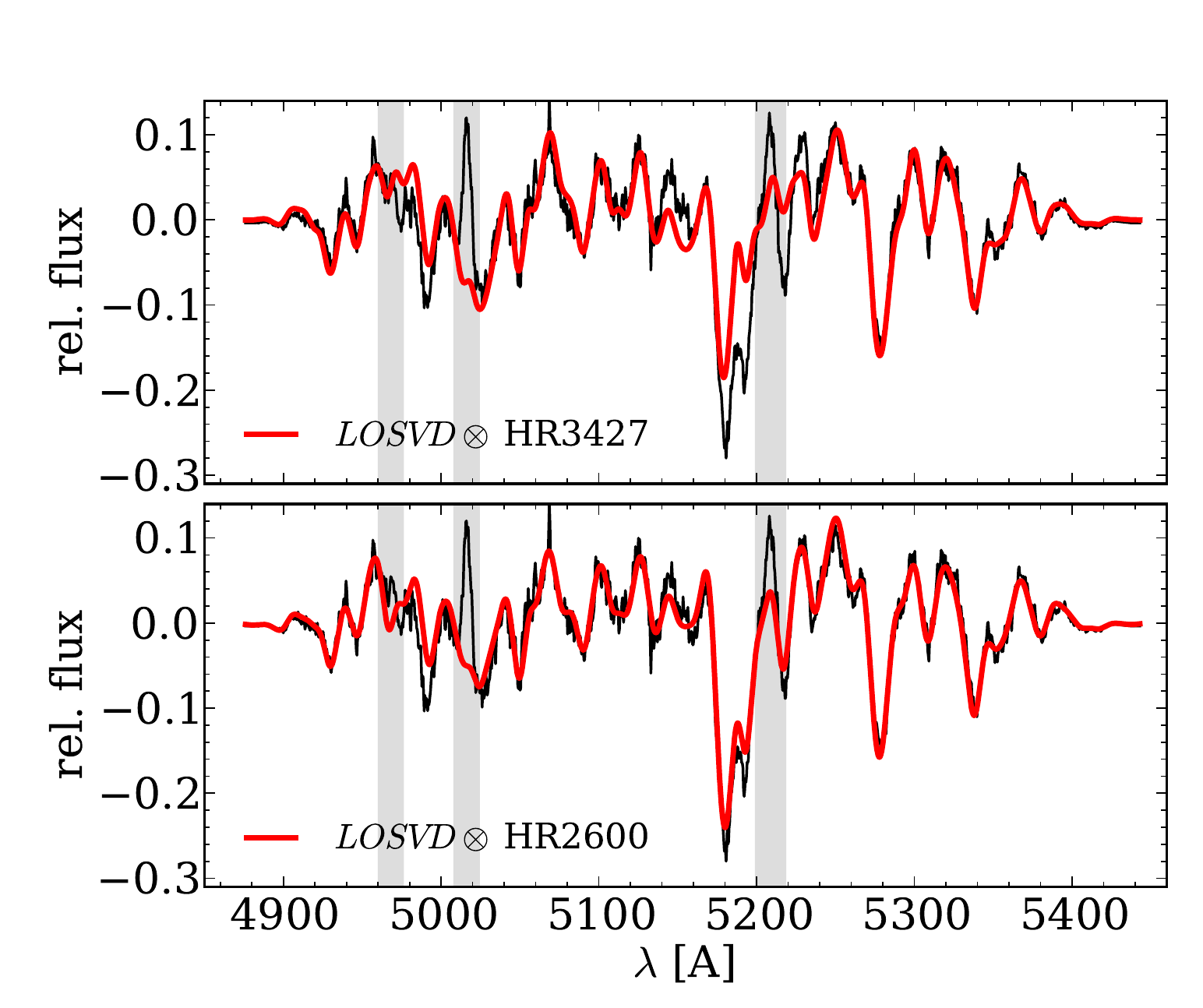}
        \end{tabular}
        \end{center}
	\caption{\label{fig:templmatch}
	Continuum removed spectrum in the central radial bin of NGC\,2841
	(black) and the broadened template spectrum (red). Grey bars mark the positions
	of the [O\,\textsc{iii}] and [N\,\textsc{i}] emission lines. {\it Upper panel:} Choosing the G8III,
	$[Fe/H]=0.16$ star HR3427 results in a notable mismatch around the Mg
	triplet region. Best fit parameters are $\sigma = 235.1 \pm 3.0$~kms$^{-1}$, $h_3 =
	0.037 \pm 0.009$, $h_4 = 0.041 \pm 0.009, \mbox{RMS} = 0.034$).  {\it Lower panel:}
	Using HR2600 (K2III, $[Fe/H]=-0.35$) results in a much better match with $\sigma
	= 241.2 \pm 3.2$~kms$^{-1}$, $h_3 = 0.022 \pm 0.009$, $h_4 = 0.048 \pm 0.009, \mbox{RMS} =
	0.024$. While FCQ finds values for the LOSVD moments that fully agree
	within the errors, the residual spectrum will look very different for those two
	cases and render the detection of weak emission lines impossible in the case of
	HR3427.}
\end{figure}
\begin{figure} 
\begin{center} 
  \includegraphics[width=0.5\textwidth]{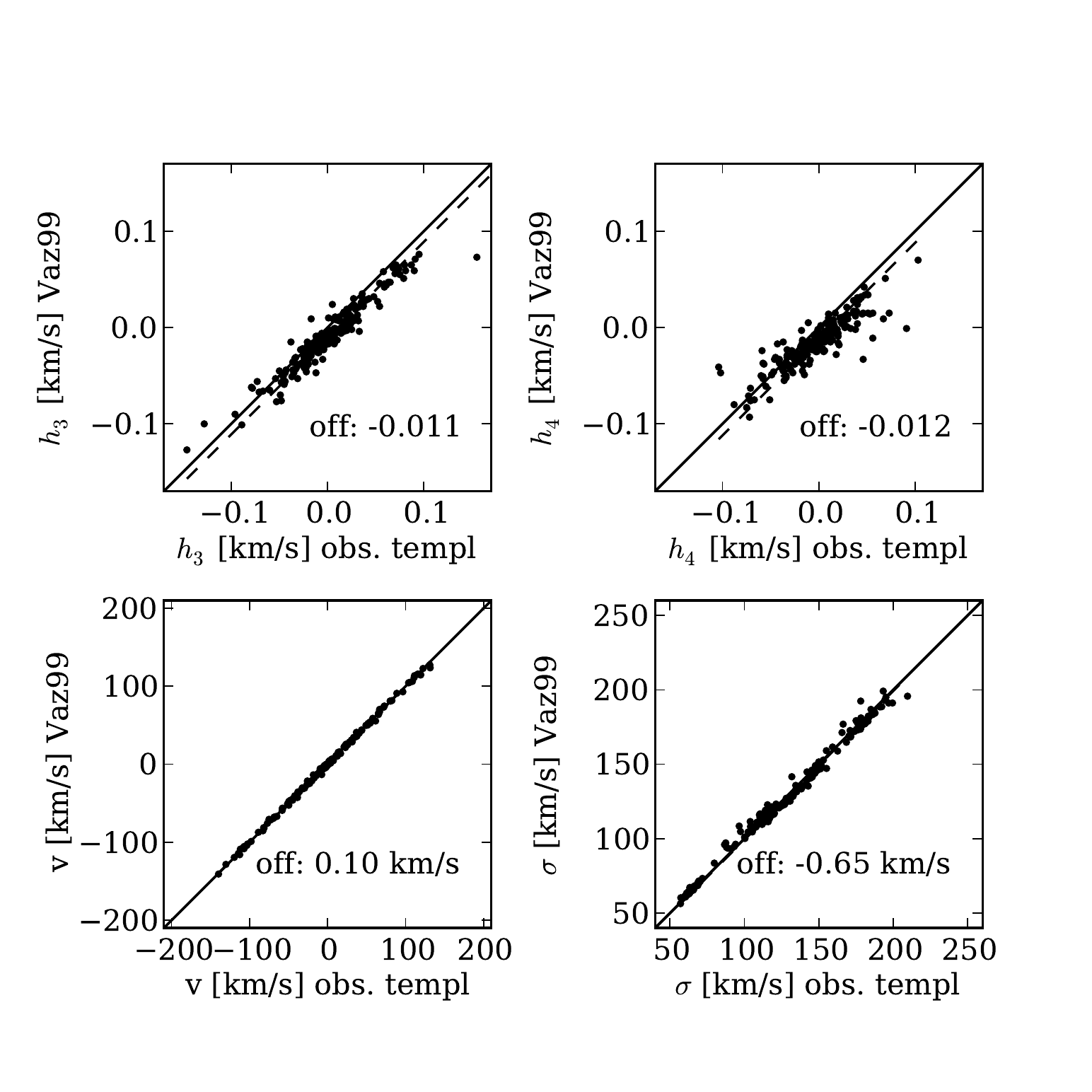} 
\end{center} 
\caption{\label{fig:templComp1A}
Moments of the LOSVD that we obtain from the kinematic extraction using
the Vazdekis SSP library vs.\ the values that we obtain for the observed
templates.  The solid lines correspond to a one-to-one correlation, the dashed
lines are actual fits to the data. 
In the case of $v$ and $\sigma$ the fitted line is covered by the one-to-one
line and not visible.
Note: We only compare galaxies with no
obvious sign of emission.
}
\end{figure} 

We run FCQ with the collection of all velocity templates. Then we choose the single
best-fitting template based on the minimum RMS between the broadened
template and the galaxy spectrum
\begin{eqnarray}
\mbox{RMS} = \int_{\lambda_1}^{\lambda_2}  \left( G(\lambda) - B(\lambda) \otimes S_i(\lambda) \right)^2 d\lambda,
\end{eqnarray}
where $\lambda_1 = $~4817~\AA~to $\lambda_2 =$~5443~\AA~is the fitted
wavelength range, $G(x)$ is the galaxy spectrum, $S_i(x)$ is the $i$-th
template spectrum, and $B(x)$ is the broadening function derived from FCQ. Note
that this is different from other algorithms such as the Maximum Penalized
Likelihood (MPL) technique of \citet{Gebhardt2000} or the Penalized
Pixel-Fitting method (pPXF) of \citet{Cappellari2004}) which fit a linear
combination of their templates. FCQ subsequently fits Gaussians with Hermite
expansions ($h_3$ and $h_4$ moments; \citealp{Gerhard93, van-der-Marel93}) to the
derived LOSVDs.  In Fig.~\ref{fig:templmatch} we show examples of fits with two
different broadened templates. While FCQ indeed finds very similar values for
the moments of the LOSVD, the quality of the template match differs
significantly in the two cases.

In Fig.~\ref{fig:templComp1A}, we compare the impact of the usage of either
just our observed templates or just the SSP library.  We compare only a subset
of galaxies (NGC\,2775, NGC\,2880, NGC\,3675, NGC\,4030, NGC\,4371 NGC\,4379,
and NGC\,7457) for which we detect no significant emission in order to assure
that the derived moments are not affected by emission.  The biases that we
introduce by adding the SSP templates to our library are generally small
($\Delta \sigma = -0.65$~kms$^{-1}$, $\Delta h_3 = -0.011$, $\Delta h_4 =
-0.012$ and much smaller than our median errors on the respective moments.

As our spectra often reach into the disk regions we deal with relatively low
velocity dispersions.  In a few cases the derived dispersions are of the order
of the instrumental resolution. The matter gets complicated by the fact that
the disk regions are also the regions of lowest surface brightness and
therefore the regions with poorest $S/N$.  It is important to understand how
reliable the derived moments are under these circumstances.

A caveat of the deconvolution in Fourier space is the amplification of high
frequency noise. Fourier-based algorithms therefore filter the signal before
the actual deconvolution step. FCQ uses the optimal Wiener filter
\citep{Brault1971, Simkin1974}.  The basic idea is to decompose the Fourier
transform of the input data into a Gaussian contribution --- the data part --
and an exponential function --- the noise part.  The {\it optimal} Wiener filter
then weighs the various signal channels according to their relative
contribution to the data part of the input signal (for details see
\citealt{Bender1990}).  While a purely Gaussian LOSVDs ought to be well
modelled by a Gaussian in Fourier space, the Gauss-Hermite moments cause higher
frequency shoulders, that are easily {\it swallowed} by the noise. 
Adjusting the filter width may recover characteristics of the LOSVD
\citep{Bender1994} at the cost of increased statistical uncertainty. 

Here we choose not to broaden the Wiener Filter as this yields better stability
against statistical deviations. But this causes biases especially at low
velocity dispersions. To correct for these biases we carry out extensive Monte
Carlo simulations on a regular parameter grid of velocity dispersion, $h_3$,
$h_4$, $S/N$ and template.  We generate artificially broadened spectra at each
grid point with 30 different noise realisations according to the input signal
to noise.  We find that the necessary corrections to $\sigma$, $h_3$ and $h_4$
are well behaved and linear functions between input and retrieved values and
independent of input template if the velocity dispersion is larger than
75~kms$^{-1}$, the signal to noise is larger than 30 per pixel and a stellar
template is used.  The SSP templates cause non-linear behaviour at small
velocity dispersions. While we still use the SSP templates to generate
broadened model spectra during the emission line fitting, the reported
kinematic values are exclusively based on stellar templates, and corrected for
biases using 
\begin{eqnarray*}
	\sigma(r) & = & a_{\sigma} \cdot \sigma^{\mbox{\tiny FCQ}}(r) + b_{\sigma} \\
	h_3(r) & = & a_{h_3} \cdot {h_3}^{\mbox{\tiny FCQ}}(r) + b_{h_3} \\
	h_4(r) & = & a_{h_4} \cdot {h_4}^{\mbox{\tiny FCQ}}(r) + b_{h_4}.
\end{eqnarray*}
Tab.~\ref{tab:sigma_corrections} and \ref{tab:h3h4_corrections} list the
corresponding parameters that we obtain from the simulations.  For velocity
dispersions below 75~kms$^{-1}$ and $S/N<30$ per pixel, we do not report values for $h_3$
and $h_4$. Further we report values of $v$ and $\sigma$ only for $S/N>20$.

We estimate statistical errors in the derived moments through Monte Carlo simulations as
described in \citet{Mehlert00}. Once the optimum LOSVD is derived by FCQ we
generate the synthetic spectra using the fitted $v$, $\sigma$, $h_3$ and
$h_4$-parameters and the best fitting stellar template. In a similar manner 
as for the bias correction, 100 different
realisations of artificial noise are then added to the spectra to reach the same
signal to noise values as in the original spectra. We then use FCQ again to
derive the kinematics on those spectra. The reported errors correspond to the 
statistical one-sigma deviations from the mean.
\begin{center}
\begin{deluxetable}{rrrrr}
\tabletypesize{\scriptsize}
\footnotesize
\tablecaption{Parameters for the linear bias corrections in velocity dispersion\label{tab:sigma_corrections}}
\tablewidth{0pt}
\tablehead{
 & \multicolumn{4}{c}{$S/N$ per pixel}\\
\colhead{} & \colhead{22.5} & \colhead{40.0} & \colhead{37.5} & \colhead{75.0} 
}
\startdata
{\bf $a_{\sigma}$} & 1.06 & 1.06 & 1.06 & 1.07 \\
{\bf $b_{\sigma}$} & -11.04 & -10.89 & -10.80 & -10.70 
\enddata
\end{deluxetable}
\end{center}

\begin{center}
\begin{deluxetable}{rrrrr}
\tabletypesize{\scriptsize}
\footnotesize
\tablecaption{Parameters for the linear bias corrections in $h_3$ and $h_4$\label{tab:h3h4_corrections}}
\tablewidth{0pt}
\tablehead{
 & & \multicolumn{3}{c}{$S/N$ per pixel}\\
\colhead{} & \colhead{$\sigma$[kms$^{-1}$]} & \colhead{30.3} & \colhead{37.5} & \colhead{75.0} 
}
\startdata
 & 75.0 & 1.3084 & 1.2947 & 1.2734 \\ 
{\bf $a_{h_3}$}  & 100.0 & 1.1281 & 1.1142 & 1.0874 \\
 & 150.0 & 1.0182 & 1.0104 & 1.0000 \\
 & 200.0 & 1.0103 & 1.0059 & 0.9988 \\
 & 250.0 & 1.0037 & 1.0008 & 0.9945 \\
 & &  & & \\
& 75.0 & 0.0002 & 0.0001 & 0.0001 \\
{\bf $b_{h_3}$} & 100.0 & 0.0003 & 0.0003 & 0.0002 \\
& 150.0 &  0.0003 & 0.0003 & 0.0003 \\
& 200.0 &  0.0009 & 0.0008 & 0.0006 \\ 
& 250.0 &  0.0010 &0.0010 & 0.0009 \\
& & &  & \\
& 75.0 & 1.8521 & 1.8088 & 1.7277 \\
{\bf $a_{h_4}$} & 100.0 & 1.4994 & 1.4500 & 1.3555\\
& 150.0 & 1.0857 & 1.0655 & 1.0307 \\
& 200.0 & 1.0407 & 1.0275 & 1.0045 \\
& 250.0 & 1.1503 & 1.1371 & 1.1089 \\
& & &  & \\ 
 & 75.0 & 0.0864 & 0.0814 & 0.0738 \\ 
{\bf $b_{h_4}$} & 100.0 & 0.0280 & 0.0240 & 0.0160 \\
 & 150.0 & 0.0165 & 0.0148 & 0.0127 \\
 & 200.0 & 0.0105 & 0.0104 & 0.0104 \\
 & 250.0 & -0.0027 & -0.0015 & 0.0006 
\enddata
\end{deluxetable}
\end{center}
%
\subsection{Emission line subtraction and gas kinematics}
\label{sec:emissionlines}
\begin{figure}
        \begin{center}
        \begin{tabular}{c}
                \includegraphics[width=0.45\textwidth]{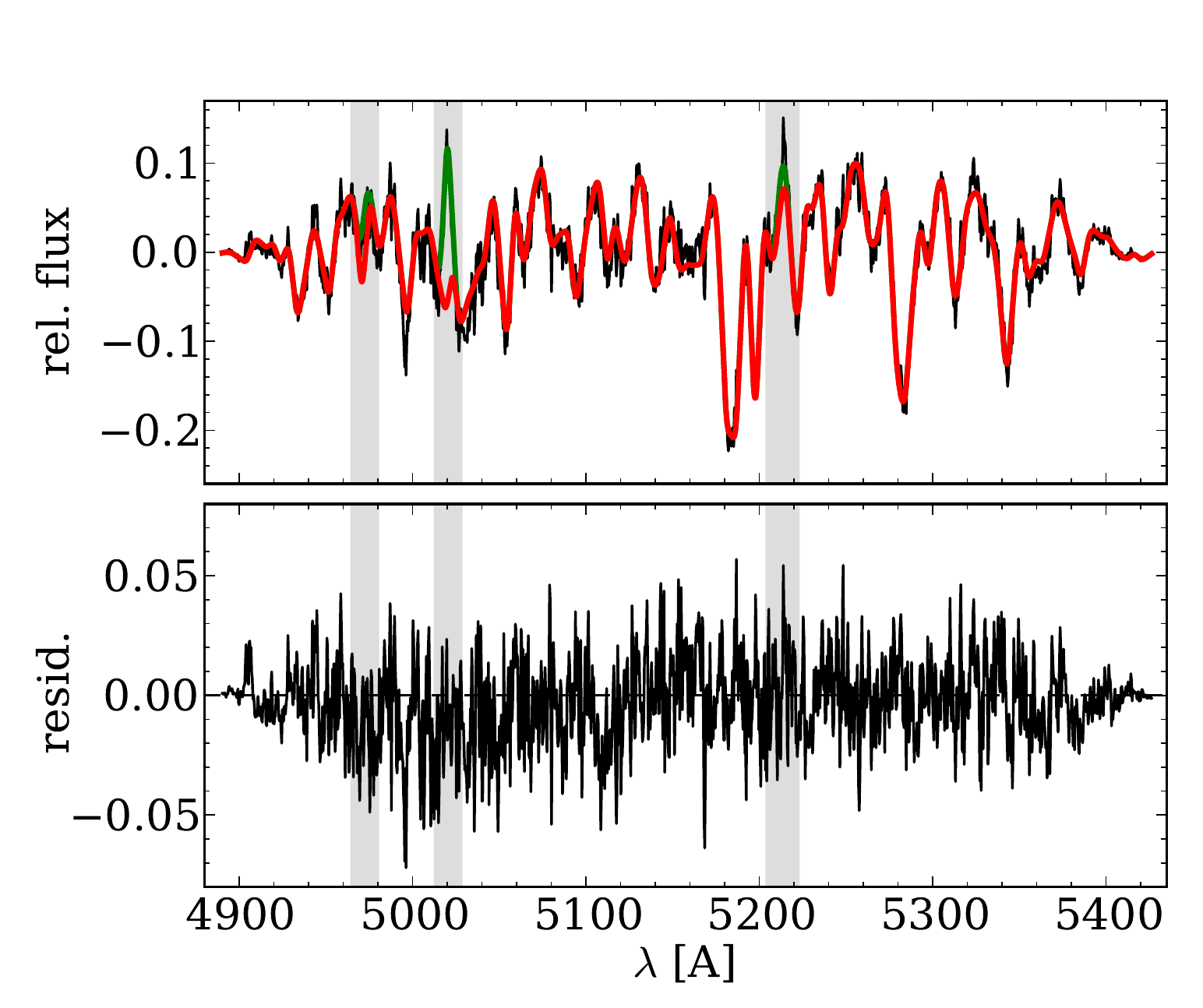} \
        \end{tabular}
        \end{center}
        \caption{Example of the nebular emission line subtraction. Here
		the gas emission was removed from one of the central spectra of NGC\,3368.
		{\it Upper panel:} After subtraction of the best fitting broadened stellar template (red)
		the algorithm finds a significant emission signal in the [O\,\textsc{iii}]
		lines and the nitrogen doublet (green). 
		The vertical bars mark the search range for emission.
		{\it Lower panel:} Residuals between observed and model spectrum after
		removal of the emission. 
		\label{fig:gasfit}} 
\end{figure}

A significant fraction of the objects in our sample show emission in $H_\beta$
(4861.32~\AA), [O\,\textsc{iii}] (4958.83~\AA~\&~5006.77~\AA) and [N\,\textsc{i}]
(5197.90~\AA~\&~5200.39~\AA). 
The nitrogen emission line lies on the red flank of the Mg triplet feature --
the most important kinematic feature in our spectral range.  While typically
weak, the nitrogen emission often significantly affects the derivation of
$h_3$-moments.  $h_3$-moments measure the asymmetric deviation from a Gaussian and
are expected to behave antisymmetrically with respect to the galaxy centre in the
case of axisymmetric systems. Deviations from this antisymmetry may hint at
contamination by nitrogen emission lines. 
We therefore decided to remove nebular emission following a similar procedure
as the GANDALF routine \citep{Sarzi2006}: We perform a first fit to the galaxy
spectrum over a larger spectral range reaching from 4820~\AA~to 5440~\AA~using 
the FCQ algorithm.
We then subtract the best fitting broadened stellar spectrum from the galaxy
spectrum and fit Gaussian functions --- using a standard least squares
algorithm --- to the residual emission. The algorithm first searches for emission in a
500~kms$^{-1}$ window around the brighter oxygen line at 5007~\AA~(the oxygen doublet
is well resolved at our instrumental resolution) red-shifted to the systemic
velocity. It fits for the three parameters of
amplitude, central velocity, and the velocity dispersion. It then goes on to
the other and generally weaker emission lines and performs a fit to their amplitude while
assuming the same central velocity and velocity dispersion as the oxygen line. 
In principle the ratio of the two oxygen emission lines is given by atomic
physics and is a constant of value 0.33.  Rather than fixing these values during the fit
we also fitted the lower amplitude line as this provided another handle on the
reliability of our method.
%
We then subtract the best fitting emission lines from the original galaxy
spectrum and repeat the FCQ multiple-template fit. The best fitting broadened
template is again subtracted from the input spectrum and the gas emission fit is
repeated on the improved difference spectrum.  An example for a spectrum that
shows signs of nebular emission is shown in the upper panel of
Fig.~\ref{fig:gasfit}. We plot the residuals between the recorded spectrum and
the broadened model spectrum after the removal of the emission in the lower
panel.  This iterative approach was found to converge very
quickly.  A fourth FCQ fit typically yields no significant change in the
derived kinematics any more.  Our reported stellar kinematics went through
three subsequent iterations of template fitting with two interleaved gas emission
removal steps.
The gas removal routine gives access to the study of line ratios
\citep{Sarzi2006} and is also a necessary step before the calculation of
absorption line indices and subsequent stellar population analysis. Both of
which will be subject to a forthcoming publication.

\subsection{Identification of pseudobulges}
\label{sec:pseudoid}
If no classification is available already from \cite{Fisher2008} or
\cite{Fisher2010}, we follow the same procedure for the identification of
pseudobulges.  We define a bulge photometrically as the  excess light over the
inwards extrapolation of the outer disk exponential luminosity profile.  The
bulge-to-disk decompositions that we adopt in \S\ref{sec:photom} allow us to
determine the bulge region of an object.  Here, we classify bulges using
close-to $V$-band HST images ($F547M$, $F555W$ \& $F606W$).  While these bands
are subject to dust obscuration they are also sensitive to an enhancement in star
formation rate, a feature commonly observed in pseudobulges \citep{Fisher2006}.
We visually inspected the HST images to see whether the bulge regions contain
disk-like structures such a nuclear spirals, nuclear bars and/or nuclear rings.
If such structure is present we call the photometric bulge a pseudo-bulge. If
there is no structure (the bulge resembles an elliptical galaxy with
a smooth light distribution), we call this bulge a classical bulge.

Weak central dust spirals that also occur frequently in elliptical galaxies
(e.g. \citealt{Storchi-Bergmann07}) --- often distinguishable from the outer
disk because they are inclined differently --- are no reason for us to call a
bulge a pseudobulge. NGC\,2841 poses an example for this situation. A dust spiral
can easily be identified in HST F438W, but it is misaligned with the outer
disk also seen only in vicinity of the nucleus.

Yet, a few objects remain for which we do not feel confident to assign a
classification based on their HST morphology: NGC\,2460, NGC\,3953, NGC\,4826 and NGC\,7217. We
treat them as unclassified throughout this work. As mentioned in
\S\ref{sec:sample}, we also do not classify the bulges of the galaxies NGC\,3593
and NGC\,7331 due to their high inclination. In Appendix~\ref{sec:individuals} we give
a detailed explanation for the bulge classification for each individual object.
\subsection{Photometry} 
\label{sec:photom}
We use the results from decompositions of surface brightness profiles
to investigate possible correlations between photometric parameters
and kinematic structure. Also the bulge-disk decompositions serve to
identify the actual bulge region of a particular galaxy.

\begin{center}
\begin{deluxetable*}{cp{4cm}clp{1.7cm}l}
\tabletypesize{\scriptsize}
\tablecaption{Data Sources\label{tab:phot_data_sources}}
\tablehead{
\colhead{No.} & \colhead{Instrument} &  \colhead{Filter} & \colhead{Scale} & \colhead{Field Of View} & \colhead{Reference} \\
\colhead{} & \colhead{} & \colhead{} & \colhead{(arcsec/pixel)} & \colhead{(arcsec)}
} 
\startdata
 1 & HST Nicmos 1                          & F160W      & 0.043  & 11 $\times$ 11   & MAST archive$^1$\\
 2 & HST Nicmos 2                          & F160W      & 0.075  & 19.2 $\times$ 19.2 & MAST archive$^1$\\
 3 & HST Nicmos 3                          & F160W      & 0.20    & 51.2 $\times$ 51.2 & MAST archive$^1$\\
 4 & 2MASS                                 & H          & 1.0    & 512  $\times$ 1024 & IPAC archive\\
 5 & 2MASS                                 & H          & 1.0    & variable size, mosaic          & \citet{Jarrett2003} \\
 6 & Spitzer IRAC                          & 3.6\,$\mu$m & 0.60    & variable size, scan          & IPAC archive$^2$\\
 7 & Spitzer IRAC (SINGS)                  & 3.6\,$\mu$m & 0.75   & variable size, scan          & SINGS$^3$\\
 8 & HST ACS/WFC                           & F814W      & 0.049   & 202  $\times$ 202  & MAST archive$^1$\\
 9 & HST WFPC2                             & F547M      & 0.10    & 80   $\times$ 80   & MAST archive$^1$\\
10 & HST WFPC2                             & F814W      & 0.10    & 80   $\times$ 80   & MAST archive$^1$\\
11 & Perkins 1.8\,m OSIRIS                 & H          & 1.5    & 412  $\times$ 412  & \citet{Eskridge2002} \\
12 & Lick 3\,m pNIC                        & K          & 0.24  & 15.4   $\times$ 15.4   & \citet{Rauscher1995} \\
13 & CTIO 1.5\,m OSIRIS                    & H          & 1.1    & 312  $\times$ 312  & \citet{Eskridge2002} \\
14 & William Herschel Telescope INGRID     & K          & 0.24  & 252  $\times$ 252  & \citet{Knapen2003} \\
15 & Calar Alto Observatory 2.2\,m  MAGIC Nicmos 3      & K      & 0.66   & 172  $\times$ 172  & \citet{Mollenhoff2001} \\
16 & Mauna Kea 0.61\,m  Nicmos256          & K        & 2.1  & 644  $\times$ 568  & \citet{Tully1996} \\
17 & UKIRT 3.8\,m IRCAM II                 & H          & 1.7   & 198  $\times$ 72   & \citet{de-Jong1994} \\
\enddata
\tablenotetext{Notes:}{
1) {\tt http://archive.stsci.edu/} 
2) {\tt http://irsa.ipac.caltech.edu/} 
3) {\tt http://data.spitzer.caltech.edu/popular/sings/}
}
\end{deluxetable*}
\end{center}

We derive surface photometry following the prescriptions in \citet{Fisher2008}
and \citet{Fisher2010}. For each galaxy, we combine multiple data sources
mostly in the infrared (but sometimes in optical bands) to obtain a final 1D
composite surface brightness profile. The different data have been calibrated
against the H-band using 2MASS magnitudes. High-resolution HST imaging is used
in the galaxy center, while wide-field images sample the outer
disk. The resulting composite profile of each galaxy is used to derive the
bulge-to-disk photometric decomposition. 
Our
method is well tested and has been used in several publications
\citep{Fisher2008,Fisher2010, Kormendy2009}.  Our principal source of data is 2MASS
$H$-band maps \citep{Skrutskie2006}. When available we use data from 2MASS
Large Galaxy Atlas \citep{Jarrett2003}. We prefer $H$-band over $K_s$ because
the 2MASS $H$-band data is more sensitive than the 2MASS $K_s$ data. For all
galaxies we use the NASA/IPAC Extragalactic Database to search for ancillary
ground based $H$-band data. We also include Spitzer 3.6~\um~data.
Finally, when available we also include high resolution $F160W$ images from
HST/NICMOS. In a few cases, where galaxies were lacking archival NICMOS data,
we use $I$ or $R$ band data instead. The high resolution data can be crucial to
accurately constraining the bulge-disk decomposition. \cite{Fisher2010}
investigates the uncertainty introduced from mixing filters in this way, it is
typically smaller than 0.1~mags, and therefore small compared to the
uncertainty in the fit. Also, \cite{Fisher2008} derives very similar S\'ersic
indices with $V$-band data as with $H$-band profiles.

We fit ellipses to all images. Isophotal fitting is carried out using
the code of \cite{BM87}. See \cite{Fisher2008} for a brief summary of
the procedure. The code returns a 2-D surface brightness profile
(including for each ellipse center, major \& minor axis size, position
angle, and mean surface brightness). We then combine all profiles into
a composite surface brightness profile.  The power of this method is
two-fold.  First, combination of surface brightness profiles allows us
to robustly identify systematic errors from point-spread-functions and
sky subtraction. Secondly the resulting composite profile has an
extremely high dynamic range in radius, which is necessary to
accurately constrain the bulge-disk decomposition (see discussion in the
Appendix of \citealp{Fisher2008} and also \citealp{Kormendy2009}). The zero points of
our profiles are matched against the 2MASS data.

We determine bulge and disk parameters by fitting each
surface brightness profile with a one-dimensional S\'ersic function
plus an exponential outer disk,
\begin{equation}
I(r)=I_0\exp\left[-(r/r_0)^{1/n} \right ] + I_d\exp\left[-(r/h) \right ]\, ,
\end{equation}
where $r$ represents the distance along the major axis,
$I_0$ and $r_0$ are the central surface brightness and
scale length of the bulge, $I_d$ and $h$ represent the central surface
brightness and scale length of the outer disk, and $n$ represents
the bulge S\'ersic index \citep{Sersic1968}. The half-light radius, $r_e$, of the bulge is
obtained by converting $r_0$,
\begin{equation}
r_e=(b_n)^nr_0,
\end{equation}
where the value of $b_n$ is a proportionality constant defined such that
$\Gamma (2 n) = 2 \gamma(2 n,b_n)$ \citep{Ciotti1991}. $\Gamma$ and $\gamma$
are the complete and incomplete gamma functions, respectively. We use the
approximation $b_n \approx 2.17 n - 0.355$ \citep{Caon1993}. We restrict our
range in possible S\'ersic indices to $n > 0.33$ to ensure that the
approximation is accurate.  Bulge and disk magnitudes are adjusted to account
for the shape of the bulge using the ellipticity profile from the isophote
fitting. 

Intermediate type galaxies are known to contain many components that
are not well described by the decomposition into a S\'ersic bulge and
exponential disk (e.g. bars, rings, nuclear star clusters).  Similar
to the outer disk, in the bulge we exclude significant non-S\'ersic
components such as nuclear bars and nuclear rings.  The inclusion of
these features can have unpredictable effects on the S\'ersic index,
depending on the relative size of the feature, and what type it
is. The appendix of \cite{Fisher2008} discusses how masking data in
the bulge will affect the decomposition. Essentially, this has the
effect of decreasing the robustness of the fit, which will be reflected
in the error bars. The parameters from the decompositions are
presented in Tab.~\ref{tab:ndecomps}
, the image data sources are described in Tab.~\ref{tab:phot_data_sources}.

The surface brightness profile of NGC\,5566 does not follow a 
typical bulge/disk profile. We publish the obtained kinematics here
but exclude this object from all further analysis.
\begin{center}
\begin{deluxetable*}{cccrrrrrrl}
\tabletypesize{\scriptsize}
\tablecaption{Bulge to disk decomposition parameters$^a$. \label{tab:ndecomps}}
\tablewidth{0pt}
\tablehead{
\colhead{Galaxy} & \colhead{Bulge}  & \colhead{$n$} &
\colhead{$\mu_e$} & \colhead{$r_e$} & \colhead{$m^{Sersic}$} &
\colhead{$\mu_0^{disk}$} & \colhead{$h$} & \colhead{$m^{disk}$} & \colhead{Data sources}\\
& \colhead{morph.} & & \colhead{mag arcsec$^{-2}$} & \colhead{arcsec} & \colhead{mag} & \colhead{mag arcsec$^{-2}$} & \colhead{arcsec} & \colhead{mag} & \colhead{} \\
 \colhead{(1)} & \colhead{(2)} & \colhead{(3)} & \colhead{(4)} & \colhead{(5)} & \colhead{(6)} & \colhead{(7)} & \colhead{(8)} & \colhead{(9)} & \colhead{(10)} 
}
\startdata
NGC\,1023 &  c &  2.52~$\pm$~0.81 &  15.76~$\pm$~1.11 &  12.35~$\pm$~4.39 &  7.15~$\pm$~1.11 &  16.62~$\pm$~0.32 &  62.20~$\pm$~6.56 &  5.66~$\pm$~0.35 & 5,10,15\nl
NGC\,2460 &  p &  3.49~$\pm$~0.32 &  18.02~$\pm$~0.46 &  12.69~$\pm$~4.40 &  9.19~$\pm$~0.46 &  16.40~$\pm$~0.28 &  11.42~$\pm$~0.98 &  9.11~$\pm$~0.31 & 2,4 \nl
NGC\,2681 &  p &  3.82~$\pm$~0.31 &  14.58~$\pm$~0.63 &  3.74~$\pm$~3.00 &  8.35~$\pm$~0.63 &  17.32~$\pm$~0.33 &  23.78~$\pm$~2.76 &  8.44~$\pm$~0.37  & 3,4,6 \nl
NGC\,2775 &  c &  3.23~$\pm$~0.93 &  17.28~$\pm$~1.02 &  15.86~$\pm$~5.50 &  8.00~$\pm$~1.02 &  17.20~$\pm$~0.61 &  41.28~$\pm$~7.91 &  7.12~$\pm$~0.67 & 4,6,10,11,14,15 \nl
NGC\,2841 &  c &  3.22~$\pm$~0.58 &  16.55~$\pm$~0.68 &  15.46~$\pm$~8.98 &  7.33~$\pm$~0.68 &  16.49~$\pm$~0.17 &  60.51~$\pm$~2.82 &  5.58~$\pm$~0.18 & 2,5,7 \nl
NGC\,2859 &  c &  2.34~$\pm$~0.65 &  16.23~$\pm$~1.08 &  $8.3 (< 21.4^b)$ &  8.52~$\pm$~1.08 &  19.11~$\pm$~0.47 &  55.17~$\pm$~8.93 &  8.41~$\pm$~0.52 & 5,6,8 \nl
NGC\,2880 &  c &  3.41~$\pm$~0.48 &  17.49~$\pm$~0.59 &  $11.7 (< 24.2^b)$ &  8.85~$\pm$~0.59 &  18.29~$\pm$~0.40 &  25.47~$\pm$~1.88 &  9.26~$\pm$~0.41 & 4,10 \nl
NGC\,2964 &  p &  1.01~$\pm$~0.34 &  15.43~$\pm$~0.51 &  2.04~$\pm$~0.50 &  11.18~$\pm$~0.51 &  16.40~$\pm$~0.18 &  16.07~$\pm$~0.65 &  8.37~$\pm$~0.19 & 2,4,6 \nl
NGC\,3031 &  c &  4.09~$\pm$~0.48 &  17.14~$\pm$~0.62 &  70.70~$\pm$~54.91 &  4.49~$\pm$~0.62 &  16.59~$\pm$~0.25 &  132.80~$\pm$~8.05 &  3.98~$\pm$~0.26 & 2,5,7\nl
NGC\,3166 &  p &  1.24~$\pm$~0.30 &  14.37~$\pm$~0.39 &  4.36~$\pm$~1.12 &  8.37~$\pm$~0.39 &  15.86~$\pm$~0.44 &  15.47~$\pm$~2.66 &  7.92~$\pm$~0.50 & 5,6,9,11\nl
NGC\,3245 &  c &  2.75~$\pm$~0.56 &  15.16~$\pm$~0.80 &  4.51~$\pm$~1.70 &  8.69~$\pm$~0.80 &  16.44~$\pm$~0.28 &  21.60~$\pm$~1.49 &  7.77~$\pm$~0.30 & 2,4,6 \nl
NGC\,3351 &  p &  1.38~$\pm$~0.74 &  15.99~$\pm$~0.60 &  8.08~$\pm$~3.10 &  8.59~$\pm$~0.60 &  17.01~$\pm$~0.44 &  49.48~$\pm$~5.13 &  6.54~$\pm$~0.47 & 2,5,7 \nl
NGC\,3368 &  p &  2.46~$\pm$~0.77 &  15.97~$\pm$~0.75 &  13.08~$\pm$~7.29 &  7.25~$\pm$~0.75 &  16.57~$\pm$~1.95 &  35.26~$\pm$~29.61 &  6.84~$\pm$~2.31 & 2,4,7\nl
NGC\,3384 &  p &  1.58~$\pm$~0.22 &  14.39~$\pm$~0.40 &  5.35~$\pm$~1.16 &  7.83~$\pm$~0.40 &  16.96~$\pm$~0.15 &  44.42~$\pm$~3.13 &  6.73~$\pm$~0.18 & 2,5,6\nl
NGC\,3521 &  c &  3.66~$\pm$~0.77 &  15.48~$\pm$~1.50 &  8.55~$\pm$~6.36 &  7.48~$\pm$~1.50 &  15.99~$\pm$~0.31 &  49.53~$\pm$~3.87 &  5.52~$\pm$~0.33 & 2,5,7 \nl
NGC\,3593 &  ? &  1.22~$\pm$~0.21 &  16.18~$\pm$~0.27 &  14.55~$\pm$~2.16 &  7.57~$\pm$~0.27 &  17.61~$\pm$~0.30 &  52.52~$\pm$~4.99 &  7.02~$\pm$~0.33 & 3,5,6,11 \nl
NGC\,3627 &  p &  1.50~$\pm$~0.58 &  14.53~$\pm$~0.59 &  4.77~$\pm$~1.52 &  8.24~$\pm$~0.59 &  16.72~$\pm$~0.15 &  65.99~$\pm$~0.15 &  5.63~$\pm$~0.16 & 3,5,7 \nl
NGC\,3675 &  p &  1.57~$\pm$~1.12 &  16.35~$\pm$~1.93 &  $9.0 (< 23.4^b)$ &  8.66~$\pm$~1.93 &  16.22~$\pm$~0.31 &  36.85~$\pm$~3.42 &  6.40~$\pm$~0.34 & 2,5,6,11 \nl
NGC\,3898 &  c &  3.22~$\pm$~0.86 &  15.93~$\pm$~1.27 &  7.63~$\pm$~2.88 &  8.24~$\pm$~1.27 &  17.25~$\pm$~0.64 &  29.00~$\pm$~4.56 &  7.94~$\pm$~0.68 & 2,5,6 \nl
NGC\,3945 &  p &  1.79~$\pm$~0.48 &  16.00~$\pm$~0.76 &  9.82~$\pm$~3.50 &  8.05~$\pm$~0.76 &  19.02~$\pm$~0.44 &  83.17~$\pm$~26.76 &  7.43~$\pm$~0.64 & 5,6,10\nl
NGC\,3953 &  ? &  2.43~$\pm$~0.68 &  17.15~$\pm$~0.96 &  12.74~$\pm$~8.03 &  8.49~$\pm$~0.96 &  17.44~$\pm$~0.15 &  66.14~$\pm$~4.73 &  6.34~$\pm$~0.19 & 2,3,5,7 \nl
NGC\,3992 &  c &  3.18~$\pm$~1.18 &  17.44~$\pm$~1.44 &  12.23~$\pm$~3.74 &  8.73~$\pm$~1.44 &  17.65~$\pm$~0.50 &  77.54~$\pm$~19.36 &  6.20~$\pm$~0.62 & 5,9,16\nl
NGC\,4030 &  p &  1.98~$\pm$~1.30 &  16.50~$\pm$~1.52 &  5.18~$\pm$~2.14 &  9.89~$\pm$~1.52 &  15.60~$\pm$~0.36 &  15.82~$\pm$~1.95 &  7.61~$\pm$~0.41 & 2,4,6,13\nl
NGC\,4203 &  c &  2.45~$\pm$~0.83 &  15.72~$\pm$~1.51 &  $6.78 (<15.9^b)$ &  8.43~$\pm$~1.51 &  17.28~$\pm$~0.37 &  30.31~$\pm$~3.53 &  7.88~$\pm$~0.41 & 4,6,10\nl
NGC\,4260 &  c &  3.68~$\pm$~0.42 &  19.08~$\pm$~0.48 &  21.49~$\pm$~11.93 &  9.08~$\pm$~0.48 &  17.02~$\pm$~0.19 &  21.69~$\pm$~1.59 &  8.34~$\pm$~0.22 & 2,4,6\nl
NGC\,4274 &  p &  1.52~$\pm$~0.24 &  15.49~$\pm$~0.28 &  5.92~$\pm$~1.10 &  8.73~$\pm$~0.28 &  17.03~$\pm$~0.16 &  46.53~$\pm$~2.92 &  6.70~$\pm$~0.19 & 5,6,8\nl
NGC\,4314 &  p &  2.72~$\pm$~0.96 &  17.12~$\pm$~1.09 &  10.53~$\pm$~6.96 &  8.82~$\pm$~1.09 &  16.70~$\pm$~0.36 &  35.08~$\pm$~2.65 &  6.98~$\pm$~0.38 & 2,4,6,14\nl
NGC\,4371 &  p &  2.21~$\pm$~1.00 &  16.66~$\pm$~1.35 &  10.97~$\pm$~6.00 &  8.37~$\pm$~1.35 &  18.10~$\pm$~0.98 &  44.59~$\pm$~13.45 &  7.86~$\pm$~1.07 & 1,4,6\nl
NGC\,4379 &  c &  2.39~$\pm$~0.55 &  16.72~$\pm$~0.69 &  6.40~$\pm$~1.97 &  9.56~$\pm$~0.69 &  16.91~$\pm$~0.37 &  13.52~$\pm$~1.48 &  9.26~$\pm$~0.40 & 1,4,8\nl
NGC\,4394 &  p &  1.58~$\pm$~0.67 &  16.28~$\pm$~0.87 &  6.10~$\pm$~1.81 &  9.43~$\pm$~0.87 &  18.40~$\pm$~0.25 &  57.75~$\pm$~6.38 &  7.60~$\pm$~0.30 & 4,6,11,12\nl
NGC\,4448 &  p &  1.19~$\pm$~0.25 &  16.43~$\pm$~0.31 &  6.70~$\pm$~1.01 &  9.52~$\pm$~0.31 &  16.85~$\pm$~0.17 &  28.42~$\pm$~0.97 &  7.59~$\pm$~0.18 & 5,6,10,11\nl
NGC\,4501 &  p &  1.25~$\pm$~1.06 &  15.43~$\pm$~1.34 &  5.31~$\pm$~1.53 &  8.99~$\pm$~1.34 &  15.81~$\pm$~0.46 &  39.26~$\pm$~3.89 &  5.85~$\pm$~0.49 & 5,6,12\nl
NGC\,4536 &  p &  1.47~$\pm$~0.35 &  14.77~$\pm$~0.62 &  3.98~$\pm$~1.18 &  8.88~$\pm$~0.62 &  17.18~$\pm$~0.15 &  32.32~$\pm$~1.82 &  7.64~$\pm$~0.17 & 2,5\nl
NGC\,4569 &  p &  2.34~$\pm$~0.97 &  15.13~$\pm$~1.59 &  4.80~$\pm$~2.33 &  8.60~$\pm$~1.59 &  16.84~$\pm$~0.32 &  61.32~$\pm$~6.42 &  5.91~$\pm$~0.36 & 2,3,5,7\nl
NGC\,4698 &  c &  2.51~$\pm$~0.53 &  15.66~$\pm$~0.66 &  5.11~$\pm$~1.41 &  8.97~$\pm$~0.66 &  17.32~$\pm$~0.34 &  34.93~$\pm$~2.76 &  7.61~$\pm$~0.36 & 2,4,6,11\nl
NGC\,4736 &  p &  1.23~$\pm$~0.30 &  13.80~$\pm$~0.41 &  7.76~$\pm$~1.65 &  6.56~$\pm$~0.41 &  14.93~$\pm$~0.35 &  26.82~$\pm$~3.85 &  5.80~$\pm$~0.41 & 3,5,7\nl
NGC\,4772 &  c &  3.03~$\pm$~0.88 &  17.73~$\pm$~1.20 &  13.49~$\pm$~8.24 &  8.84~$\pm$~1.20 &  18.87~$\pm$~0.55 &  71.96~$\pm$~16.04 &  7.59~$\pm$~0.64 & 4,11\nl
NGC\,4826 &  ? &  3.93~$\pm$~0.88 &  16.93~$\pm$~1.06 &  28.93~$\pm$~10.26 &  6.25~$\pm$~1.06 &  16.50~$\pm$~0.27 &  67.95~$\pm$~6.86 &  5.34~$\pm$~0.31 & 2,5,7\nl
NGC\,5055 &  p &  1.71~$\pm$~1.03 &  17.35~$\pm$~1.36 &  27.88~$\pm$~16.08 &  7.16~$\pm$~1.36 &  16.57~$\pm$~0.38 &  68.70~$\pm$~5.53 &  5.40~$\pm$~0.40 & 3,5,7\nl
NGC\,5248 &  p &  1.29~$\pm$~0.45 &  16.47~$\pm$~0.67 &  $0.7 ( <4.9^b)$ &  8.74~$\pm$~0.67 &  17.43~$\pm$~0.28 &  43.99~$\pm$~3.72 &  7.21~$\pm$~0.30 & 2,5,6\nl
NGC\,7177 &  p &  2.03~$\pm$~0.52 &  16.35~$\pm$~0.59 &  7.02~$\pm$~3.17 &  9.08~$\pm$~0.59 &  16.42~$\pm$~0.31 &  15.45~$\pm$~0.73 &  8.49~$\pm$~0.32 & 2,4,6,17\nl
NGC\,7217 &  ? &  3.20~$\pm$~1.03 &  17.04~$\pm$~1.21 &  13.41~$\pm$~9.35 &  8.13~$\pm$~1.21 &  16.50~$\pm$~0.64 &  28.54~$\pm$~6.60 &  7.23~$\pm$~0.72 & 2,4,6,11\nl
NGC\,7331 &  ? &  2.85~$\pm$~1.02 &  16.20~$\pm$~1.30 &  16.63~$\pm$~9.90 &  6.88~$\pm$~1.30 &  16.97~$\pm$~0.47 &  61.61~$\pm$~9.04 &  6.03~$\pm$~0.52 & 3,5,7\nl
NGC\,7743 &  p &  3.66~$\pm$~0.52 &  15.15~$\pm$~0.89 &  2.37~$\pm$~1.04 &  9.94~$\pm$~0.89 &  17.14~$\pm$~0.22 &  22.27~$\pm$~1.96 &  8.41~$\pm$~0.25 & 2,4,6\nl
\enddata
\tablenotetext{Notes:}{
1) Galaxy name. 
2) Bulge classification: c = classical, p = pseudobulge, ? = not classified.
3) Bulge S\'ersic index.
4) Bulge surface brightness at $r_e$.
5) Bulge effective radius along the major axis.
6) Bulge apparent magnitude.
7) Disk central surface brightness.
8) Disk scale length.
9) Disk apparent magnitude.
10) Image data sources, see Tab.\,\ref{tab:phot_data_sources}.\\
{\it Comments:} 
a) These photometric decompositions are based on infrared and optical data, but
calibrated against the H-band. b) The error on the effective radius is
comparable to or larger than the value itself. We list the value that is
preferred by the fit and the upper limit.
}
\end{deluxetable*}
\end{center}
\subsection{Bulge Radius}
Here we are particularly interested in the kinematic properties of the bulge
regions of our observed galaxies. Of course the derived LOSVDs will always be
the light weighted average of all components (bulge, disk, bar) along a
particular line of sight through the galaxy. But the photometric bulge to disk
decompositions allow us to determine within which radius the bulge should
dominate.  We define the {\it bulge radius} $r_b$ along the major axis as the
radius where the light contribution of the photometric bulge component exceeds
the light contribution of the disk component by 25\%:
\begin{eqnarray} 
  I_{0}^{bulge} \exp\left(-(\frac{r_{b}}{r_0})^{\frac{1}{n}} \right) = 
   1.25 \cdot I_{0}^{disk} \exp{\left(-\frac{r_{b}}{h}\right) } 
\end{eqnarray}
where $I_{0}^{bulge}$ is the central surface brightness of the bulge component,
$I_{0}^{disk}$ is the central surface brightness of the disk component,
$r_0$ and $h$ are the scale lengths of the bulge and disk components and
$n$ is the S\'ersic index (see previous Section for the relation
between $r_0$ and $r_e$).
One might argue that the bulge effective radius $r_e$ is a more natural choice
as $r_b$ of course is dependent on the disk parameters. But we find that in a
number of galaxies $r_e$ actually lies in a region that is dominated by disk
light. The choice of 25\% is a compromise between the desire to be reasonably
dominated by the bulge component on the one hand and still wanting to maintain
a sufficient number of resolution elements within the bulge radius on
the other. The values for the bulge radius are listed in Tab.~\ref{tab:params}.
In Fig.\ \ref{fig:kprofiles} we indicate the location of the bulge radius 
though a dashed vertical line.   
\section{Results}
\label{sec:results}
\subsection{Kinematic profiles and comparison with literature}
Tab.~\ref{tab:data_example} gives an example of the format of the measured
stellar kinematic moments as function of the distance from the center of the
galaxy.  The full listing is available electronically\footnote{\tt
http://cds.u-strasbg.fr/}.
In the Appendix~\ref{sec:kprofiles} we plot the kinematic profiles.  When
available, we also plot data from the literature for comparison.  Integral
Field Spectroscopic data from SAURON is available for some of the galaxies in
our sample. In those cases we create pseudo long-slit data through
interpolation of the SAURON  $v$, $\sigma$, $h_3$, and $h_4$ maps along a slit
aperture with a position angle corresponding to our observation.  In general
the agreement of our data with the published values is acceptable.

In a few cases such as NGC\,4203 a difference between the previously published 
data and ours are explained by the difference in the observed position angle.
 
\cite{Bertola1995} find somewhat larger velocity dispersions for NGC\,4379 than
we do. Formally their instrumental dispersion should allow to
resolve the 80~kms$^{-1}$ --- 118~kms$^{-1}$ that we find for the dispersion in the bulge. 

\citet{Dumas2007} finds larger velocity dispersions in the cases of NGC\,3351 and
NGC\,5248 than we do. The dispersion of those objects is probably too low to be
resolved by their instrumental dispersion of $\approx$ 110.~kms$^{-1}$. 

\cite{Vega-Beltran2001} find systematically lower velocity dispersions for
NGC\,2841 than we do. However, we also plot data from \cite{Heraudeau1998} which
are in excellent agreement with ours.  

The SAURON data for NGC\,4698 \citep{Falcon-Barroso2006} suggest a somewhat larger
velocity dispersion over our whole observed range than we find. They also find
negative $h_3$ moments on the east side.  The dispersion of this galaxy
($\approx$~140~kms$^{-1}$) should be well resolved by SAURON and such the difference
remains somewhat mysterious but small. 
\begin{center}
\begin{deluxetable*}{lcccccc}
\footnotesize
\tablecaption{Format example of the measured stellar kinematics.\label{tab:data_example}}
\tablewidth{0pt}
\tablehead{
\colhead{Galaxy} & \colhead{PA} & \colhead{r} & \colhead{v} & \colhead{$\sigma$} & \colhead{$h_3$} & \colhead{$h_4$}\\ 
\colhead{} & \colhead{[deg]} & \colhead{["]} & \colhead{[kms$^{-1}$]} & \colhead{[kms$^{-1}$]} & \colhead{} & \colhead{}\\    
\colhead{(1)} & \colhead{(2)} & \colhead{(3)} & \colhead{(4)} & \colhead{(5)} & \colhead{(6)} & \colhead{(7)}    
} 
\startdata
NGC\,1023 &	87 &	51.46 &	186.55 $\pm$	2.71 &	92.16 $\pm$	2.55 &	-0.067 $\pm$	0.017 &	-0.037 $\pm$	0.015\\
\enddata
\tablenotetext{Notes:}{
1) Identifier. 
2) Observed position angle.
3) Distance from the center (positive: east; negative: west).
4) Velocity relative to systemic velocity.
5) Velocity dispersion.
6) Gauss-Hermite $h_3$ moment.
7) $h_4$ moment.\\
The full listing is available electronically{\tt http://cds.u-strasbg.fr/}.
}
\end{deluxetable*}
\end{center}

\subsection{Signatures of bars in velocity profiles}
In our sample, \nbarred~out of \ngal~of the galaxies are classified as barred
or as hosting an oval. Bars and ovals will affect the observed kinematics and
their presence should be reflected in the moments of the observed LOSVD.
\citet{Bureau2005} use $N$-body simulations to derive diagnostics for the
presence of bars in edge-on disks. They find that {\it double-hump} rotation
curves, plateaus and shoulders in velocity dispersion, and correlation of $h_3$
moments with velocity in contrast to the usually-seen anti-correlation
are indicators for the presence of a bar.  The {\it double-hump} describes a
rotation curve that first rises quickly with radius, reaches a local maximum
then drops slightly and starts rising again towards larger radii. We do see
similar features in a number of our galaxies even though they are not observed
edge-on (e.g. see the rotation curves for NGC\,2841, NGC\,3351, and NGC\,3384 in
Fig.\ \ref{fig:kprofiles}). The signature is not always strong enough to form an
actual local minimum after the fast inner rise. Instead, in some cases we
observe shelves: the rise in velocity stagnates for a certain radial range
but becomes larger again before finally flattening out (e.g. NGC\,1023 and NGC\,3627 in
Fig.\ \ref{fig:kprofiles}).
    
Out of \nbarred\ barred galaxies (including \noval~ovals), 20 do show such features.
However our data do not extend very far into the disk region in many of the
objects in our sample; also visibility may be inhibited by the coarse
spatial binning of some of our spectra. Further, this diagnostic tool was
developed for edge-on systems, so it is likely that we  miss bar signatures in
the velocity profiles. However, 9 of the 16 non-barred galaxies show either
shelves or double humps which may be an indication that those systems actually do
host a bar that is not readily seen photometrically.

\subsection{Central velocity dispersions}
We calculate the central velocity dispersion of the galaxies in our sample by
averaging the major axis dispersion within a tenth of the effective bulge
radius $r_e$ that we obtain from the photometric decomposition. The values for
the central dispersions are given in Tab.~\ref{tab:params}. 
The quoted errors correspond to the formal errors of the derived mean
within $r_e/10$. 

\begin{figure*}
        \begin{center}
        \begin{tabular}{c}
        \includegraphics[width=0.7\textwidth]{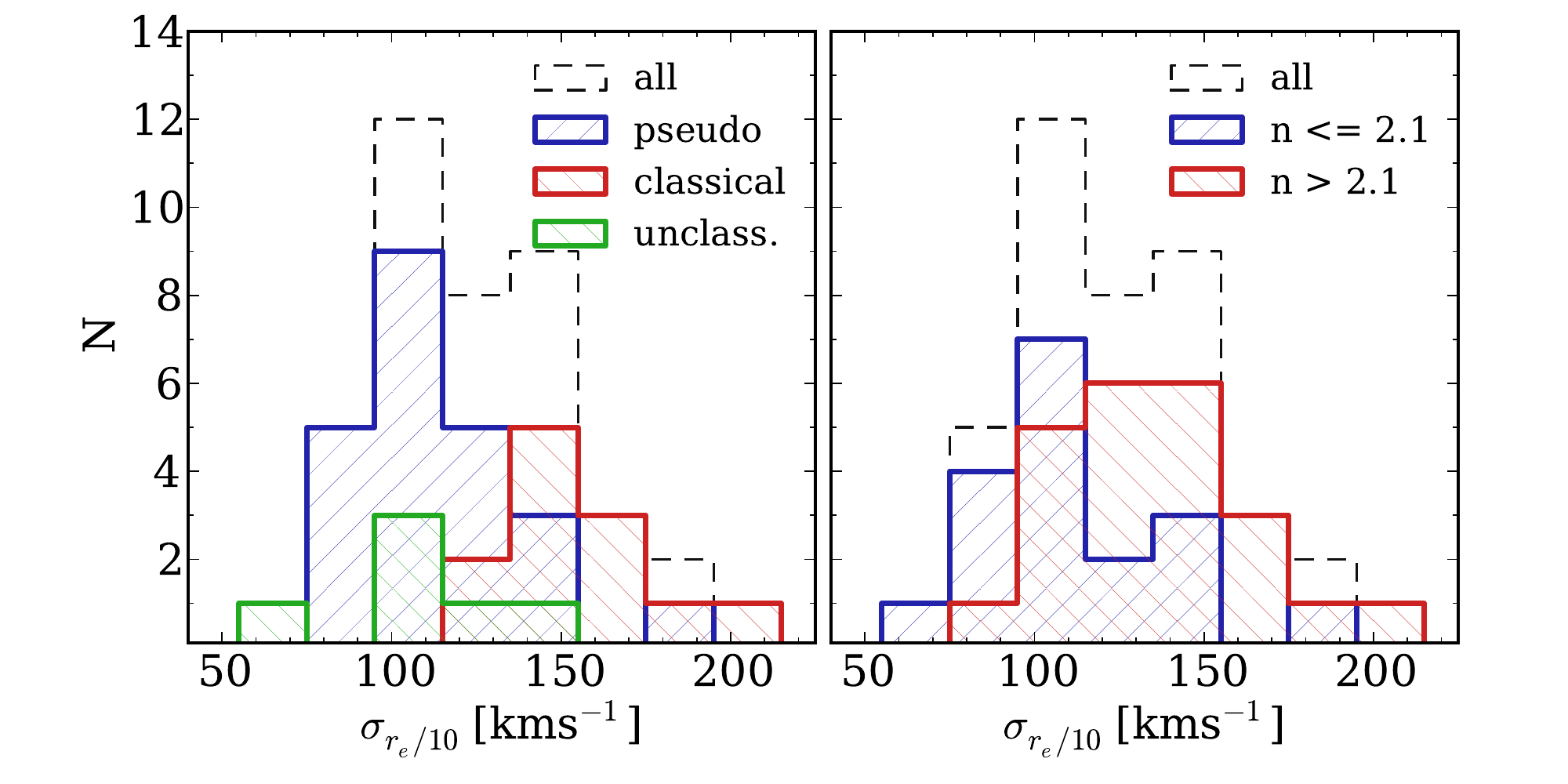}
        \end{tabular}
        \end{center}
	\caption{\label{fig:sigc_hists}
	Histograms of the central velocity dispersions.  The left panel discriminates
	bulge types by morphology, the right panel discriminates them by their S\'ersic index.
	}
\end{figure*}

In Fig.~\ref{fig:sigc_hists} we show corresponding histograms of the central
dispersions. In the left panel we discriminate bulge types based on their
morphology, in the right panel we discriminate by S\'ersic index.  There is
significant overlap between the distributions of velocity dispersions for the
classical and pseudobulges.  Nonetheless, it is clear that, in our sample,
pseudobulges have on average lower velocity dispersions.  We find in our sample
that classical bulges become exceedingly rare below central velocity
dispersions of 100~kms$^{-1}$.  However, we caution that our sample is not
volume limited.  
%
\subsection{Velocity dispersion gradients}
\label{sec:struct_vs_kin}
Inspection of the individual rotation curves reveals a wide variety of
structures; however, in particular the shape of the velocity dispersion profile
seems to fall into two rough classes.
\begin{figure*} 
\begin{center} 
\begin{tabular}{cc}
  \includegraphics[width=0.4\textwidth]{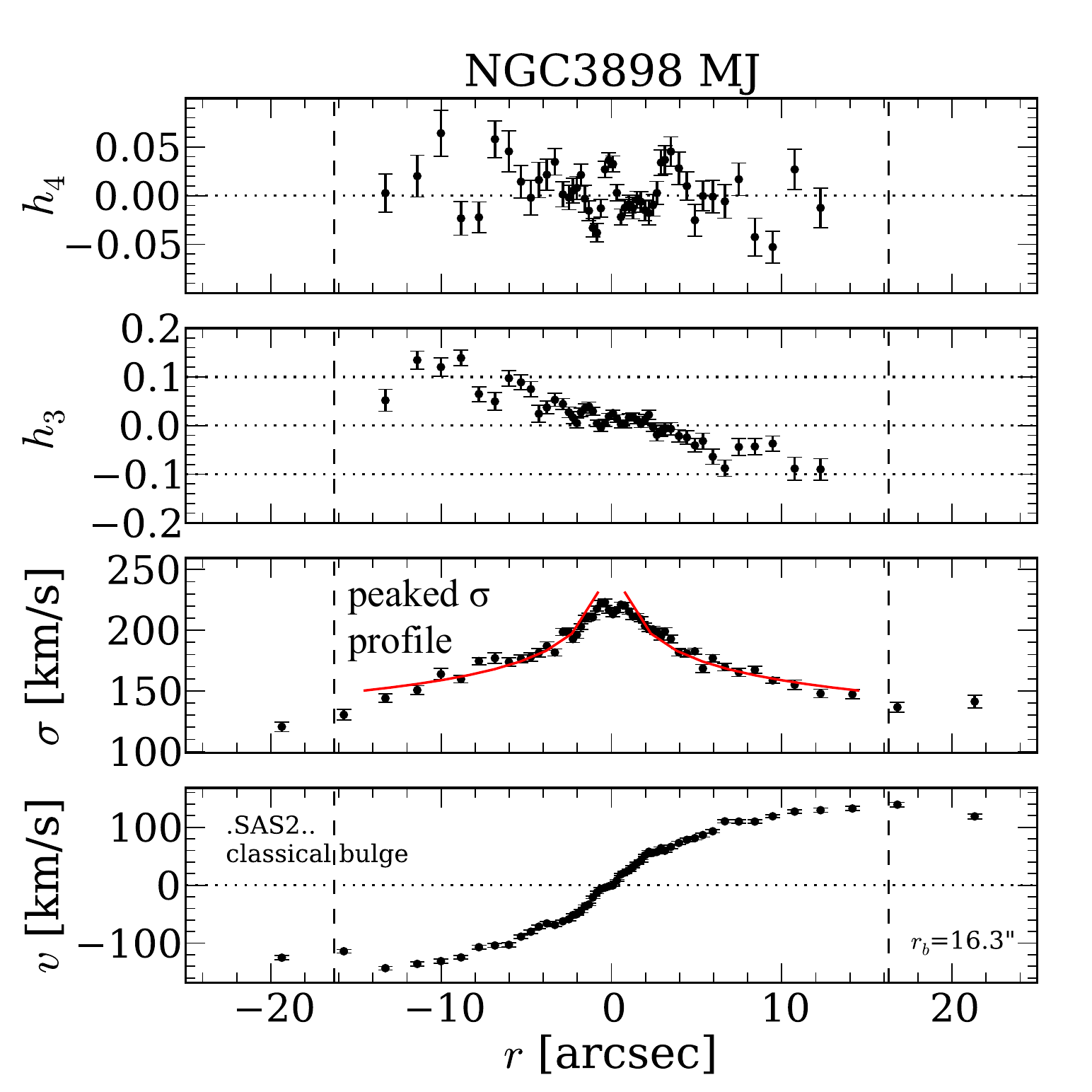}
  \includegraphics[width=0.4\textwidth]{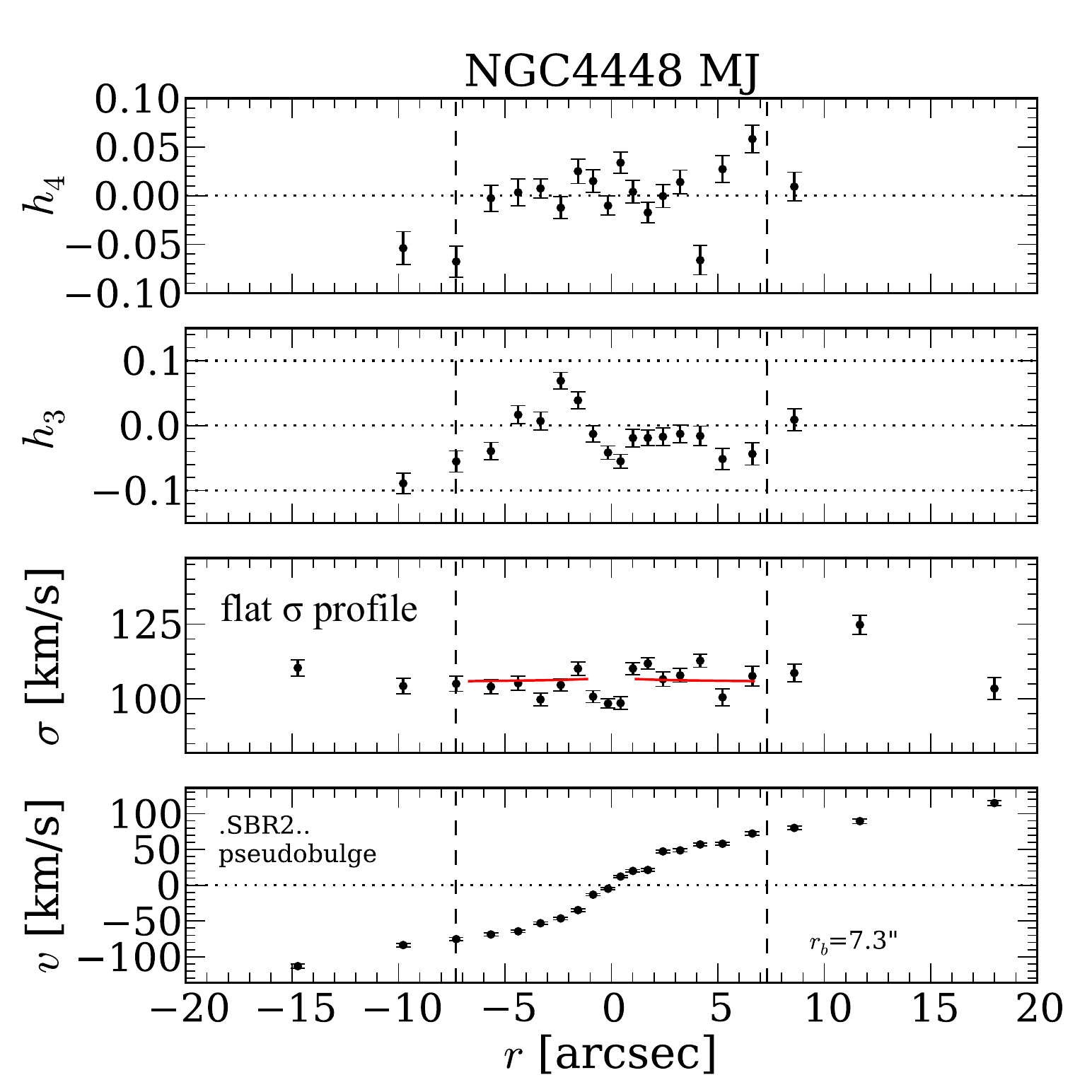}
\end{tabular} 
\end{center} 
\caption{Major axis kinematic profiles for NGC\,3898 and NGC\,4448.  Positive radii
are east of the galaxy center.  We plot from bottom to top the rotational
velocity, velocity dispersion, $h_3$ and $h_4$ moments.  The curvature of the
red lines correspond to the derived logarithmic slope of the dispersion
profile, they are scaled to match the depicted profile. Their extend indicates
the radial range which is taken into account
for the derivation of the slope (see text)
.
\label{fig:twoprofiles}} 
\end{figure*}
In Fig.~\ref{fig:twoprofiles} we show the kinematic profiles for the two
galaxies NGC\,3898 and NGC\,4448 from our sample. Depicted are the velocity, the
velocity dispersion, as well as the $h_3$ and $h_4$ moments of the
Gauss-Hermite expansion of the LOSVDs.  Dashed lines indicate the bulge radius
from the photometric decomposition.  While in the case of NGC\,3898 the velocity
dispersion rises all the way to the centre, NGC\,4448 has a relatively flat
dispersion profile within the bulge radius.

Similarly to \citet{Fisher1997} we examine,
the logarithmic slope of the velocity dispersion within the bulge radius and
call it $\gamma$. We derive the slope point-wise and then take the average, i.e.
\begin{eqnarray} 
\label{eq:gamma}
  \gamma = <\frac{ \Delta \log(\sigma) }{ \Delta \log(r)}>\mid_{r_{min}<r<r_b}, 
\end{eqnarray} 
where $r_{min}$ always excludes the inner FWHM of the seeing of the particular
observation and in some cases is chosen larger to exclude central features like
nuclear regions of enhanced star formation (see
Appendix~\ref{sec:individuals}).  In Fig.~\ref{fig:twoprofiles} we also
overplot lines which correspond to the derived $\gamma$ values.  Further, in
order to avoid a dependence of the slope on the particular binning scheme of
each kinematic dataset, we use a different binning for the purpose of
determining $\gamma$: we bin radially in 5 equally-sized bins in $log(r)$. In
cases where the resulting bins do not all at least contain one data point, we
use our previous bins.

An alternative to the presented method is using the ratio of the averaged
velocity dispersions within two annuli within the bulge radius
\begin{eqnarray} 
\label{eq:delta} 
\delta =
\frac{<\sigma>\mid_{r_{min}<r<r_b/3} }{<\sigma>\mid_{r_b/3<r<r_b} }
\end{eqnarray} 
as proxy for the slope. The choice of $r_b/3$ as cut radius for
the two different annuli is somewhat arbitrary, but we do not find a strong
dependence of our results on the specific radius chosen. 
Both values for the slope, $\gamma$ and $\delta$, are reported 
in Tab.~\ref{tab:params}. 

We find that all bulges which are classified as pseudobulges indeed show
flattened velocity dispersion profiles or even sigma drops
(e.g.\ NGC\,3351, NGC\,3368, and NGC\,3627 in Fig.\ \ref{fig:kprofiles}).
The
dispersion profiles of many pseudobulges are sometimes slightly asymmetric. On
the other, hand a majority of the classical bulges show centrally peaked
velocity dispersion profiles
(e.g.\ NGC\,1023, NGC\,2841, NGC\,2880, and NGC\,3245 in Fig.\ \ref{fig:kprofiles}).

\begin{figure}
        \begin{center}
        \begin{tabular}{c}
        \includegraphics[width=0.45\textwidth]{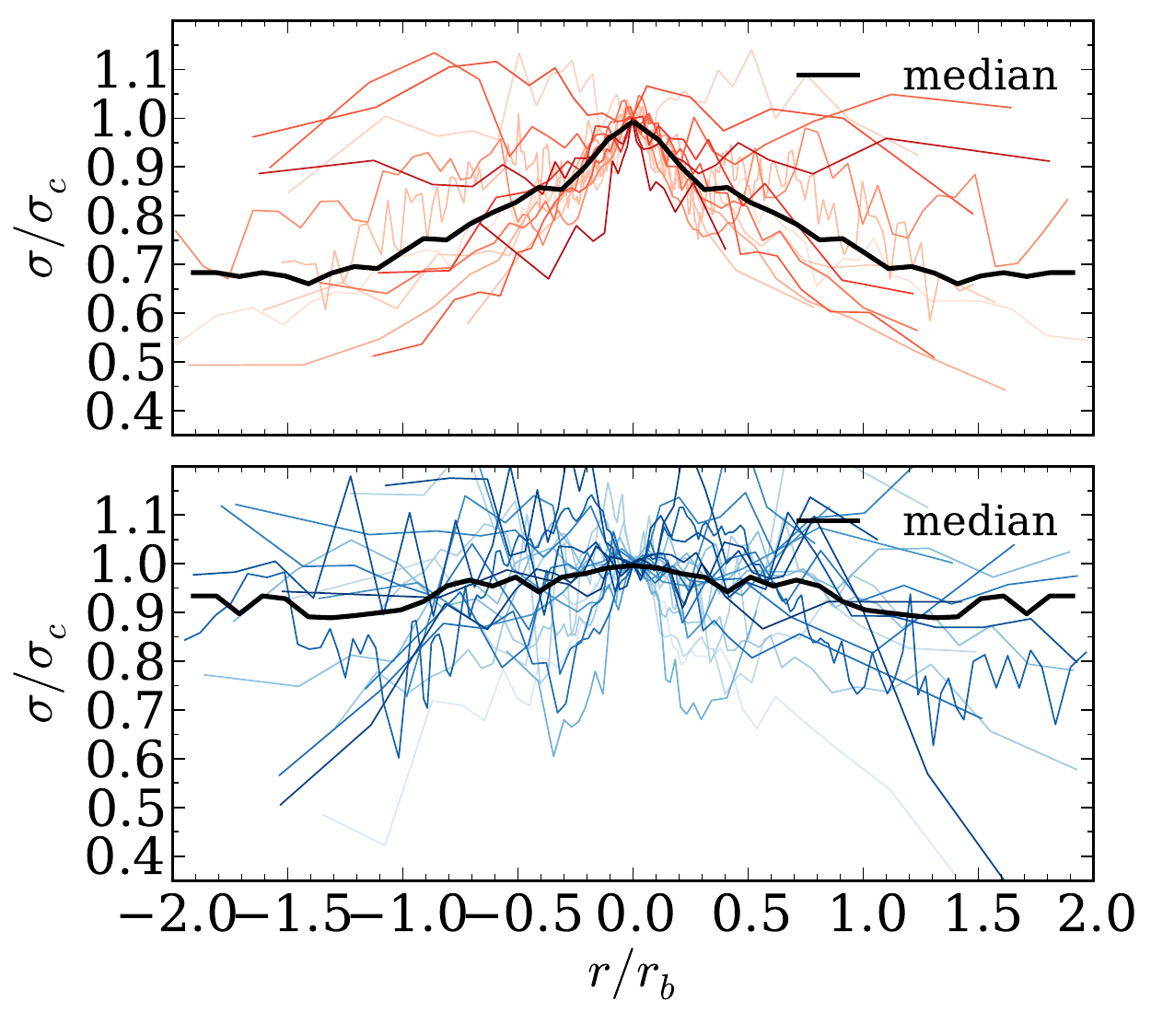}\\
        \includegraphics[width=0.45\textwidth]{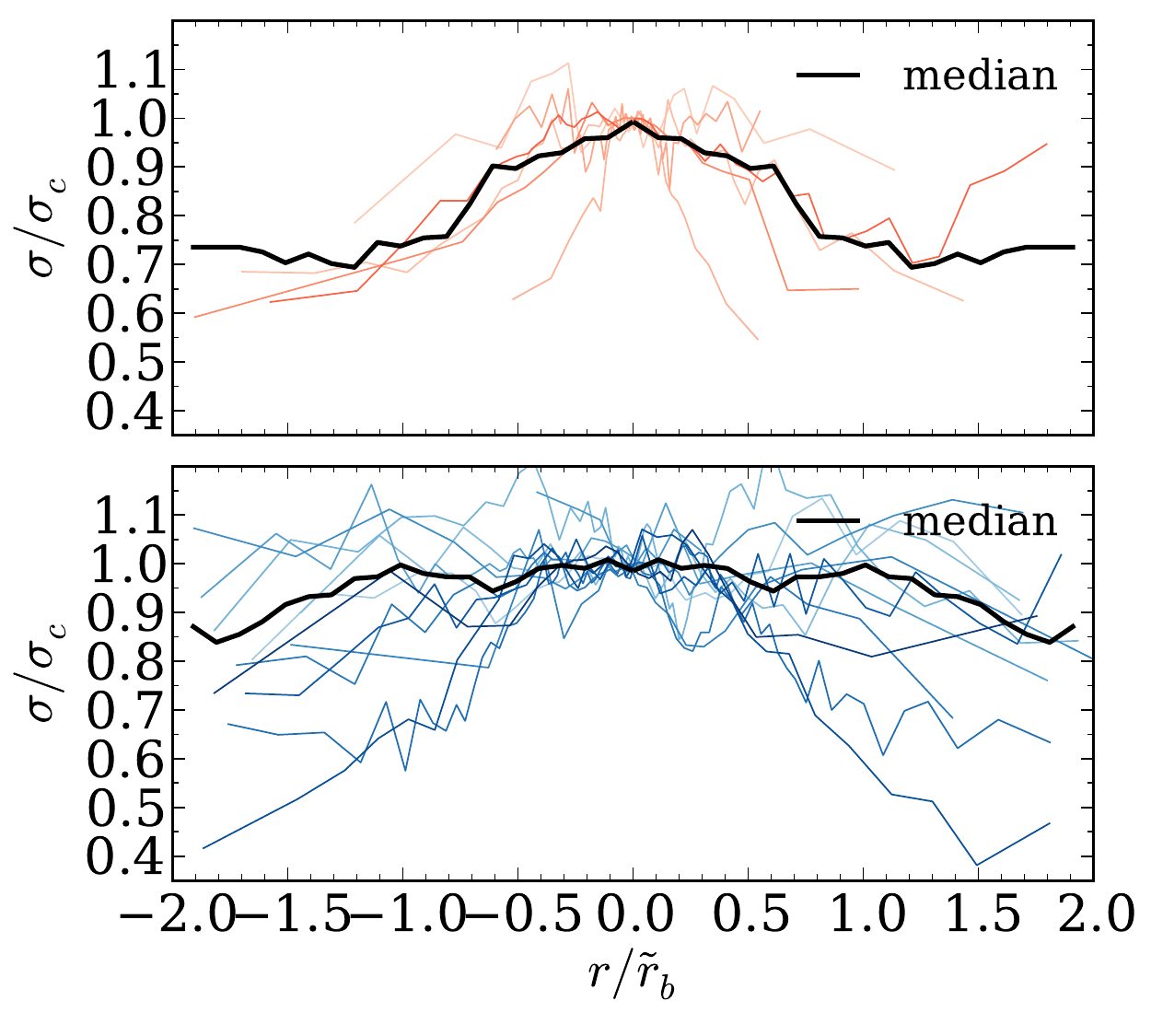}
        \end{tabular}
        \end{center}
	\caption{\label{fig:overplots}
	Major (top panels) and minor axis (bottom panels) velocity dispersion
	profiles, normalized by central velocity dispersion and bulge radius. Profiles
	of classical bulges are plotted in red in the respective upper panels, 
	those of pseudobulges in blue in the corresponding lower panels. 
	Different color shades correspond to different galaxies. The
	thick black line shows the median profile for all bulges in one panel. The
	bulge radii for the minor axis profiles have been corrected
	using the mean bulge ellipticity according to 
	$\tilde r_{b} = \left( 1 - <\epsilon> \right) \cdot r_{b}$.
	}
\end{figure}

Fig.~\ref{fig:overplots} summarizes this finding qualitatively, where
we plot the velocity dispersion profiles along the major and minor
axes for all our bulges separated by bulge type, normalized by central
dispersion and bulge radius. We do not plot bulges that were left
unclassified. For this plot we adjust the bulge radius that was
obtained from a major axis profile by the mean ellipticity in the
bulge region. Whilst not as clear, partly due to the lower number
of profiles, but partly probably also due to the subtleties of
choosing a correct radius for the normalisation, we again find that
classical bulges tend to show centrally rising velocity dispersions.

\begin{figure*}
        \begin{center}
        \begin{tabular}{c}
                \includegraphics[width=0.8\textwidth]{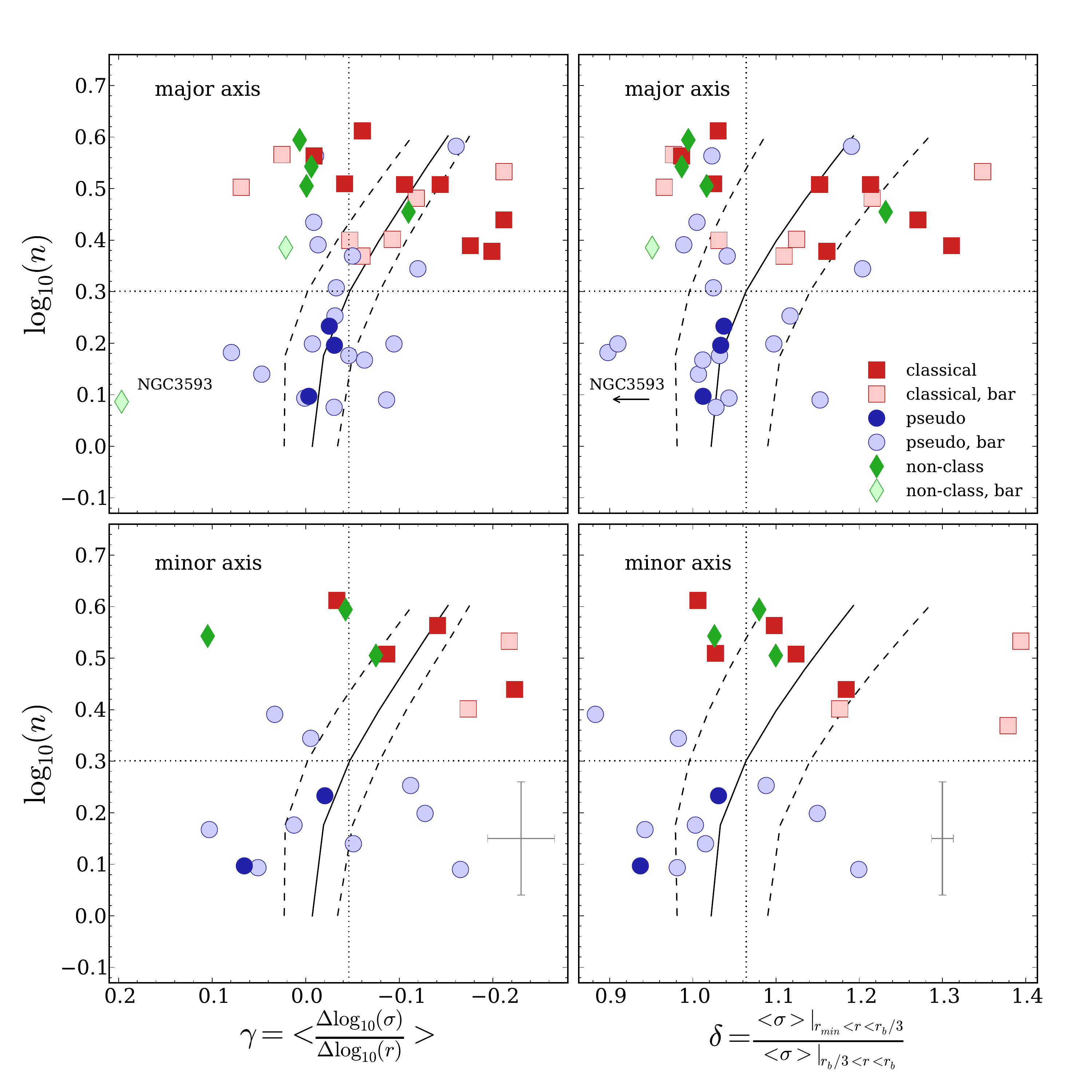} \\ 
        \end{tabular}
        \end{center}
	\caption{\label{fig:struct_vs_kin}
	S\'ersic index $n$ is shown as a function of both metrics for the
	flatness of the velocity dispersion profile, $\gamma$ and
	$\delta$, respectively. 
	{\it Upper left:} Shows the major axis logarithmic slope of
	the velocity dispersion.  
	Red squares and blue circles correspond to classical
	and pseudobulges respectively.  Open symbols label barred galaxies. The green
	diamonds represent unclassified objects.  The black solid line shows the
	respective behaviour of the isotropic models in \citet{Ciotti1991}, here the
	slopes were calculated in the radial range $r_e/10 < r < r_e$.  
	The dashed
	lines show the $\gamma$ values one would obtain by choosing the outer integration
	radius 50\% smaller or larger.  The horizontal line marks a S\'ersic index of
	two. The vertical dotted lines mark $\gamma = -0.046$ and $\delta = 1.06$,
	the respective values that the isotropic models take for a S\'ersic index of 2.
	{\it Upper right:} Shows the ratio of the averaged velocity dispersion
	in two different annuli. NGC\,3593 falls far to the
	left with $\delta = 0.7$ (see discussion in \S\ref{sec:discussion} and Appendix \ref{sec:individuals}). 
	{\it Lower panels:} Same for the minor axis dispersions.
	The radii were adjusted according the mean
	bulge ellipticity $\tilde r_{b} = \left( 1 - <\epsilon> \right) \cdot r_{b}$.
	The error bars correspond to the typical errors in the derived
	quantities, they also apply to the upper panels. 
	}
\end{figure*}

In Fig.~\ref{fig:struct_vs_kin} we now plot the S\'ersic index from the
photometric decomposition as a function of both metrics for the slope of the
velocity dispersion. Similar to the distributions of central velocity
dispersion, there is significant overlap in profile slope. 
Nonetheless, the
bulges with large values of S\'ersic index tend to have steeply decaying
dispersions profiles. 
Similarly the bulges with low S\'ersic indices more
commonly have flat dispersion profiles. This result is true for both the
logarithmic slope of dispersion, and the dispersion ratio.

The increasing slope of velocity dispersion with S\'ersic index is not fully
unexpected. For instance, \citet{Ciotti1991} describes a series of models for
isotropic and spherical galaxies which have a surface brightness profiles that
follow a S\'ersic law. He gives projected velocity dispersion profiles for his
models and outside of the very central regions ($r > r_e/10$), and for S\'ersic
indices larger than one, the slope is a monotonically increasing function of
$n$. We calculate slopes and $\sigma$ ratios for these profiles in a similar
manner as we did for our data.  One caveat of this exercise is that our
definition of a bulge radius is not applicable in the case of the one-component
models. Also, we have to chose an inner cut radius for the fit as the models
feature central sigma drops in the case of small $n$.  Sigma drops are an
observed phenomenon (e.g.~\citealt{Falcon-Barroso2006}), but our spatial
resolution is typically not fine enough to resolve these. We somewhat
arbitrarily fit for $\gamma$ in the radial range of $r_e/10 < r < r_e$ and
calculate $\sigma$ ratios for $r_e/10 < r < r_e/3$ and $r_e/3 < r < r_e$. Note
that our effective bulge radii are on average 15\% smaller than the definition
of the bulge radius that we use throughout the work.  In
Fig.~\ref{fig:struct_vs_kin} we overplot the obtained values as a black line.
The dashed lines show the range of values one would obtain by choosing 50\%
larger or smaller outer cut radii for the integration.  While the spherical and
isotropic galaxies are a very simplistic model for the variety of bulges in our
sample, one can see that the general trends are reproduced, however a more
detailed dynamical modelling is needed to confirm this result. 

In the major axis plots all unbarred pseudobulges fall below or very close to
$\gamma = - 0.05$ and $\delta = 1.06$ (the corresponding values of the isotropic
models for $ n = 2$) and only one unbarred classical bulge falls below $\gamma
= - 0.05$.  However 3 out of 8 unbarred classical bulges do fall significantly
below $\delta = 1.09$ suggesting that $\gamma$ is more successful in
discriminating bulge types.  Again this picture is complicated further once
barred galaxies are taken into account. The additional component of a bar seems
to lead towards flatter dispersion profiles. 

%
\subsection{Influence of seeing on velocity dispersion}
The seeing disk and the width of the slit will {\it smear} the observed
velocities and can create increases in the observed line of sight velocity
dispersion. This effect is commonly known as {\it slit smearing}.  All data
presented here were observed with a slit width of 1~arcsecond. The effect of
slit smearing on the velocity dispersion is therefore expected to be negligible
compared to the effect caused by the seeing ($> 1.2$~arcseconds in all cases).
At least two galaxies NGC\,3384 and NGC\,3521 do show peaks in velocity dispersion
in the central arcseconds 
(see Fig.\ \ref{fig:kprofiles})
. In both galaxies the velocity profile also rises
rapidly in the centre.  We test whether this rapid rise in combination with the
seeing may be responsible for the observed dispersion peak. We model the point
spread function (PSF) with a Gaussian of the same FWHM. We then calculate the
standard deviation of the velocity which is weighted by the PSF amplitude at
all radii and subtracted the result from the observed velocity dispersion.  In
this simple one-dimensional model the PSF smearing does generate a central peak
which is of similar size and amplitude as the observed one. We cannot rule out
the possibility that the central peaks of NGC\,3384 and NGC\,3521 can be explained
through PSF-smearing alone.  We however refrain from correcting the presented
velocity dispersions as an accurate correction has to include the knowledge of
a high resolution luminosity profile and a more rigorous, 2-dimensional
modelling of the PSF.  We rather exclude the central peaks from the further
analysis.
%
\subsection{Distribution of $h_3$ and $h_4$ moments}
\begin{figure}
        \begin{center}
        \begin{tabular}{c}
                \includegraphics[width=0.45\textwidth]{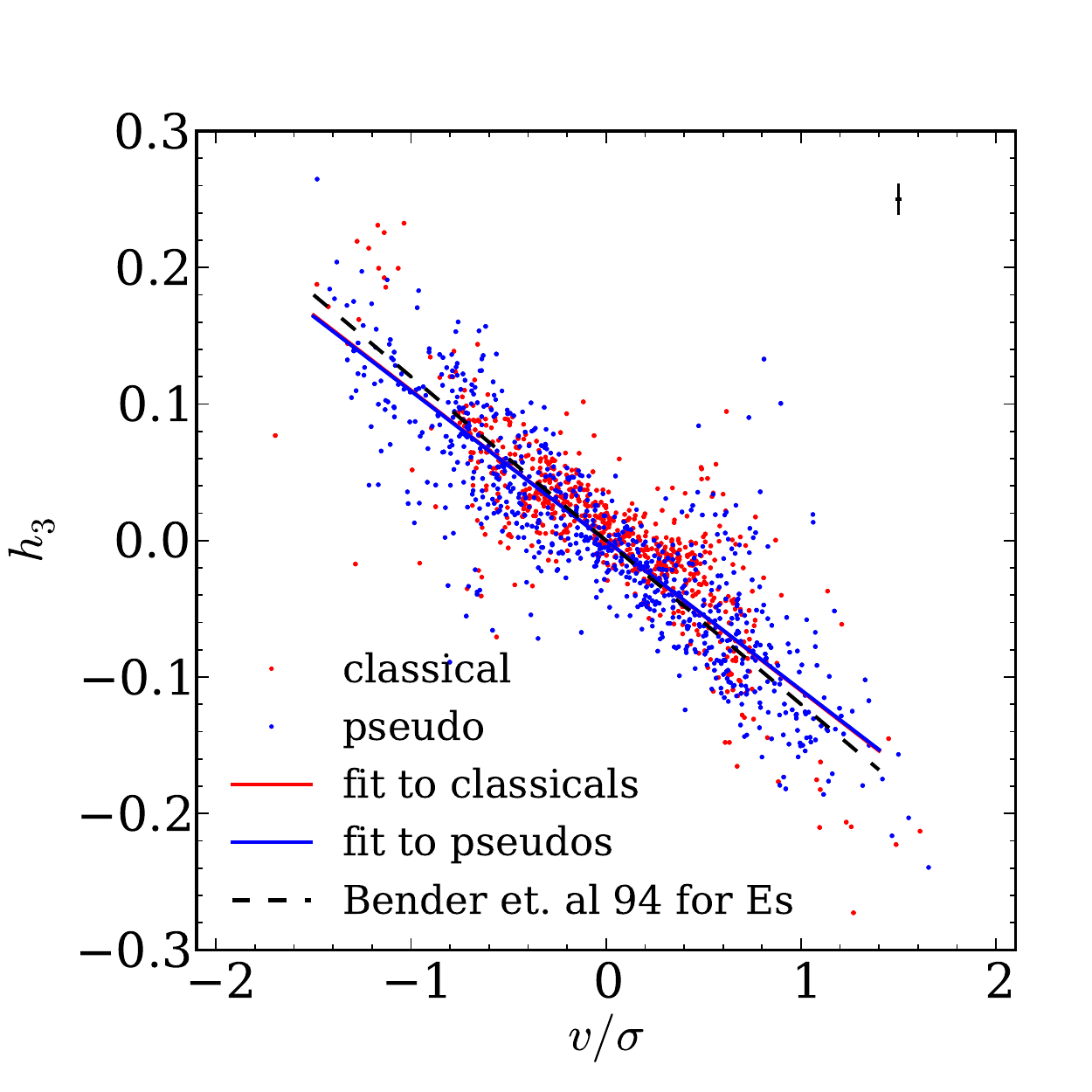} \\ 
        \end{tabular}
        \end{center}
	\caption{\label{fig:h3_vs_v_over_sig}
	Local correlation between $h_3$ and v/$\sigma$ along the major axis
	for the galaxies in our sample.  We only plot points for which the
	error in $h_3$ is lower than 0.05. Plotted are all galaxies for which
	the bulge was classified either as classical (red) or a pseudobulge (blue).  
	Typical error bars are shown in the upper right of the diagram.
	The red and blue lines correspond to the 
	fitted linear correlations for the classical and pseudobulges respectively.
	The dashed black line represents the value for the correlation that \citet{Bender1994}
	obtained for their sample of early types.
	} 
\end{figure}
As $h_3$ measures the asymmetric deviation from a purely Gaussian distribution
it detects lower velocity tails of the velocity distribution along the line of
sight. Such tails arise naturally in disks \citep{Binney87}. \citet{Bender1994}
found that local $h_3$ and local $v/\sigma$ are strongly anti-correlated with a
slope of $-0.12$ in their sample of elliptical galaxies. \citet{Fisher1997}
finds a similar anti-correlation in the inner regions of his lenticular
galaxies but also sees that, for a number of his objects, at values of
$v/\sigma \approx 1$ the anti-correlation turns, at least briefly but
abruptly, into a correlation.

\begin{figure*}
  \begin{center}
  \begin{tabular}{c}
	  \includegraphics[width=0.7\textwidth]{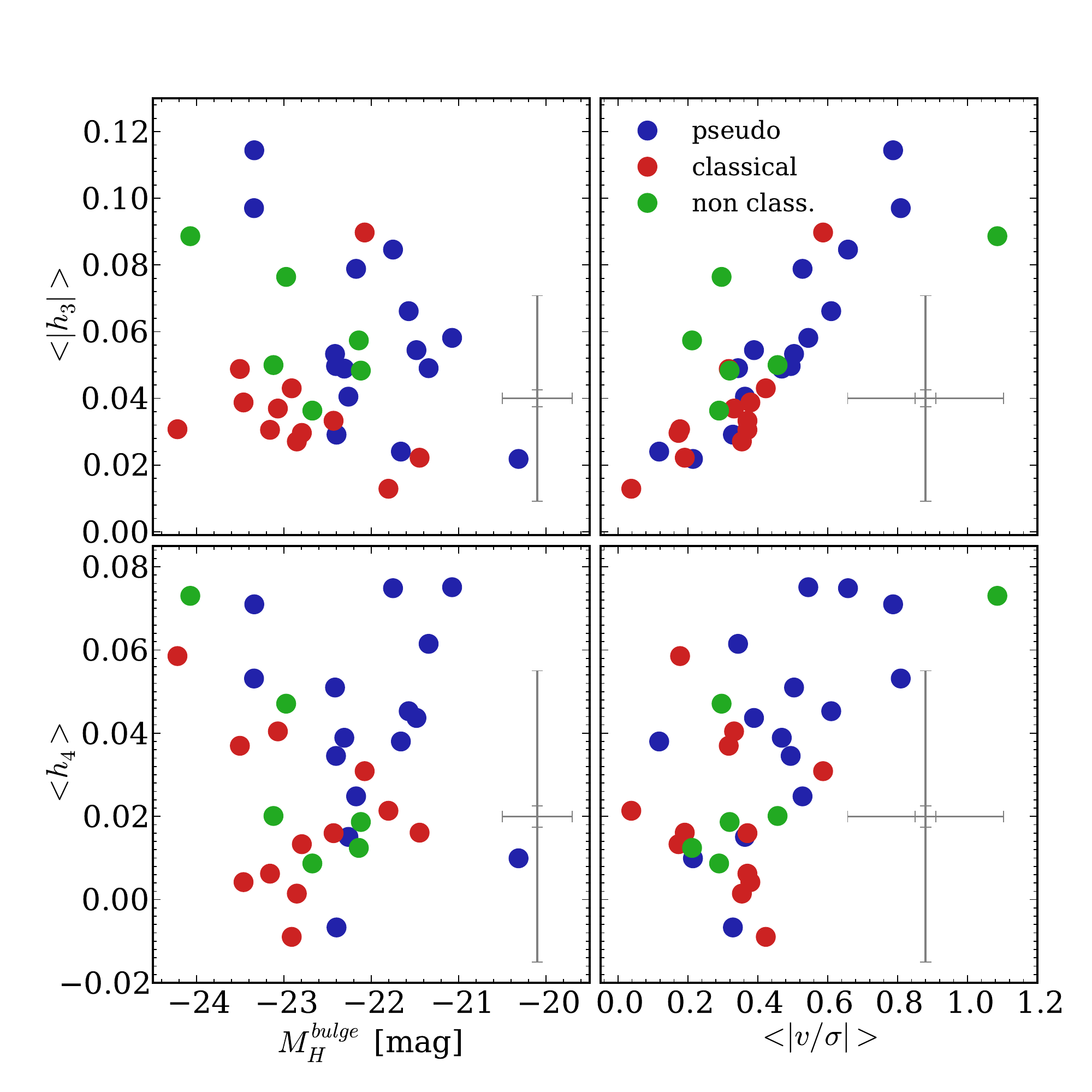} \\ 
  \end{tabular}
  \end{center}
  \caption{\label{fig:h3h4_corr}
  Major axis correlations between bulge averaged Gauss-Hermite moments
  $<h_3>$ and $<h_4>$, bulge luminosities and $<v/\sigma>$. {\it Upper left:}
  $<h_3>$ as function of bulge $H$-band magnitude. 
  Red circles represent classical bulges,
  blue circles are pseudobulges, green circles represent unclassified bulges.
  {\it Upper right:} $<h_3>$ as function of $<v/\sigma>$. No inclination
  corrections were applied. 
  {\it Lower left:} $<h_4>$ as function of bulge magnitude.  
  {\it Lower right:} $<h_4>$ as function of $<v/\sigma>$.
  Representative error bars are displayed in the lower right hand part of each panel.
  In the case of the bulge magnitude they correspond to the typical error. For
  all other plotted quantities the larger error bars correspond to the typical RMS
  scatter of that quantity within the considered radial range, whereas the smaller
  error bar corresponds to the formal error of the derived average.
  } 
\end{figure*}

We reproduce the plot for the local correlation of $h_3$ and $v/\sigma$ from
\citet{Bender1994} for our sample in Fig.~\ref{fig:h3_vs_v_over_sig} and
color-code pseudobulges in blue and classical bulges in red. We find that the
same correlation is reproduced in our intermediate type galaxies. The $h_3$ moments
are generally anti-correlated with $v/\sigma$ out to $v/\sigma \approx 0.5$,
irrespective of bulge type. A linear fit to the complete set of data points
gives a slope of ($h_3 = (-0.106 \pm 0.001)\cdot v/\sigma)$.  
Separate fits to the subsample of classical bulges and pseudobulges give
values that are indistinguishable within the errors. A Kolmogorov-Smirnov test
\citep{Smirnov39,Press02} for the median values of $v \cdot \sigma^{-1}
\cdot h_3^{-1}$ within individual galaxies yields a probability of 76\% for
the hypothesis that the classical and the pseudobulges stem from the same
distribution. This local correlation is reproduced in the mean values for the
bulge region (Fig.~\ref{fig:h3h4_corr}). 

We further test for a possible correlation with $H$-band bulge magnitude (see
Fig.~\ref{fig:h3h4_corr}) and the bulge averaged value $<h_3>$.  We do not see
any correlation between bulge luminosity and $<h_3>$. 

%

The $h_4$ moment of the Gauss-Hermite expansion measures the symmetric
deviation from a Gaussian distribution. Negative $h_4$ describe a more boxy,
centrally flattened distribution, more positive values describe centrally
peaked distributions with extended wings.
The averaged $h_4$ moments in the bulges are generally close to zero, the
median for the complete sample of major axis spectra is 0.03 with a standard
deviation of 0.046. 
None of our bulges show obvious dips in the $h_4$ profile
as the ones described by \citet{Debattista2005,Mendez-Abreu2008}.  But this
diagnostic for boxy-peanut shape bulges only applies to low inclinations ($i <
30 ^\circ $), given that the inclinations of most of our galaxies is larger
than $30 \Deg$ (41 out of \ngal) this is not further surprising.  However,
fourteen galaxies show a double peak in the $h_4$ profile within the bulge
region
(e.g.\ NGC\,1023, NGC\,3031, NGC\,3945, and NGC\,7331 in Fig.\ \ref{fig:kprofiles})
. This is typically seen in combination with a rapid increase of the
rotational velocity and relatively strong $h_3$ moments. From our Monte Carlo
simulations described in \S\ref{sec:kin_derive} we can rule out that the
observed peaks are a result of a degeneracy between $h_3$ and $h_4$ moments
in the fit.

We find no correlation between the averaged $<h_4>$ moments and the bulge
luminosities. However, while the error bars are large, 
larger $h_4$ moments seem to be found in bulges with larger averaged 
$v/\sigma$ (Fig.~\ref{fig:h3h4_corr}). 
\citet{Bender1994} also discuss the possibility of a similar trend in their
subsample of rotationally flattened galaxies.

There is a mild indication that pseudobulges and classical bulges show different
distributions in the average $<h_3>$ and $<h_4>$ moments.  For pseudobulges we
find an average value for $<h_3>$ of 0.06 with an RMS scatter of 0.03 while
for classical bulges the mean value of $<h_3>$ is 0.04 with a scatter or 0.03.
A KS test and a Student's two-tailed t-test for two independent samples yield a
probability of 0.3\% and 3\%, respectively, for the two subsamples to stem from
the same distribution.  For $<h_4>$ we find and average of 0.04 with a scatter
of 0.02 in the pseudobulges, and 0.02 with a scatter of 0.02 in the classical
bulges. Here the KS test finds a 5\% probability for the null hypothesis while
the t-test yields a 1.5\% probability.  As both the $h_3$ and the $h_4$ moment
are affected by the inclination and the scatter is large, this trend has to be
taken with caution.  A larger sample and kinematic modelling will be needed to
confirm if this is a signature of systematic different anisotropies in the two
classes of bulges.

%
\begin{center}
\begin{deluxetable*}{lrrrrrrrrrr}
\tabletypesize{\scriptsize}
\tablecaption{Structural and kinematic parameters. \label{tab:params}}
\tablewidth{0pt}
\tablehead{
\colhead{Galaxy} 
&\colhead{$r_{\mu_{b}=\mu_{d}}$}
&\colhead{$r_b$}
&\colhead{$<\epsilon_b>$} 
&\colhead{$<\epsilon_d>$}
&\colhead{$\sigma_{r_e/10}$}
&\colhead{$\gamma_{MJ}$} 
&\colhead{$\delta_{MJ}$}
&\colhead{$\gamma_{MN}$} 
&\colhead{$\delta_{MN}$}
&\colhead{$\frac{<v^2>}{<\sigma^2>}$}
\\
\colhead{} 
&\colhead{arcsec} 
&\colhead{arcsec} 
&\colhead{} 
&\colhead{} 
&\colhead{kms$^{-1}$} 
&\colhead{} 
&\colhead{}
&\colhead{}
&\colhead{}
&\colhead{}
\\
\colhead{(1)} &\colhead{(2)} &\colhead{(3)} &\colhead{(4)} &\colhead{(5)} &\colhead{(6)} &\colhead{(7)} &\colhead{(8)} &\colhead{(9)} &\colhead{(10)} &\colhead{(11)}
\\
}
\startdata
NGC\,1023 & 	21.3 & 	19.0 & 	0.22 & 	0.55 & 	212.9~$\pm$~5.2 & 	-0.09 & 	1.12 & 	-0.17 & 	1.176 & 	0.165	\nl
NGC\,2460 & 	8.3 & 	6.6 & 	0.19 & 	0.25 & 	111.4~$\pm$~3.5 & 	-0.01 & 	0.99 & 	0.10 & 	1.026 & 	0.358	\nl
NGC\,2681 & 	14.6 & 	13.2 & 	0.11 & 	0.20 & 	112.5~$\pm$~1.3 & 	-0.16 & 	1.19 & 	 \nodata  & 	 \nodata  & 	 \nodata$^a$ 	\nl
NGC\,2775 & 	19.4 & 	16.8 & 	0.10 & 	0.16 & 	173.9~$\pm$~13.7 & 	-0.04 & 	1.02 & 	1.03 & 	1.027 & 	0.133	\nl
NGC\,2841 & 	17.4 & 	15.2 & 	0.22 & 	0.49 & 	222.2~$\pm$~19.3 & 	-0.11 & 	1.15 & 	-0.09 & 	1.124 & 	0.207	\nl
NGC\,2859 & 	30.0 & 	27.6 & 	0.16 & 	0.22 & 	176.8~$\pm$~5.4 & 	-0.06 & 	1.11 & 	-2.04 & 	1.379 & 	0.678	\nl
NGC\,2880 & 	26.7 & 	22.7 & 	0.20 & 	0.35 & 	142.2~$\pm$~5.3 & 	-0.21 & 	1.35 & 	-0.22 & 	1.394 & 	0.496	\nl
NGC\,2964$^b$  & 	3.4 & 	3.1 & 	0.15 & 	0.28 & 	88.4~$\pm$~1.3 & 	 \nodata  & 	 \nodata  & 	 \nodata  & 	 \nodata  & 	0.117	\nl
NGC\,3031 & 	72.0 & 	61.3 & 	0.24 & 	0.43 & 	157.5~$\pm$~13.6 & 	-0.06 & 	1.03 & 	-0.03 & 	1.006 & 	0.400	\nl
NGC\,3166 & 	9.9 & 	9.1 & 	0.39 & 	0.25 & 	151.4~$\pm$~6.1 & 	0.00 & 	1.04 & 	0.05 & 	0.981 & 	1.365	\nl
NGC\,3245 & 	9.5 & 	8.5 & 	0.20 & 	0.44 & 	225.2~$\pm$~8.3 & 	-0.21 & 	1.27 & 	-0.22 & 	1.184 & 	0.300	\nl
NGC\,3351 & 	14.2 & 	12.9 & 	0.16 & 	0.24 & 	90.0~$\pm$~4.2 & 	0.05 & 	1.01 & 	-0.05 & 	1.015 & 	0.778	\nl
NGC\,3368 & 	23.3 & 	20.4 & 	0.17 & 	0.34 & 	122.5~$\pm$~6.6 & 	-0.01 & 	0.99 & 	0.03 & 	0.883 & 	0.574	\nl
NGC\,3384 & 	15.5 & 	14.4 & 	0.20 & 	0.34 & 	150.3~$\pm$~2.4 & 	-0.09 & 	1.10 & 	-0.13 & 	1.150 & 	0.521	\nl
NGC\,3521 & 	12.2 & 	10.8 & 	0.35 & 	0.45 & 	129.5~$\pm$~2.9 & 	-0.01 & 	0.99 & 	-0.14 & 	1.098 & 	0.905	\nl
NGC\,3593 & 	32.1 & 	29.5 & 	0.49 & 	0.62 & 	62.3~$\pm$~3.1 & 	0.20 & 	0.70 & 	 \nodata  & 	 \nodata  & 	0.065	\nl
NGC\,3627 & 	11.7 & 	10.9 & 	0.27 & 	0.51 & 	116.1~$\pm$~3.9 & 	-0.05 & 	1.03 & 	0.01 & 	1.003 & 	0.267	\nl
NGC\,3675 & 	9.8 & 	8.5 & 	0.29 & 	0.49 & 	114.7~$\pm$~5.3 & 	-0.03 & 	1.03 & 	 \nodata  & 	 \nodata  & 	0.363	\nl
NGC\,3898 & 	17.7 & 	15.7 & 	0.25 & 	0.41 & 	219.0~$\pm$~8.3 & 	-0.14 & 	1.21 & 	 \nodata  & 	 \nodata  & 	0.295	\nl
NGC\,3945 & 	33.3 & 	31.0 & 	0.19 & 	0.17 & 	183.1~$\pm$~5.4 & 	-0.03 & 	1.12 & 	-0.11 & 	1.088 & 	1.062	\nl
NGC\,3953 & 	16.6 & 	14.6 & 	0.26 & 	0.48 & 	110.6~$\pm$~3.1 & 	0.02 & 	0.95 & 	 \nodata  & 	 \nodata  & 	0.178	\nl
NGC\,3992 & 	14.9 & 	13.2 & 	0.22 & 	0.49 & 	144.2~$\pm$~9.5 & 	0.07 & 	0.97 & 	 \nodata  & 	 \nodata  & 	0.406	\nl
NGC\,4030$^b$ & 	3.6 & 	3.0 & 	0.11 & 	0.19 & 	102.9~$\pm$~4.5 & 	 \nodata  & 	 \nodata  & 	 \nodata  & 	 \nodata  & 	0.213	\nl
NGC\,4203 & 	16.4 & 	14.7 & 	0.11 & 	0.11 & 	170.1~$\pm$~3.6 & 	-0.18 & 	1.31 & 	 \nodata  & 	 \nodata  & 	0.249	\nl
NGC\,4260 & 	9.0 & 	7.3 & 	0.21 & 	0.53 & 	143.8~$\pm$~14.3 & 	0.03 & 	0.98 & 	 \nodata  & 	 \nodata  & 	0.131	\nl
NGC\,4274 & 	12.3 & 	11.3 & 	0.40 & 	0.34 & 	106.9~$\pm$~5.3 & 	0.08 & 	0.90 & 	 \nodata  & 	 \nodata  & 	0.842	\nl
NGC\,4314 & 	10.0 & 	8.6 & 	0.12 & 	0.45 & 	123.3~$\pm$~5.1 & 	-0.01 & 	1.00 & 	 \nodata  & 	 \nodata  & 	0.825	\nl
NGC\,4371 & 	25.5 & 	22.9 & 	0.29 & 	0.33 & 	125.8~$\pm$~5.0 & 	-0.12 & 	1.20 & 	0.00 & 	0.983 & 	0.426	\nl
NGC\,4379 & 	10.0 & 	8.6 & 	0.11 & 	0.20 & 	121.0~$\pm$~4.6 & 	-0.20 & 	1.16 & 	 \nodata  & 	 \nodata  & 	0.183	\nl
NGC\,4394 & 	15.2 & 	14.1 & 	0.12 & 	0.37 & 	80.0~$\pm$~3.1 & 	-0.01 & 	0.91 & 	 \nodata  & 	 \nodata  & 	1.451	\nl
NGC\,4448 & 	9.6 & 	8.5 & 	0.26 & 	0.43 & 	98.5~$\pm$~3.7 & 	-0.03 & 	1.03 & 	 \nodata  & 	 \nodata  & 	0.446	\nl
NGC\,4501 & 	7.0 & 	6.2 & 	0.19 & 	0.45 & 	144.2~$\pm$~4.9 & 	0.00 & 	1.01 & 	0.07 & 	0.937 & 	0.390	\nl
NGC\,4536 & 	10.9 & 	10.1 & 	0.39 & 	0.47 & 	98.1~$\pm$~3.3 & 	-0.06 & 	1.01 & 	0.10 & 	0.942 & 	0.724	\nl
NGC\,4569 & 	10.6 & 	9.6 & 	0.32 & 	0.57 & 	114.4~$\pm$~0.9 & 	-0.05 & 	1.04 & 	 \nodata  & 	 \nodata  & 	0.525	\nl
NGC\,4698 & 	11.9 & 	10.7 & 	0.20 & 	0.27 & 	139.3~$\pm$~10.4 & 	-0.05 & 	1.03 & 	 \nodata  & 	 \nodata  & 	0.008	\nl
NGC\,4736 & 	15.6 & 	14.2 & 	0.11 & 	0.17 & 	107.0~$\pm$~2.3 & 	-0.09 & 	1.15 & 	-0.17 & 	1.199 & 	1.330	\nl
NGC\,4772 & 	26.3 & 	23.5 & 	0.06 & 	0.42 & 	144.5~$\pm$~8.1 & 	-0.12 & 	1.22 & 	 \nodata  & 	 \nodata  & 	0.060	\nl
NGC\,4826 & 	29.5 & 	25.4 & 	0.23 & 	0.42 & 	95.7~$\pm$~6.4 & 	0.01 & 	1.00 & 	-0.04 & 	1.080 & 	0.375	\nl
NGC\,5055 & 	22.0 & 	18.3 & 	0.26 & 	0.39 & 	106.1~$\pm$~8.6 & 	-0.03 & 	1.04 & 	-0.02 & 	1.031 & 	0.471	\nl
NGC\,5248 & 	17.0 & 	15.4 & 	0.23 & 	0.37 & 	78.4~$\pm$~2.5 & 	 \nodata  & 	 \nodata  & 	 \nodata  & 	 \nodata  & 	0.620	\nl
NGC\,5566$^c$ & 	\nodata  & 	 \nodata  & 	 \nodata  & 	 \nodata  & 	 148.9~$\pm$~2.2$^d$  & 	 \nodata  & 	 \nodata  & 	 \nodata  & 	 \nodata  & 	 \nodata 	\nl
NGC\,7177 & 	10.0 & 	8.6 & 	0.17 & 	0.32 & 	115.3~$\pm$~4.8 & 	-0.03 & 	1.02 & 	 \nodata  & 	 \nodata  & 	0.508	\nl
NGC\,7217 & 	13.2 & 	11.2 & 	0.05 & 	0.10 & 	141.1~$\pm$~12.7 & 	0.00 & 	1.02 & 	-0.08 & 	1.100 & 	0.168	\nl
NGC\,7331 & 	29.4 & 	26.0 & 	0.39 & 	0.59 & 	123.6~$\pm$~13.0 & 	-0.11 & 	1.23 & 	 \nodata  & 	 \nodata  & 	1.520	\nl
NGC\,7743 & 	6.2 & 	5.6 & 	0.11 & 	0.31 & 	84.6~$\pm$~2.4 & 	-0.01 & 	1.02 & 	 \nodata  & 	 \nodata  & 	0.069	\nl
\enddata
\tablenotetext{Notes:}{
Structural and kinematic parameters for the galaxies in our sample.
1) Target name. 
2) Radius of equal bulge and disk surface brightness.
3) Adopted bulge radius for this study.
4) Mean apparent bulge ellipticity.
5) Mean apparent disk ellipticity.
6) Central velocity dispersion averaged within one-tenth of the bulge effective radius.
7) Slope of major axis velocity dispersion profile.
8) Major axis ratio of the averaged velocity dispersion within the annulus $r < r_b/3$ to averaged dispersion within $r_b/3 < r < r_b$.
9) Slope of minor axis velocity dispersion profile.
10) Minor axis ratio of velocity dispersion.
11) $<v^2>/<\sigma^2>$ \citep{Binney05} for the inclination corrected velocity.\\
{\it Comments:} 
a) The low inclination of this galaxy prevents us from deriving an inclination corrected velocity
and hence ${<v^2>}/{<\sigma^2>}$. 
b) The bulge is not sufficiently resolved to calculate the slopes of the velocity dispersion.
c) Surface brightness does follow a typical bulge/disk profile. We do not decompose the profile and only present the kinematics data here.
d) No decomposition, this is the innermost value.
}
\end{deluxetable*}
\end{center}

%
\subsection{Extreme moments and multiple kinematic components}
\label{sec:kindecomp}
\begin{figure}
        \begin{center}
        \begin{tabular}{c}
        \includegraphics[width=0.45\textwidth]{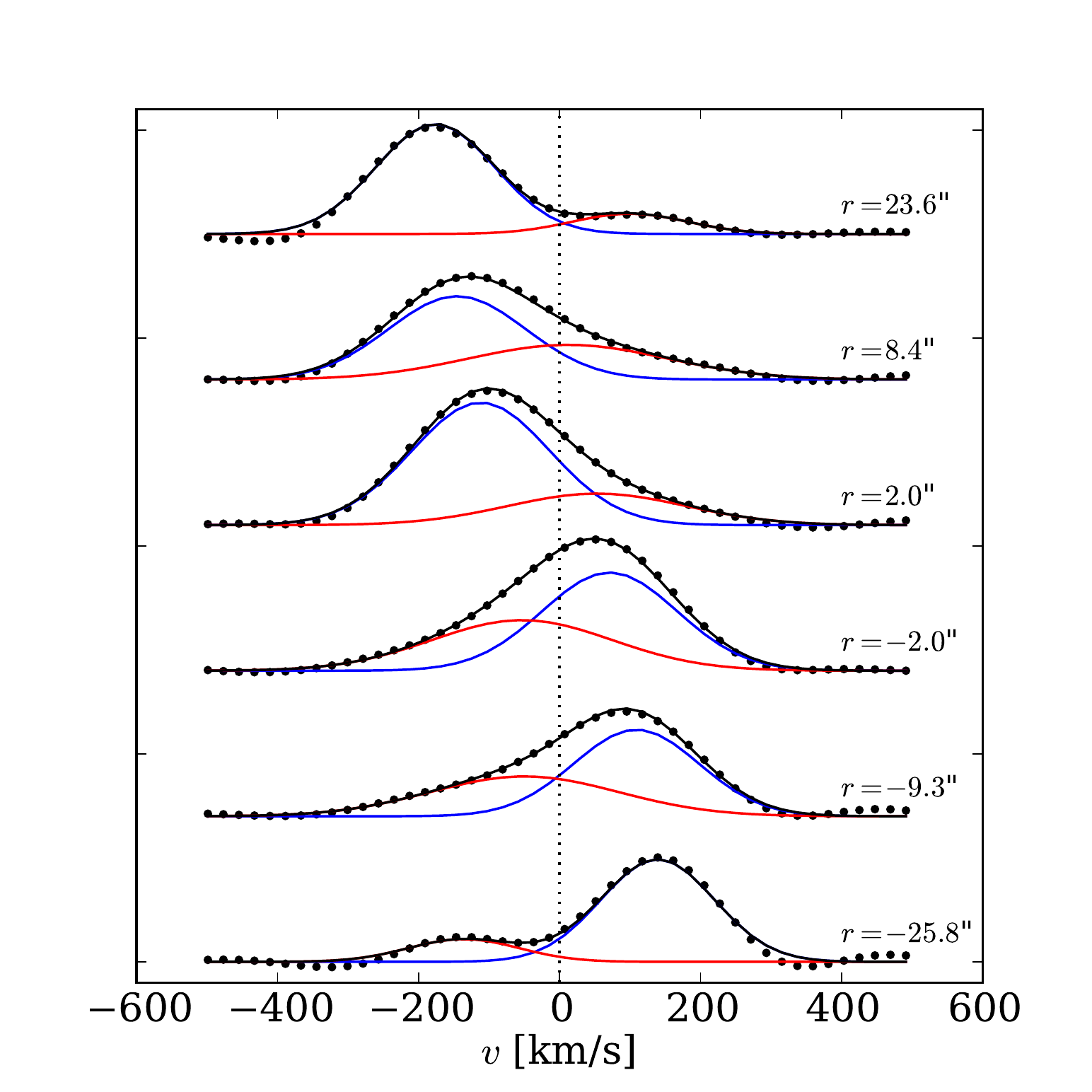}
        \end{tabular}
        \end{center}
	\caption{\label{fig:n3521_losvds}
	Example for the double-Gaussian decompositions for NGC\,3521.  The FCQ
	derived full line of sight velocity distribution for five selected radii is
	plotted in black.  The two-Gaussian kinematic decompositions plotted in red and
	blue. 
	}
\end{figure}

Five galaxies show extreme $h_3$ and $h_4$ moments. The most extreme case,
NGC\,3521, (a classical bulge) exhibits values of $h_3$ and $h_4$ as large as 0.24
and 0.35, respectively (see Fig.\ \ref{fig:kprofiles}).
NGC\,3945, NGC\,4736, NGC\,7217 (all pseudobulges), and NGC\,7331
(unclassified
) show values of
$h_3$ and $h_4$ of up to 0.2. The LOSVD is poorly reproduced by a Gauss-Hermite
expansion at the radii of such extreme higher moment values (see Fig.\ \ref{fig:n3521_losvds}).
The reason lies in the existence of a secondary kinematic component in all those
cases. For NGC\,3521 this has been reported by \citet{Zeilinger2001} who attributed
the counter-rotating stellar component to the presence of a bar.
The two-component nature of NGC\,7217 was discovered before by
\citet{Merrifield1994}. They suggest that the second component is the result of
an extended period of accretion with intermittent change of angular momentum of
the infalling material. \citet{Prada1996} reported a counter-rotating bulge in
NGC\,7331.

Two more systems in our sample, NGC\,2841 \citep{Bertola1999b} and NGC\,3593
\citep{Bertola1996}, were reported to host counter-rotating components, and a
kinematically decoupled component was found in NGC\,4698 \citep{Corsini1999,
Bertola1999, Falcon-Barroso2006} see \citet{Pizzella2004} for a review.
NGC\,3593 is the only galaxy in our sample for which the rotation curve itself
already reveals counter rotation through a twist --- the rotation changes sign
with respect to the systemic velocity at a radius of about 20~arcseconds (see Fig.\
\ref{fig:kprofiles} and Appendix\ \ref{sec:individuals}). The case is similar
whilst not as pronounced for NGC\,4698 where the rotation curve becomes very flat
towards the center (see Fig.\ \ref{fig:kprofiles}).
However, in these cases the secondary component does not result in unusually
strong $h_3$ and $h_4$ moments in our data.

In an attempt of a fairer treatment of their complexity we decomposed the
FCQ-derived LOSVDs into two separate Gaussian components in a similar manner to
\cite{Scorza1995,Zeilinger2001}. We the Metropolis-Hastings algorithm
\citep{Hastings1970, Press2007} to infer the parameters and error bars. Before
the fit, the spectra are binned to a minimum $S/N$ of 75 per pixel. We run four
simultaneous chains for each radial bin. The step width is tuned to achieve a
25\% acceptance ratio and after convergence the first half of the chain is
discarded (clipped).  The run is aborted if the chains do not converge after
100.000 steps.  

\begin{figure*}
        \begin{center}
        \begin{tabular}{cc}
        \includegraphics[width=0.4\textwidth]{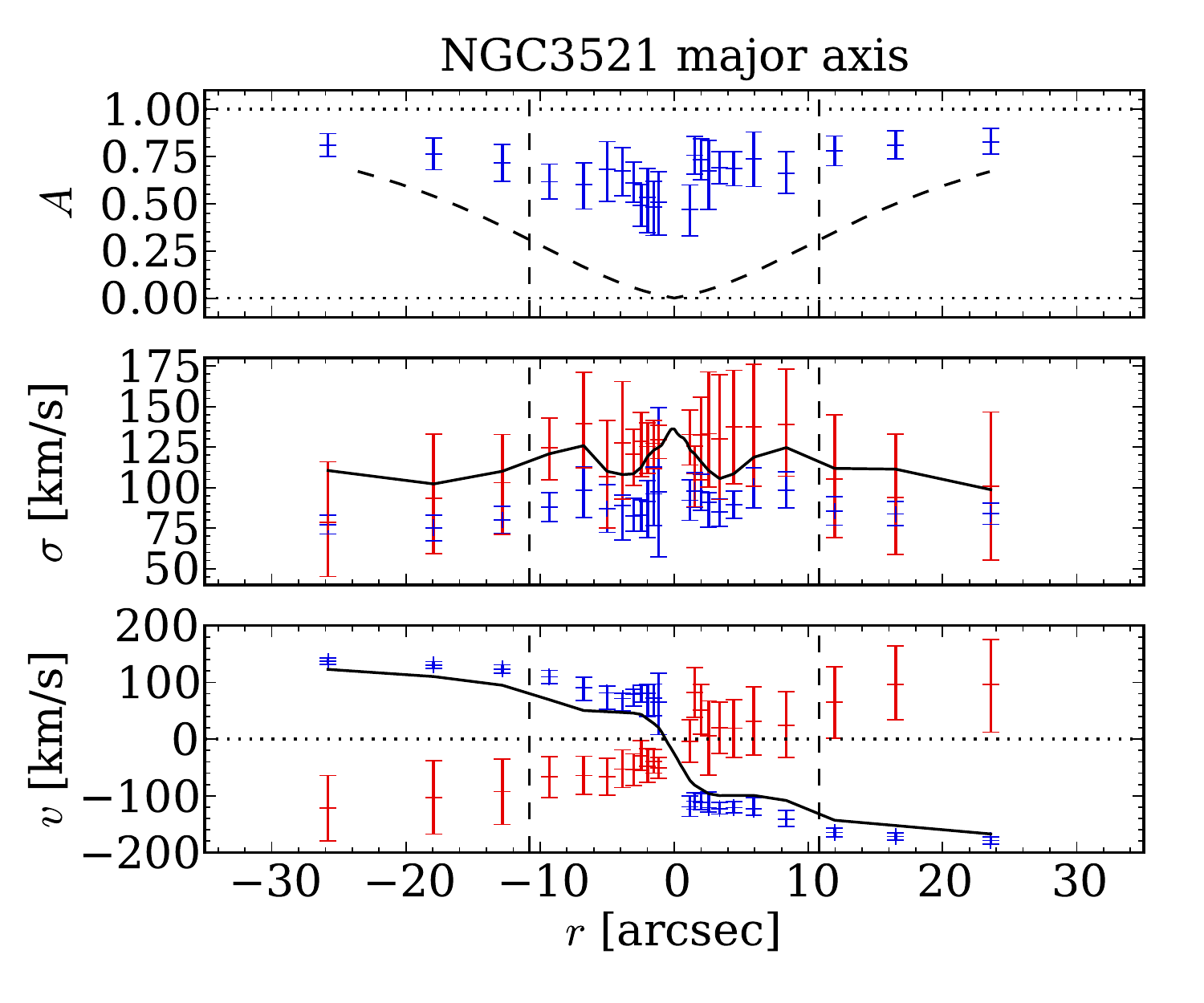}&
        \includegraphics[width=0.4\textwidth]{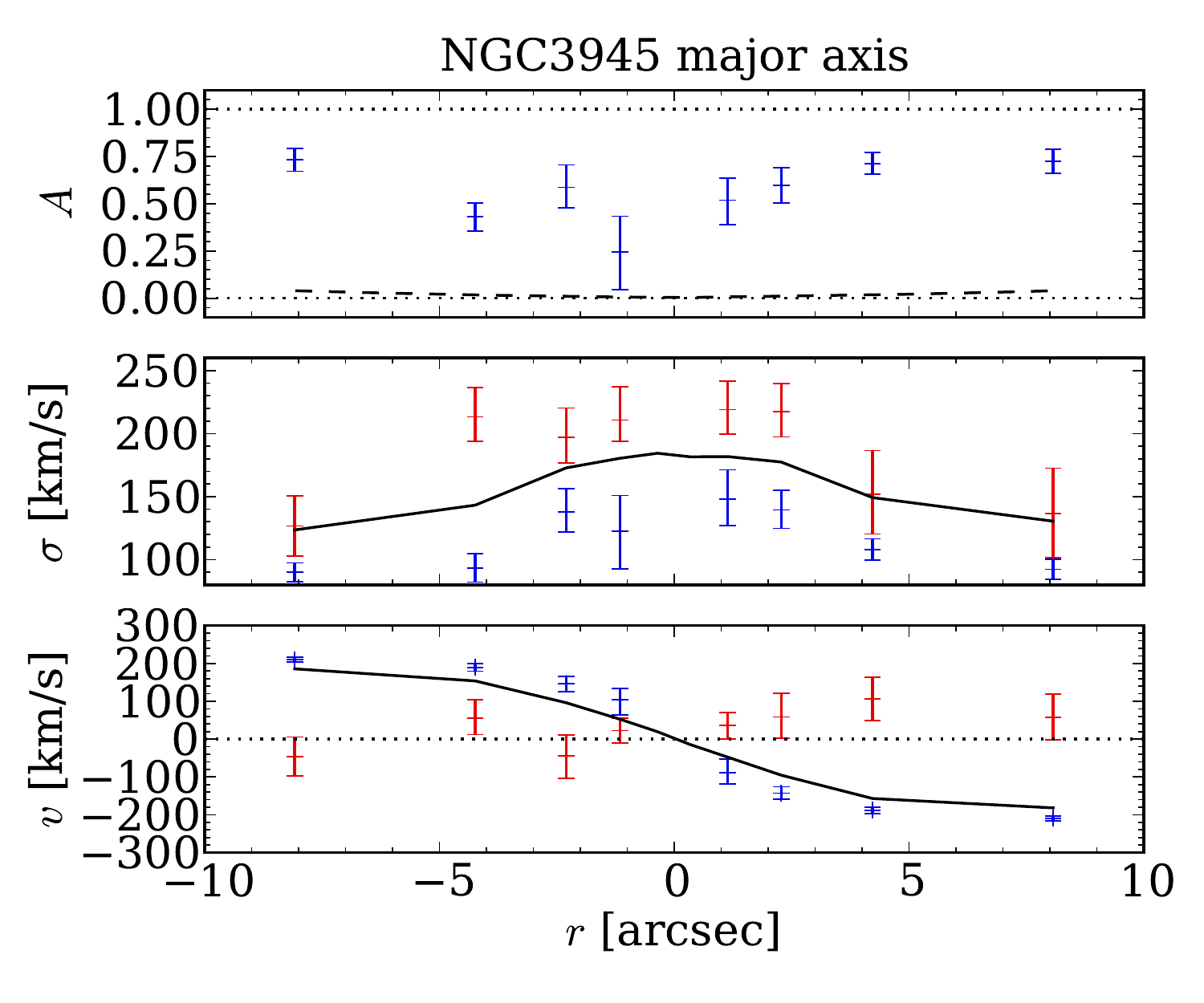}\\
        \includegraphics[width=0.4\textwidth]{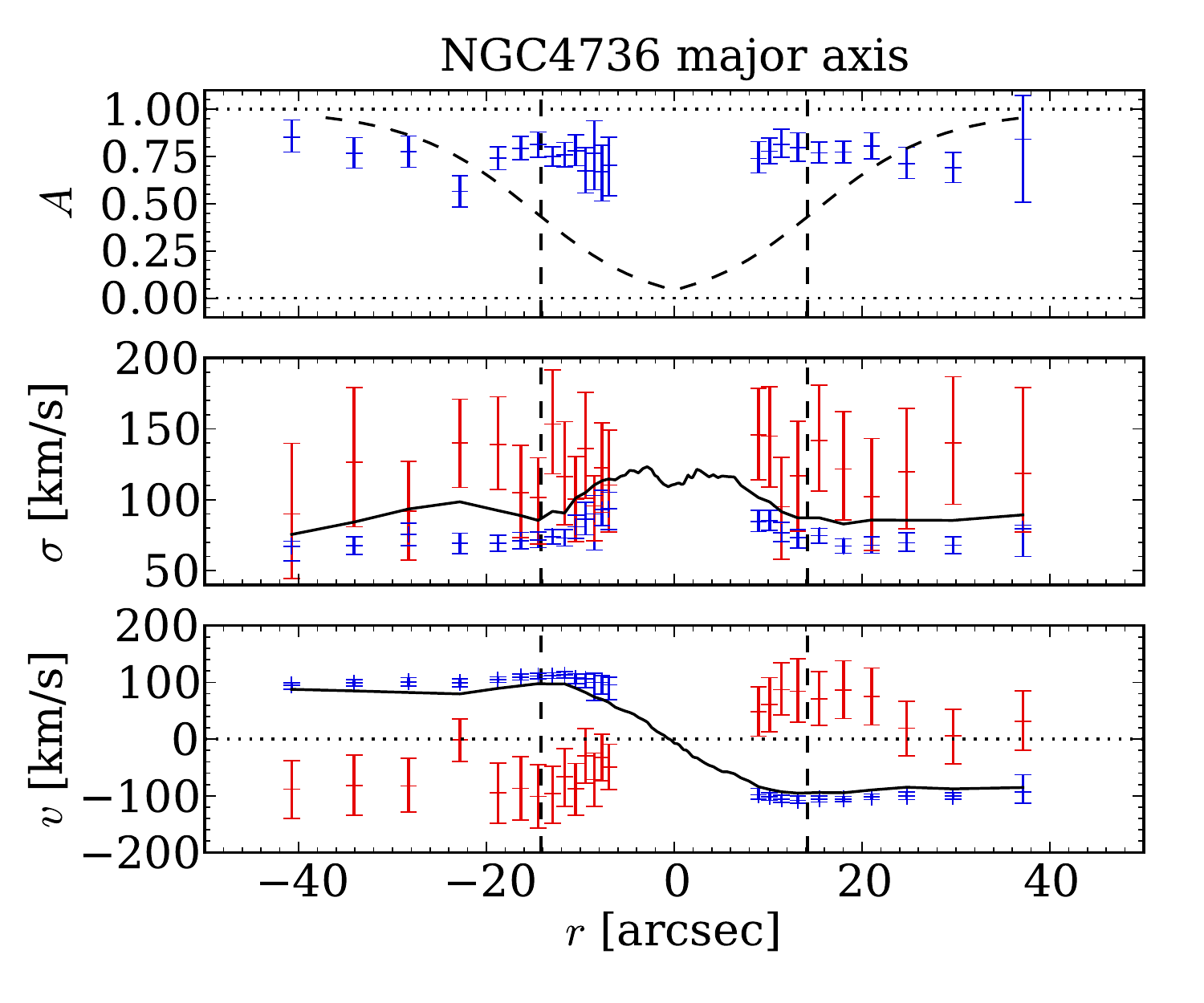}&
        \includegraphics[width=0.4\textwidth]{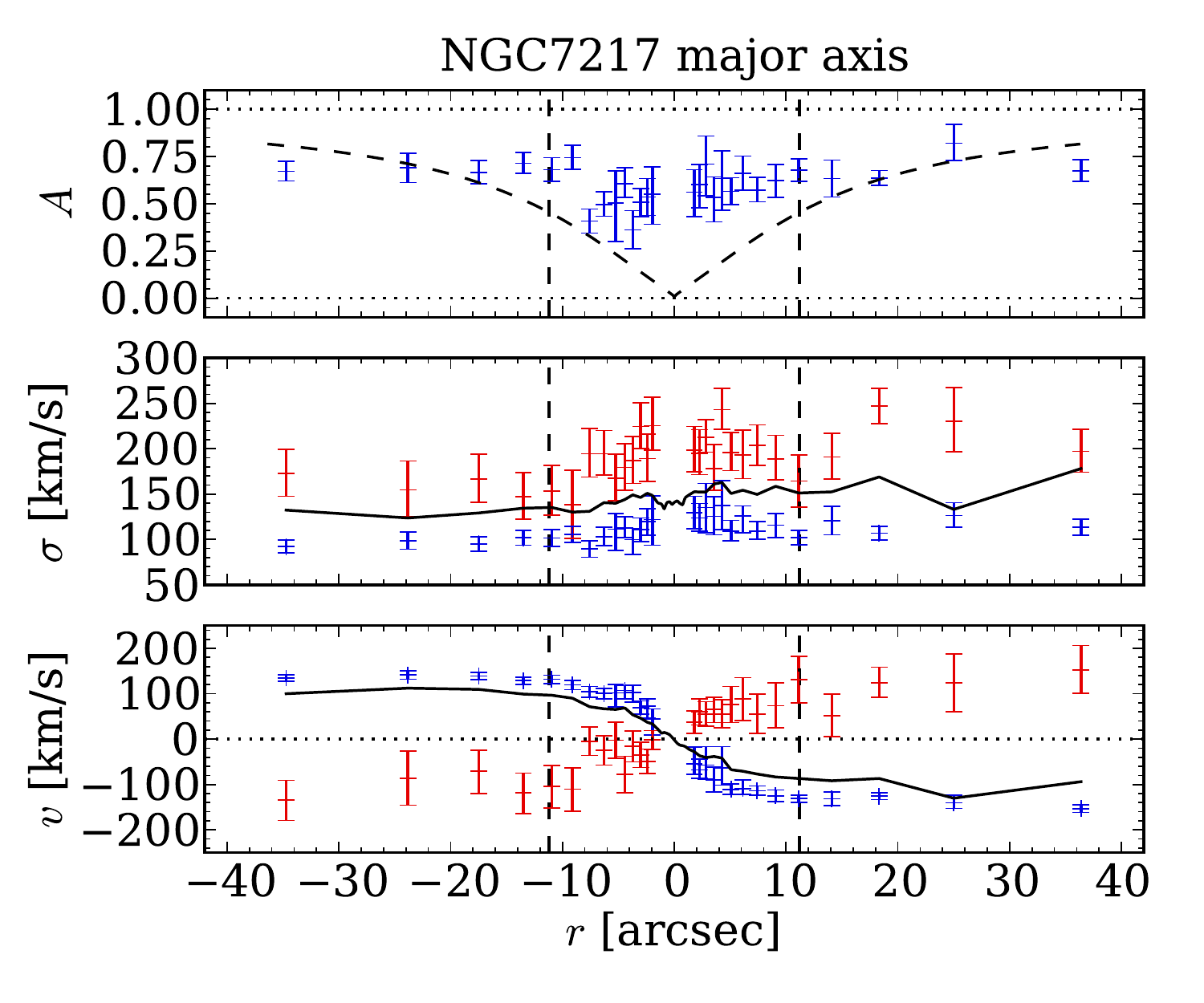}\\
        \includegraphics[width=0.4\textwidth]{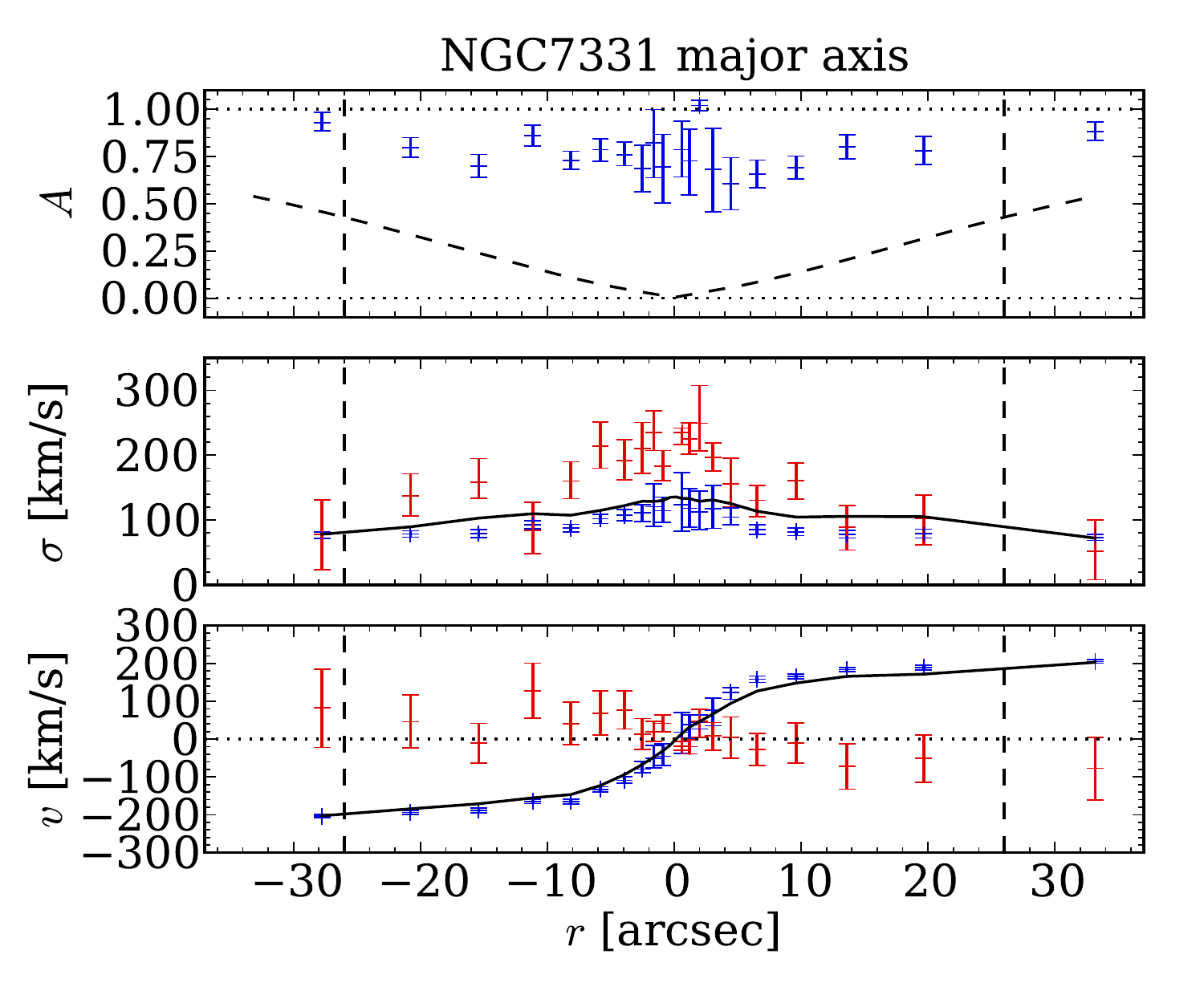}
        \end{tabular}
        \end{center}
	\caption{\label{fig:kinDecomp}
	\scriptsize
	Kinematic decompositions of the major axis data of NGC\,3521, NGC\,3945,
	NGC\,4736, NGC\,7217 and NGC\,7331.  
	In a procedure similar to \cite{Scorza1995,Zeilinger2001} 	
	we fit two Gaussians to the FCQ derived full LOSVDs at
	all radii.  Blue and red error bars show mean velocity and dispersion of the
	two Gaussians where the Metropolis-Hastings algorithm chains achieved
	convergence. The length of the error bars represent the 20\% and 80\%
	quantiles.  Black curves indicate the FCQ fitted moments of a single
	Gauss-Hermite expansion of the LOSVD. Here the error bars are comparable to the
	thickness of the line. The upper panel shows the relative weight of the fast
	component with respect to the total light. 
	The dashed curve shows the disk to total ratio from the photometric
	decomposition. Vertical lines indicate the bulge
	radius. 
	}
\end{figure*}

Fig.~\ref{fig:kinDecomp} shows the result of this decomposition.  The plotted
values are the maximum-likelihood values, and the end of the error bars mark
the 20\% and 80\% quantiles in all four chains after clipping.  Central values
with strong degeneracies between the parameter sets are omitted.  In all five
galaxies we do find significant second components under the assumption that
individual components are purely Gaussian.  In Tab.~\ref{tab:kindecomp} we list
the integrated fractions of light in the two different kinematic components and
compare those to the values that one would expect from the photometric
decomposition. 

\begin{center}
\begin{deluxetable}{clcc}
\tablecaption{Light fractions in kinematic subcomponent. \label{tab:kindecomp}}
\tablewidth{0pt}
\tablehead{\colhead{galaxy} & \colhead{morph.}  & \colhead{disk light} &\colhead{light in fast component}\\ 
\colhead{} &\colhead{} &\colhead{[\%]} &\colhead{[\%]}\\ 
\colhead{(1)} &\colhead{(2)} &\colhead{(3)} &\colhead{(4)} 
}
\startdata
NGC\,3521 &	SAB(rs)bc	&	34	&	44 \\
NGC\,3945 &	(R)SB$^+$(rs)	&	 2	&	34 \\
NGC\,4736 &	(R)SA(r)ab	&	55	&	50 \\
NGC\,7217 &	(R)SA(r)ab	&	55	&	38 \\
NGC\,7331 &	SA(s)b		&	26	&	47 
\enddata
\tablenotetext{notes:}{
Comparison of the amount of light in the lower dispersion component to the light
that one would expect from the photometric decomposition. We integrate the
amplitudes of the Gaussian fits to the two components over all radii shown in
Fig.~\ref{fig:kinDecomp}.  
1) Galaxy. 
2) Morphology (RC3). 
3) Integrated light in the disk component from the extrapolation of the outer disk exponential profile. 
4) Integrated light in the lower dispersion component. 
}
\end{deluxetable}
\end{center}
%
%
\section{Discussion}
\label{sec:discussion}
\subsection{Dichotomous dispersion profiles of classical and pseudobulges}
It is commonly assumed that the bulge light --- typically determined from a
bulge-disk decomposition --- represents a dynamically hot component.  Yet, it
has been known for a long time that rotation-supported bulges exist
\citep{Kormendy1982b}. Also, many bulges have lower central velocity
dispersions than expected from the \cite{Faber1976} relation \citep{KK04}.
\citet{Falcon-Barroso2006,Ganda2006} find that several galaxies in the SAURON
survey have centers that are dynamically colder than the surrounding disks.
Thus, it is now clear that not all galaxies fit the picture that a
bulge is a dynamically hot component.

There is an observed dichotomy in bulge properties including S\'ersic index,
bulge morphology, star formation \& ISM properties, and optical color
\citep{Carollo1997, Gadotti2001, KK04, Fisher2006, Fisher2008, Fisher2010}.
Recently, \citet{Fisher2010} showed that the dichotomous properties in S\'ersic
index, morphology, and ISM properties are consistent. Furthermore they show that
bulges of different type occupy different regions in the projection of fundamental
plane properties, thus indicating that there are very likely two physically
distinct classes of bulges. Are dynamics part of this dichotomy?

The high spectral resolution of 39~kms$^{-1}$ of our data, enables us
to recover dispersions out into the disk regions in many of our
targets --- a feature uncommon to many similar surveys.  We extract
LOSVDs using the FCQ algorithm with additional procedures to account
for nebular emission and template mismatch. We recover $v$, $\sigma$,
$h_3$, and $h_4$ moments of a Gauss-Hermite model of the LOSVDs as
function of radius.

We observe a great variety of shapes of kinematic profiles
(see Fig.\ \ref{fig:kprofiles} and 
their detailed description in Appendix\ \ref{sec:individuals}).
Similarly to \citet{Falcon-Barroso2006} and \citet{Ganda2006}, we find that it
is not necessarily true that the center of a bulge has the highest observed
velocity dispersion (e.g.\ NGC\,3593). In our sample only $\sim$1/3 of the
galaxies have centrally peaked velocity dispersion profiles (like NGC\,3898).
Many galaxies have roughly flat velocity dispersion profiles. In these
galaxies there is no apparent transition in velocity dispersion from the bulge
to the disk region unlike in the stellar surface brightness profile
(e.g.\ NGC\,4448 and NGC\,5055).

It is interesting to note that to the radial extent of our data, in many of these
cases the disk velocity dispersion is as high as the central velocity
dispersion of the galaxy.  For example in NGC\,4448 the disk velocity dispersion
remains above 100~kms$^{-1}$.  Therefore some disks of spiral galaxies are not
necessarily cold stellar systems over the radii that we cover in this study.

From minor axis data of 19 S0 to Sbc bulges \citep{Falcon-Barroso2003} showed
that higher ellipticity bulges have shallower velocity dispersion profiles
profiles. If pseudobulges appear photometrically flattened then one might
expect that the steepness of the dispersion profile should correlate with bulge
type.
We show in Fig.~\ref{fig:overplots} that the shape of the velocity dispersion
profile correlates very well with bulge type. Galaxies with classical bulges
have centrally peaked profiles.  Galaxies with pseudobulges have, on average,
flat dispersion profiles.  We have attempted to quantify this using the
logarithmic derivative of the velocity dispersion as function of radius
(Eq.~\ref{eq:gamma}) and also the ratio of dispersions at different radii
(Eq.~\ref{eq:delta}). We find that pseudobulges and classical bulges occupy
different regions in the parameter space of logarithmic derivative of velocity
dispersion and S\'ersic index (see Fig.~\ref{fig:struct_vs_kin}) in a way that
is not inconsistent with models of dynamically isotropic systems
\citep{Ciotti1991}.

It is important to note that the dynamics of a few galaxies are not well
described by a simple monotonic trend of velocity dispersion with radius; we
stress that for the purpose of this paper we are interested in the bulk
properties of the distribution of stellar dynamics.  The great variety of
shapes in dispersion profiles that we observe
(for a detailed description see Appendix\ \ref{sec:individuals})
is likely to be a consequence of
the fact that there are multiple ways to heat galactic disks, for example
through mergers
\citep{van-Albada1982, Quinn1993, Eliche-Moral2006, Hopkins2008},
bars \citep{Saha2010}, and other disk
instabilities \citep{Sellwood1993, Combes1990}.  Even under the strong
assumption that classical and pseudobulges are dynamically distinct, it does not
seem plausible that any simple description of the kinematic profile cleanly
separates classical from pseudobulges.

A few classical bulges
in barred galaxies
such as NGC\,3992 do not seem to fit this general picture.  However, all those
galaxies are barred and it is conceivable that bars may distort the kinematic
profile of a classical bulge as they vertically heat the disk they reside in
\citep{Gadotti05, Saha2010}.  Central velocity dispersions lie higher by a
factor of two than at the bulge radius in the most extreme cases in our sample,
e.g., in NGC\,1023, NGC\,3898, and NGC\,4203.  If bars raise velocity
dispersions by a factor of up to four as suggested by \citet{Saha2010}, the
signature of a central dispersion peak can of course be easily washed out.

NGC\,4826 has extreme amounts of dust in its central region --- hence also its
name {\it black eye} or {\it evil eye} galaxy.  The bulge was consequently
classified as pseudobulge by \citet{Fisher2008}.  It stands out however, as it
has a relatively high S\'ersic index of $3.9 \pm 0.88$ for a pseudobulge.  The
V-band value of $n = 3.94\pm0.68$ \citep{Fisher2008} agrees well. The bulge
radius of 25.4~arcseconds seems small once the kinematic data are taken into
account. The velocity dispersion starts rising at about 50~arcseconds already,
which corresponds to the radius of the final flattening of the rotation curve.
It is conceivable that the large amount of dust in its center, which is
easily also visible in the infrared, may affect the decomposition. If one was
to take the value of 50~arcseconds as the bulge radius then the $\gamma$ value
would become -0.33 (0.01 before) and the sigma ratio would take a value of
1.29 (1.08 before).  This would move NGC\,4826 significantly further to the
right in both plots into the region occupied by classical bulges, in much
better agreement with the S\'ersic index. Further the disk of NGC\,4826 is
relatively free of dust and actually resembles an S0.  We hypothesize that
the unusual morphology is a result of a recent merging event. A satellite may
have fallen into an S0-like disk and brought in dust and triggered star formation.
This hypothesis is supported by the existence of two counter rotating gaseous
disks observed by \citet{Braun1992, Braun1994}. We labelled NGC\,4826
as non-classified throughout this work.
NGC\,3593 has a very large value of $\gamma$ and a very small sigma ratio, i.e.
it falls far to the left in both diagrams.  This is a result of the strong
depression in velocity dispersion in the bulge region. NGC\,3593 is the only
galaxy in the sample where we observe actual counter rotation in the sense of a
change of sign of the mean rotational velocity in the bulge region. The small
S\'ersic index of 0.81 supports the picture that the bulge region is dominated
by a kinematically cold and distinct but luminous disk \citep{Bertola1996}.  
NGC\,2681 is classified as a pseudobulge by morphology and yet has a relatively
large S\'ersic index of $n = 3.8$. Further, it has a centrally peaked dispersion
profile with $\gamma = -0.16$ and $\delta = 1.19$. This agreement of photometric
structure and dispersion slope prompts us to reassess the morphological
classification.  While the disk shows relatively little amounts of dust, a high
contrast dust spiral within in the bulge easily seen in HST~F555W, offers
a clear sense of rotation. Also the spiral is not obviously misaligned with
the outer disk. This galaxy may represent a prototypical case for the
breakdown of the morphological classification scheme. However, it
shows multiple bars --- possibly three \citep{Erwin99} --- and hence the central
heating may also be a consequence of its complicated dynamical structure.
NGC\,3521 has a seemingly a relatively flat dispersion slope with values 
of $\gamma = -0.01$ and $\delta = 0.99$. We discuss this object at the end of
the next Section.

\subsection{Rotational support}
In order to study the level of rotational support of a stellar system, it has
become common practice to study its location in the $v_{max}/<\sigma>$
vs.~$\epsilon$ diagram \citep{Illingworth77,Binney87, Kormendy93} --- where
$v_{max}$ measures the maximum rotational velocity and $<\sigma>$ the averaged
velocity dispersion within a certain radius, and $\epsilon$ the system's
ellipticity.  One can also directly compare \mbox{$v_{max}/<\sigma>$} to the
expected values of an oblate-spheroidal system with isotropic velocity
dispersion.  For instance, \citet{Kormendy1982a} define the anisotropy
parameter \mbox{$(v/\sigma)^*=(v_{max}/<\sigma>)/\sqrt{\epsilon/(1-\epsilon)}$}
as a measure for the rotational support of a stellar system. Values of
$(v/\sigma)^* \approx 1$ point towards a support by rotation whereas values $<
1$ indicate support by anisotropy. Those measures involve the ellipticity of
the system which is typically subject to relatively large uncertainties,
especially when measured for galaxies which are dominated by large quantities
of dust.  Here, we decide to rather examine the local $({v_{corr}}/\sigma)(r)$,
i.e.\ as a function of radius, and the averaged values of
$<{v_{corr}}^2>/<\sigma^2>$ \citep{Binney05} across the bulge region, $v_{corr}
= v_{obs}/\sin(i)$ is the inclination corrected velocity at a given radius. We
use inclinations from Hyperleda (see Tab.~\ref{tab:sample}). We apply no
further correction to the velocity dispersion.

\begin{figure}
        \begin{center}
        \begin{tabular}{cc}
                \includegraphics[width=0.4\textwidth]{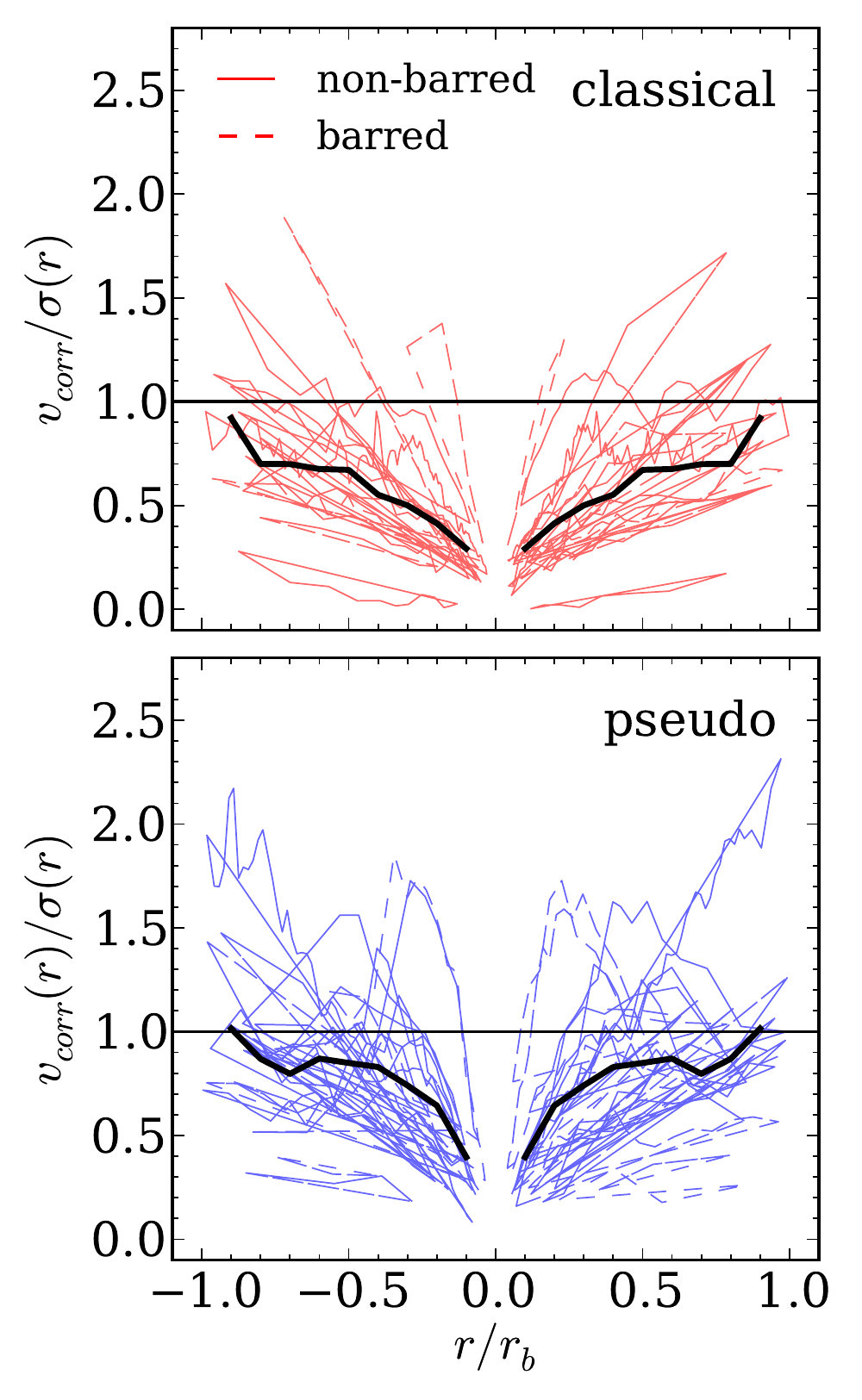} 
        \end{tabular}
        \end{center}
	\caption{\label{fig:vsig}
		Local $(v_{corr}/\sigma)(r)$ along the major axis, 
		the radii are normalized by bulge radius.
		The velocities are corrected for inclination through $1/sin(i)$. 
		Classical bulges are plotted in the upper panel, pseudobulges
		in the lower panel. The solid black line marks the median
		of all profiles.} 
\end{figure}

\begin{figure}
        \begin{center}
        \begin{tabular}{cc}
                \includegraphics[width=0.4\textwidth]{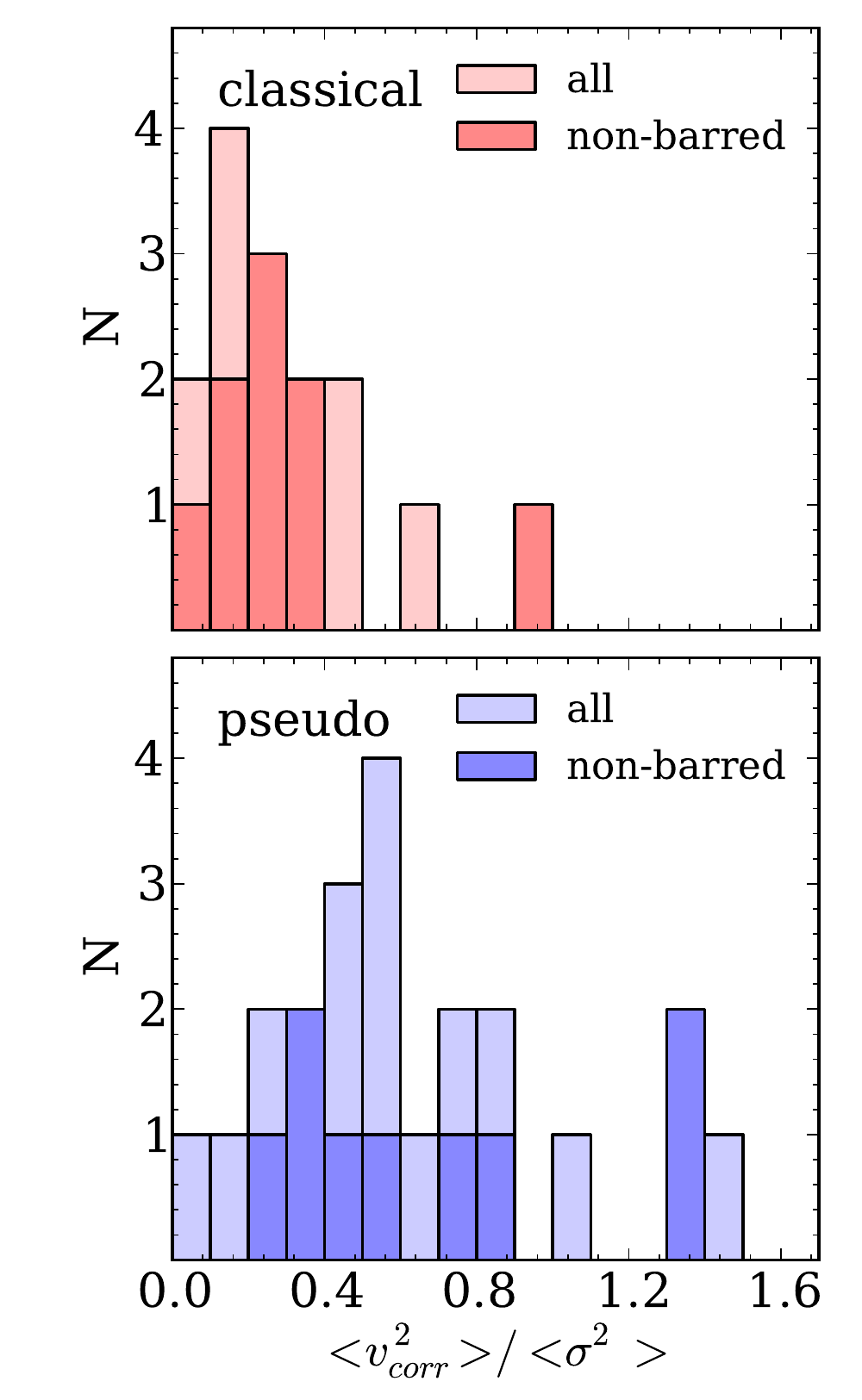} 
        \end{tabular}
        \end{center}
	\caption{\label{fig:vsig_hist}
		Histograms for the averaged values of $<v^2>/<\sigma^2>$ \citep{Binney05}.
		The velocities are corrected for inclination through $1/sin(i)$. 
		Classical bulges are plotted in the upper panels, pseudobulges
		in the lower panels.} 
\end{figure}

In Fig.~\ref{fig:vsig} we plot $(v_{corr}/\sigma)(r)$ separately for classical
and pseudobulges. We further plot histograms of the bulge-averaged quantities
in Fig.~\ref{fig:vsig_hist}.  Again we normalize the radii by the bulge radius
and exclude the central seeing FWHM from the analysis. While there is
significant overlap between the two sub samples, pseudobulges are biased
towards larger $(v_{corr}/\sigma)(r)$. This is especially seen in the
histograms for the averaged values. 
A Kolmogorov-Smirnov test \citep{Smirnov39,Press02} yields a
probability of 0.8\% (1.9\%) for the classical and the pseudobulges in
the full (non-barred) sample to stem from the same distribution. A
Student's two-tailed t-test for two independent samples yields a
probability of 0.7\% (2.5\%) for the classical and the pseudobulges in
the full (non-barred) sample.
This result supports a picture of
increased rotational support of pseudobulges that was originally described by
\citet{Kormendy93} and discussed in detail in \citet{KK04}, see also
\citet{Kormendy2008}.

\begin{figure*}
        \begin{center}
        \begin{tabular}{c}
        \includegraphics[width=0.8\textwidth]{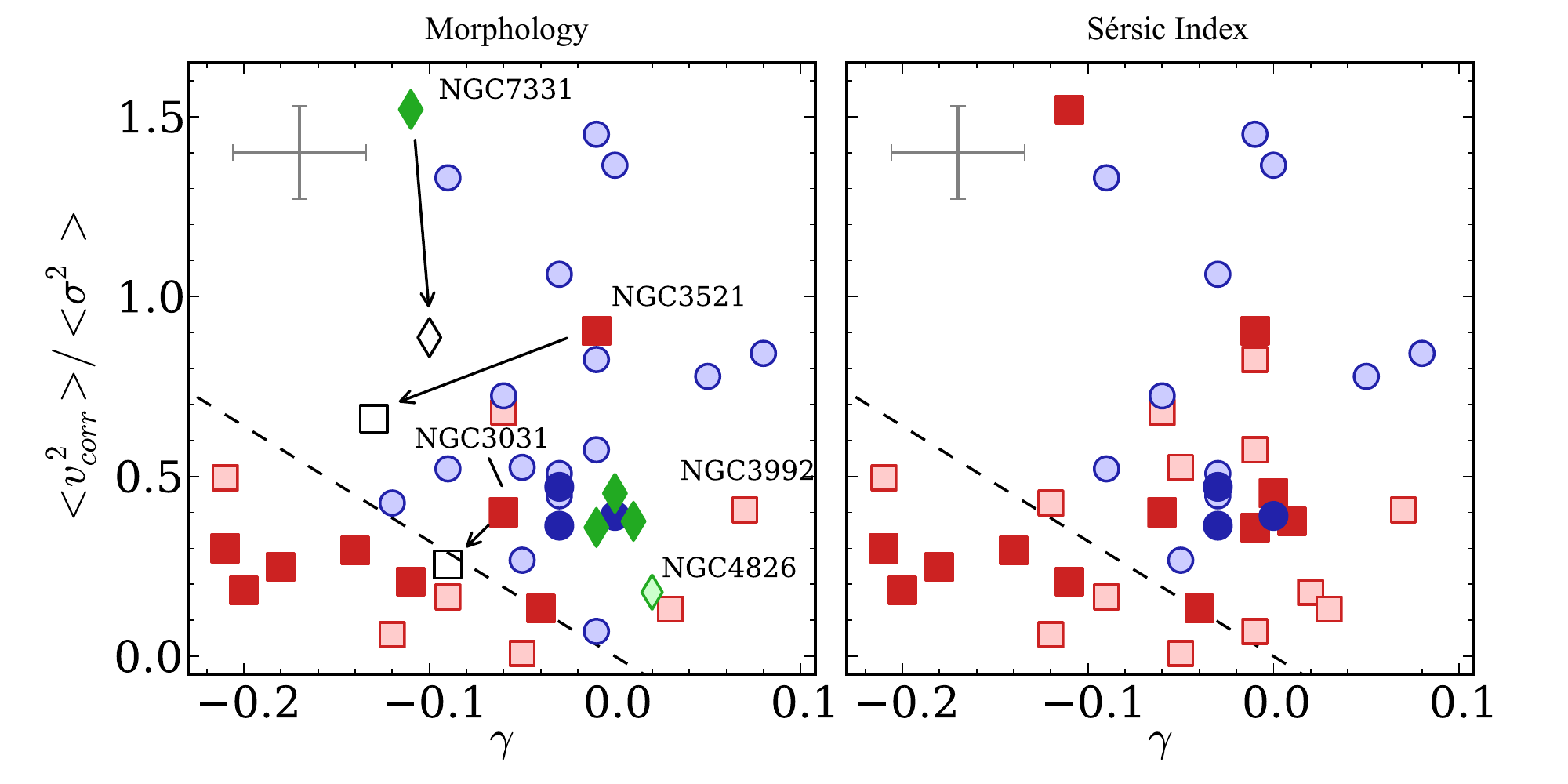}
        \end{tabular}
        \end{center}
	\caption{\label{fig:vsig_vs_gamma}
	The bulge averaged quantity $<v^2>/<\sigma^2>$ as function of the slope of
	the velocity dispersion, $\gamma$.  In the left panel we discriminate bulge
	types based on morphology.  Pseudobulges are colored blue, classical bulges red,
	and bulges that we do not classify are colored green.  Light-shaded symbols represent
	barred galaxies or galaxies hosting an oval.  In the right panel we
	discriminate by S\'ersic index (blue $n\leq2.1$, red otherwise).  For
	NGC\,3521, NGC\,3031, and NGC\,7331 we remeasure $\gamma$ changing the outer
	radius cut to isolate only the central dispersion peak as described in the
	text.  The corresponding new locations are marked as black open symbols.  The
	dashed line is drawn to contain all the pseudobulges. 
	}
\end{figure*}

While the average values of $\gamma$ and $<v^2>/<\sigma^2>$ are different for
classical and pseudobulges, neither of the two quantities separates the bulge
types. In Fig.~\ref{fig:vsig_vs_gamma} we combine both and plot
$<v^2>/<\sigma^2>$ against the logarithmic slope of the velocity dispersion,
$\gamma$, for the bulges in our sample. In the left panel we discriminate
bulges morphologically and in the right based on the bulge S\'ersic index. The
dashed line is drawn to contain all the pseudobulges. Classical bulges both
tend to have low $<v^2>/<\sigma^2>$ and steeper negative slopes than
pseudobulges. 
There are several galaxies with large S\'ersic index that fall dynamically into
the region of pseudobulges. We have discussed the possibility that bars mask or destroy
the kinematic signature of a classical bulge. However two non-barred high-S\'ersic index
galaxies remain (NGC\,3521 and NGC\,7331; NGC\,3521 also has classical
bulge morphology). Both galaxies, have centrally peaking velocity
dispersion profiles but they also show a rise in velocity dispersion
at larger radius (see Appendix~\ref{sec:kprofiles}), yet still within the
radius of the bulge. This is also a feature that is prominently seen in the
dispersion profile of NGC\,3031, but in this case the effect on the position in
the $<v^2>/<\sigma^2>$~vs.~$\gamma$ plane is not as dramatic.  This behaviour
is not what we observe in pseudobulges, which have flat dispersion profiles.
None of these three galaxies violate the general dichotomy observed in
Fig.~\ref{fig:overplots}. Therefore it is likely that their outlier location is
due to a failure in the machinery of measuring dynamical quantities. NGC\,7331
and NGC\,3521 both show signs of counter rotating components and NGC\,3031 is well
known to be interacting. It is possible in each of these galaxies that an
outside mechanism is superimposing extra kinematic structure that is visible
as an outside rise in the velocity dispersion profile of the bulge.  We
remeasure $\gamma$ changing the outer radius cut isolating only the central
dispersion peak. In the case of NGC\,3031 this moves the measured dynamical
quantities into the region of parameter space that is only occupied by
classical bulges; NGC\,3521 and NGC\,7331 both move significantly closer.

Fig.~\ref{fig:vsig_vs_gamma} illustrates agreement between
kinematic diagnostics of the bulge dichotomy with structural and morphological
indicators of bulge types.  

\subsection{Multiple kinematic components}
We observe counter rotation seen as secondary components in the full shape
of the LOSVD in five systems
(NGC\,3521, NGC\,3945, NGC\,4736, NGC\,7217, and NGC\,7331; see \S\ref{sec:kindecomp}).
It is striking how clearly the LOSVDs can be decomposed into a low-dispersion and a
high-dispersion component in all these cases. It is tempting to
interpret the latter as the bulge and the former as the disk.
However, if we plot the local disk-to-total ratios as obtained from the
photometric decomposition over the values obtained from the kinematic
decomposition (upper panels of Fig.~\ref{fig:kinDecomp}), then we
see that the disk contribution from the photometry falls short in all cases.
Therefore, the observed low-dispersion component within the bulge region is not
simply the extension of the outer disk as more light contributes to this
component as one would expect from the extrapolation of the outer disk
exponential profile alone.
It is important to point out that a Gauss-Hermite distribution with
moderate $h_3$ moments can be modelled rather well by two individual Gaussians.
$h_3$ moments, though, occur naturally in disks \citep{Binney87}
and are not a signature of an actual second component. Only in cases where the
second component is clearly seen as a second peak in the LOSVD it is safe to
assume that actually two distinct components contribute.

The observation of a counter rotating component is very interesting in the
light of the findings of \cite{Eliche-Moral2011}. They use collisionless
$N$-body simulations to study the characteristics of inner components such as
inner disks and inner rings that were formed though a minor merger.  For this,
they simulate a number of mergers with different mass ratios and orbits. In
general, while all their mergers formed an inner component supported by
rotation, none of their mergers produced a significant bar. In their simulations
all mergers with satellites on retrograde orbits do form a counter rotating
component while none of them leads to actual counter rotation in the sense of a
change of sign of the mean rotational velocity. A central increase of velocity
dispersion and strong $h_3$ and $h_4$ moments are observed in a majority of
their models. They further find, however, that $v/\sigma$ and $h_3$ are generally
anti-correlated within the inner components throughout all of their
simulations. We see all these features reflected in the aforementioned
galaxies. Of particular interest to us is the double peak in $h_4$ that their
model {\it M6P1Rb} produced. Such double peaks are very pronounced in NGC\,3945
and NGC\,4736, but are also visible in the case of NGC\,3521 and NGC\,7331. In our small
sample 3 out of the 5 systems for which counter rotation is observed are not
barred, we think that a merging event is a likely formation scenario.
%
\section{Summary}
\label{sec:summary}
In this paper we present kinematic profiles for the major axis of 45
intermediate type (S0-Scd) galaxies. Our survey differs from other similar surveys in
that we are able to resolve lower velocity dispersions which allows us to study
kinematic features in cold systems like disks and pseudobulges. We combine
these data with bulge-to-disk decompositions of the stellar light.

We find that bulges that have increased rotational support, as measured by larger
values of $<v^2>/<\sigma^2>$, are likely to have lower S\'ersic indices and show
disk-like morphology.

Classical bulges on average tend to have higher central velocity dispersions
than pseudobulges. In our sample the lowest central velocity dispersion in a
galaxy with evidence for a classical bulge through a S\'ersic index of 3.7 is
$\sigma_{r_e/10} = 85 \pm 2$~kms$^{-1}$ (NGC\,7743).  

We observe --- for the first time --- a systematic agreement between the
shape of the velocity dispersion profile and the bulge type as indicated by the
S\'ersic index. Classical bulges have centrally peaked velocity dispersion
profiles while pseudobulges in general have flat dispersion profiles and even at
times show drops in the central velocity dispersion. We confirm that this
correlation holds true if visual morphology is used for the bulge
classification instead of the S\'ersic index, as it is expected from the
good correlation between bulge morphology and S\'ersic index \citep{Fisher2010}.

We observe that the disk regions of some of our galaxies have not always a low
velocity dispersion. In some galaxies the velocity dispersion remains above
$100$~kms$^{-1}$ well into the region where the disk dominates the light.
 
We confirm the previously described multicomponent nature of the full LOSVD in
NGC\,3521 \citep{Zeilinger2001}, NGC\,7217 \citep{Merrifield1994} and NGC\,7331
\citep{Prada1996} and find two additional systems --- namely NGC\,3945 and
NGC\,4736 -- with signatures of multiple kinematic components.  They become
apparent through a secondary peak or pronounced shoulder in the full line of
sight velocity distributions (LOSVD).  We present double-Gaussian
decompositions which show the presence of a counter rotating stellar component
in all these systems.

As in elliptical galaxies \citep{Bender1994}, we find a correlation of $h_3$
and $v/\sigma$, both locally as well as in the bulge-averaged quantities.  We
observe no correlation of the higher moments with bulge luminosity, however we
find a weak correlation between the average values of $h_4$ and $v/\sigma$.

Through examination of the figures in Appendix~\ref{sec:kprofiles} it is clear
that the kinematic profiles of bulge-disk galaxies commonly contain
substructure.  Furthermore non-axisymmetric features in the stellar structure
such a bars make understanding the kinematics of these galaxies more difficult.
Future progress 
will require 2D-methods capable of resolving low velocity dispersions commonly
found in pseudobulges. We are currently executing such a survey using the
VIRUS-W spectrograph \citep{Fabricius2008}.
%
\acknowledgments
{\small
We would like to thank Luca Ciotti for providing velocity dispersion profiles
for the isotropic models presented in his 1991 paper. 
We wish to thank Jesus Falcon Barroso of the SAURON collaboration who 
made kinematic maps of a number of the galaxies available to us in various
formats which allowed for a detailed comparison of our data.  
We also wish to acknowledge the help of Gaelle Dumas who provided SAURON
data from her 2007 paper.
We thank Peter Erwin for many fruitful discussions and his valuable comments.
We would also like to express our gratefulness to the efforts of the observing
staff at the Hobby-Eberly Telescope (HET). Over the years of the duration of
this survey they have constantly provided high quality data, give very valuably
background information and have always been very helpful and approachable
concerning technical aspects of the data. 
The Hobby-Eberly Telescope is a joint project of the University of Texas
at Austin, the Pennsylvania State University, Stanford University,
Ludwig-Maximilians-Universitaet Muenchen, and Georg-August-Universitaet
Goettingen.  The HET is named in honor of its principal benefactors, William P.
Hobby and Robert E. Eberly.
The Marcario Low Resolution Spectrograph is named for Mike Marcario of High
Lonesome Optics who fabricated several optics for the instrument but died
before its completion. The LRS is a joint project of the Hobby-Eberly Telescope
partnership and the Instituto de Astronomia de la Universidad Nacional Autonoma
de Mexico.
The grism E2 used for these observations has been bought through the DFG grant
BE1091/9-1.
This work was supported by the SFB-Transregio 33 The Dark Universe by the
Deutsche Forschungsgemeinschaft (DFG).
This research has made use of the NASA/IPAC Extragalactic Database (NED) which
is operated by the Jet Propulsion Laboratory, California Institute of
Technology, under contract with the National Aeronautics and Space
Administration.
Some/all of the data presented in this paper were obtained from the
Multimission Archive at the Space Telescope Science Institute (MAST).  STScI is
operated by the Association of Universities for Research in Astronomy, Inc.,
under NASA contract NAS5-26555. Support for MAST for non-HST data is provided
by the NASA Office of Space Science via grant NAG5-7584 and by other grants and
contracts.
We acknowledge the usage of the HyperLeda database (http://leda.univ-lyon1.fr)
\citep{Paturel2003}.
This publication makes use of data products from the Two Micron All Sky Survey,
which is a joint project of the University of Massachusetts and the Infrared
Processing and Analysis Center/California Institute of Technology, funded by
the National Aeronautics and Space Administration and the National Science
Foundation.
This work is based [in part] on observations made with the Spitzer Space
Telescope, which is operated by the Jet Propulsion Laboratory, California
Institute of Technology under a contract with NASA.
Finally, we thank the anonymous referee for the careful reading and the
provided input that helped to improve this publication significantly.
}
\bibliographystyle{apj}

%
\newpage
\appendix
\section{Kinematic profiles} 
\label{sec:kprofiles}
\setcounter{figure}{15}
\begin{figure}[h]
        \begin{center}
        \begin{tabular}{lll}
	\begin{minipage}[b]{0.185\textwidth}
	\includegraphics[viewport=0 55 390 400,width=\textwidth]{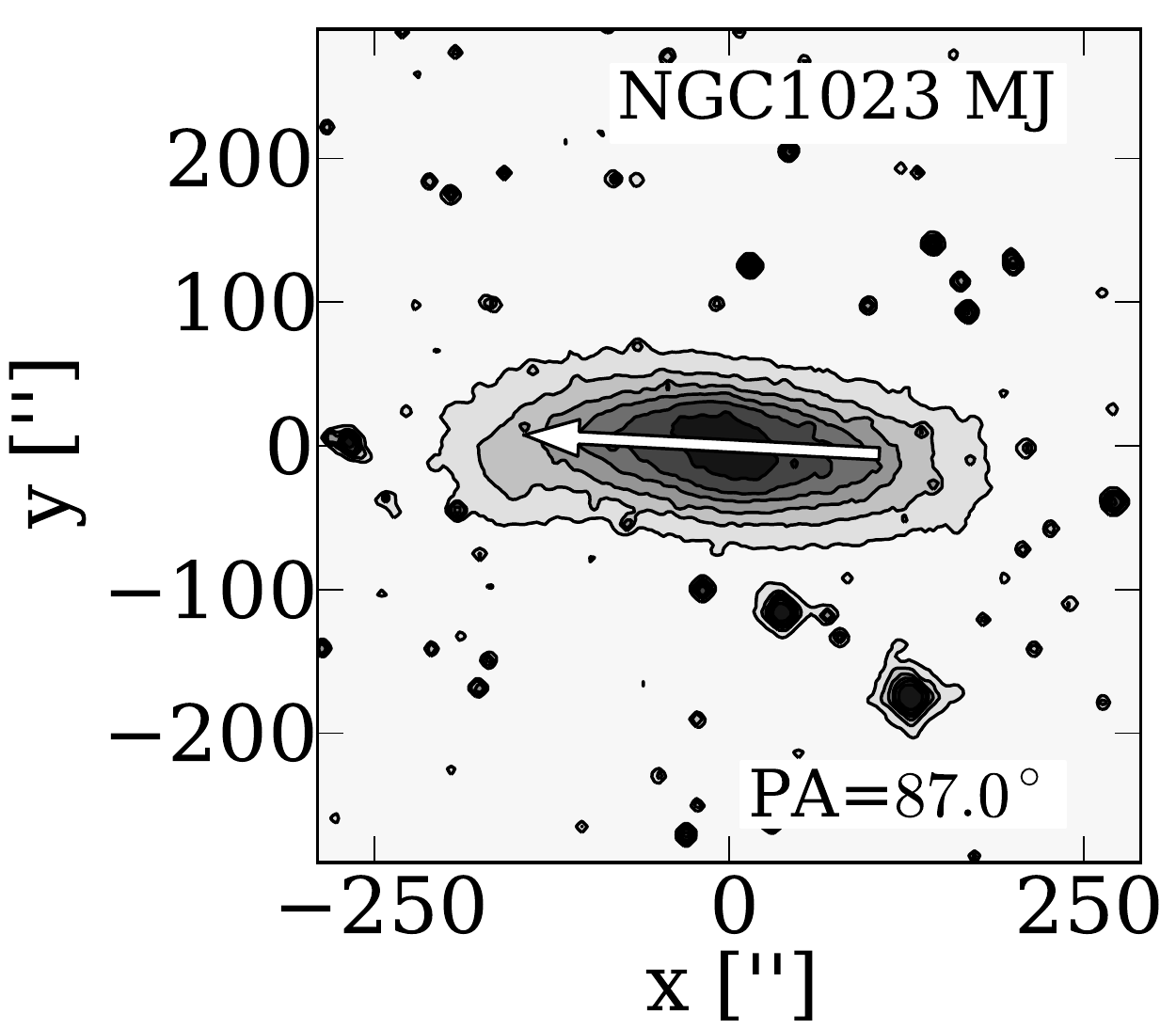} \\
	\includegraphics[viewport=0 55 390 400,width=\textwidth]{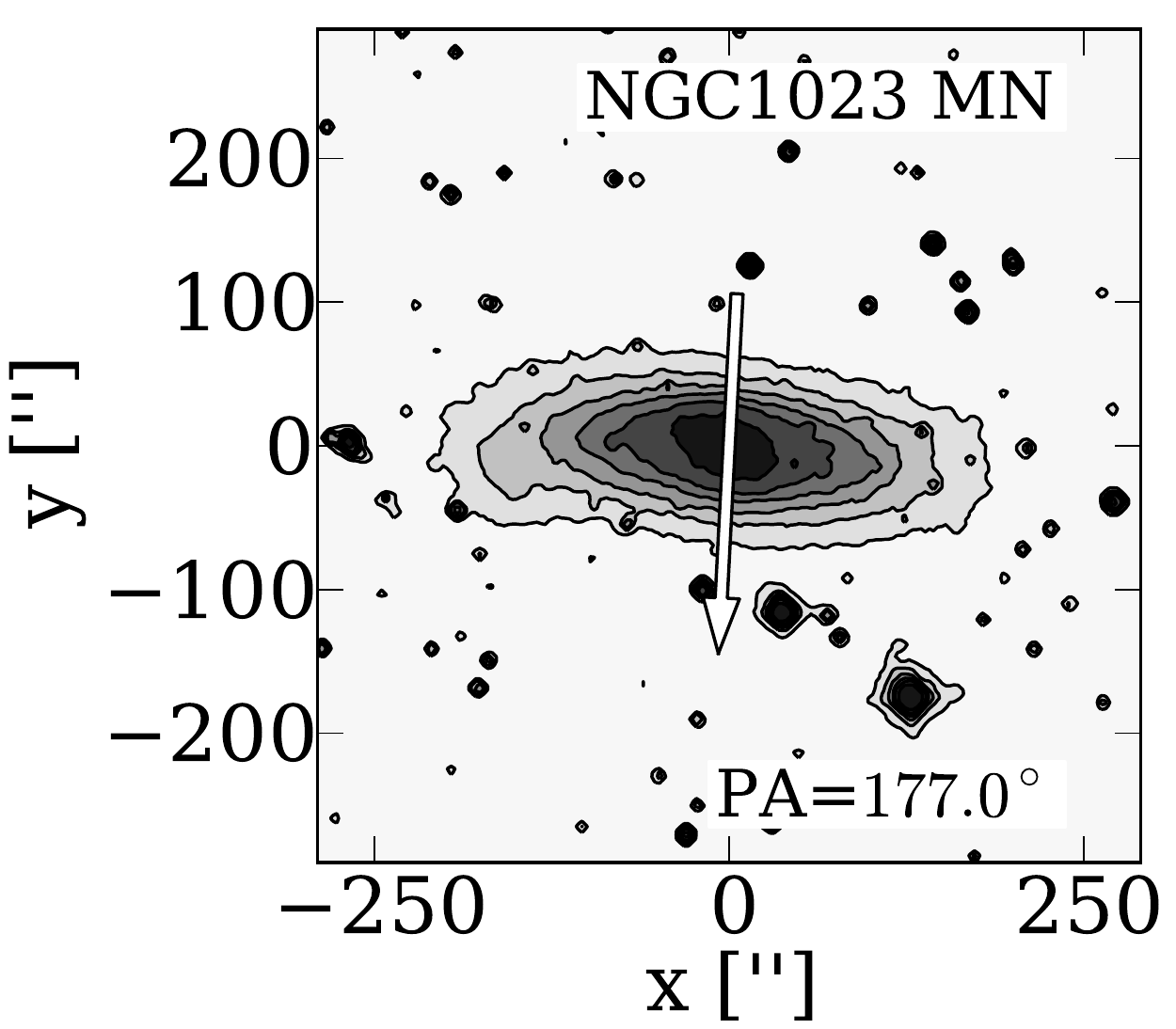}
	\end{minipage} & 
	\includegraphics[viewport=0 50 420 400,width=0.35\textwidth]{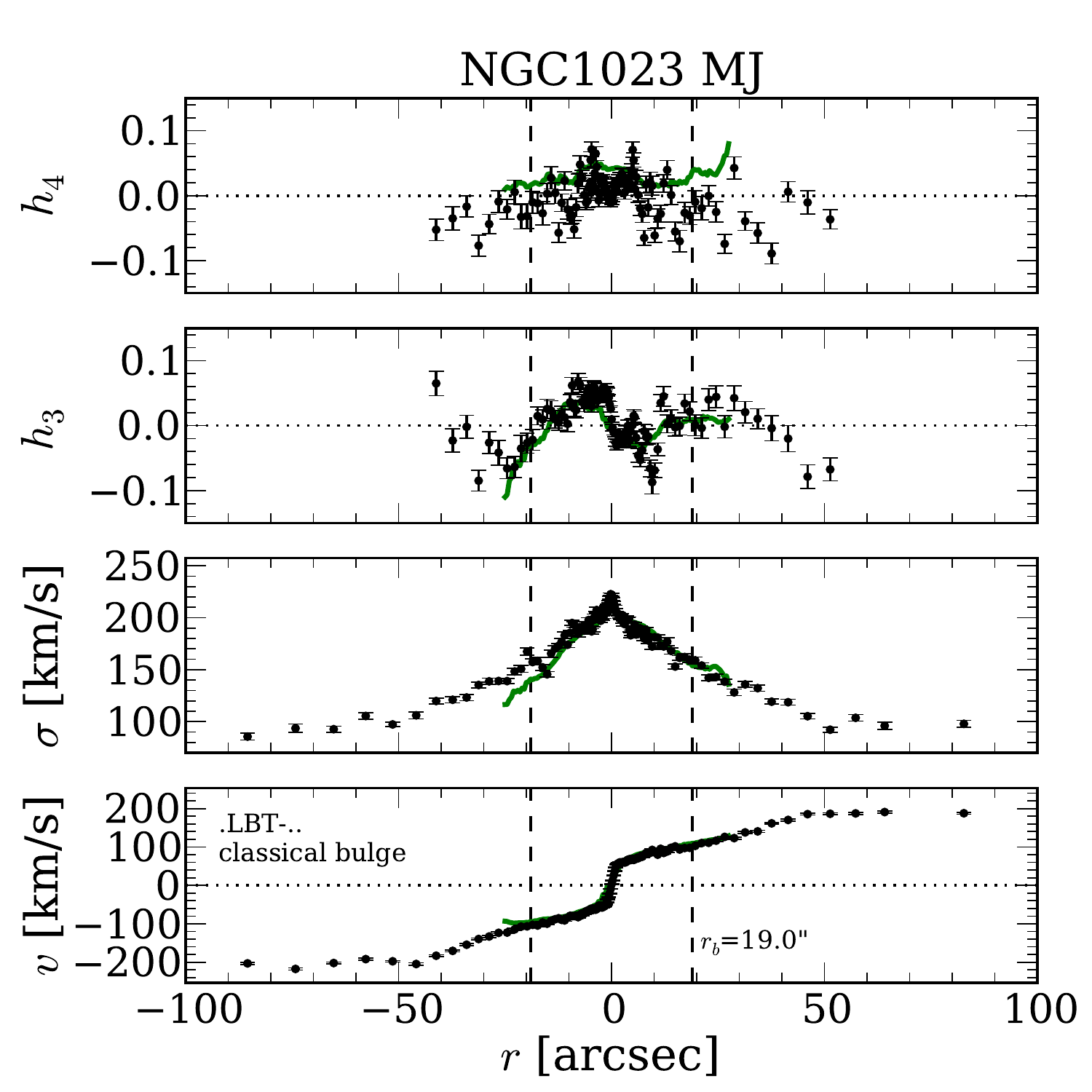} &
	\includegraphics[viewport=0 50 420 400,width=0.35\textwidth]{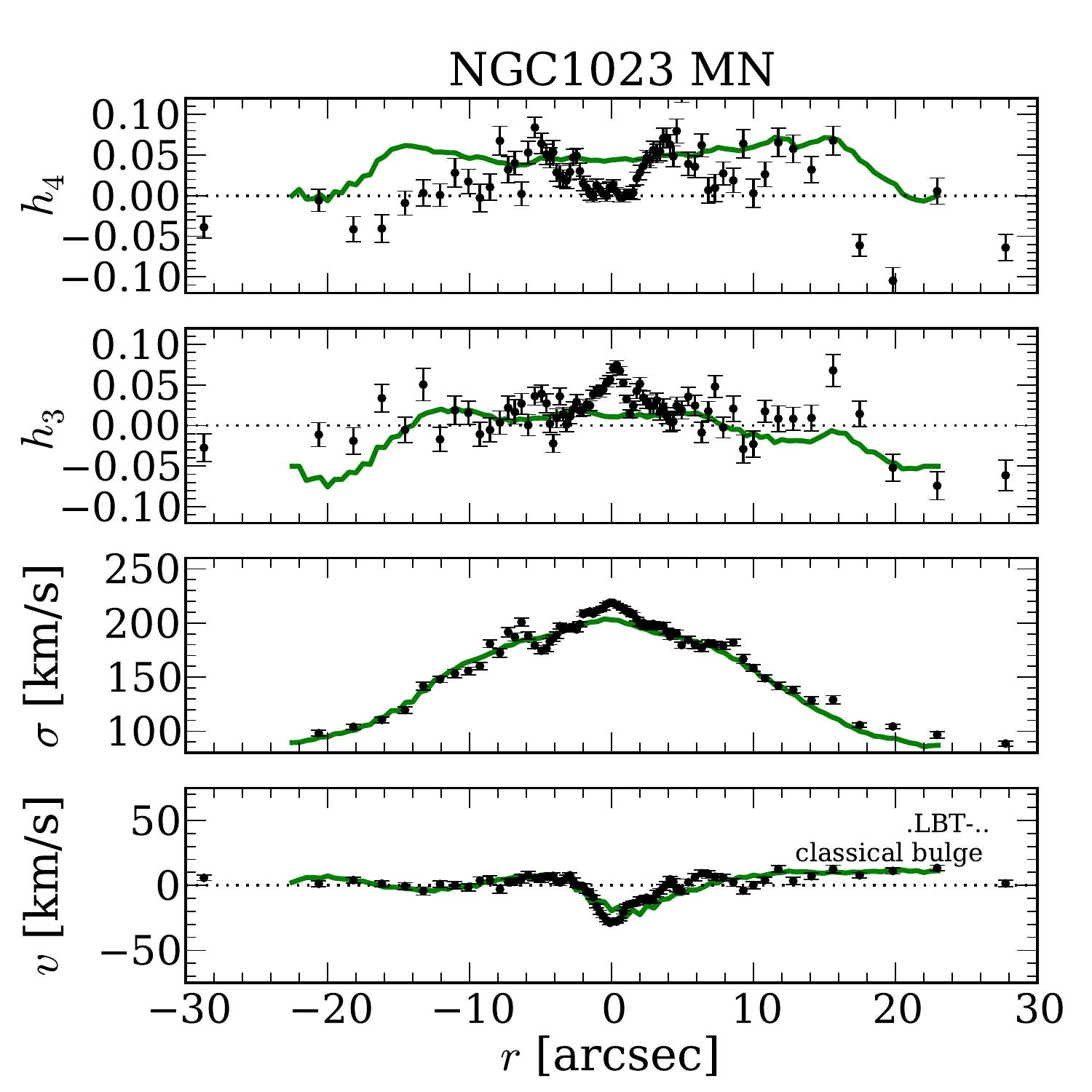}\\
        \end{tabular}
        \end{center}
        \caption{\label{fig:kprofiles}\small
	Major and minor axis kinematic profiles for NGC\,1023. The slit position is
	indicated as arrow on a Digital Sky Survey image on the left. Positive radii
	are east of the galaxy center. We plot from bottom to top the rotational
	velocity, velocity dispersion, $h_3$ and $h_4$ moments. Vertical dashed lines
	indicate the bulge radius. We plot SAURON results of \citet{Emsellem2004}
	in green. We matched our minor axis velocities to the SAURON velocity map by
	allowing an offset of the slit position. A 2~arcseconds offset to the west
	yielded the smallest residuals between the SAURON velocities and ours.
	}
\end{figure}
\setcounter{figure}{15}
\begin{figure}[h]
        \begin{center}
        \begin{tabular}{lll}
	\begin{minipage}[b]{0.185\textwidth}
	\includegraphics[viewport=0 55 390 400,width=\textwidth]{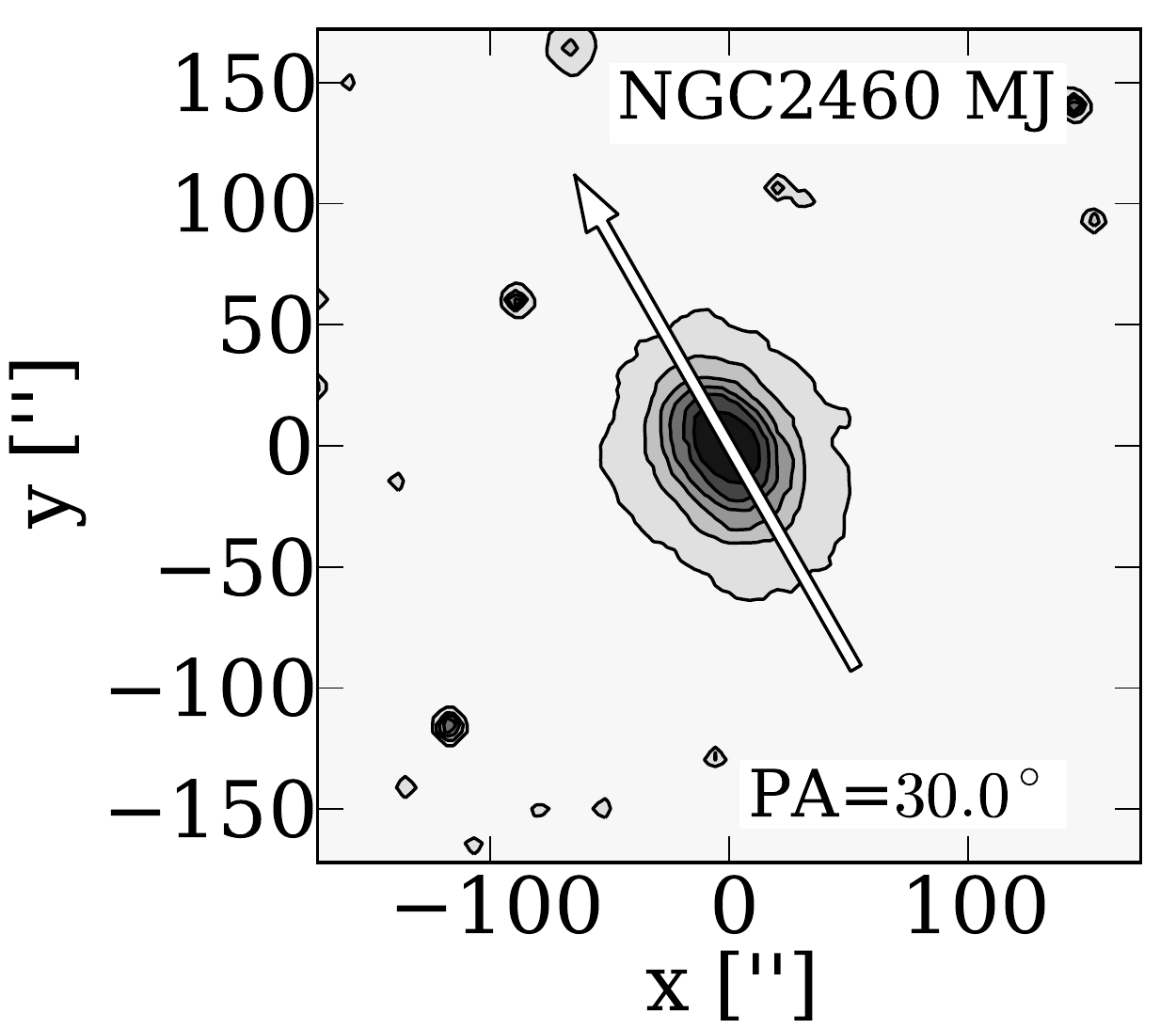} \\
	\includegraphics[viewport=0 55 390 400,width=\textwidth]{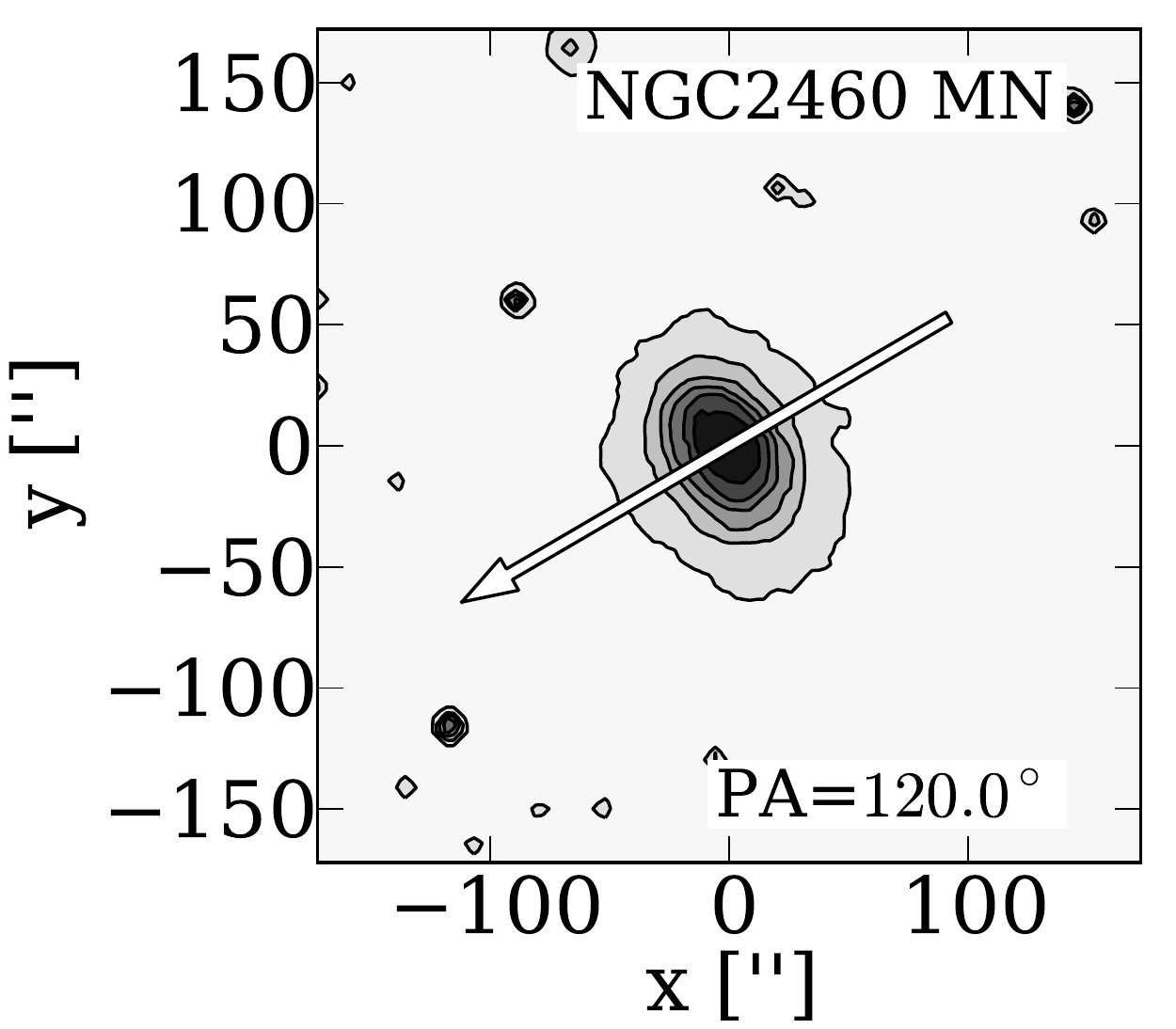}
	\end{minipage} & 
	\includegraphics[viewport=0 50 420 400,width=0.35\textwidth]{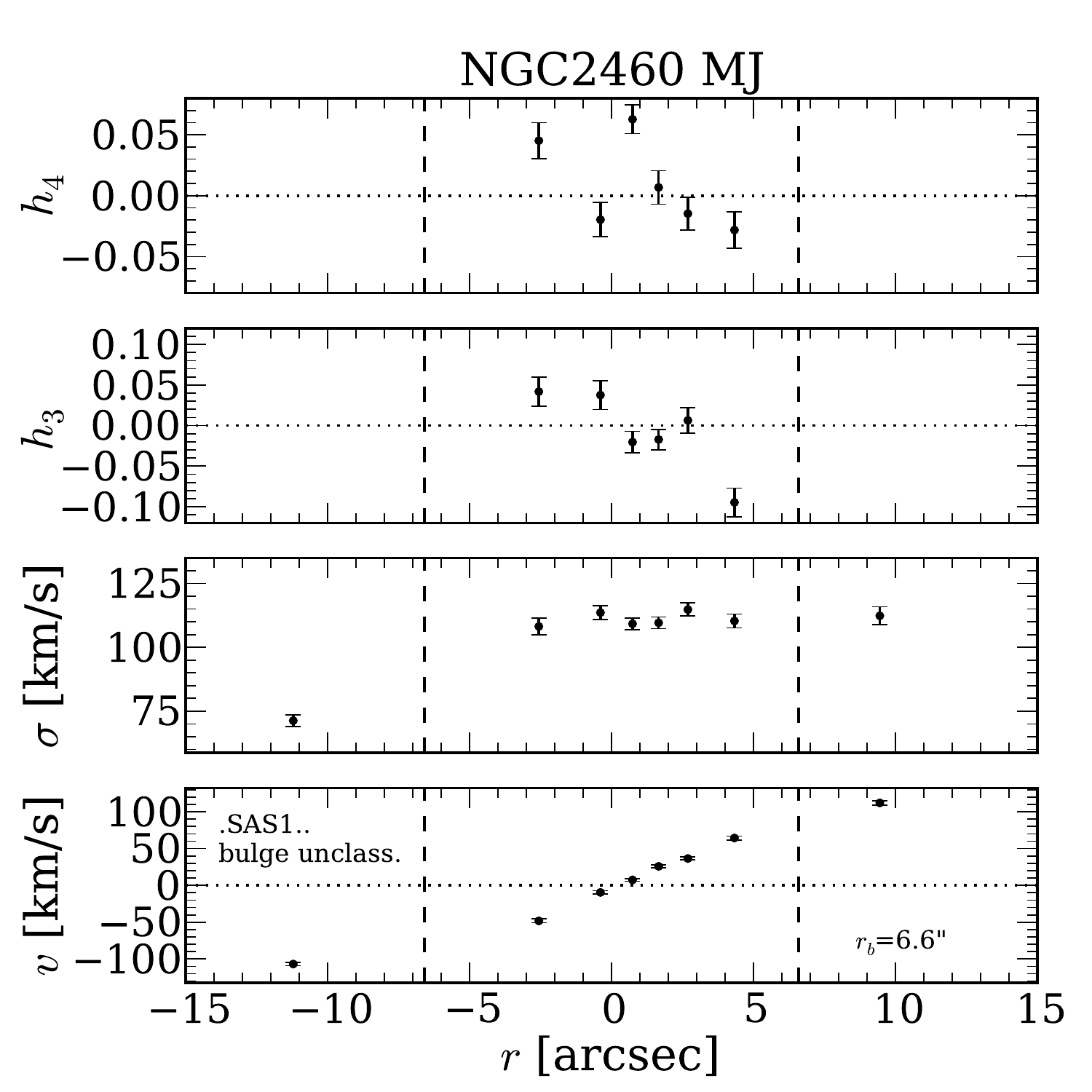} &
	\includegraphics[viewport=0 50 420 400,width=0.35\textwidth]{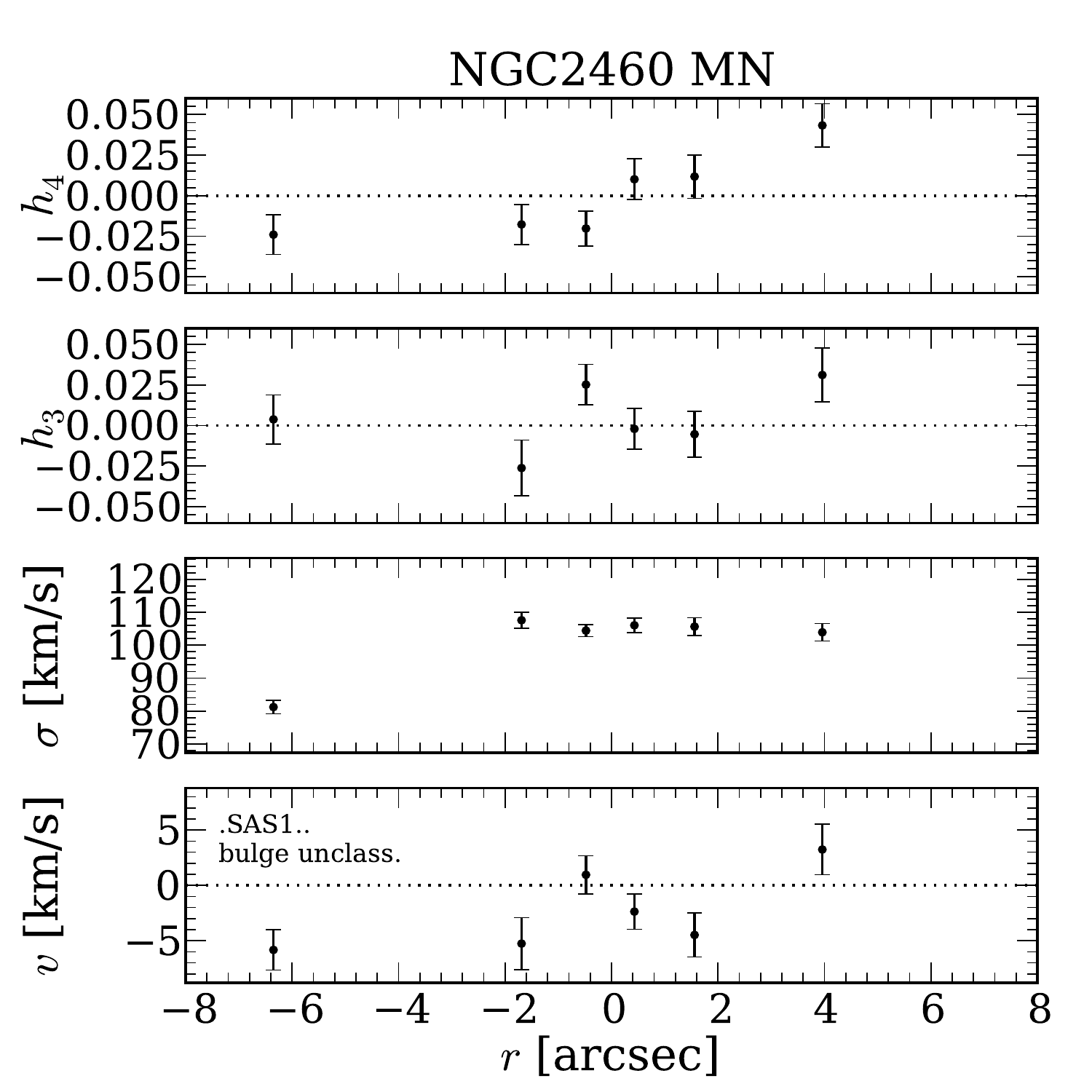}\\
        \end{tabular}
        \end{center}
        \caption{{\it continued --}\small  Major and minor axis kinematic profiles for NGC\,2460.}
\end{figure}
\clearpage
\setcounter{figure}{15}
\begin{figure}
        \begin{center}
        \begin{tabular}{lll}
        \begin{minipage}[b]{0.185\textwidth}
        \includegraphics[viewport=0 55 390 400,width=\textwidth]{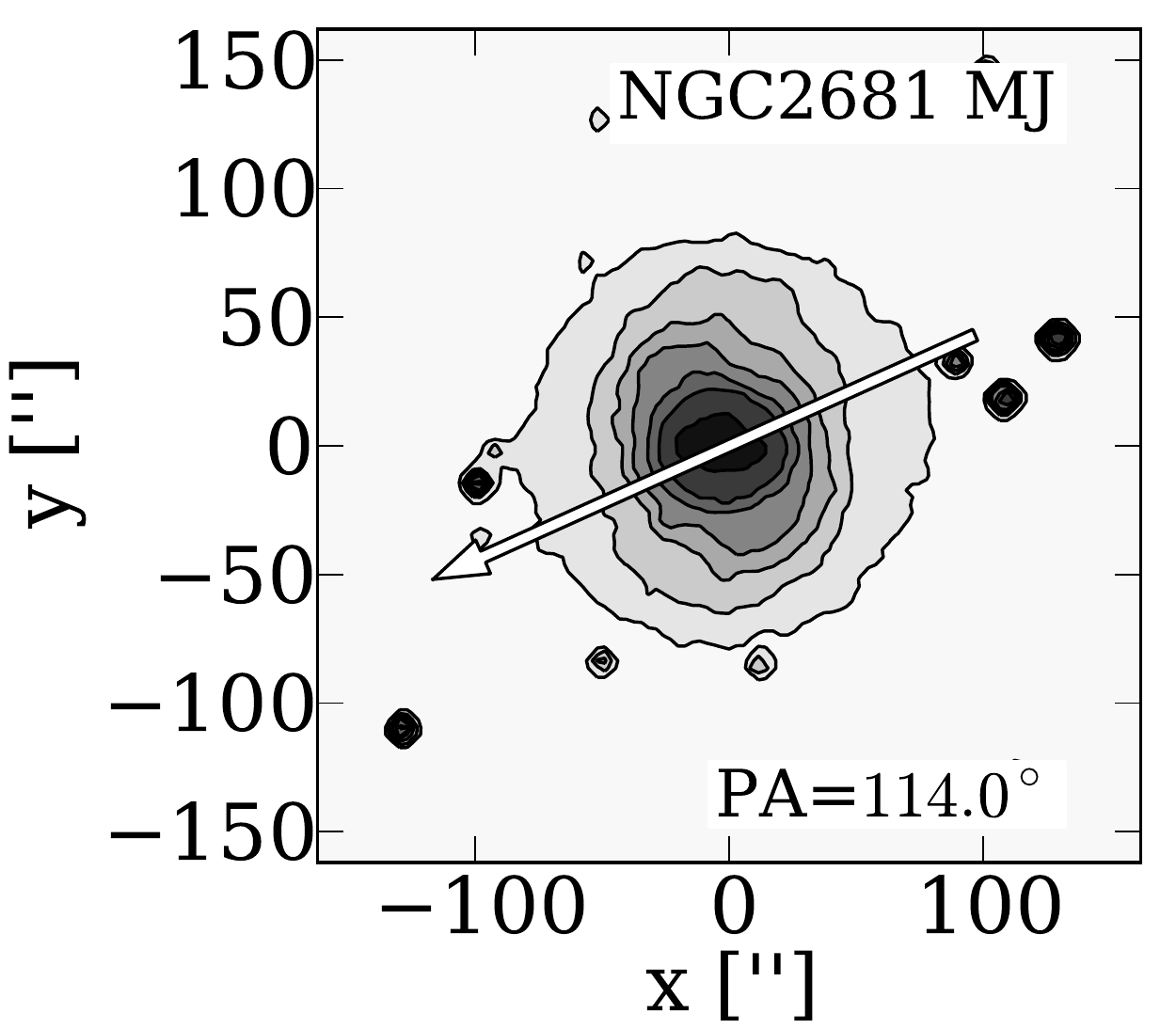}\\
        \includegraphics[viewport=0 55 390 400,width=\textwidth]{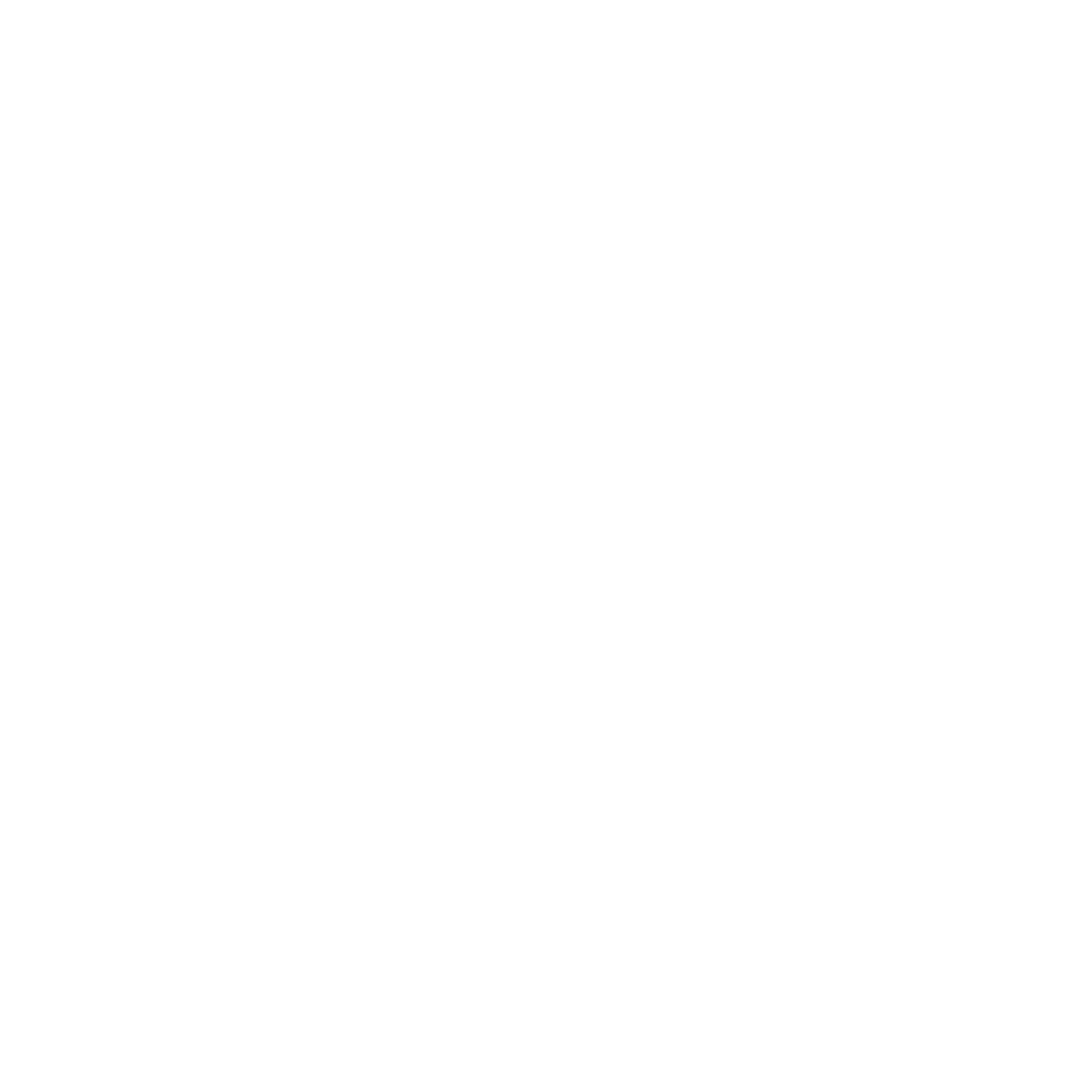}
        \end{minipage} &
        \includegraphics[viewport=0 50 420 400,width=0.35\textwidth]{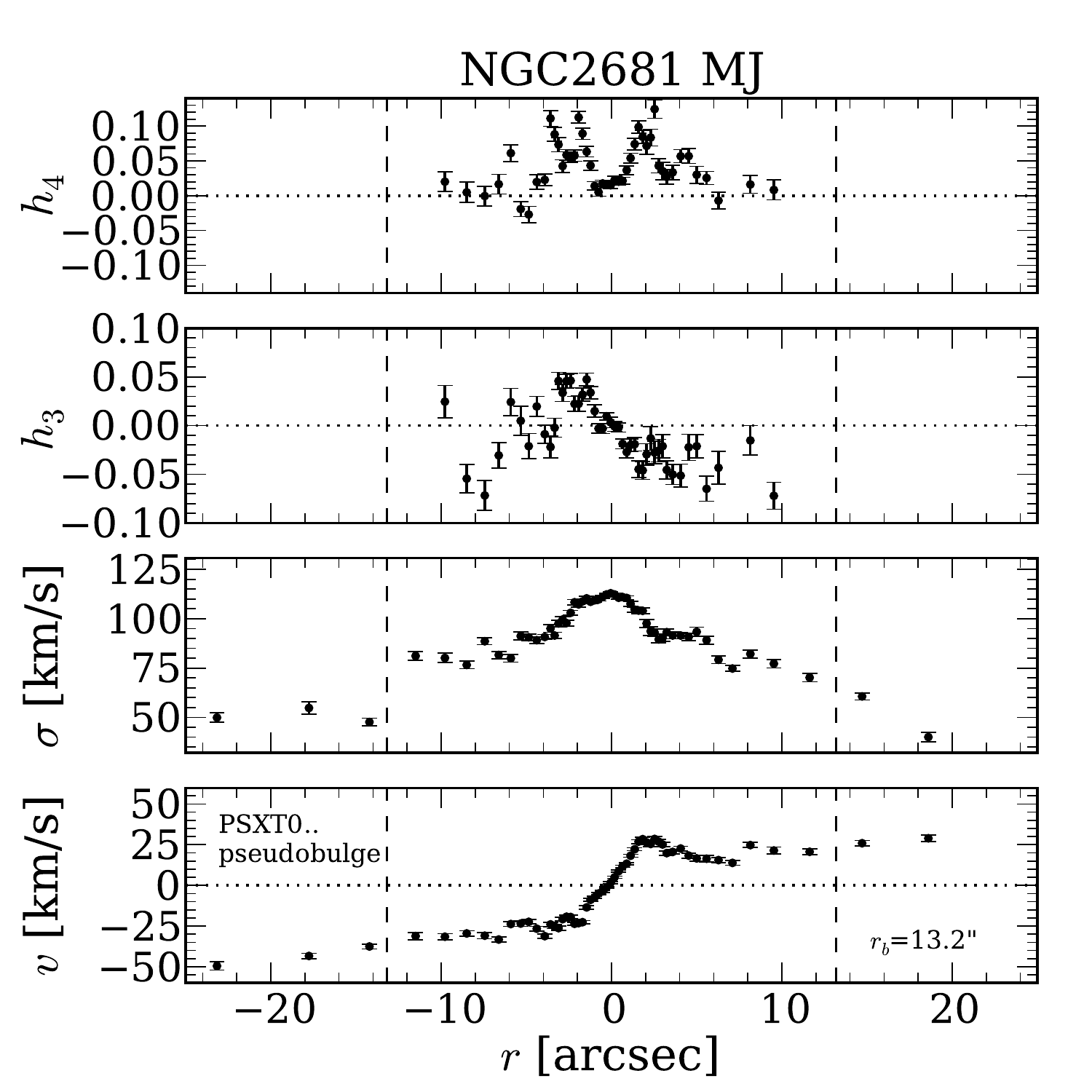}&
        \includegraphics[viewport=0 50 420 400,width=0.35\textwidth]{empty}
        \end{tabular}
        \end{center}
        \caption{{\it continued --}\small Major axis kinematic profile for NGC\,2681. 
        }
\end{figure}
\setcounter{figure}{15}
\begin{figure}
        \begin{center}
        \begin{tabular}{lll}
	\begin{minipage}[b]{0.185\textwidth}
	\includegraphics[viewport=0 55 390 400,width=\textwidth]{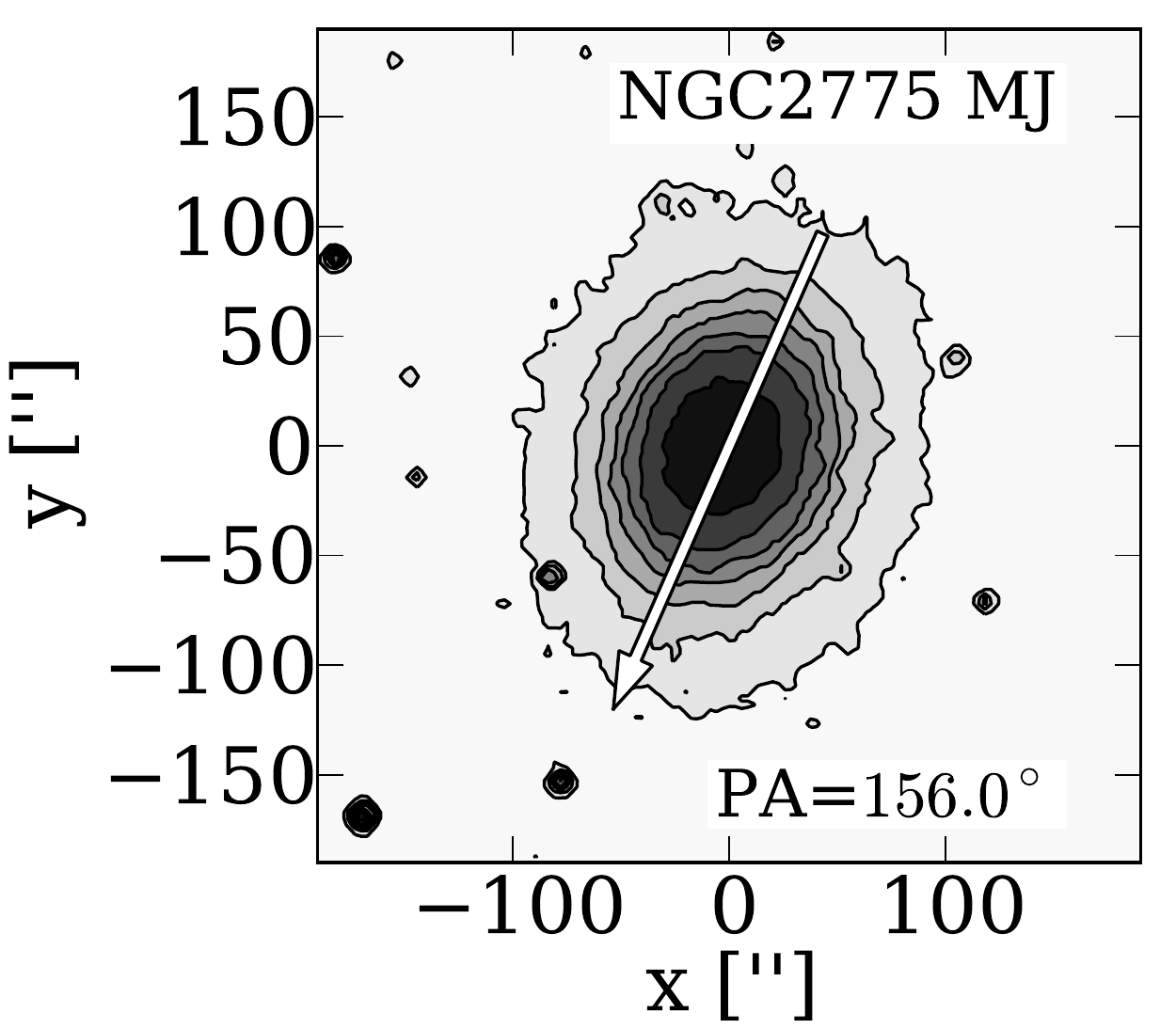} \\
	\includegraphics[viewport=0 55 390 400,width=\textwidth]{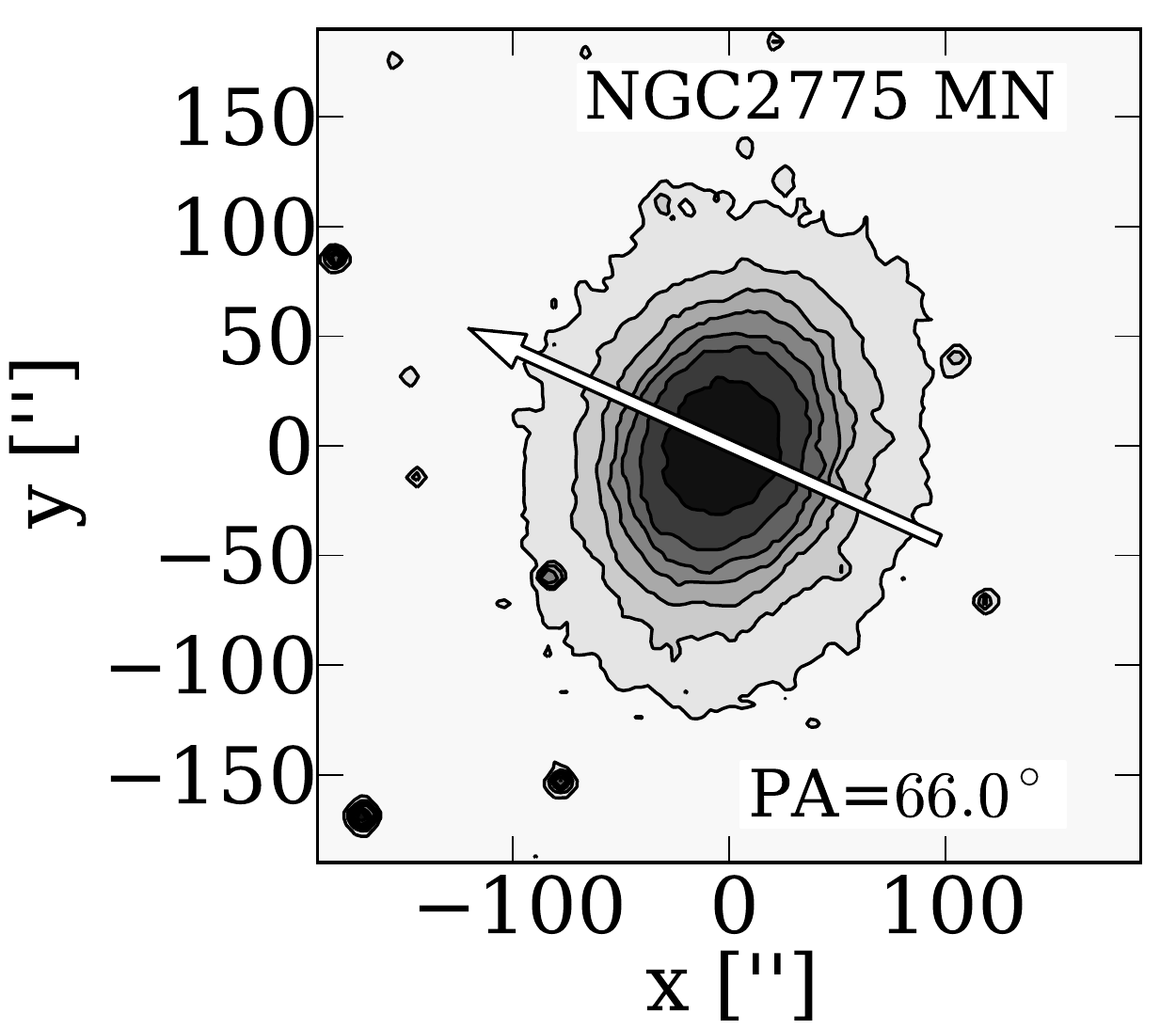}
	\end{minipage} & 
	\includegraphics[viewport=0 50 420 400,width=0.35\textwidth]{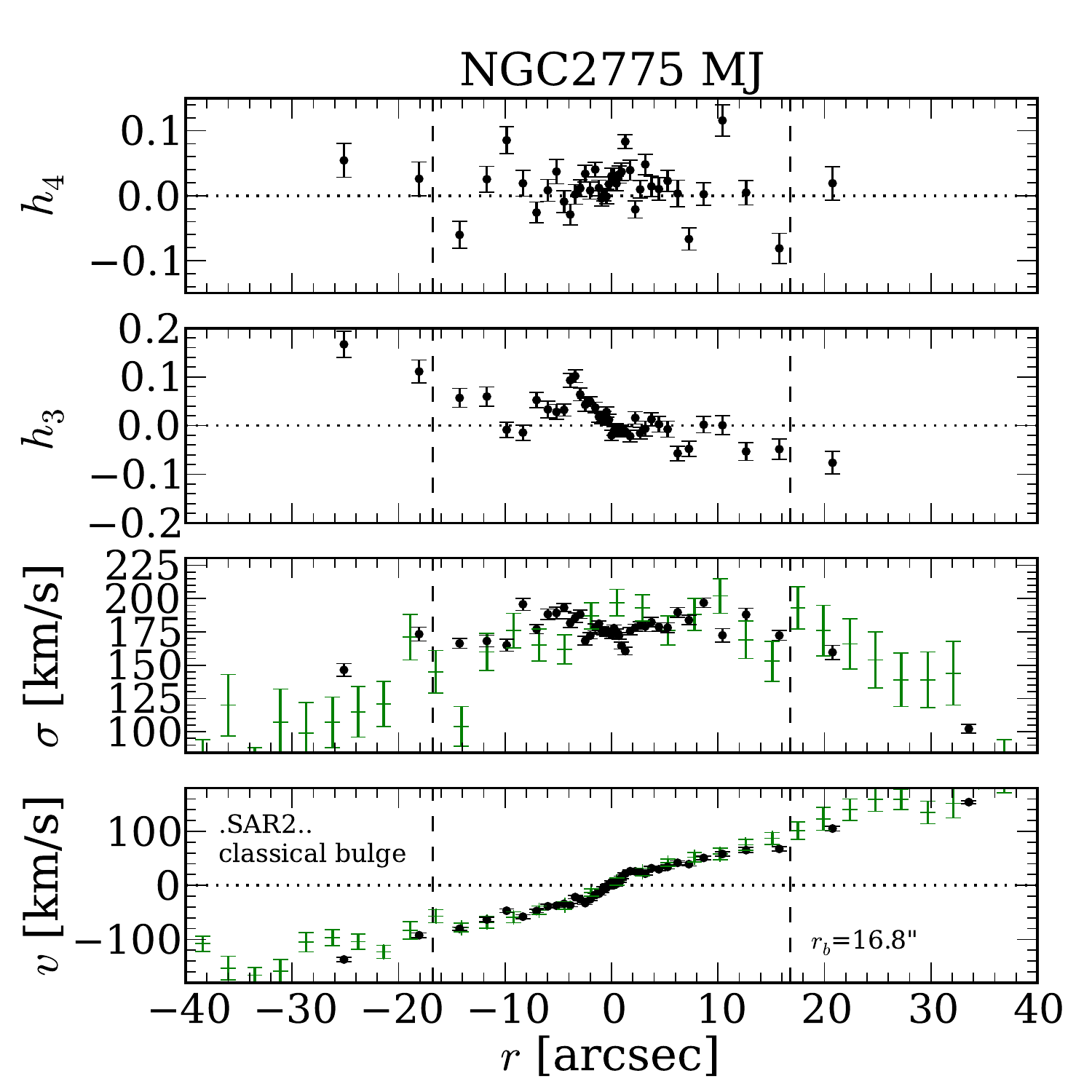} &
	\includegraphics[viewport=0 50 420 400,width=0.35\textwidth]{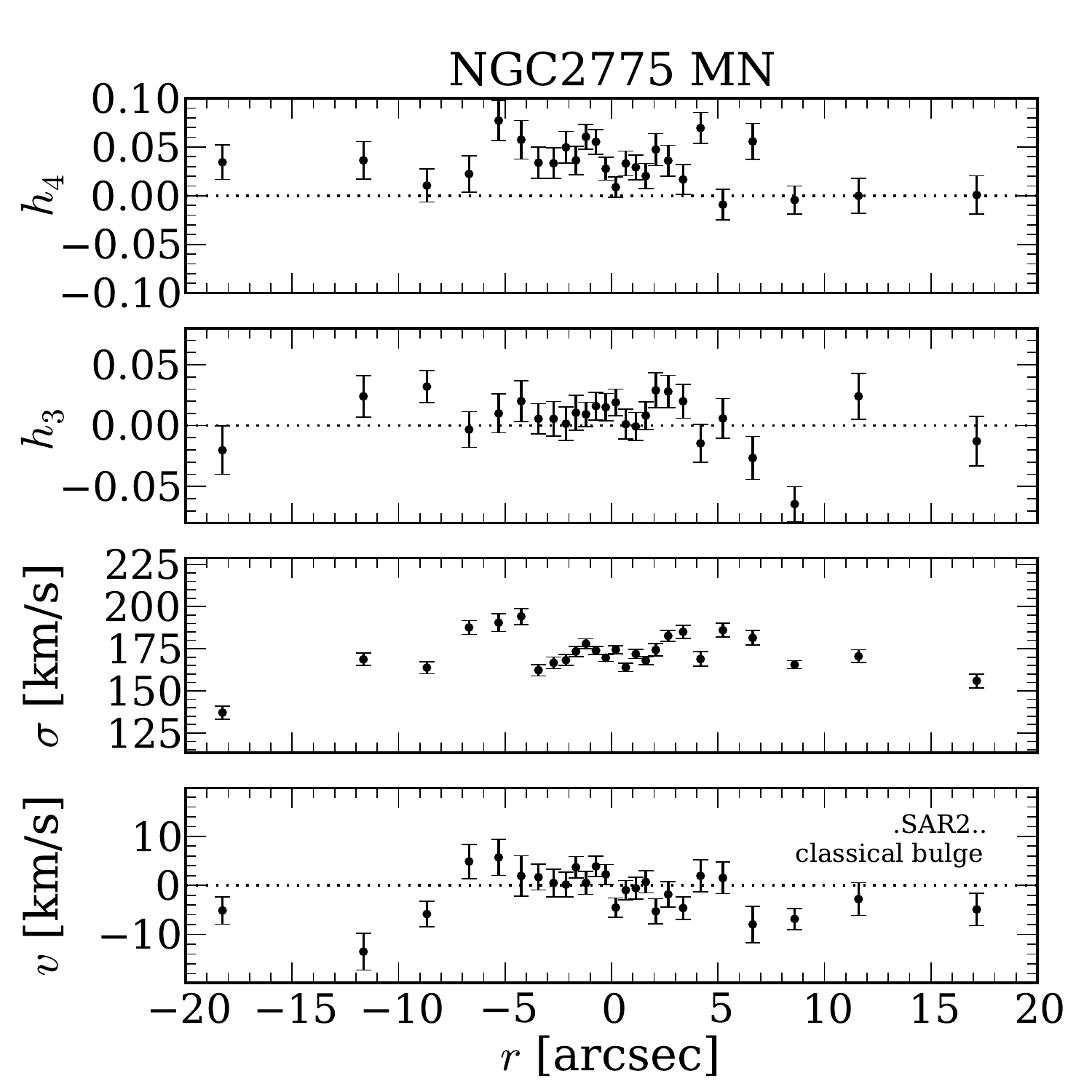}\\
        \end{tabular}
        \end{center}
        \caption{{\it continued --}\small Major and minor axis kinematic profiles for NGC\,2775.
	We plot results of \citet{Corsini1999} in green.}
\end{figure}
\setcounter{figure}{15}
\begin{figure}
        \begin{center}
        \begin{tabular}{lll}
	\begin{minipage}[b]{0.185\textwidth}
	\includegraphics[viewport=0 55 390 400,width=\textwidth]{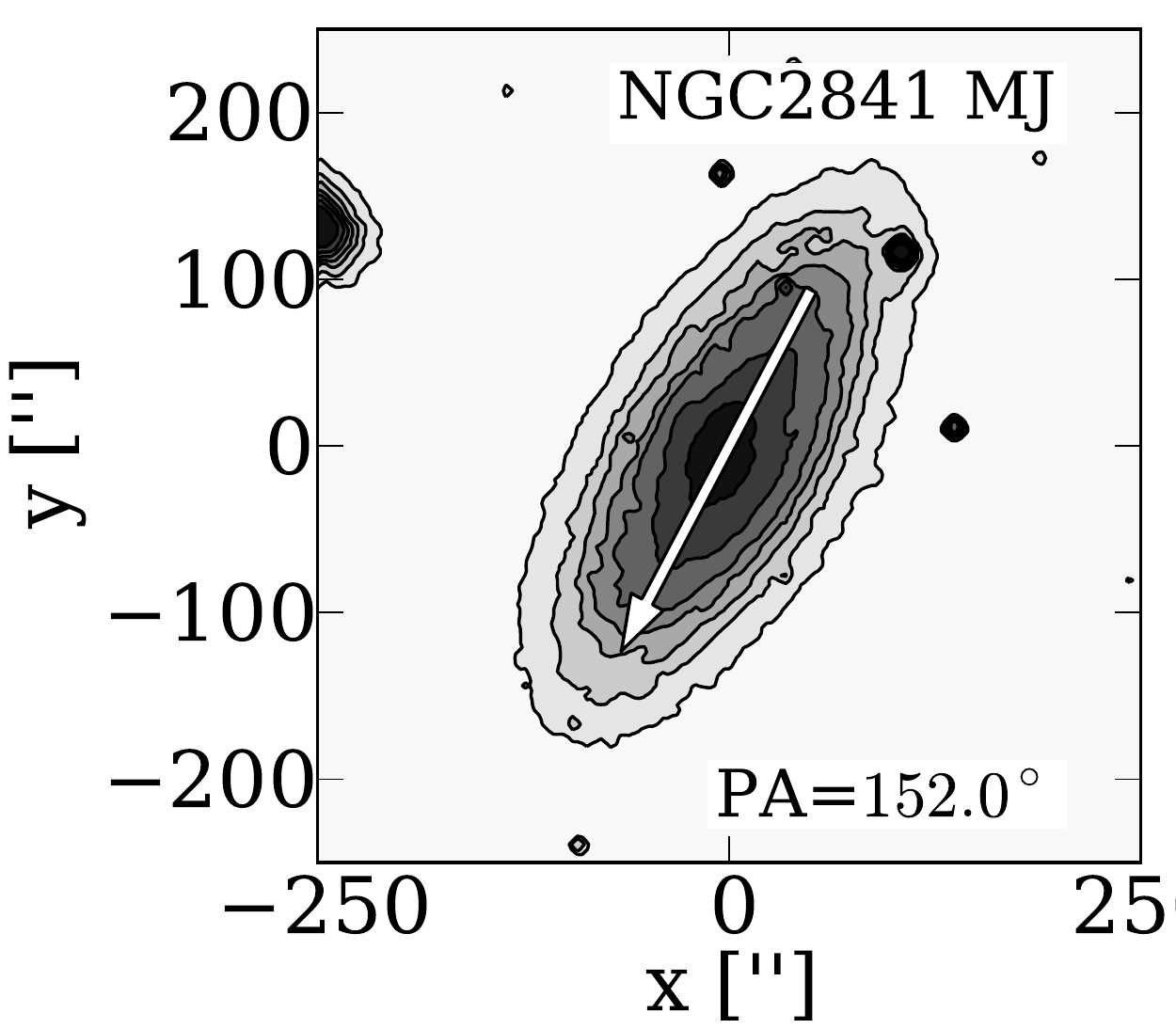} \\
	\includegraphics[viewport=0 55 390 400,width=\textwidth]{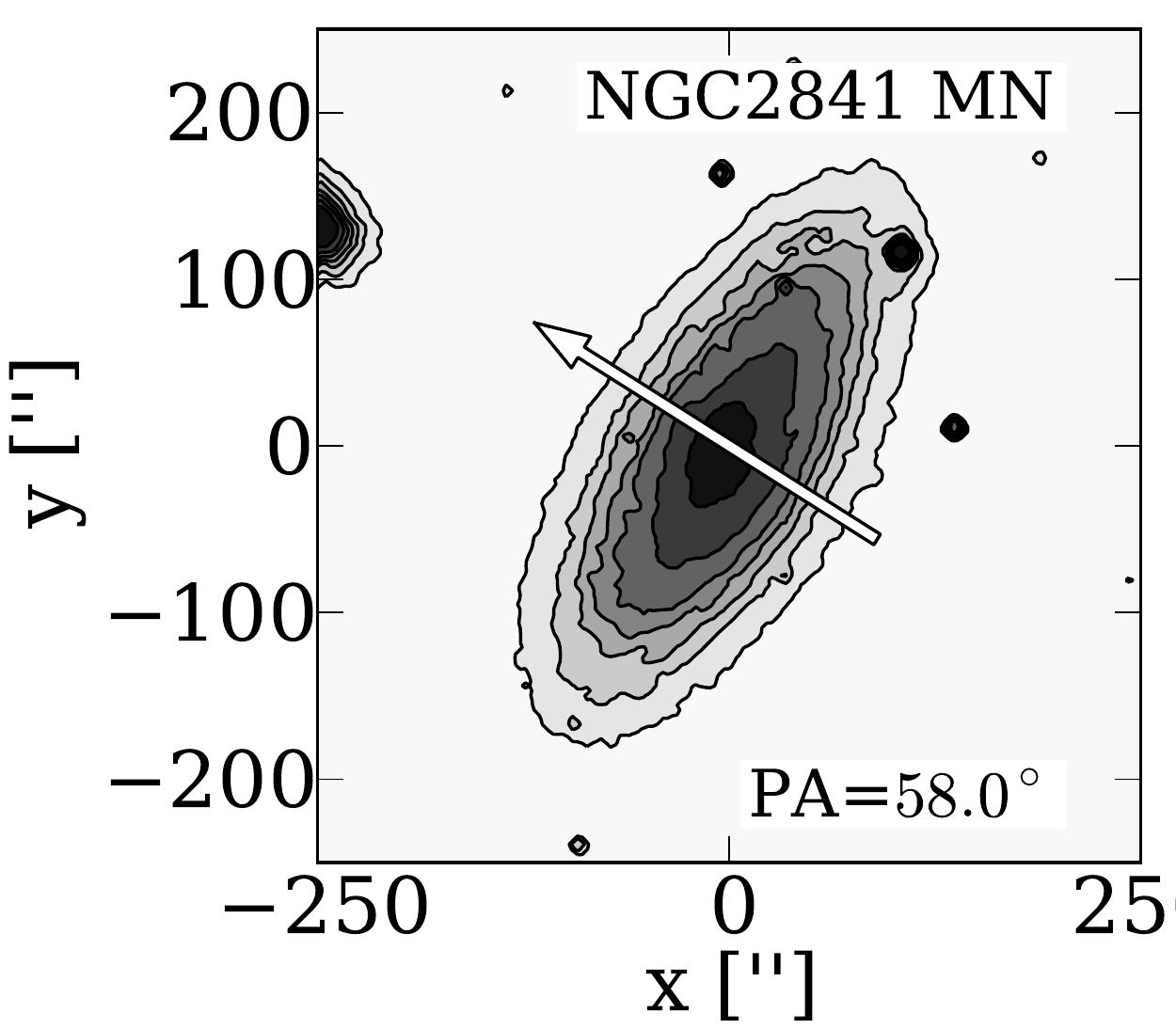}
	\end{minipage} & 
	\includegraphics[viewport=0 50 420 400,width=0.35\textwidth]{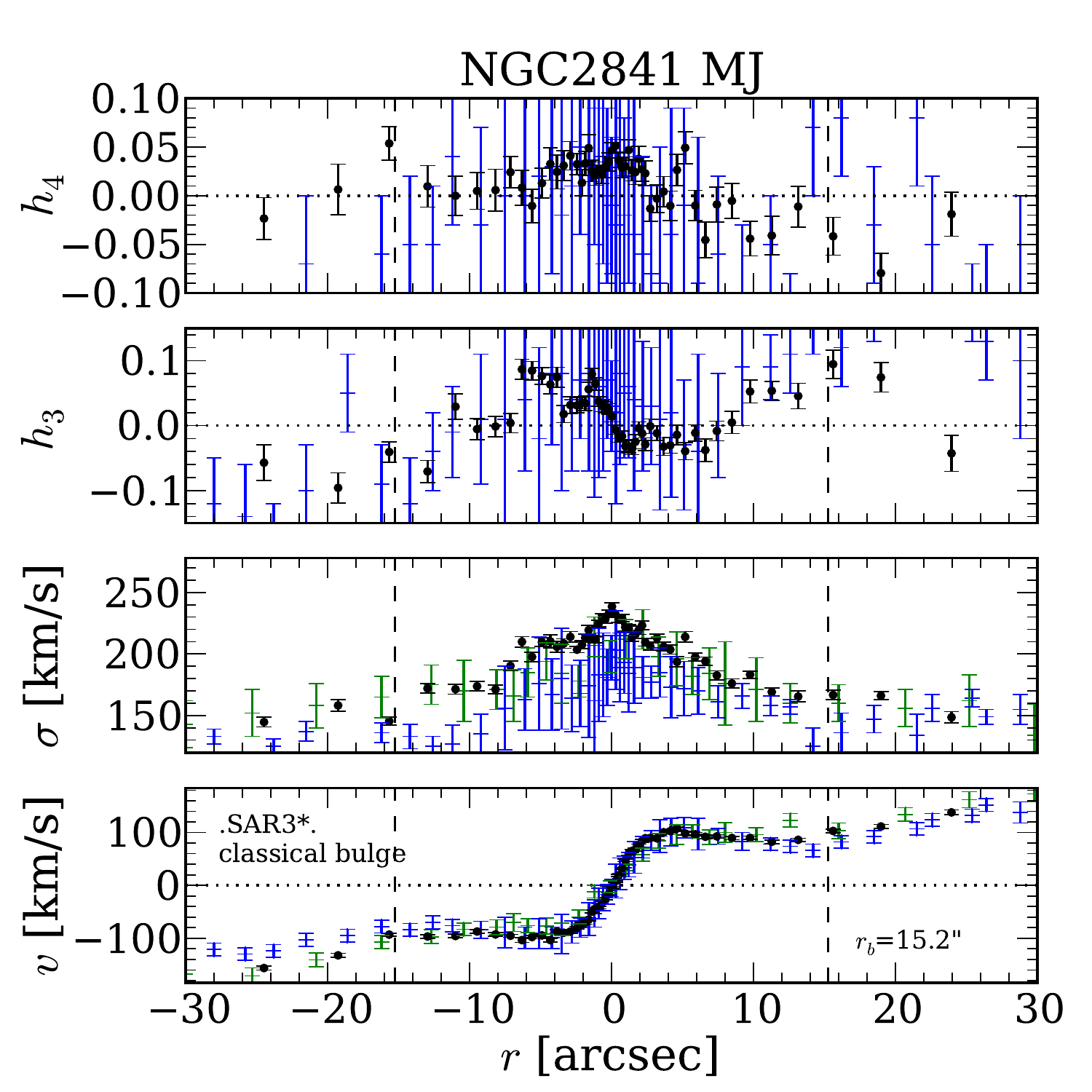} &
	\includegraphics[viewport=0 50 420 400,width=0.35\textwidth]{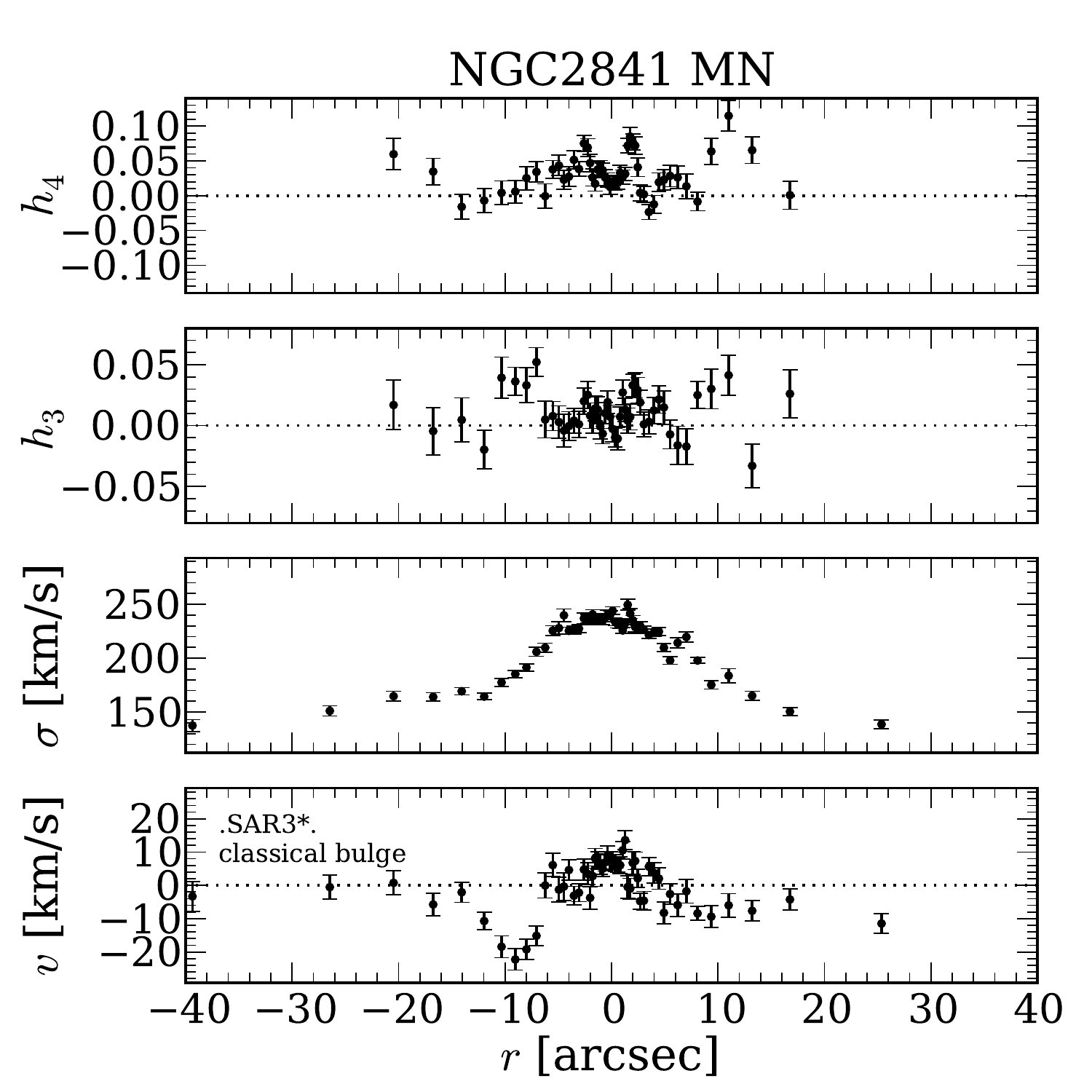}\\
        \end{tabular}
        \end{center}
        \caption{{\it continued --}\small Major and minor axis kinematic profiles for NGC\,2841.
	We plot results of \citet{Heraudeau1998} in green and those of \citet{Vega-Beltran2001} in blue.}
\end{figure}
\setcounter{figure}{15}
\begin{figure}
        \begin{center}
        \begin{tabular}{lll}
	\begin{minipage}[b]{0.185\textwidth}
	\includegraphics[viewport=0 55 390 400,width=\textwidth]{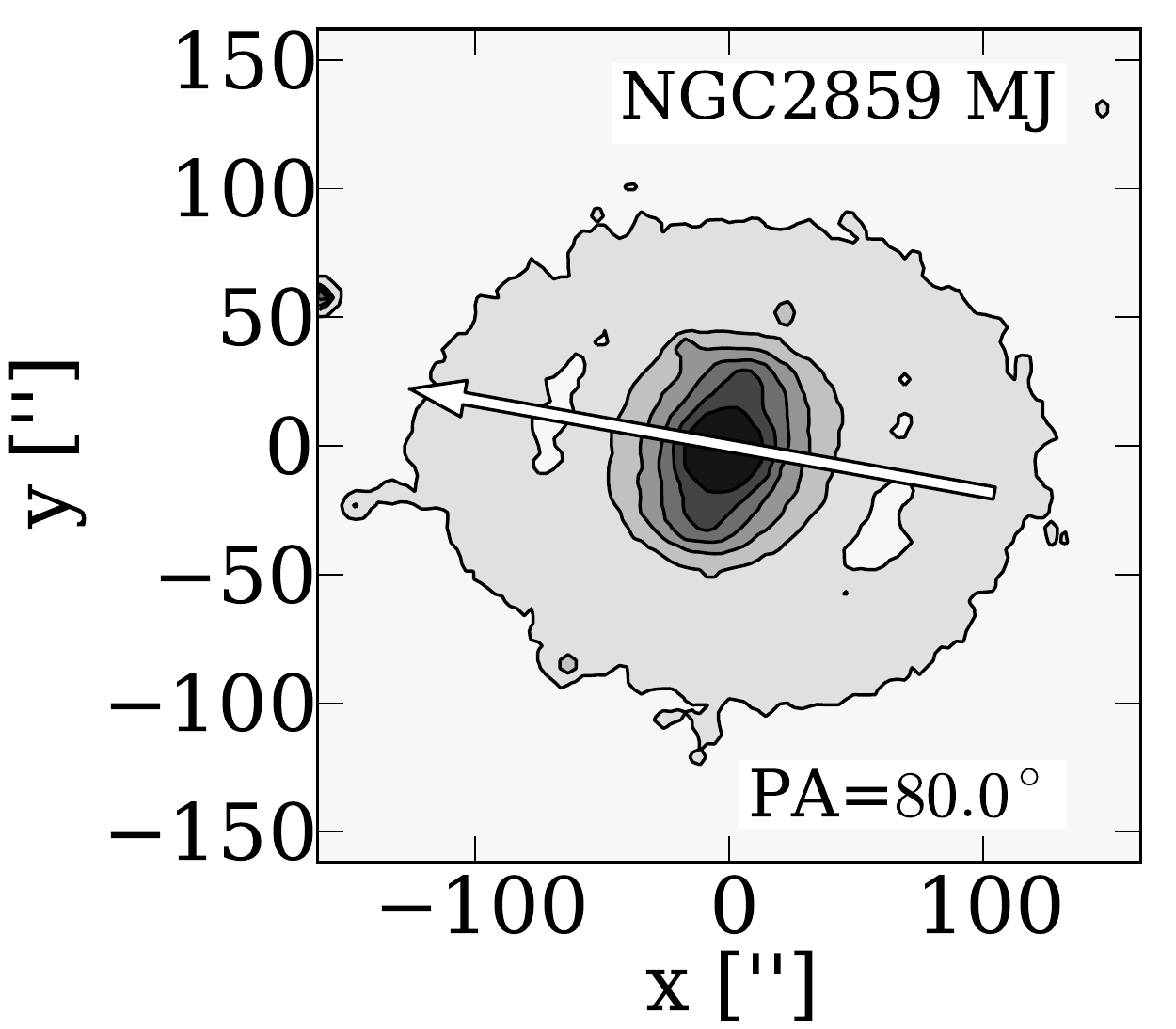} \\
	\includegraphics[viewport=0 55 390 400,width=\textwidth]{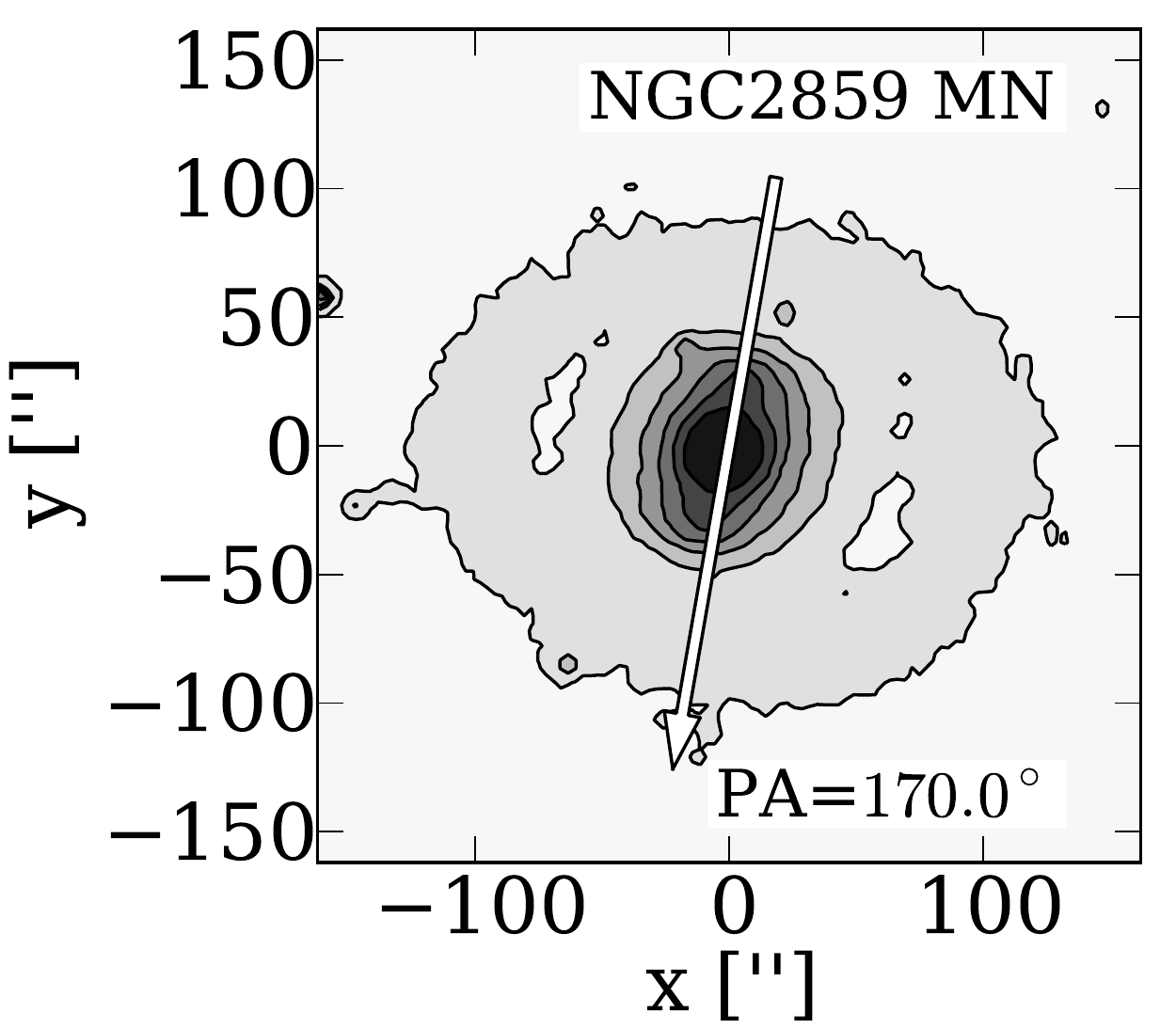}
	\end{minipage} & 
	\includegraphics[viewport=0 50 420 400,width=0.35\textwidth]{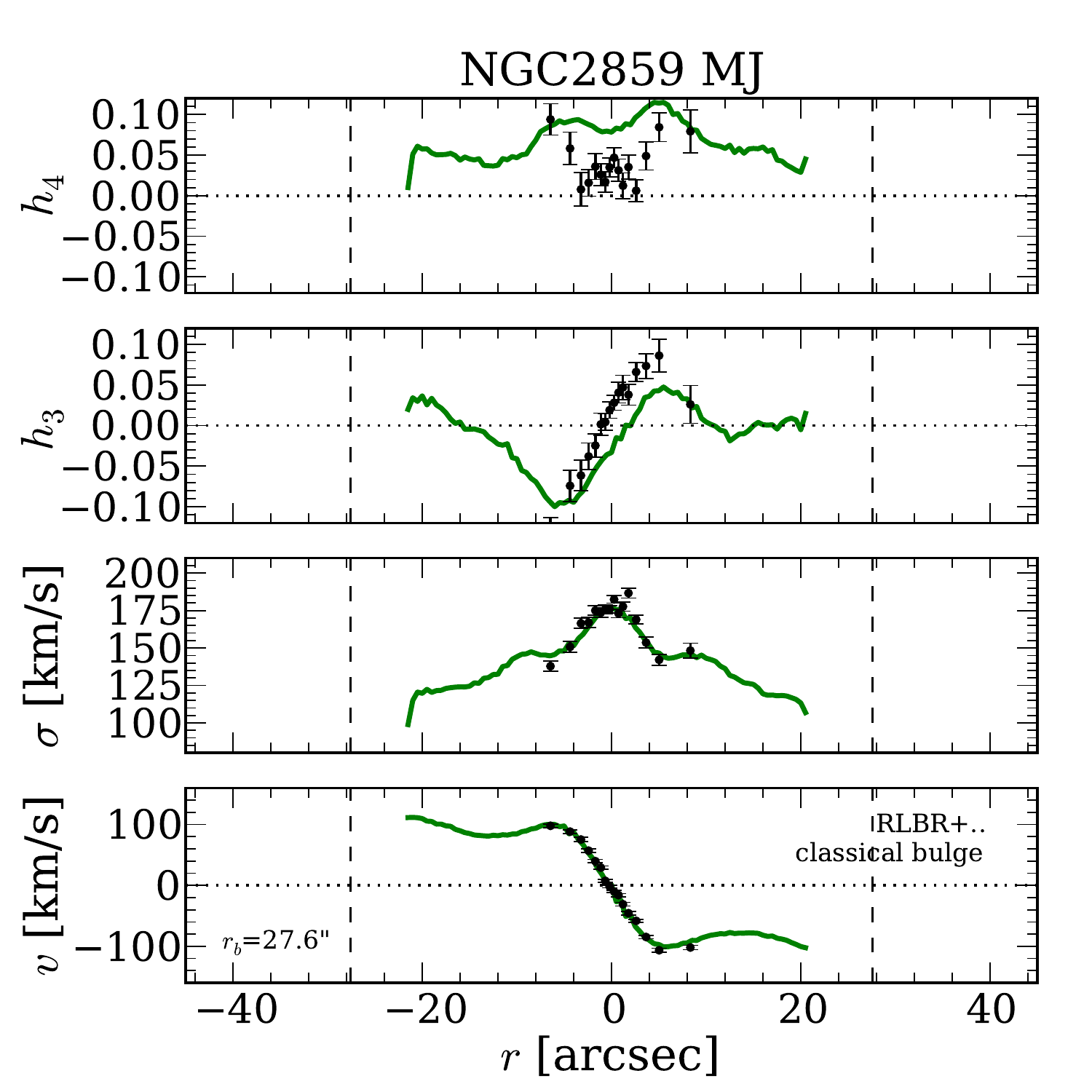} &
	\includegraphics[viewport=0 50 420 400,width=0.35\textwidth]{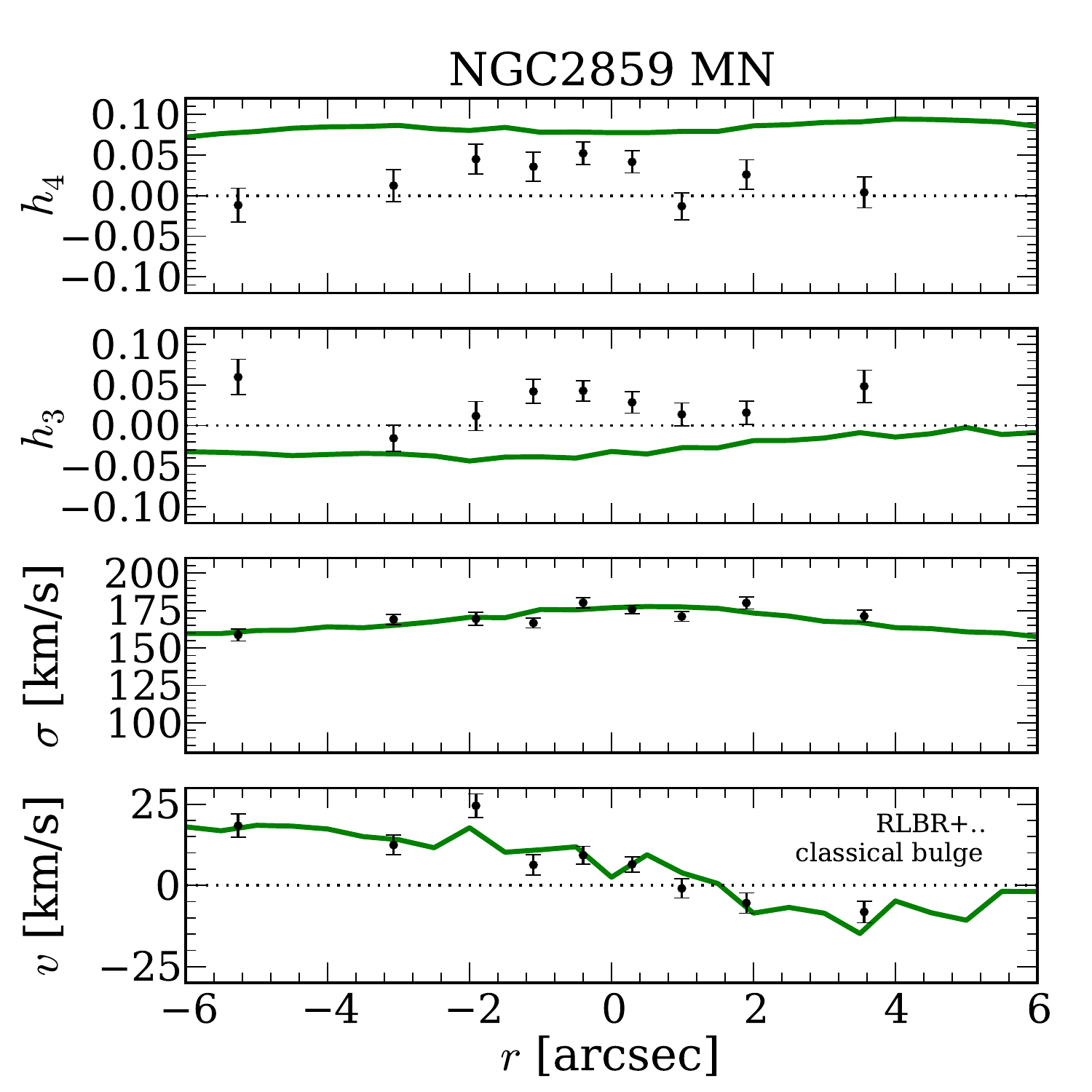}\\
        \end{tabular}
        \end{center}
        \caption{{\it continued --}\small Major and minor axis kinematic profiles for NGC\,2859.
	We plot the SAURON results \citep{de-Lorenzo-Caceres2008} in green.}
\end{figure}
\setcounter{figure}{15}
\begin{figure}
        \begin{center}
        \begin{tabular}{lll}
	\begin{minipage}[b]{0.185\textwidth}
	\includegraphics[viewport=0 55 390 400,width=\textwidth]{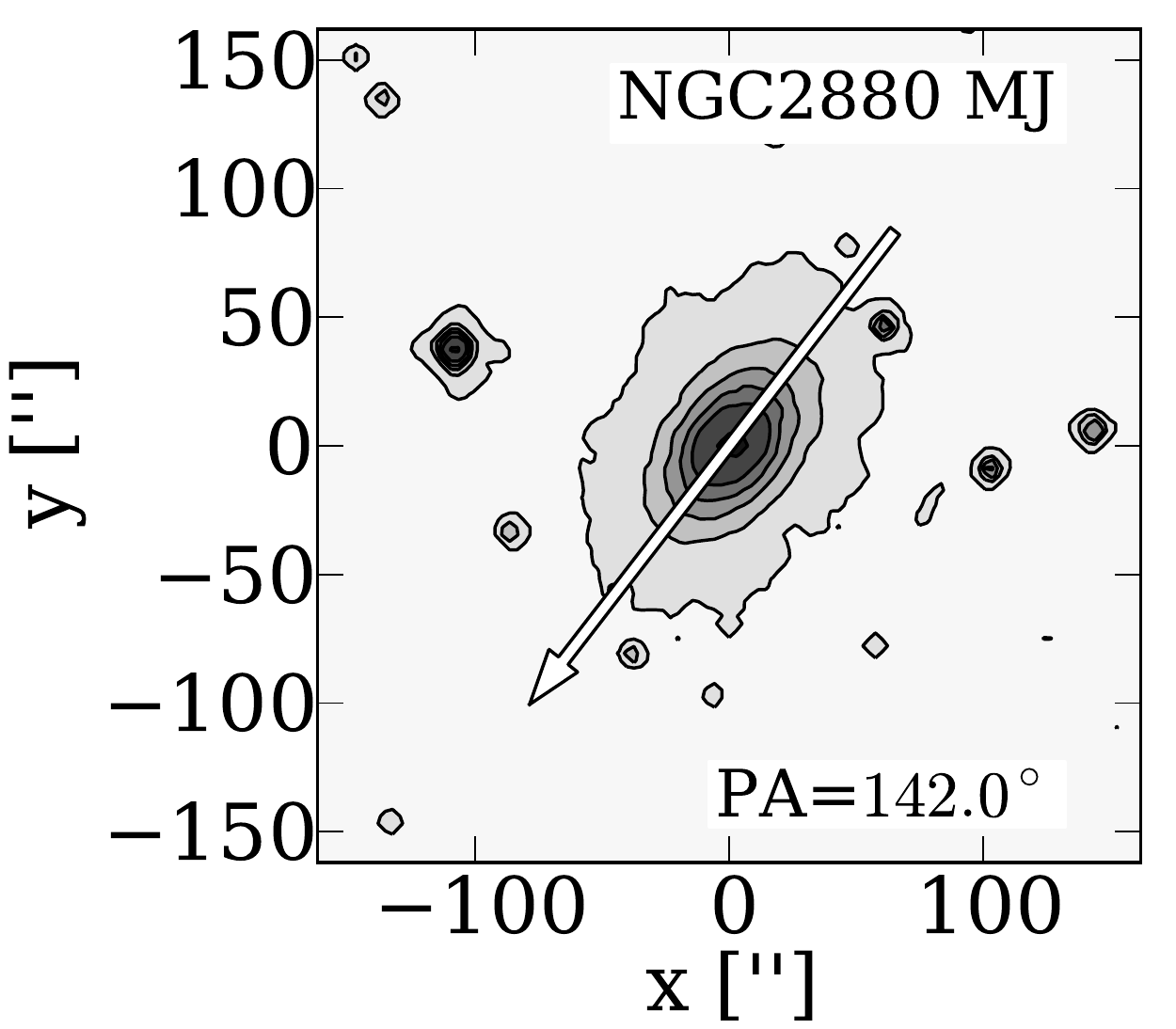} \\
	\includegraphics[viewport=0 55 390 400,width=\textwidth]{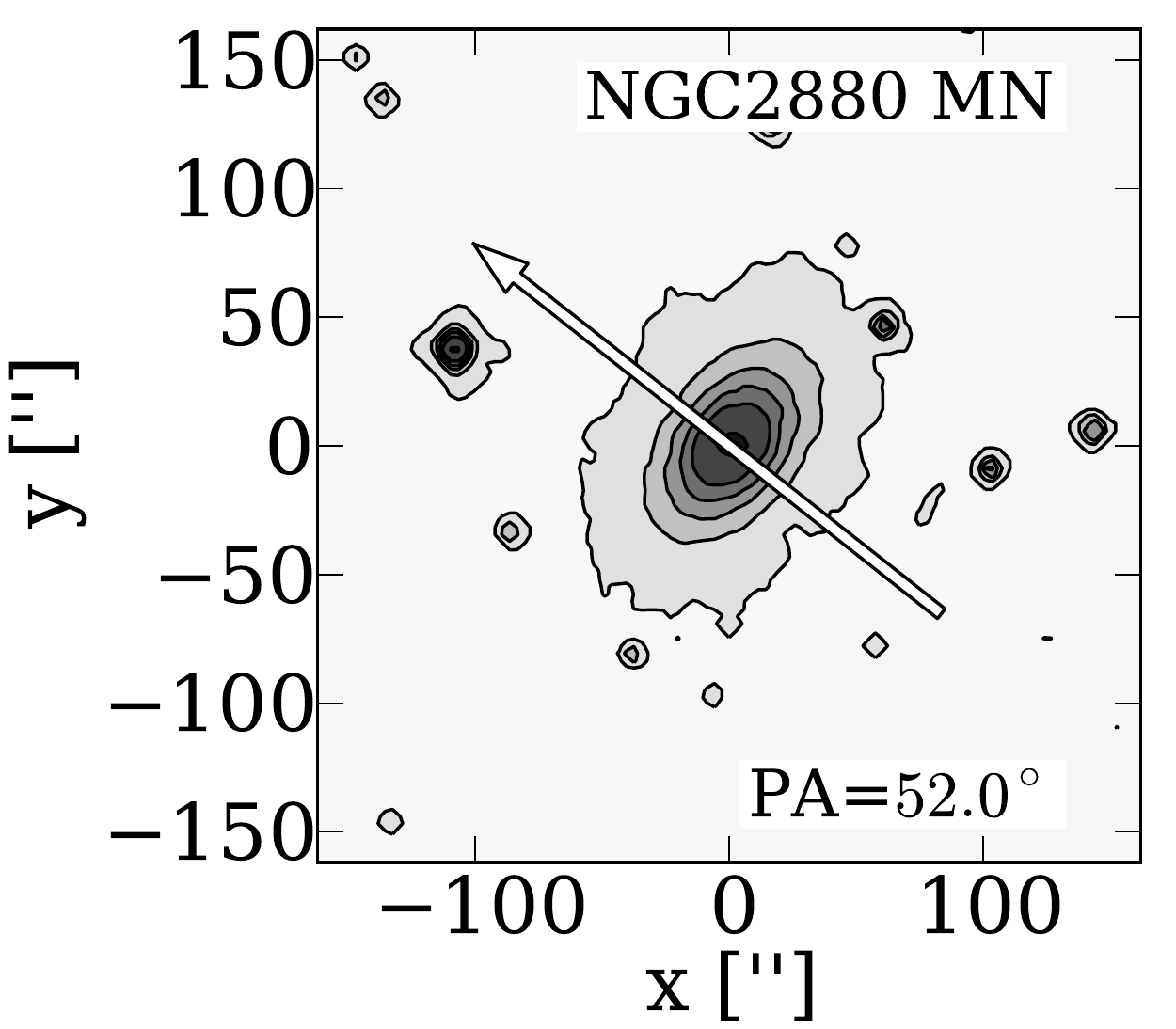}
	\end{minipage} & 
	\includegraphics[viewport=0 50 420 400,width=0.35\textwidth]{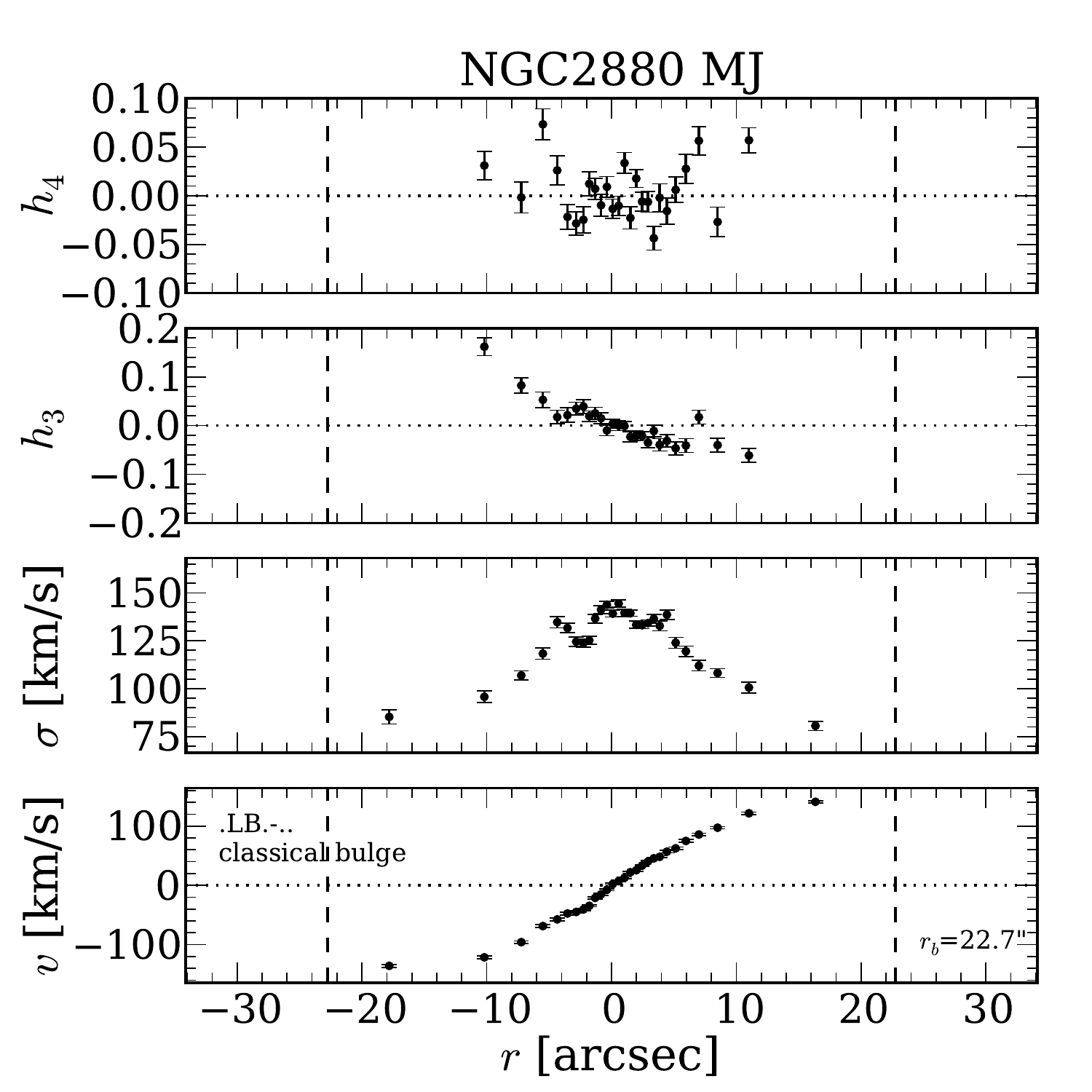} &
	\includegraphics[viewport=0 50 420 400,width=0.35\textwidth]{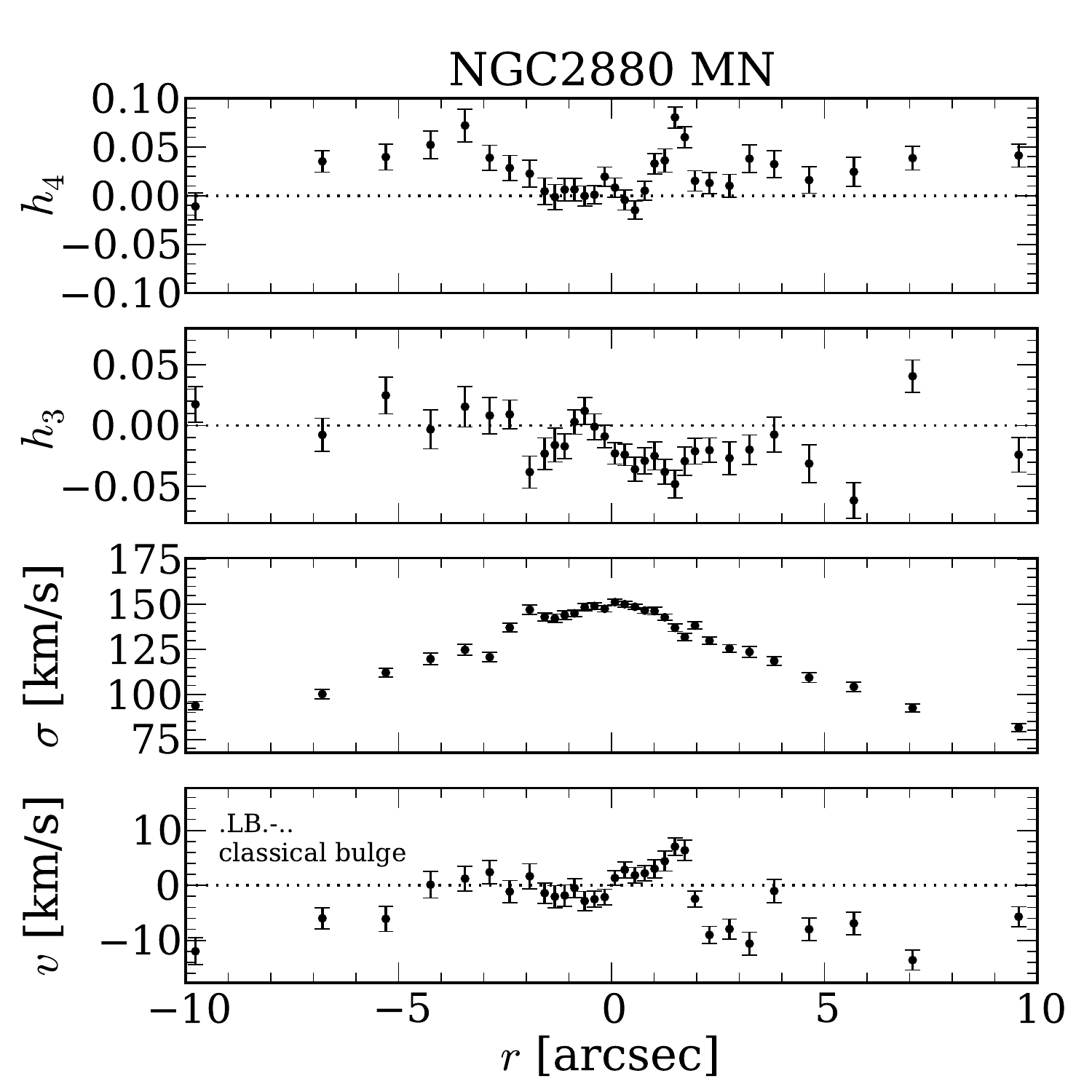}\\
        \end{tabular}
        \end{center}
        \caption{{\it continued --}\small Major and minor axis kinematic profiles for NGC\,2880.}
\end{figure}
\setcounter{figure}{15}
\begin{figure}
        \begin{center}
        \begin{tabular}{lll}
	\begin{minipage}[b]{0.185\textwidth}
	\includegraphics[viewport=0 55 390 400,width=\textwidth]{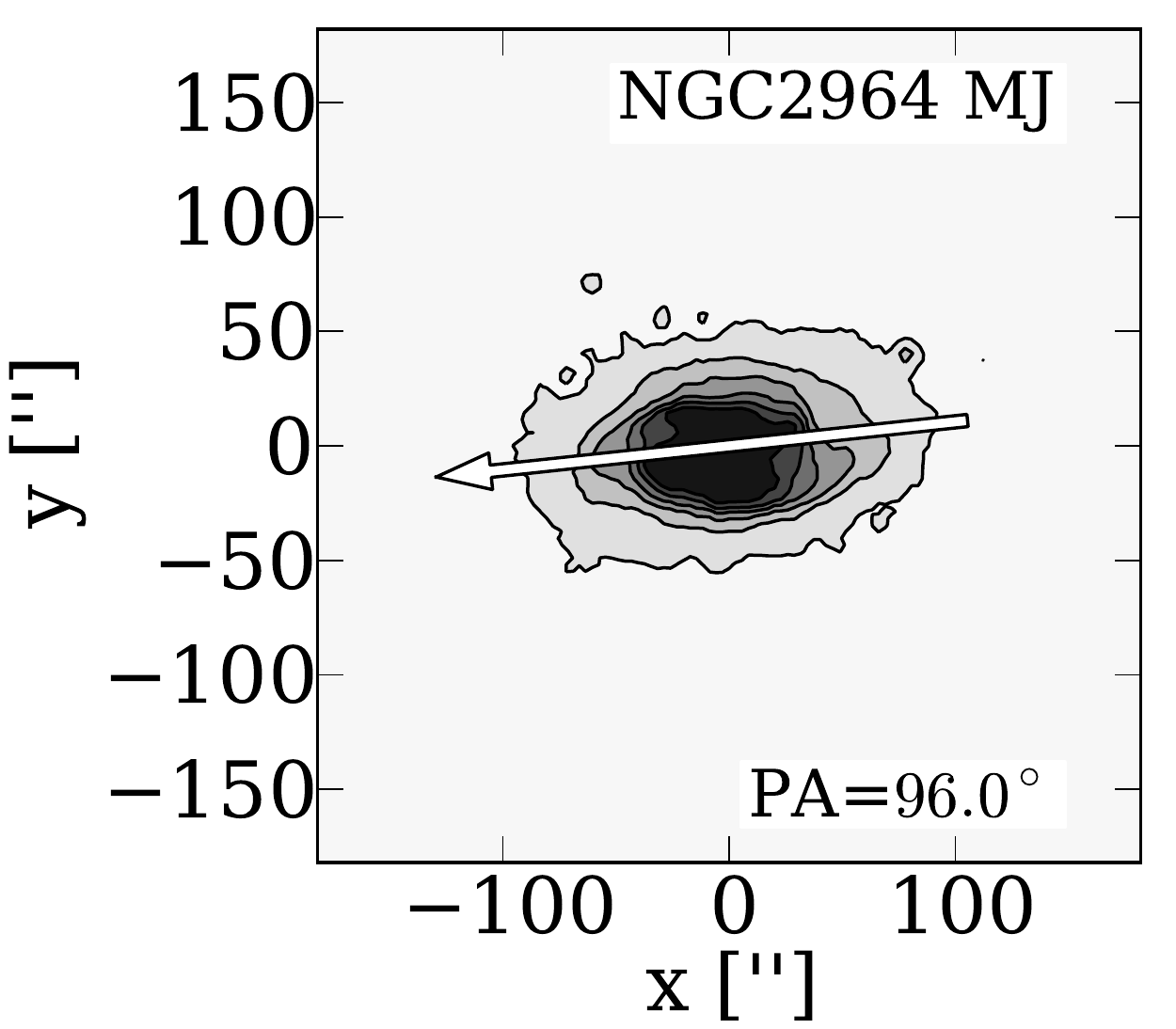} \\
	\includegraphics[viewport=0 55 390 400,width=\textwidth]{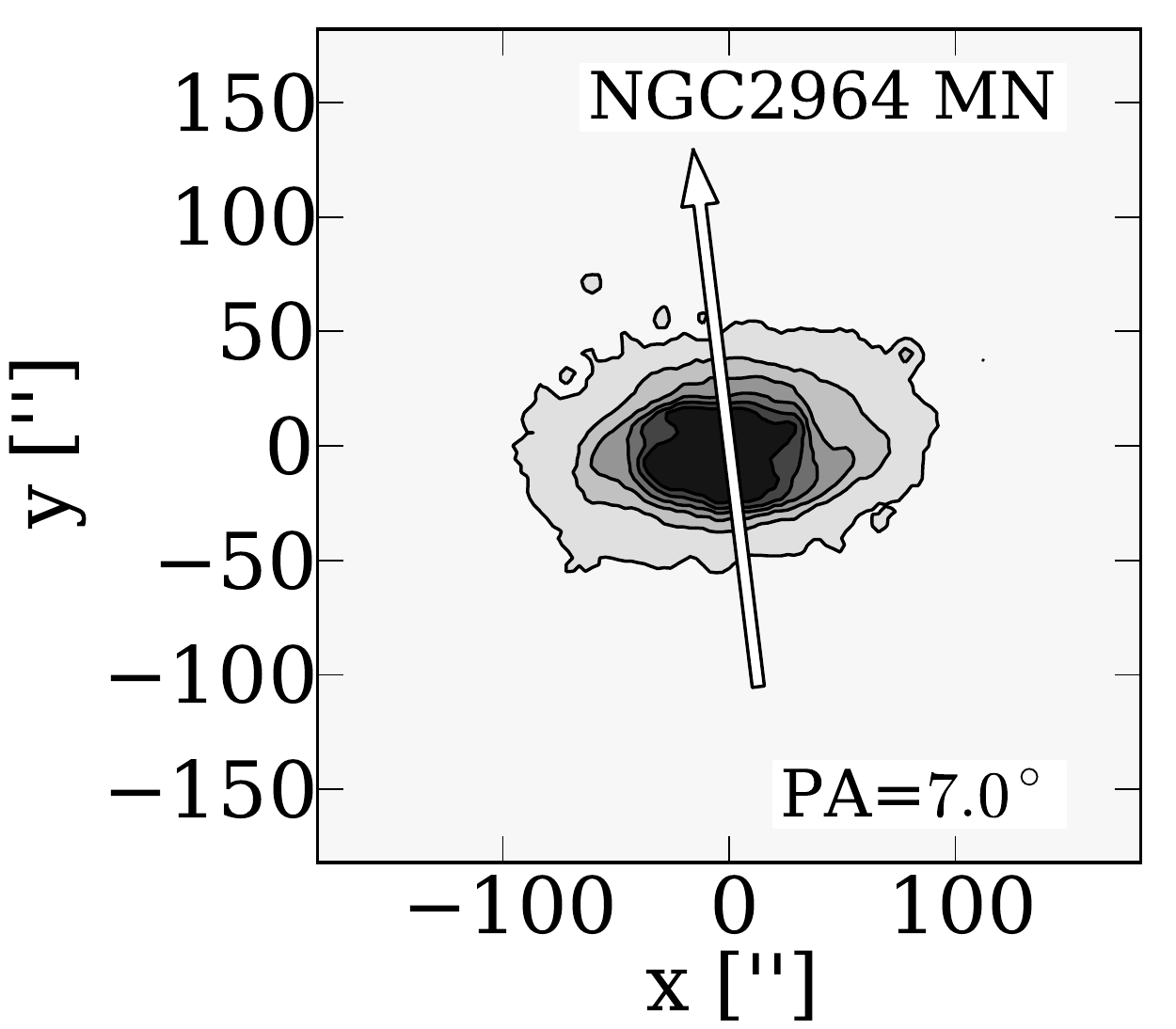}
	\end{minipage} & 
	\includegraphics[viewport=0 50 420 400,width=0.35\textwidth]{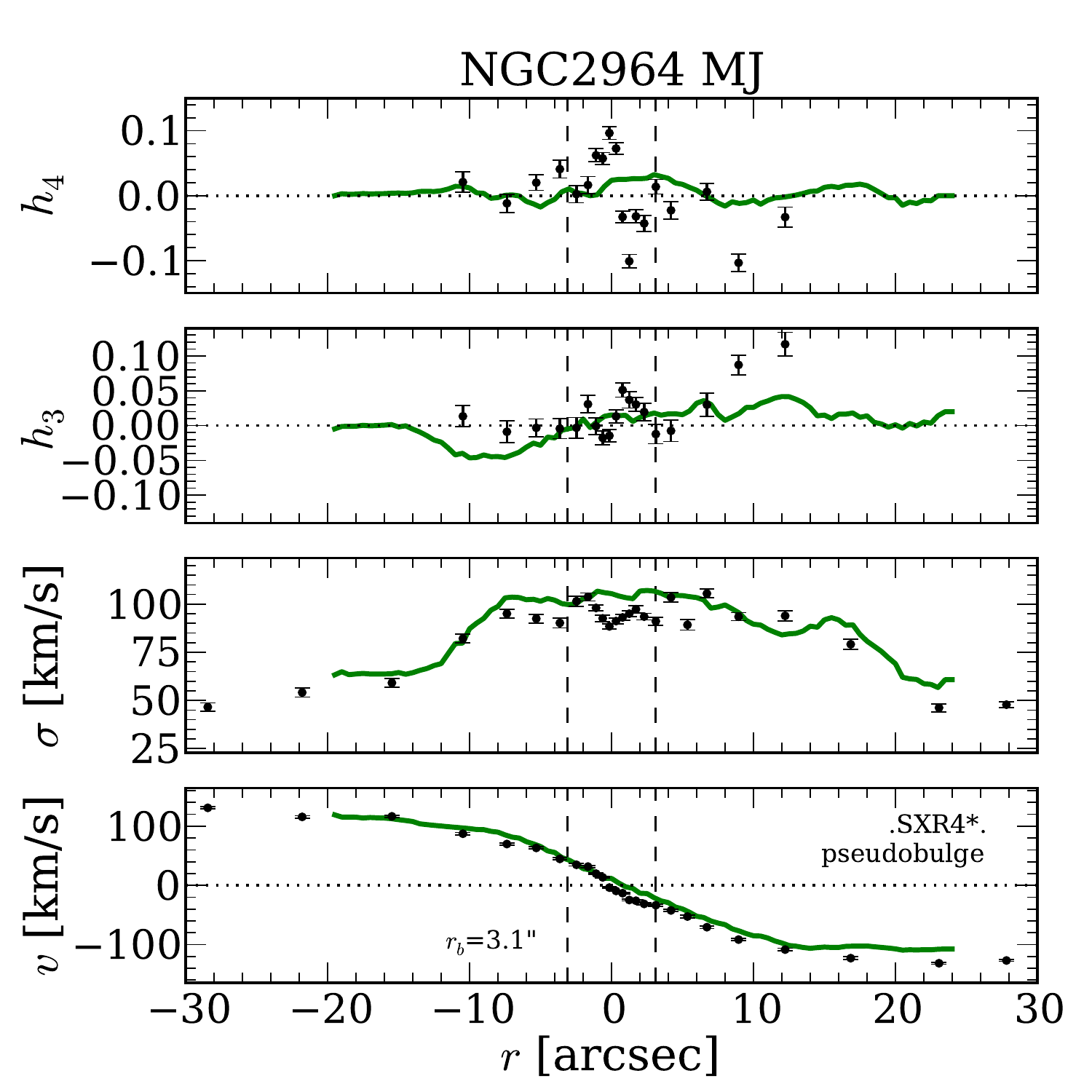} &
	\includegraphics[viewport=0 50 420 400,width=0.35\textwidth]{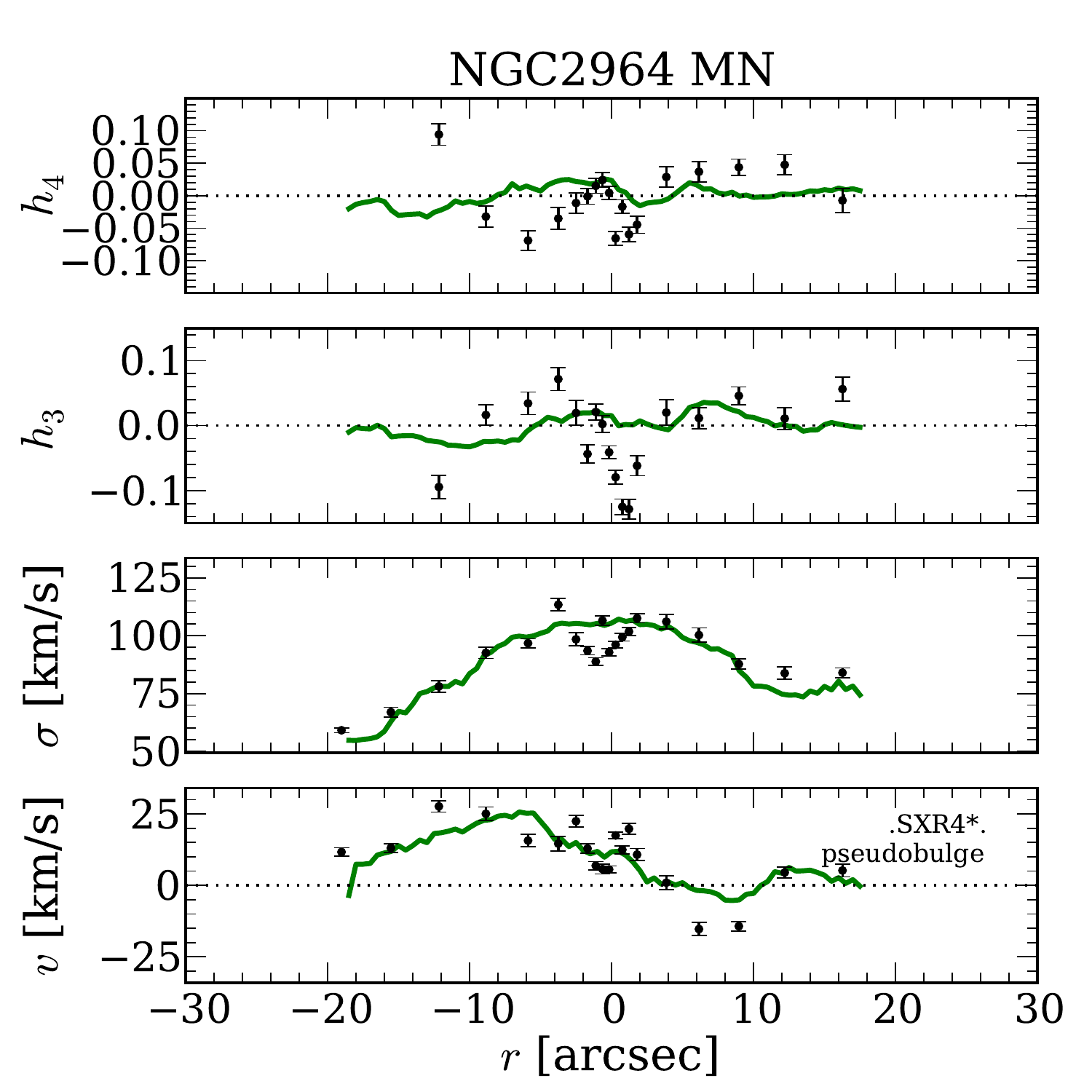}\\
        \end{tabular}
        \end{center}
        \caption{{\it continued --}\small Major and minor axis kinematic profiles for NGC\,2964.
	We plot the SAURON results \citep{Ganda2006} in green.}
\end{figure}
\clearpage
\setcounter{figure}{15}
\begin{figure}
        \begin{center}
        \begin{tabular}{lll}
	\begin{minipage}[b]{0.185\textwidth}
	\includegraphics[viewport=0 55 390 400,width=\textwidth]{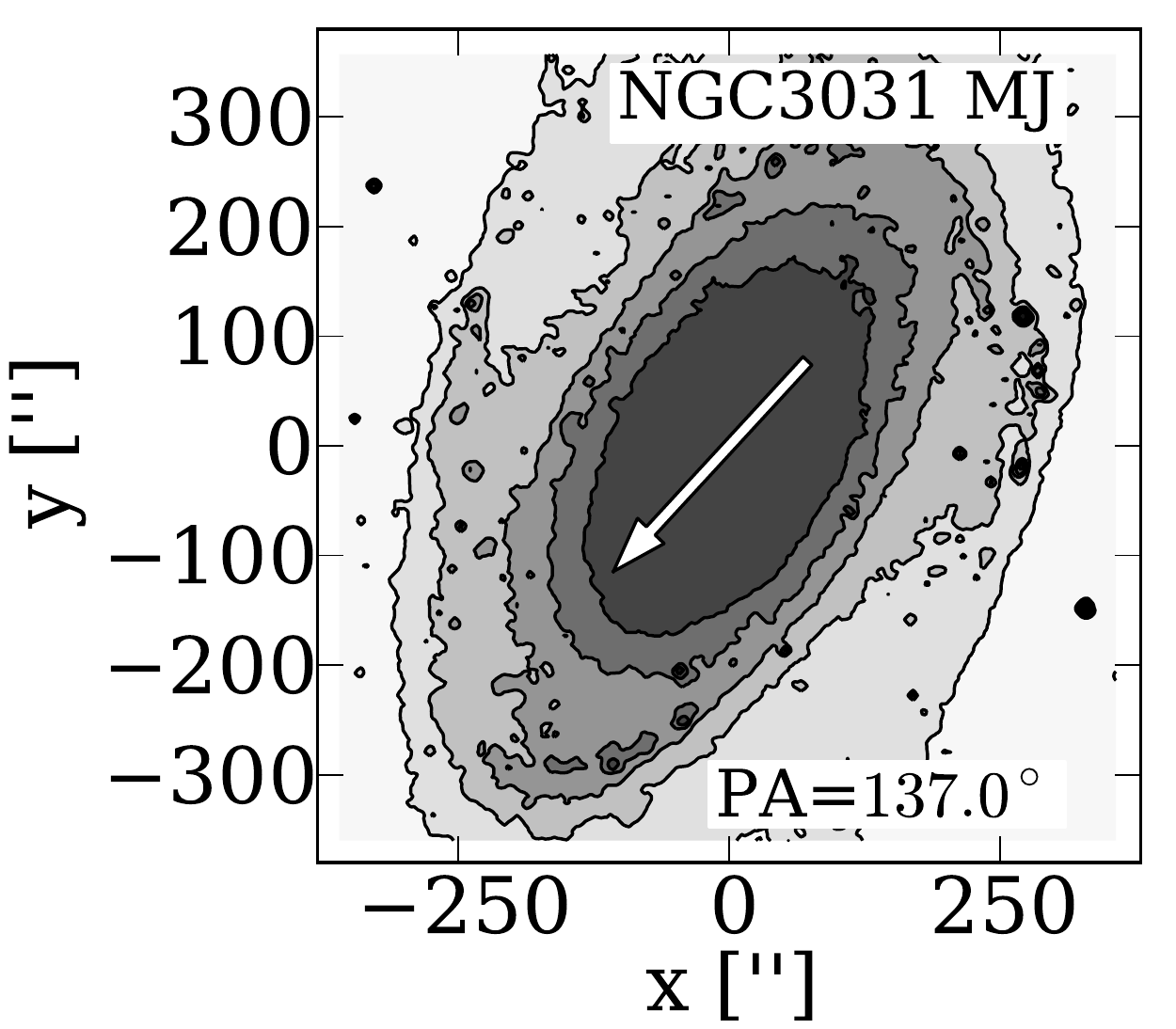} \\
	\includegraphics[viewport=0 55 390 400,width=\textwidth]{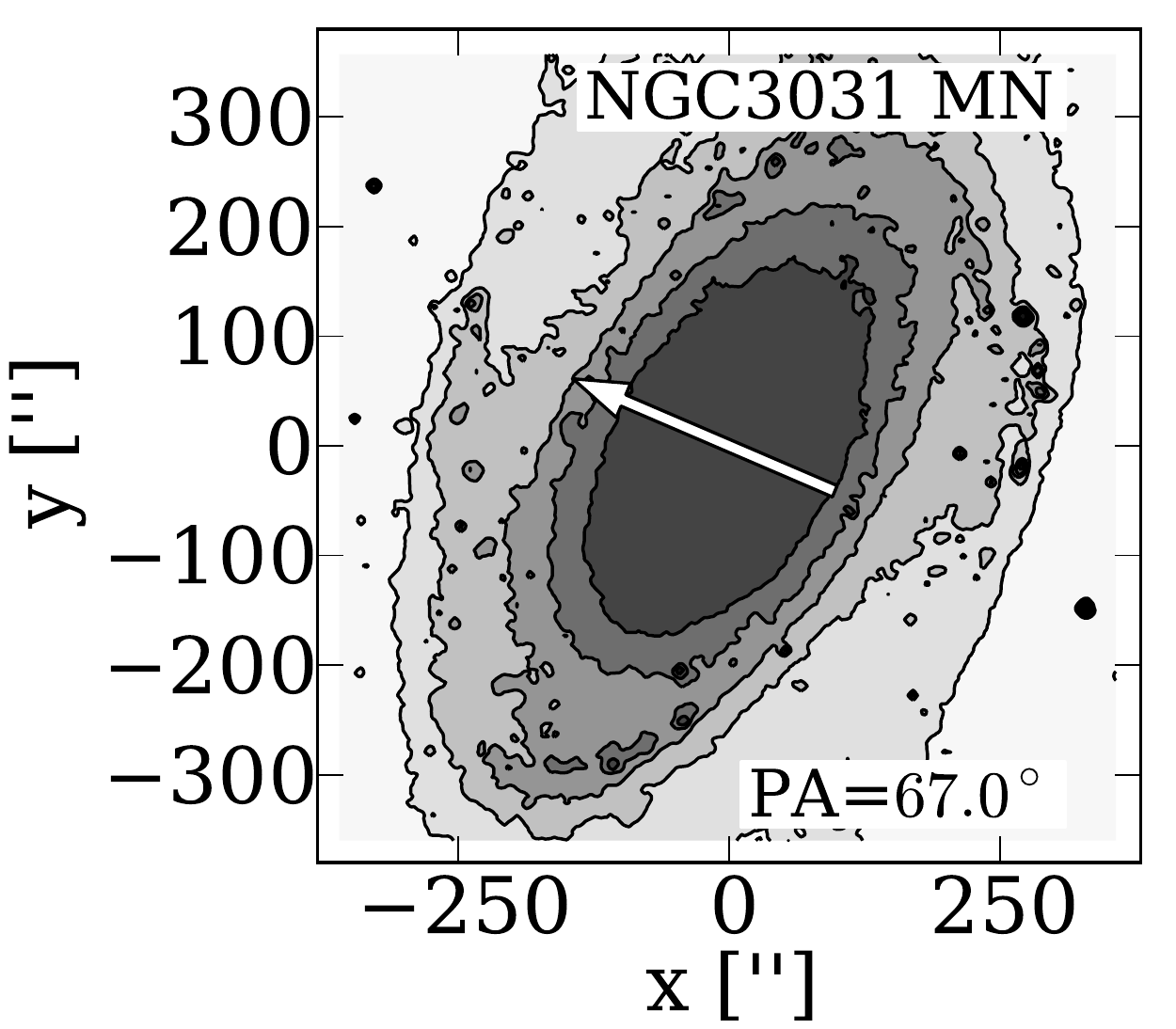}
	\end{minipage} & 
	\includegraphics[viewport=0 50 420 400,width=0.35\textwidth]{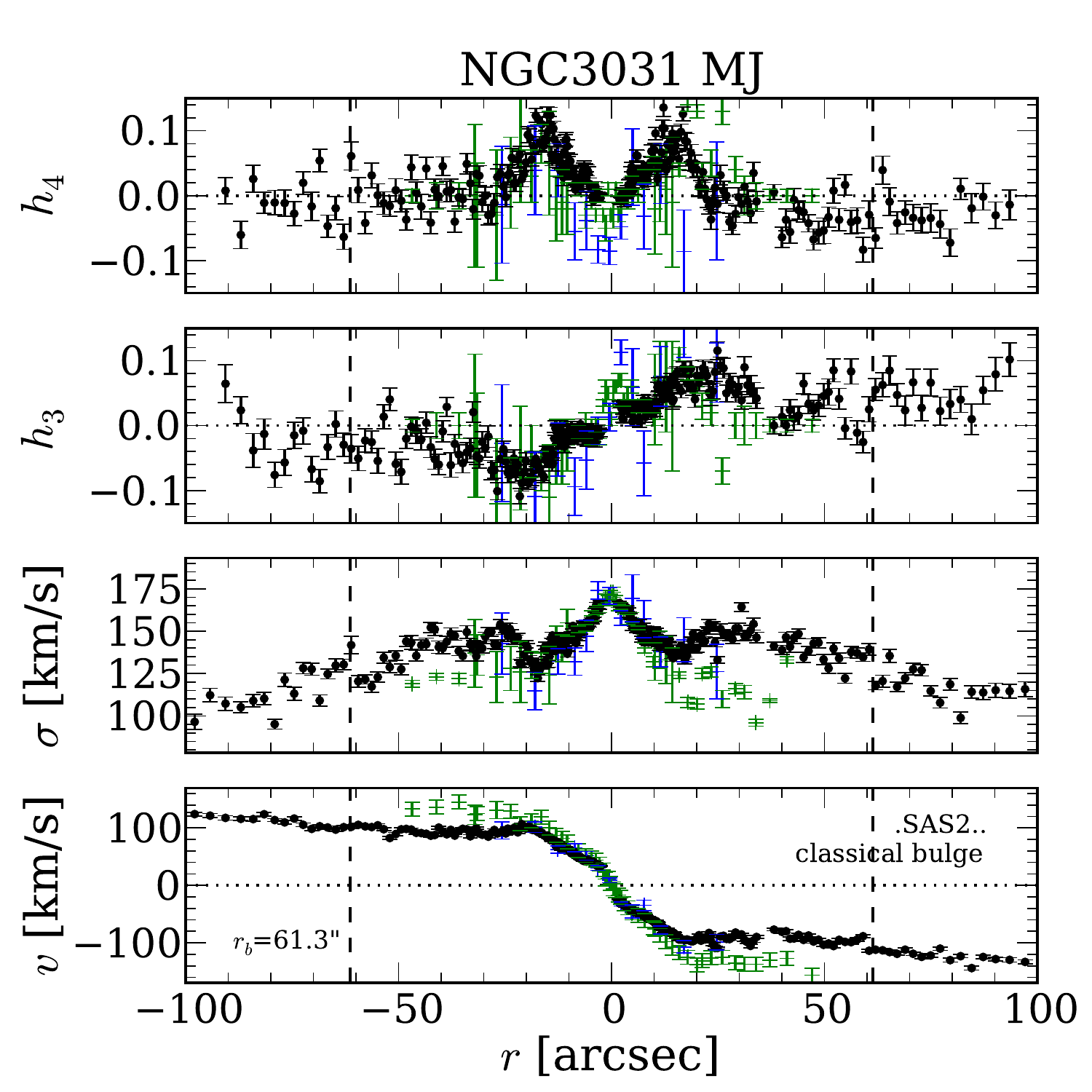} &
	\includegraphics[viewport=0 50 420 400,width=0.35\textwidth]{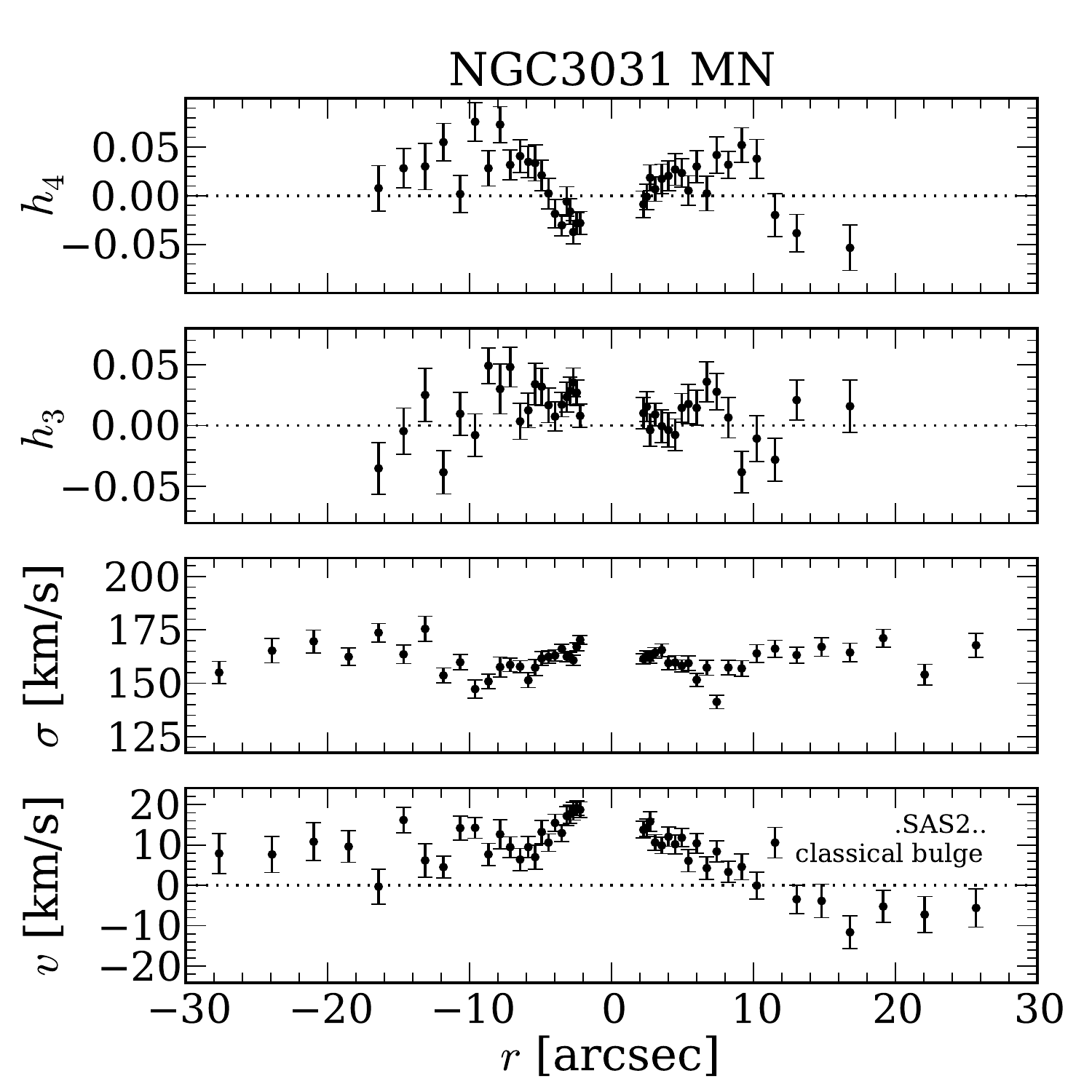}\\
        \end{tabular}
        \end{center}
	\caption{{\it continued --}\small Major and minor axis kinematic profiles for
	NGC\,3031.  Prominent central emission features connected to a liner type
	activity prevent us from deriving the central kinematics ($r \pm 2$ arcseconds)
	reliably. We do not publish moments within the affected radius.  We plot
	results of \citet{Vega-Beltran2001} in green and those of \citet{Bender1994} in
	blue.  The data of \citet{Vega-Beltran2001} were taken with a slit position
	angle of 157\Deg\ whereas we observed at a position angle of 137\Deg\ which is
	responsible
	for the offset seen in velocity. 
	}
\end{figure}
\setcounter{figure}{15}
\begin{figure}
        \begin{center}
        \begin{tabular}{lll}
	\begin{minipage}[b]{0.185\textwidth}
	\includegraphics[viewport=0 55 390 400,width=\textwidth]{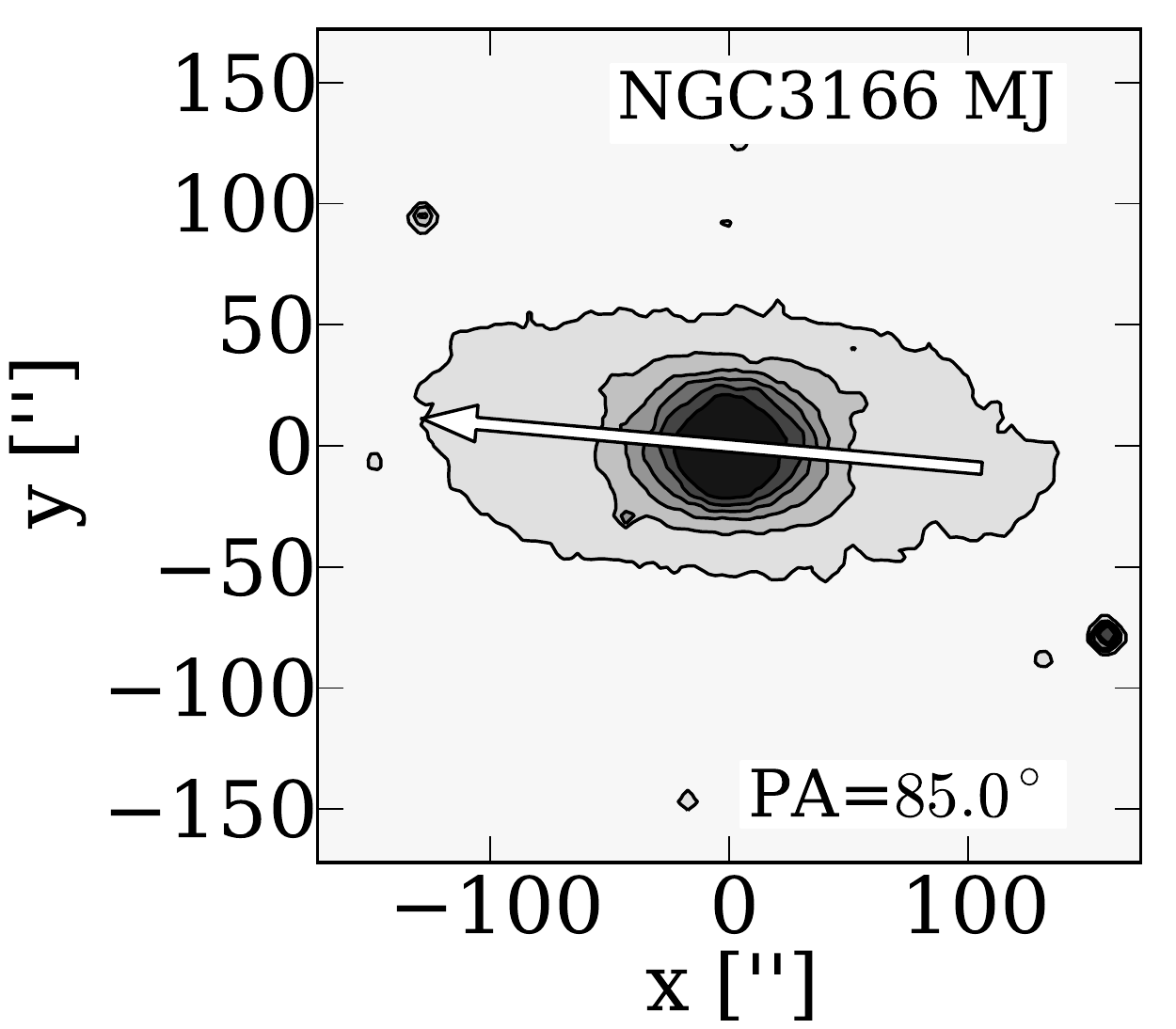} \\
	\includegraphics[viewport=0 55 390 400,width=\textwidth]{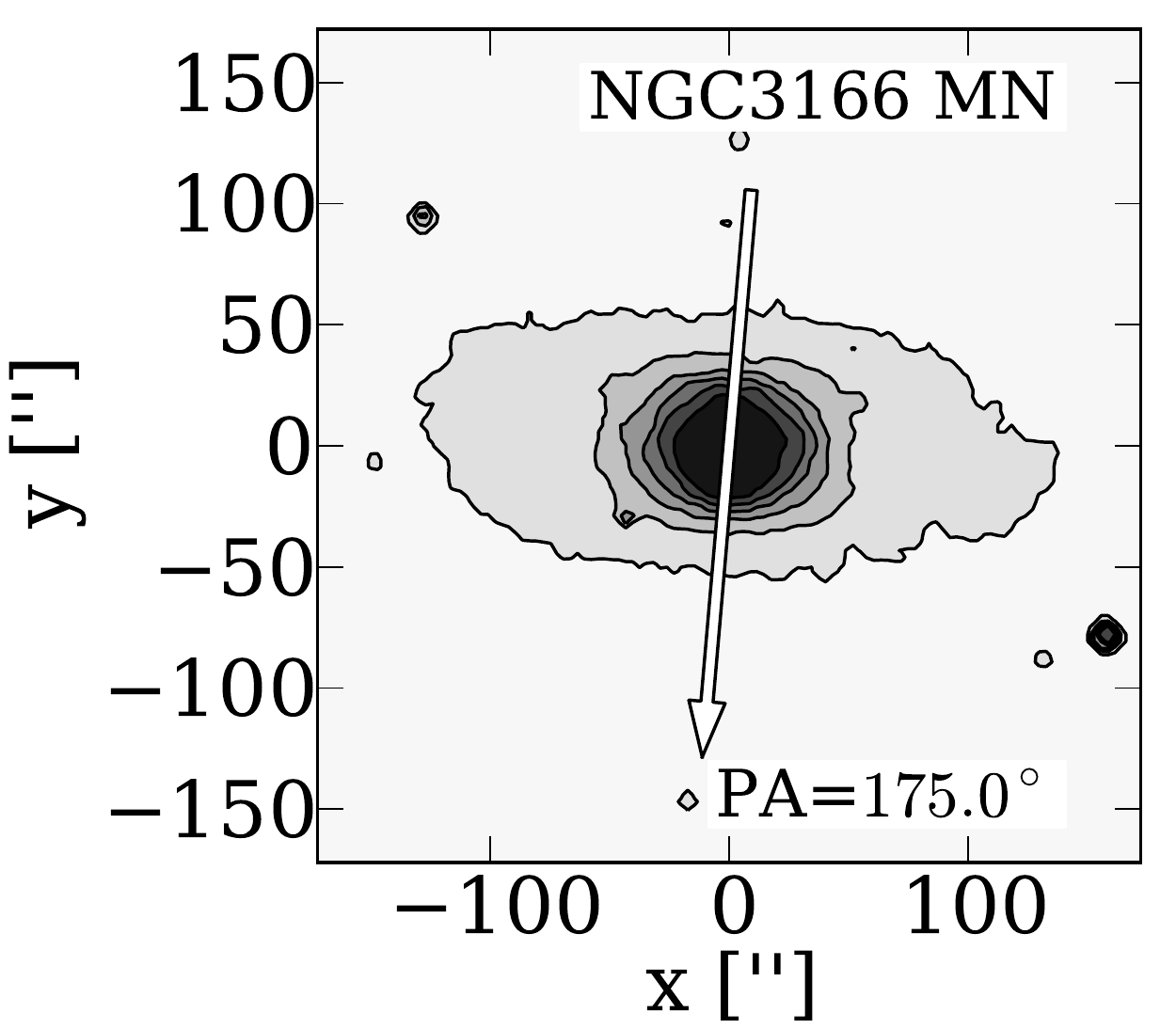}
	\end{minipage} & 
	\includegraphics[viewport=0 50 420 400,width=0.35\textwidth]{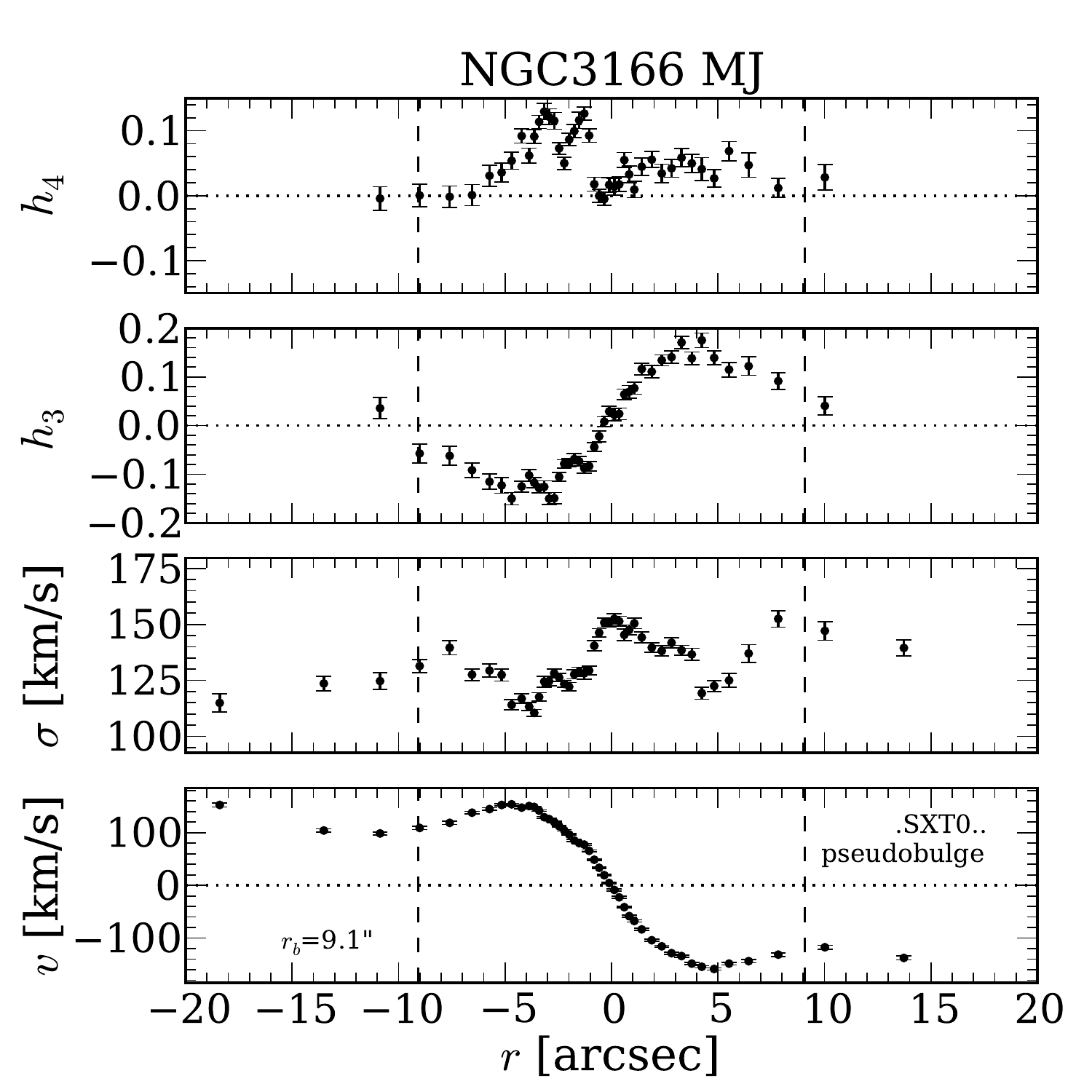} &
	\includegraphics[viewport=0 50 420 400,width=0.35\textwidth]{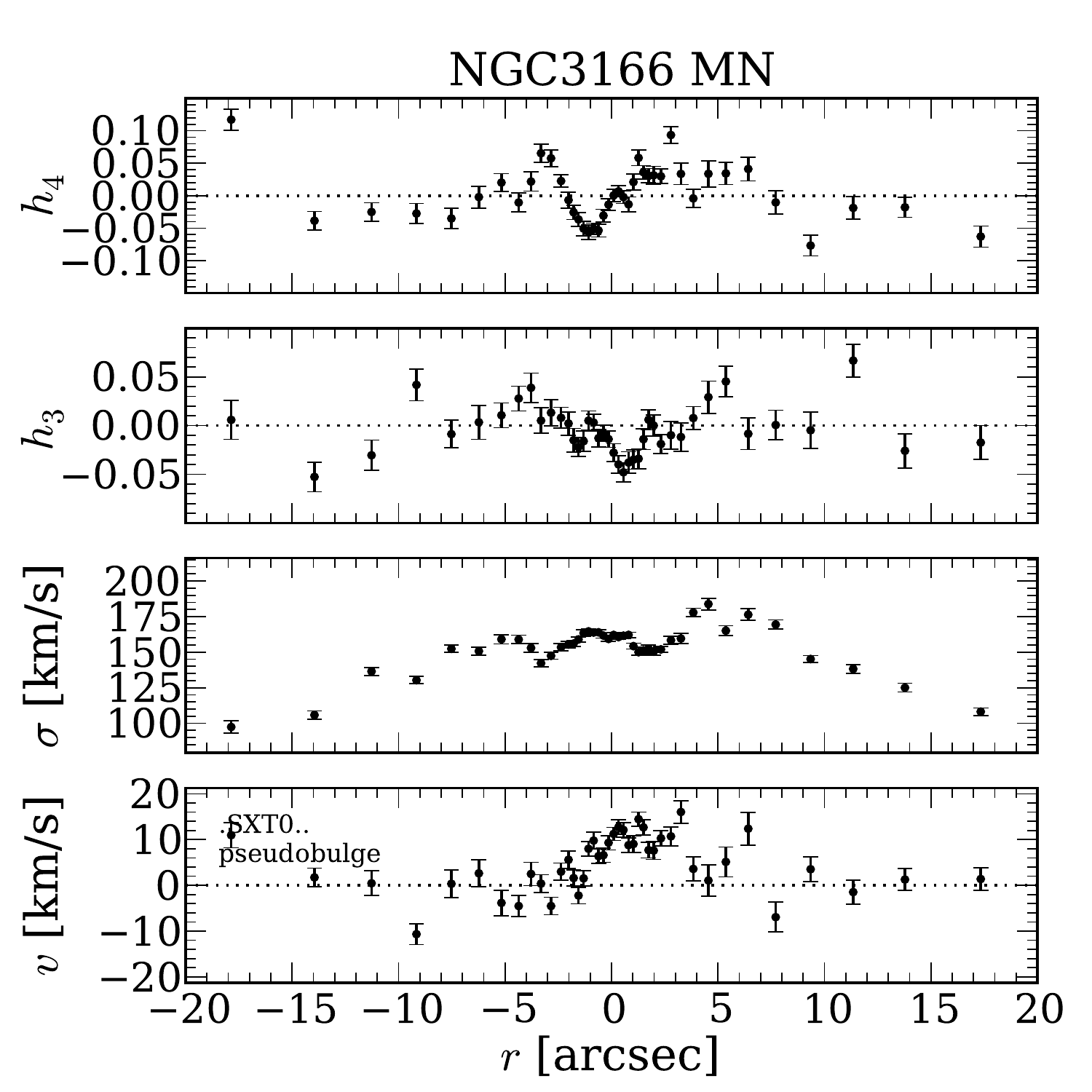}\\
        \end{tabular}
        \end{center}
        \caption{{\it continued --}\small Major and minor axis kinematic profiles for NGC\,3166.}
\end{figure}
\setcounter{figure}{15}
\begin{figure}
        \begin{center}
        \begin{tabular}{lll}
	\begin{minipage}[b]{0.185\textwidth}
	\includegraphics[viewport=0 55 390 400,width=\textwidth]{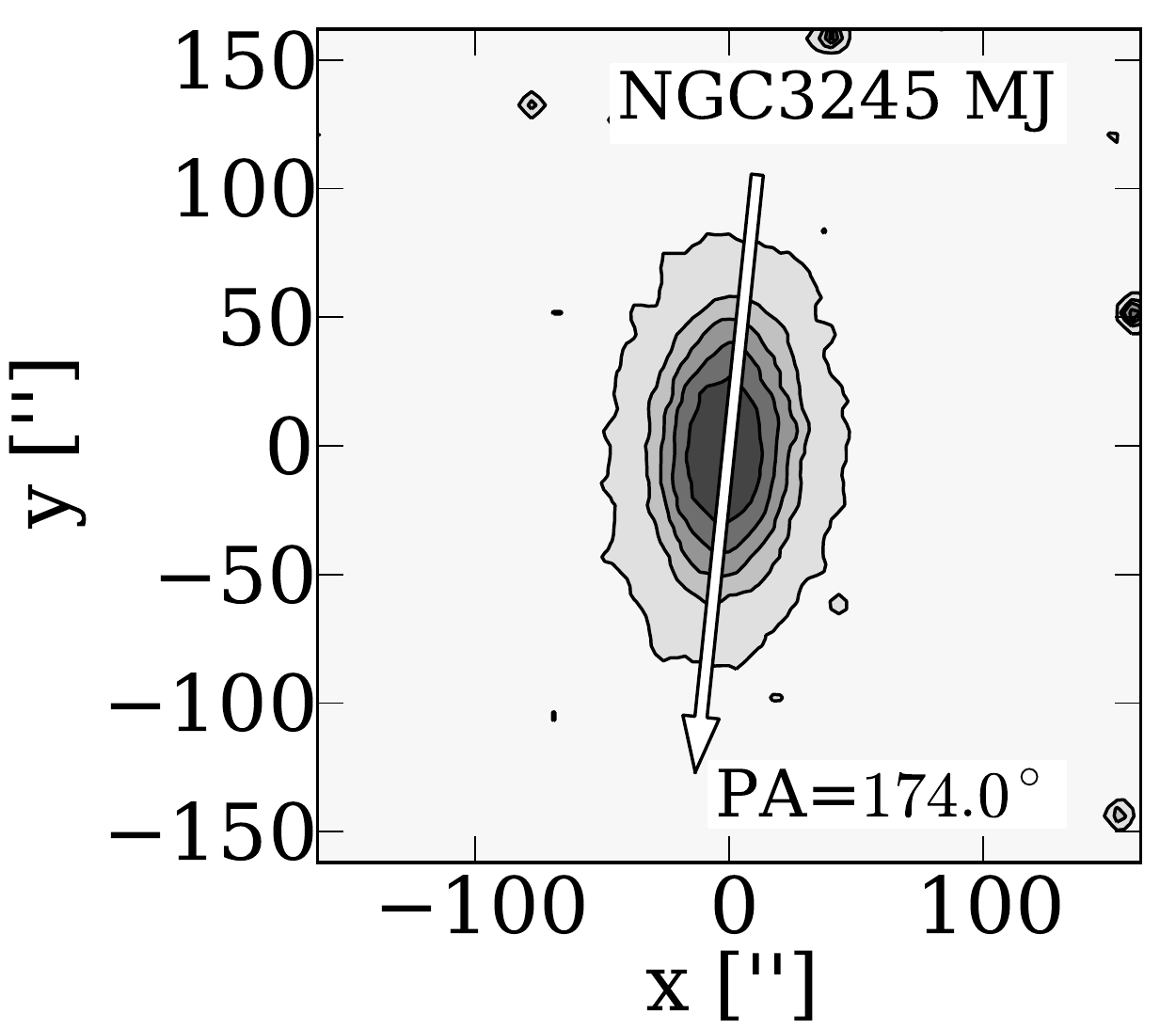} \\
	\includegraphics[viewport=0 55 390 400,width=\textwidth]{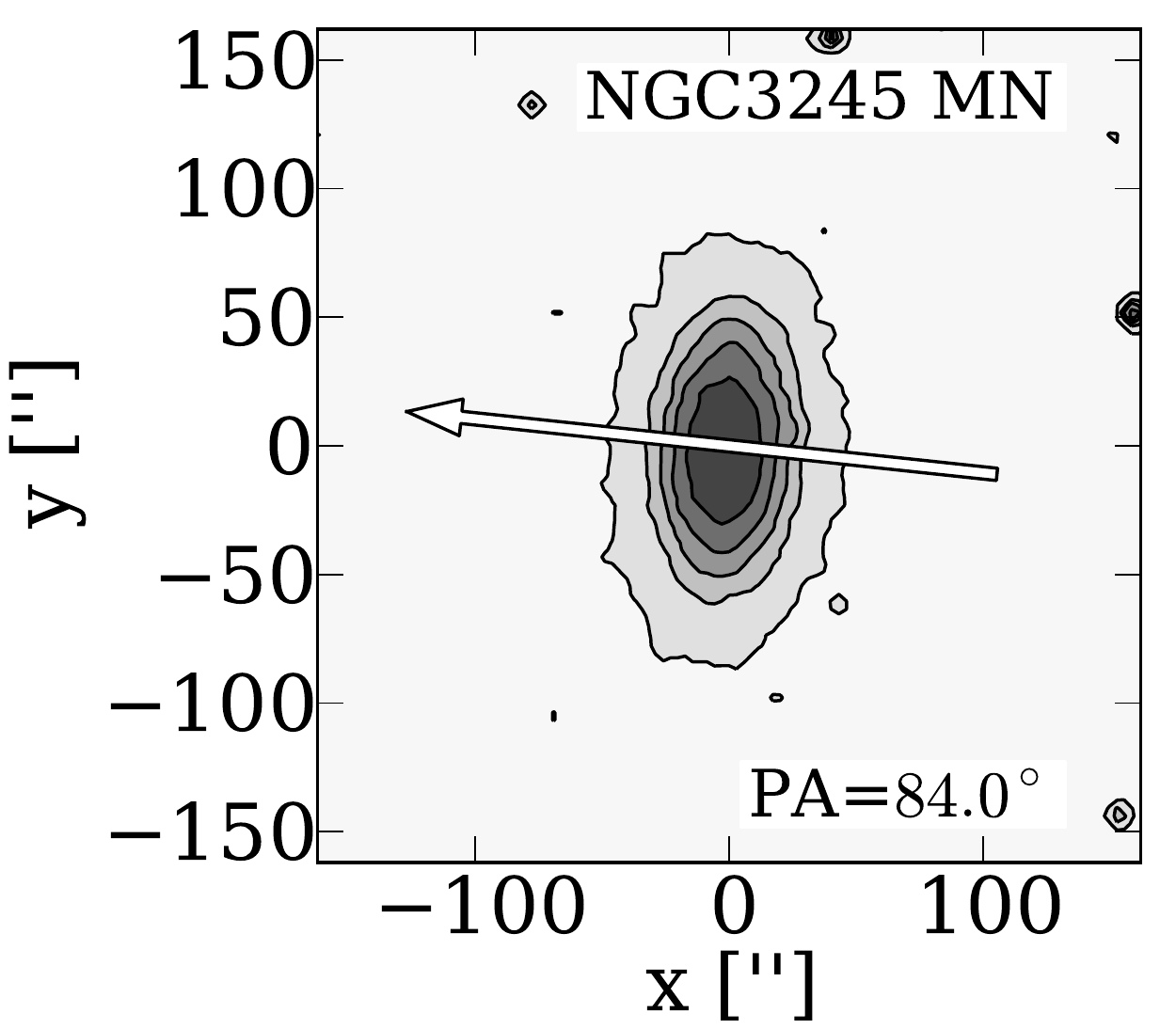}
	\end{minipage} & 
	\includegraphics[viewport=0 50 420 400,width=0.35\textwidth]{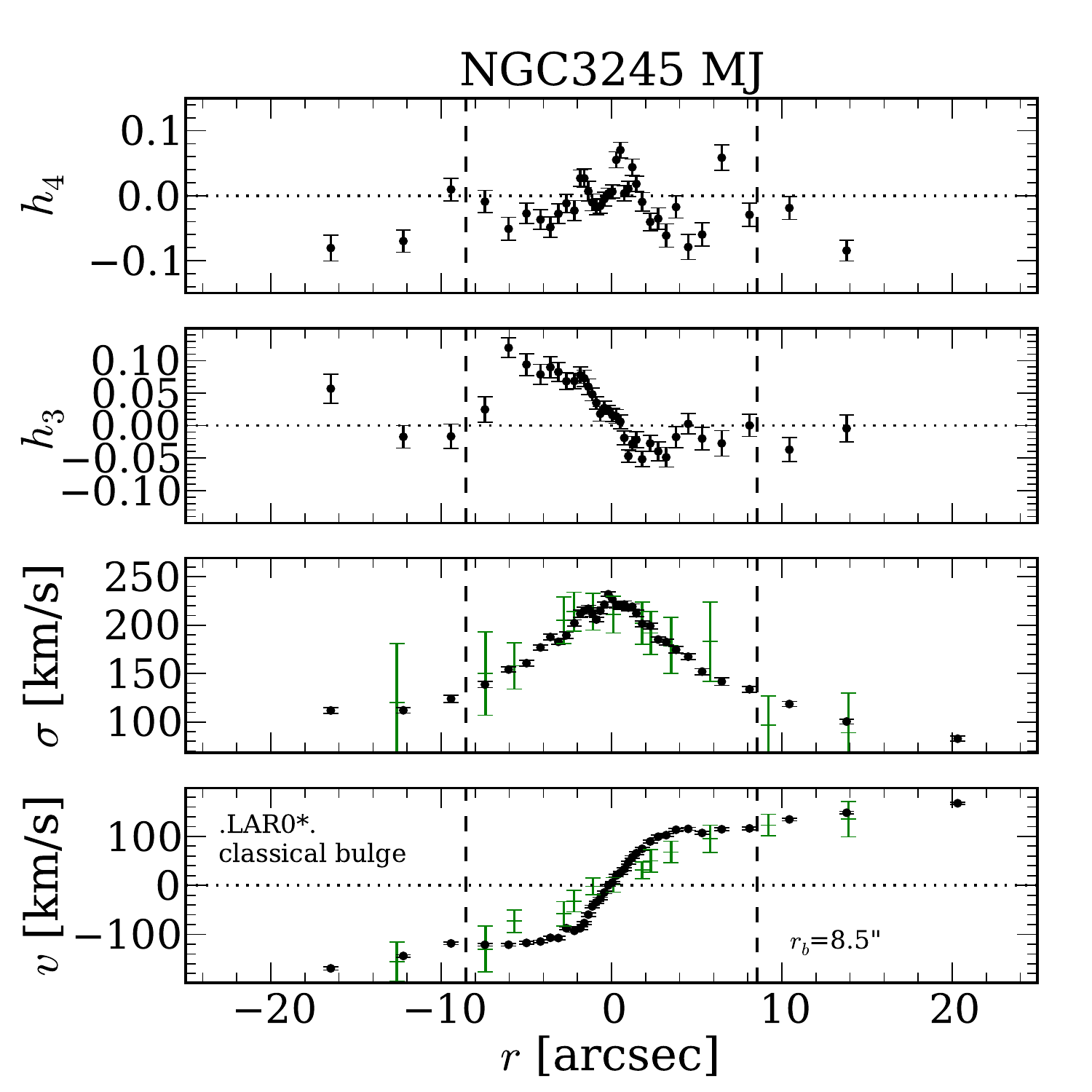} &
	\includegraphics[viewport=0 50 420 400,width=0.35\textwidth]{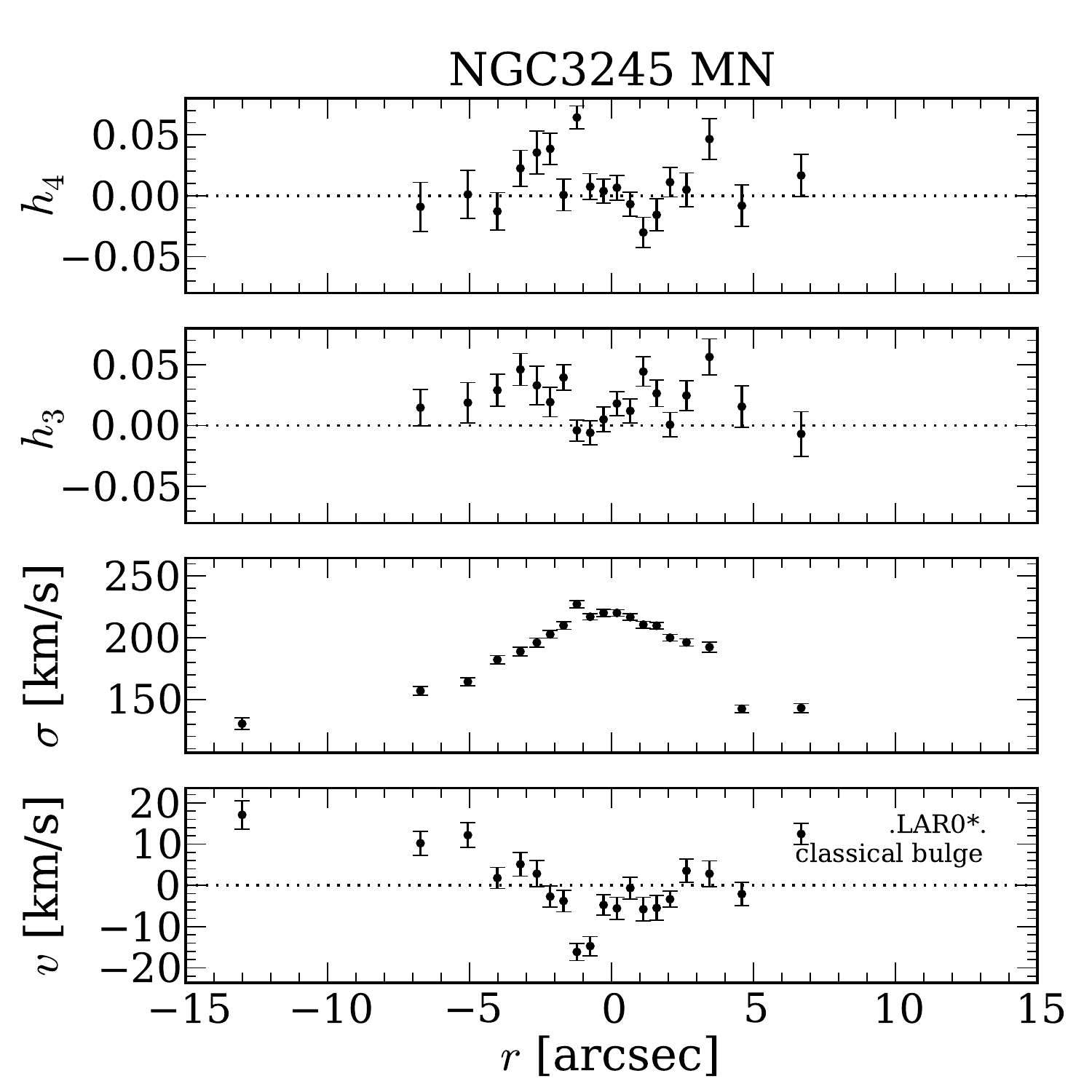}\\
        \end{tabular}
        \end{center}
        \caption{{\it continued --}\small Major and minor axis kinematic profiles for NGC\,3245.
	We plot results of \citet{SP98} in green.}
\end{figure}
\setcounter{figure}{15}
\begin{figure}
        \begin{center}
        \begin{tabular}{lll}
	\begin{minipage}[b]{0.185\textwidth}
	\includegraphics[viewport=0 55 390 400,width=\textwidth]{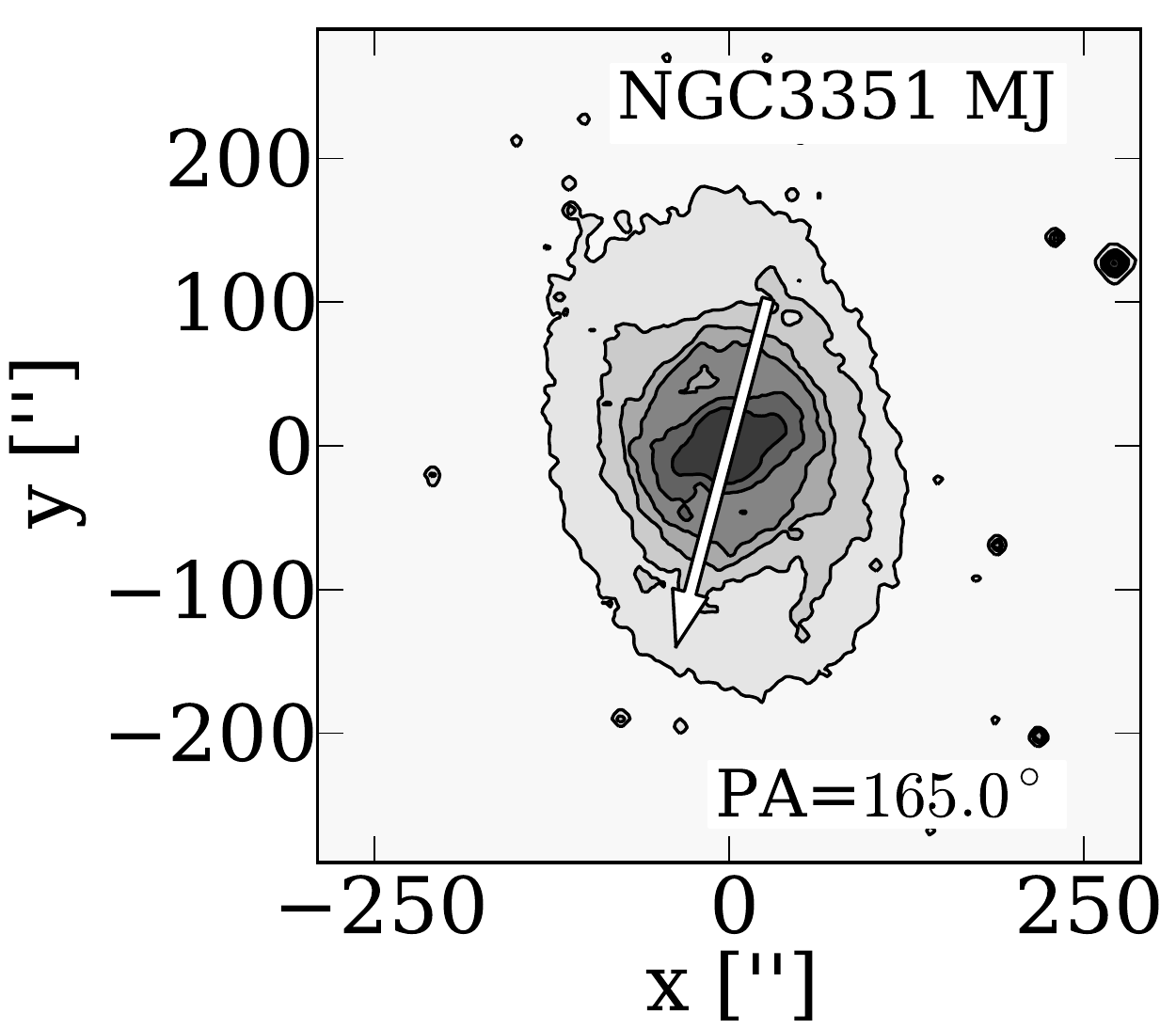} \\
	\includegraphics[viewport=0 55 390 400,width=\textwidth]{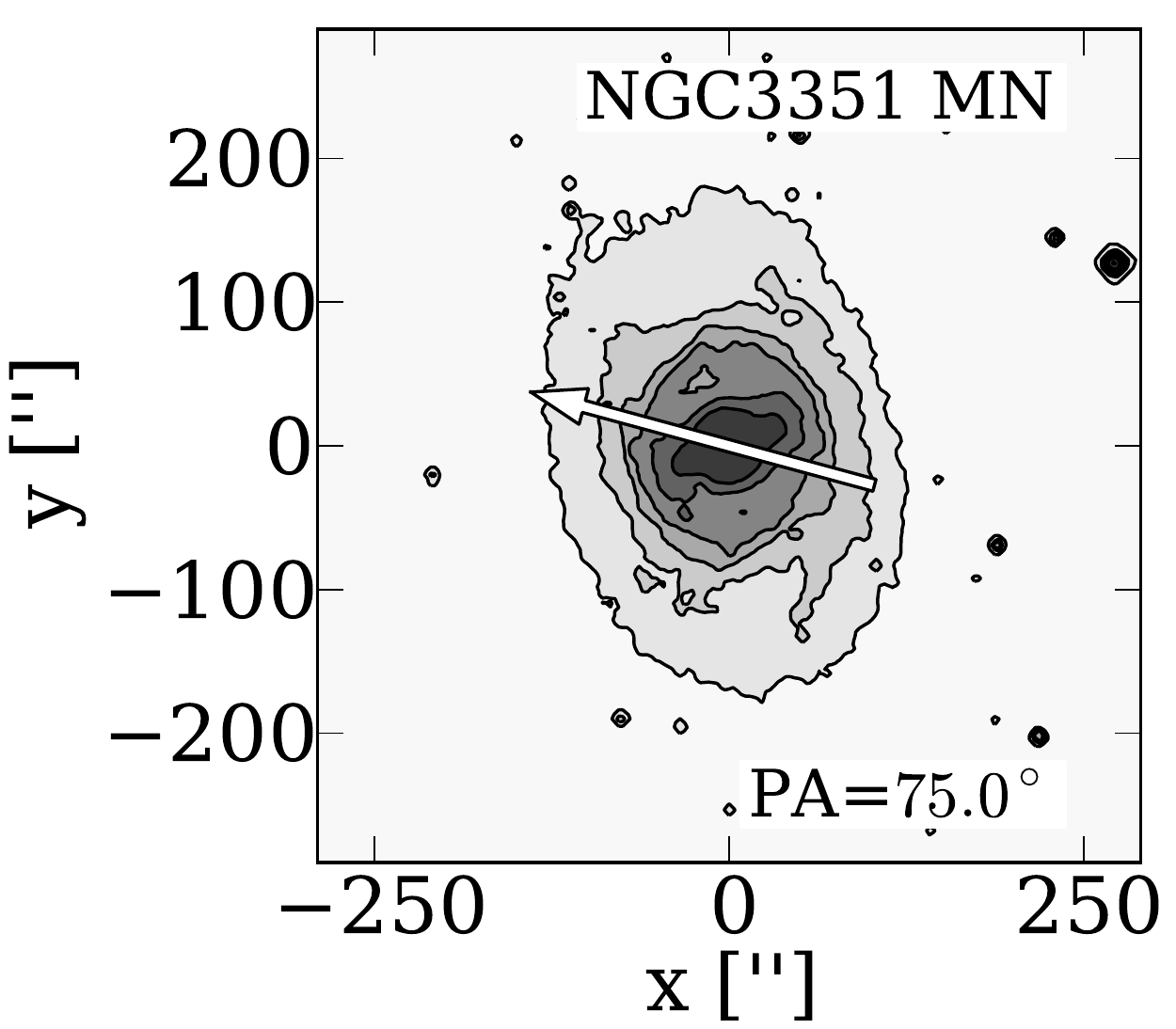}
	\end{minipage} & 
	\includegraphics[viewport=0 50 420 400,width=0.35\textwidth]{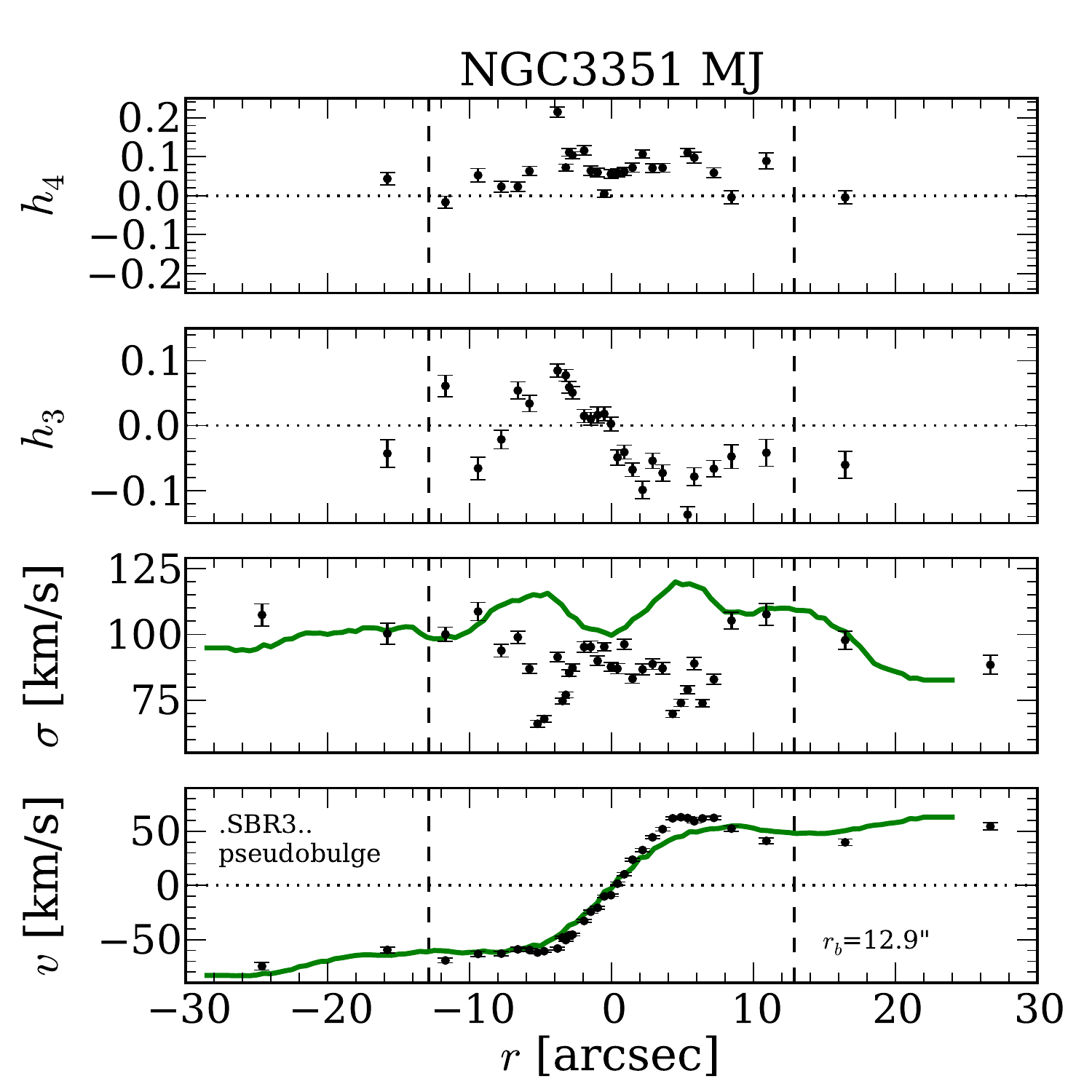} &
	\includegraphics[viewport=0 50 420 400,width=0.35\textwidth]{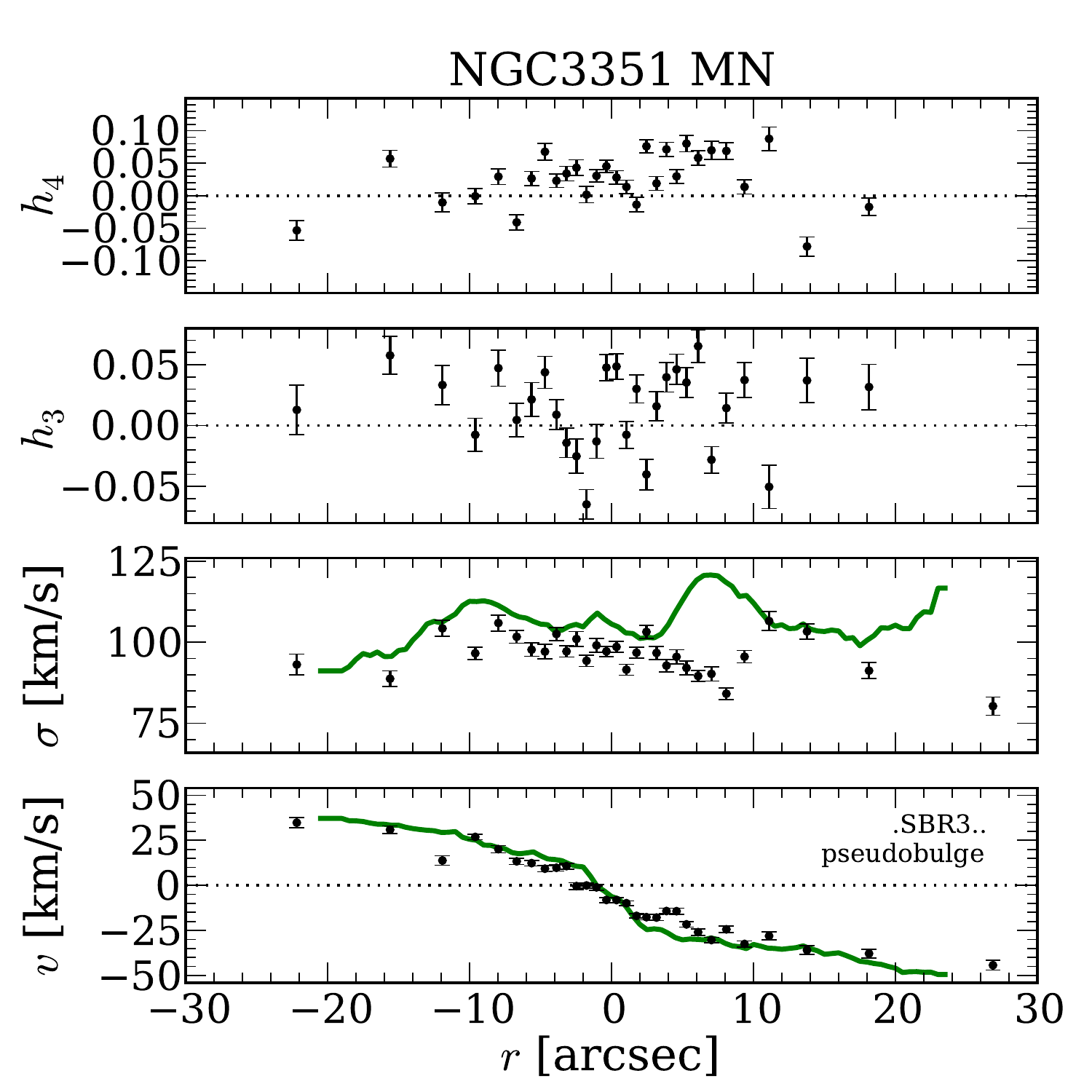}\\
        \end{tabular}
        \end{center}
        \caption{{\it continued --}\small Major and minor axis kinematic profiles for NGC\,3351. We plot
	SAURON results of \citet{Dumas2007} in green.}
\end{figure}
\setcounter{figure}{15}
\begin{figure}
        \begin{center}
        \begin{tabular}{lll}
	\begin{minipage}[b]{0.185\textwidth}
	\includegraphics[viewport=0 55 390 400,width=\textwidth]{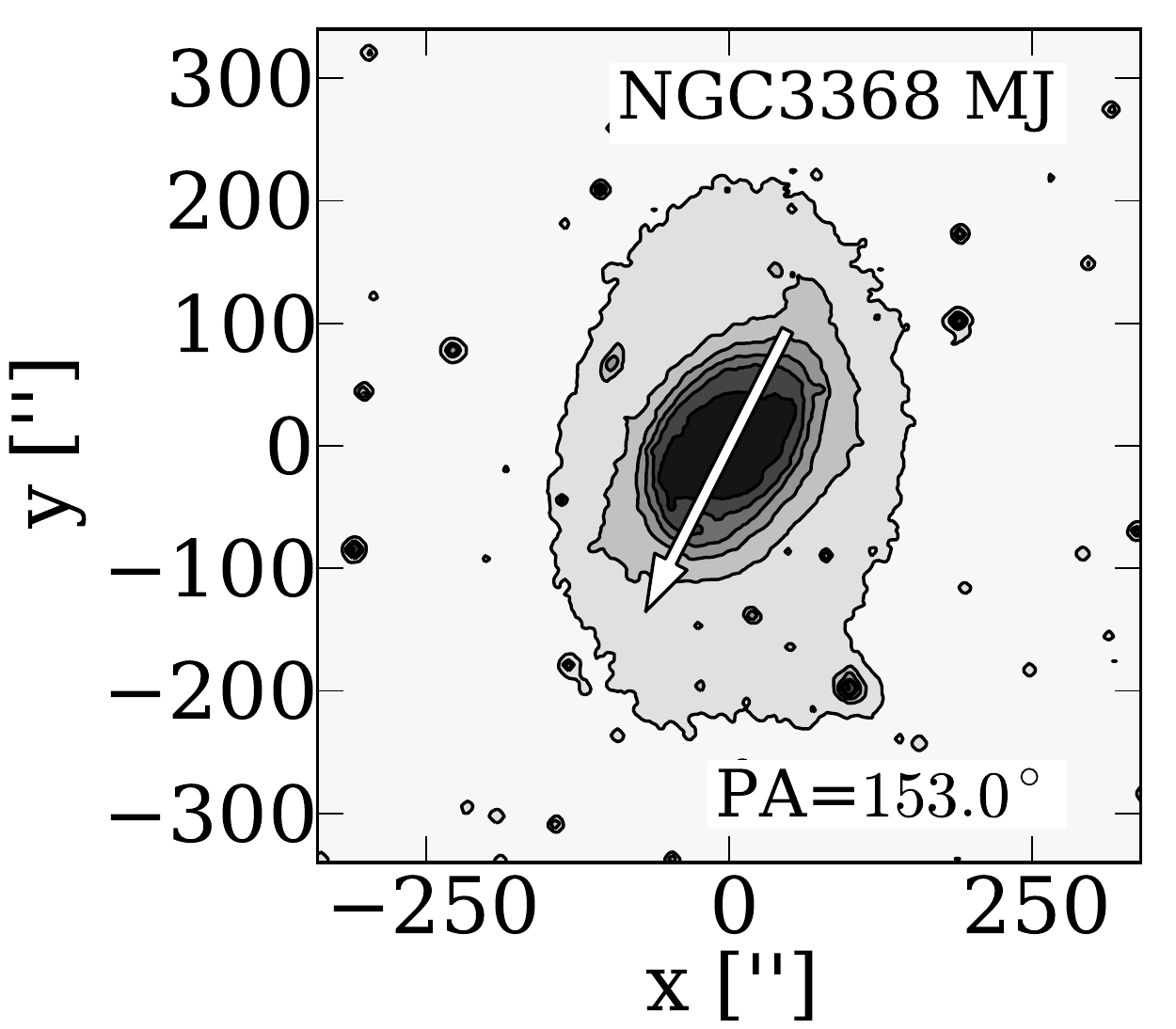}\\
	\includegraphics[viewport=0 55 390 400,width=\textwidth]{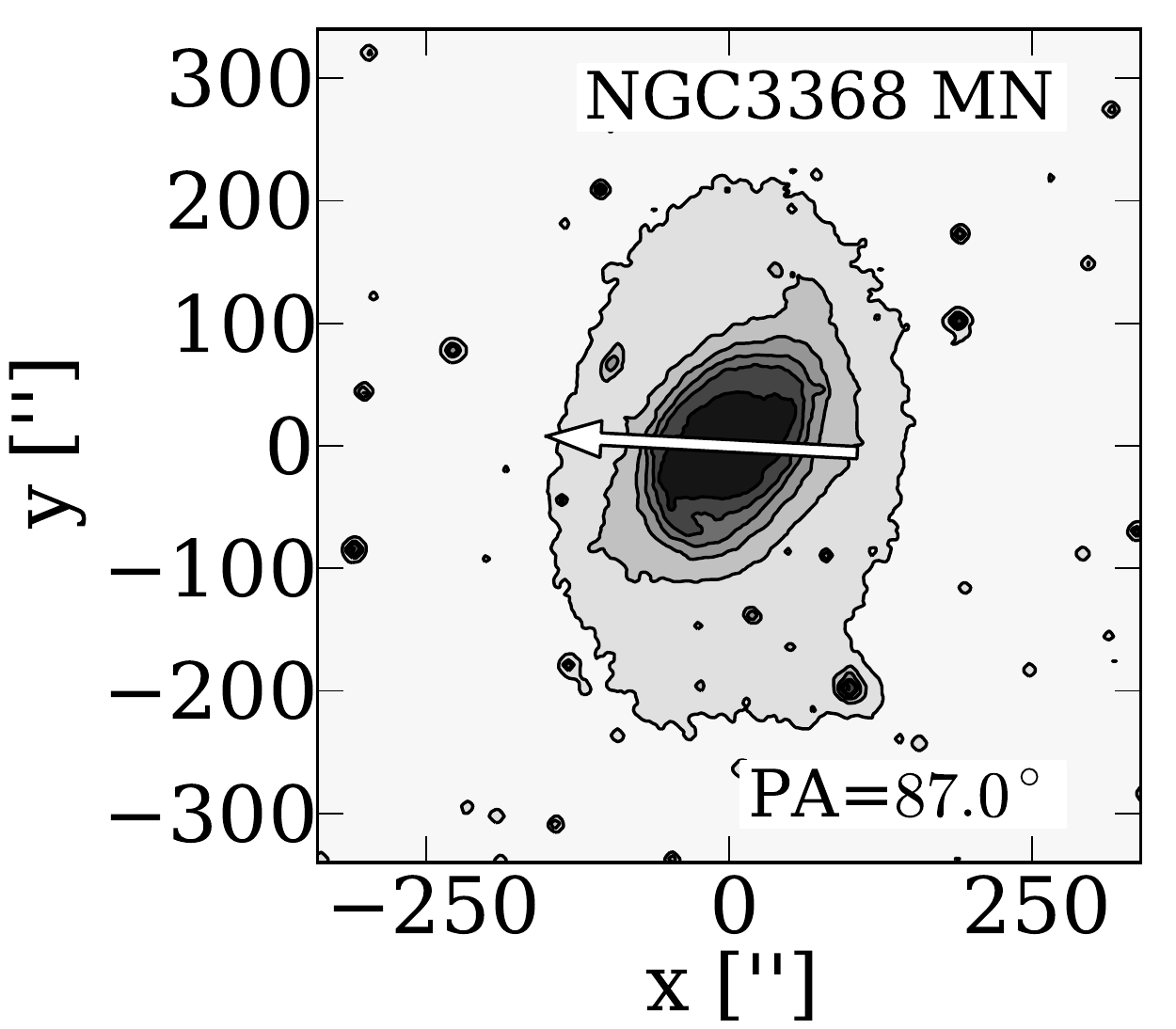}
	\end{minipage} & 
	\includegraphics[viewport=0 50 420 400,width=0.35\textwidth]{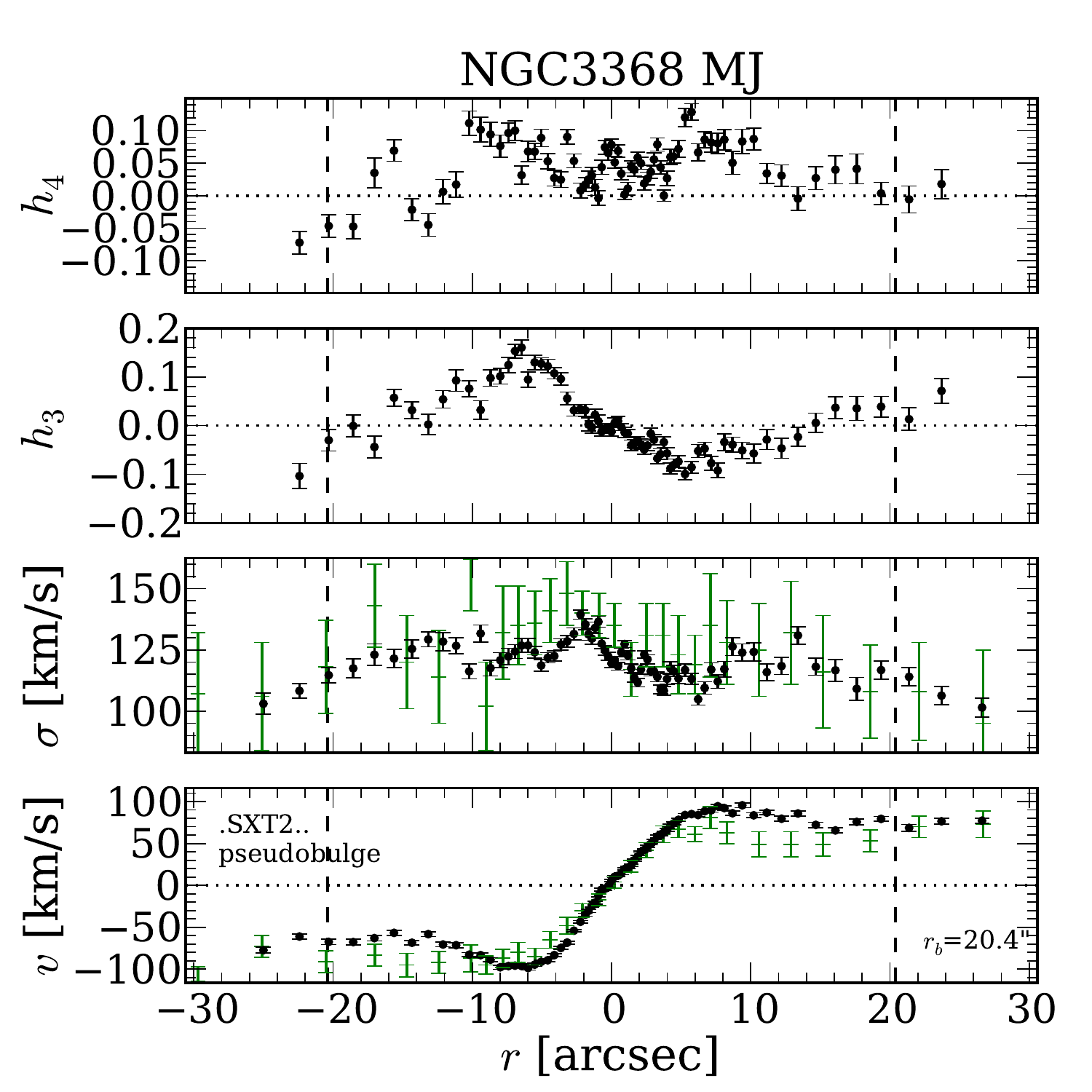} &
	\includegraphics[viewport=0 50 420 400,width=0.35\textwidth]{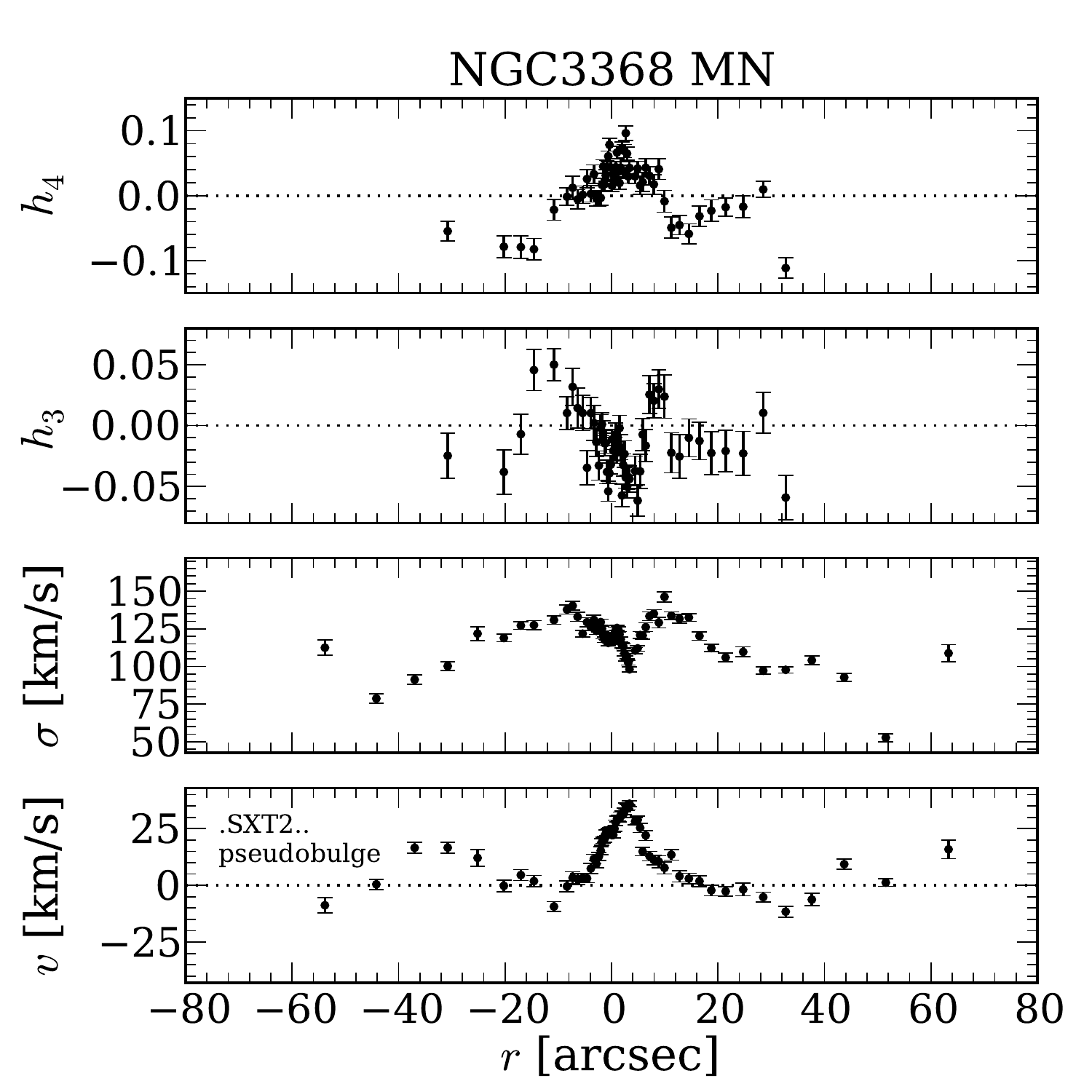}\\
        \end{tabular}
        \end{center}
        \caption{{\it continued --}\small Major and minor axis kinematic profiles for NGC\,3368.
	We plot data from \citet{Heraudeau1999} in green.} 
\end{figure}
\setcounter{figure}{15}
\begin{figure}
        \begin{center}
        \begin{tabular}{lll}
	\begin{minipage}[b]{0.185\textwidth}
	\includegraphics[viewport=0 55 390 400,width=\textwidth]{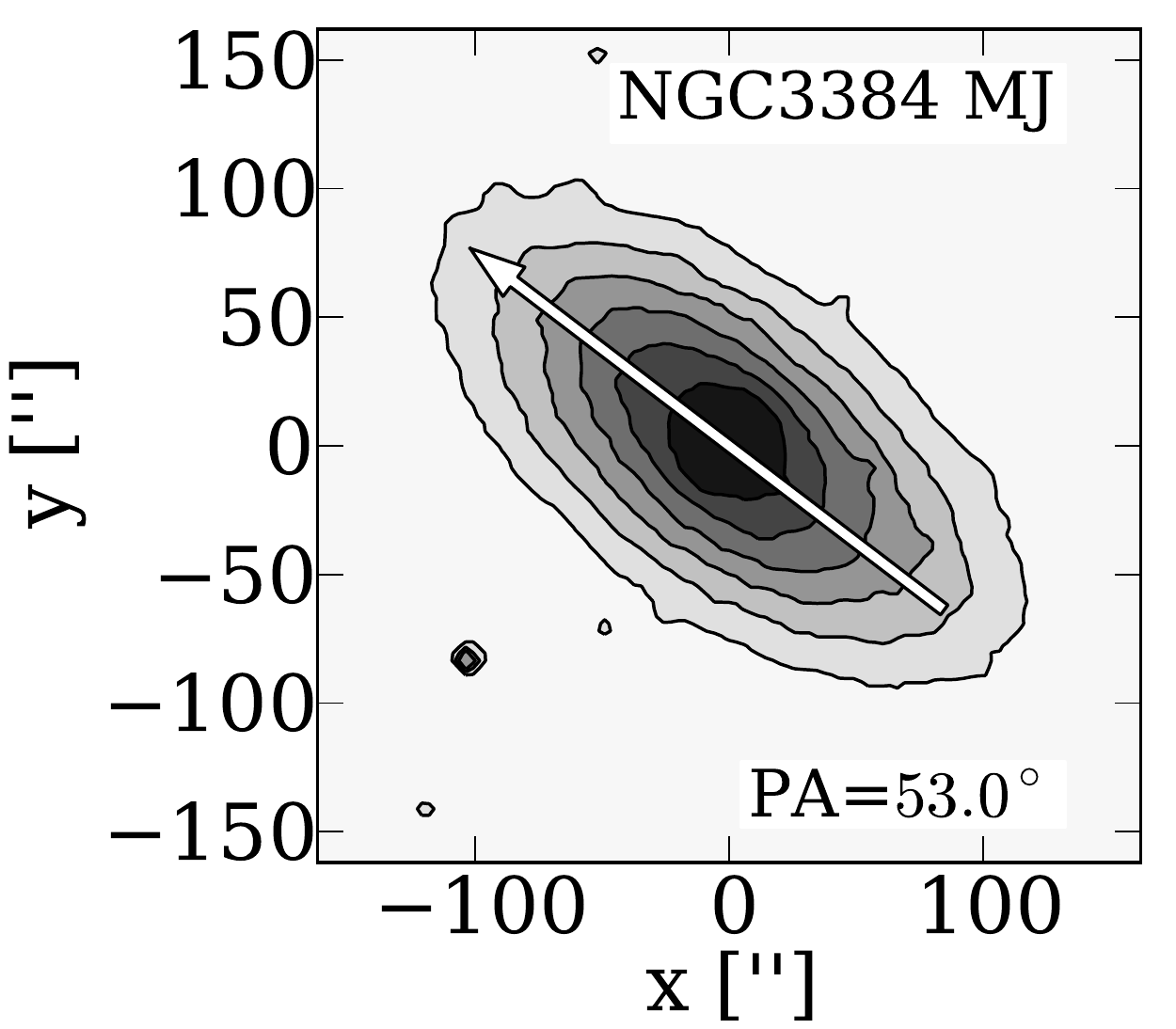}\\
	\includegraphics[viewport=0 55 390 400,width=\textwidth]{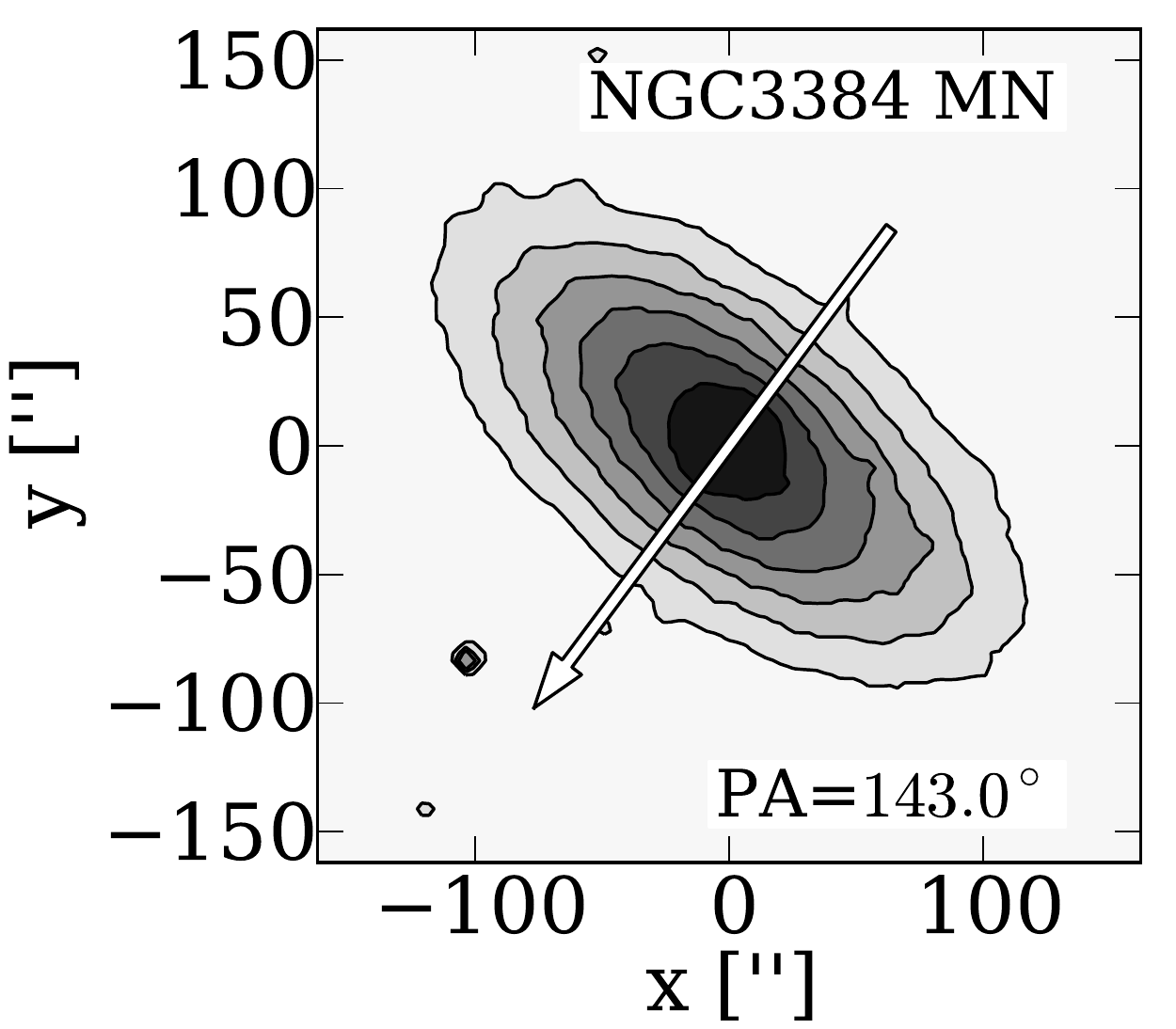}
	\end{minipage} & 
	\includegraphics[viewport=0 50 420 400,width=0.35\textwidth]{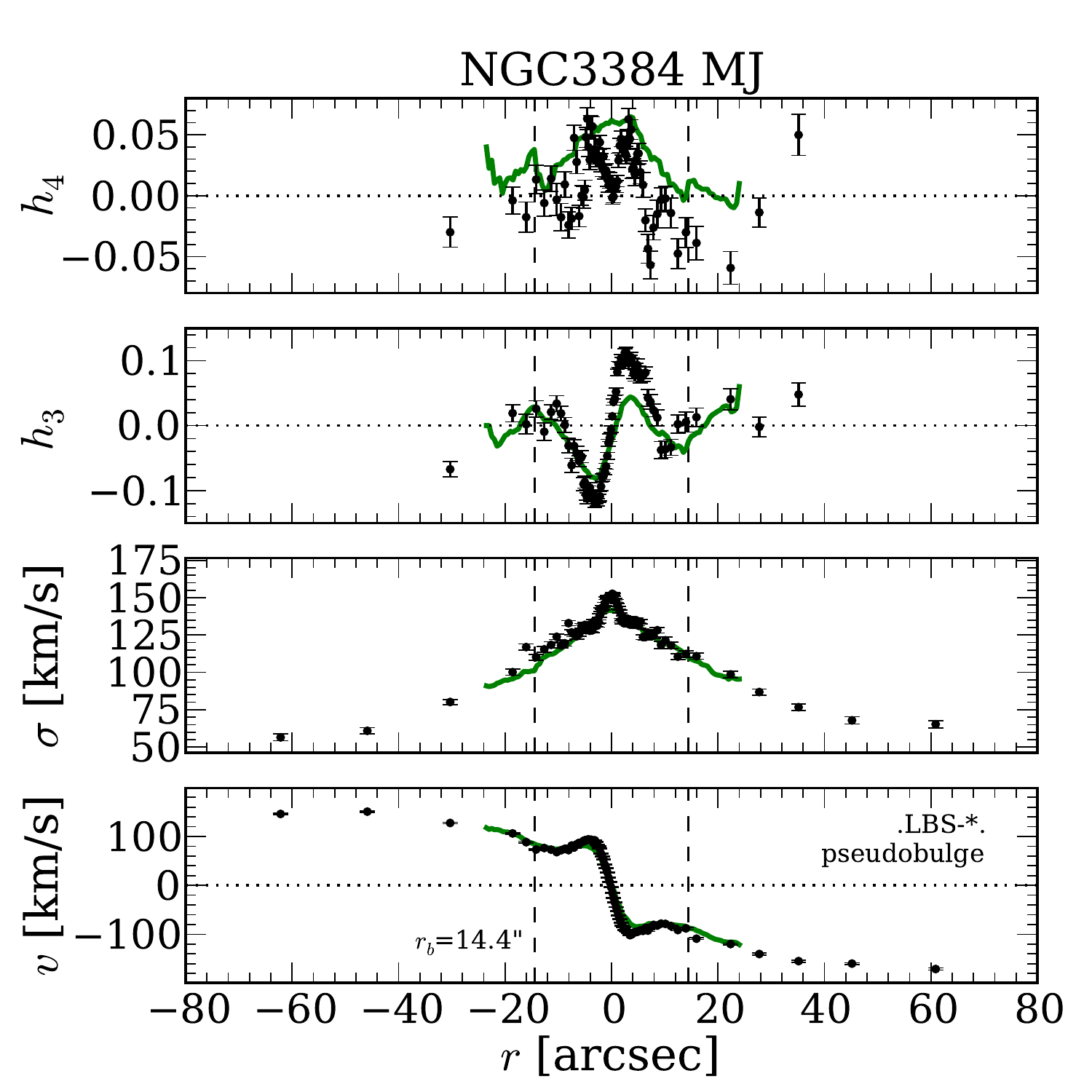} &
	\includegraphics[viewport=0 50 420 400,width=0.35\textwidth]{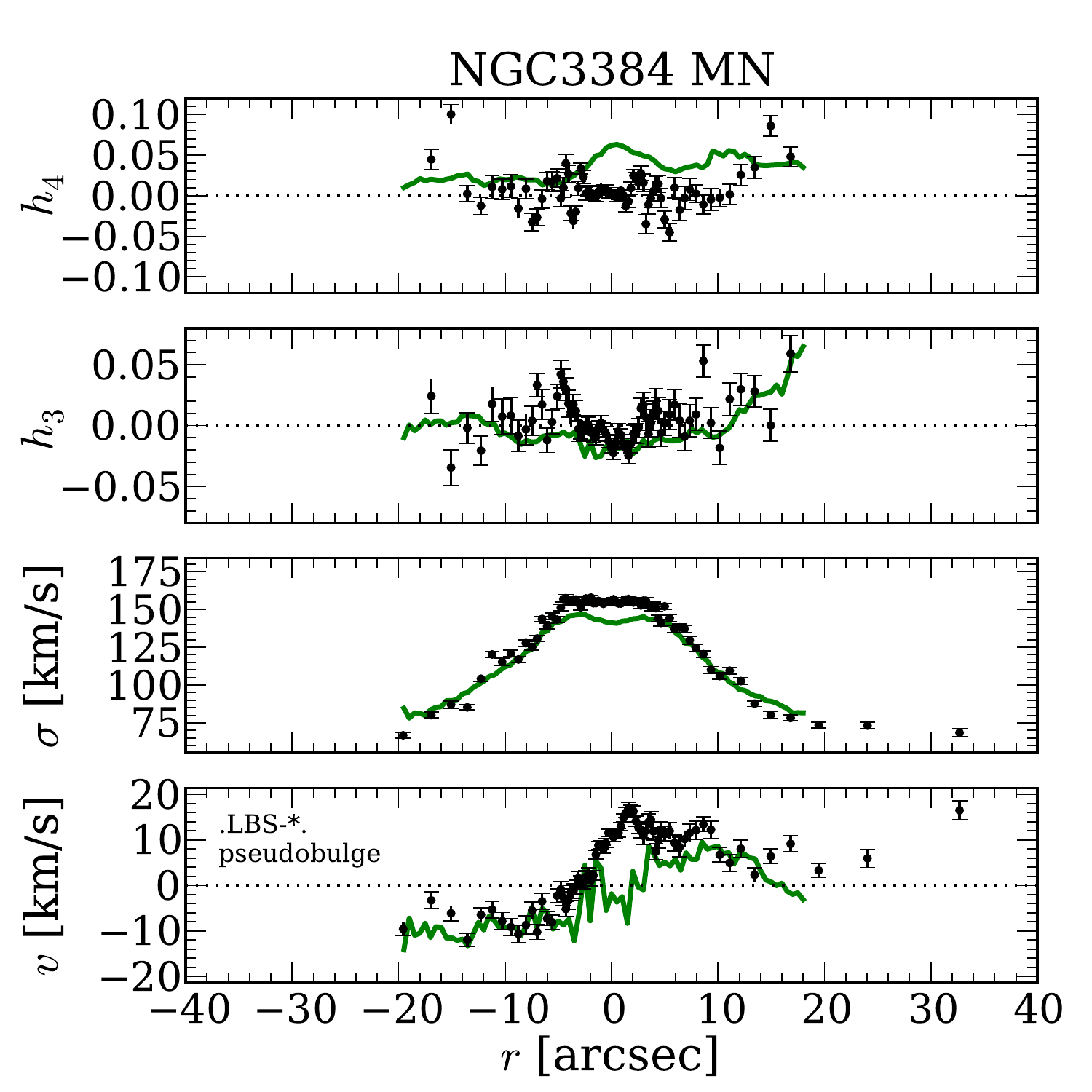}\\
        \end{tabular}
        \end{center}
        \caption{{\it continued --}\small Major and minor axis kinematic profiles for NGC\,3384.
	We plot SAURON results of \citet{Emsellem2004} in green.}
\end{figure}
\clearpage
\setcounter{figure}{15}
\begin{figure}
        \begin{center}
        \begin{tabular}{lll}
	\begin{minipage}[b]{0.185\textwidth}
	\includegraphics[viewport=0 55 390 400,width=\textwidth]{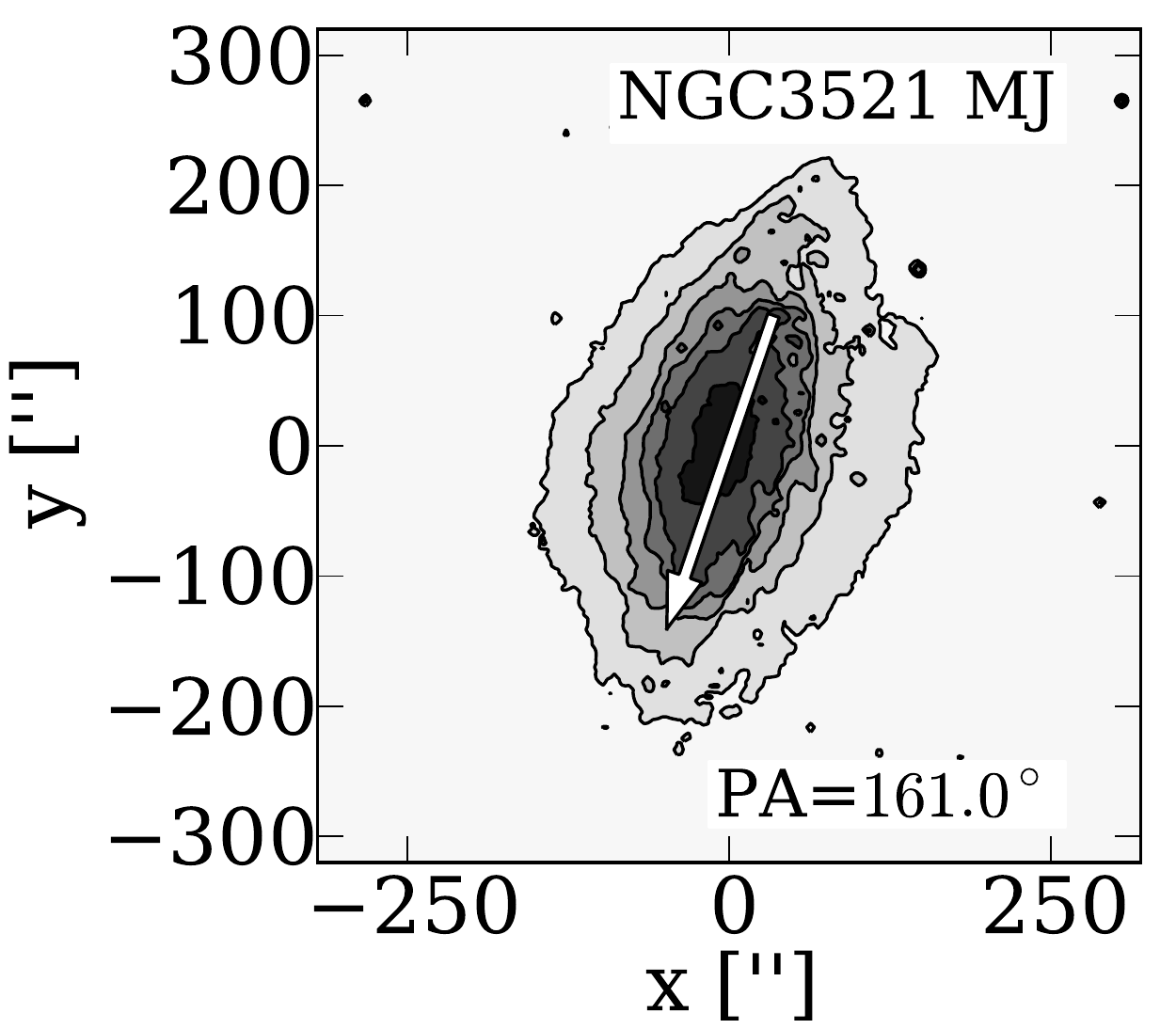}\\
	\includegraphics[viewport=0 55 390 400,width=\textwidth]{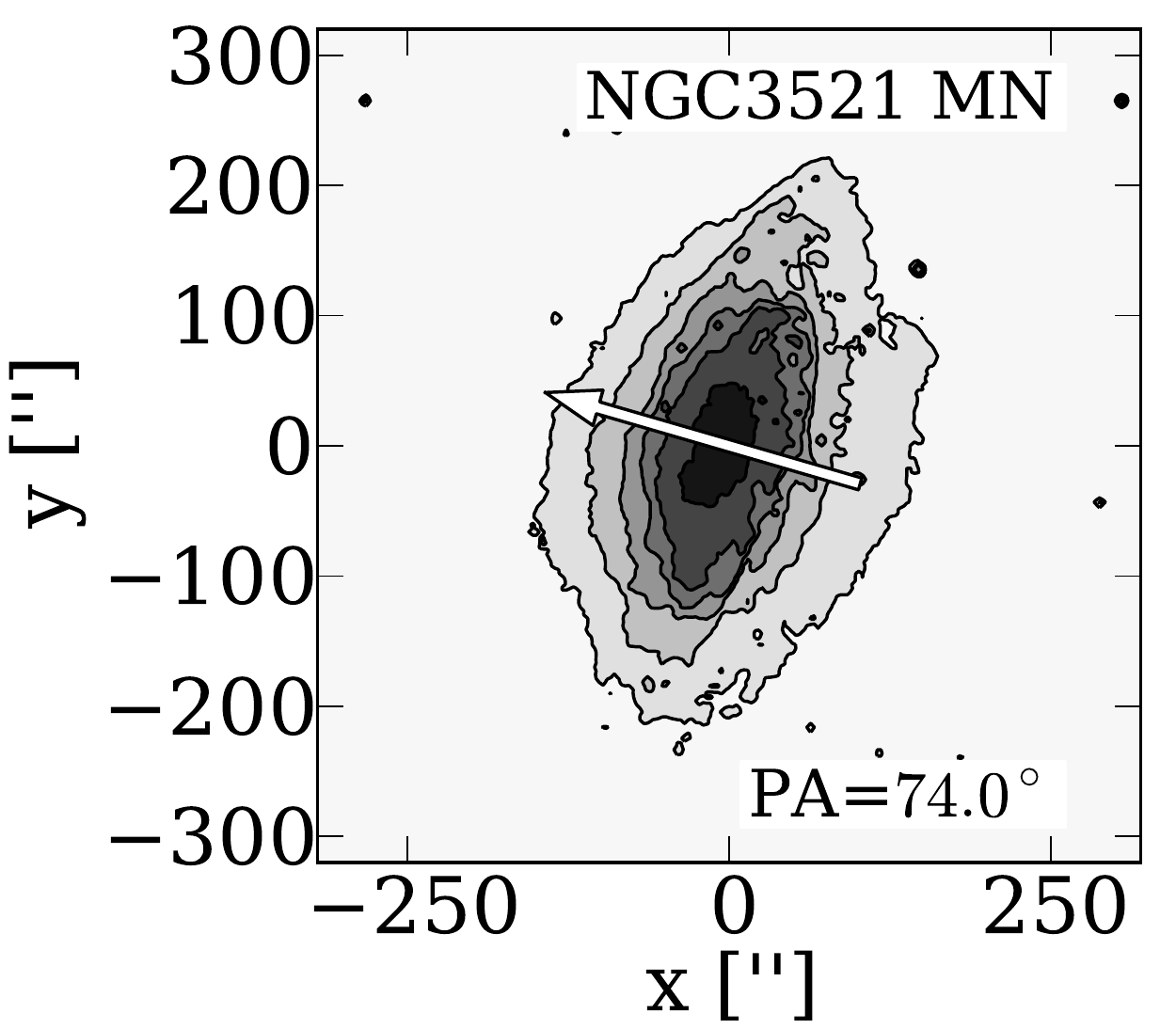}
	\end{minipage} & 
	\includegraphics[viewport=0 50 420 400,width=0.35\textwidth]{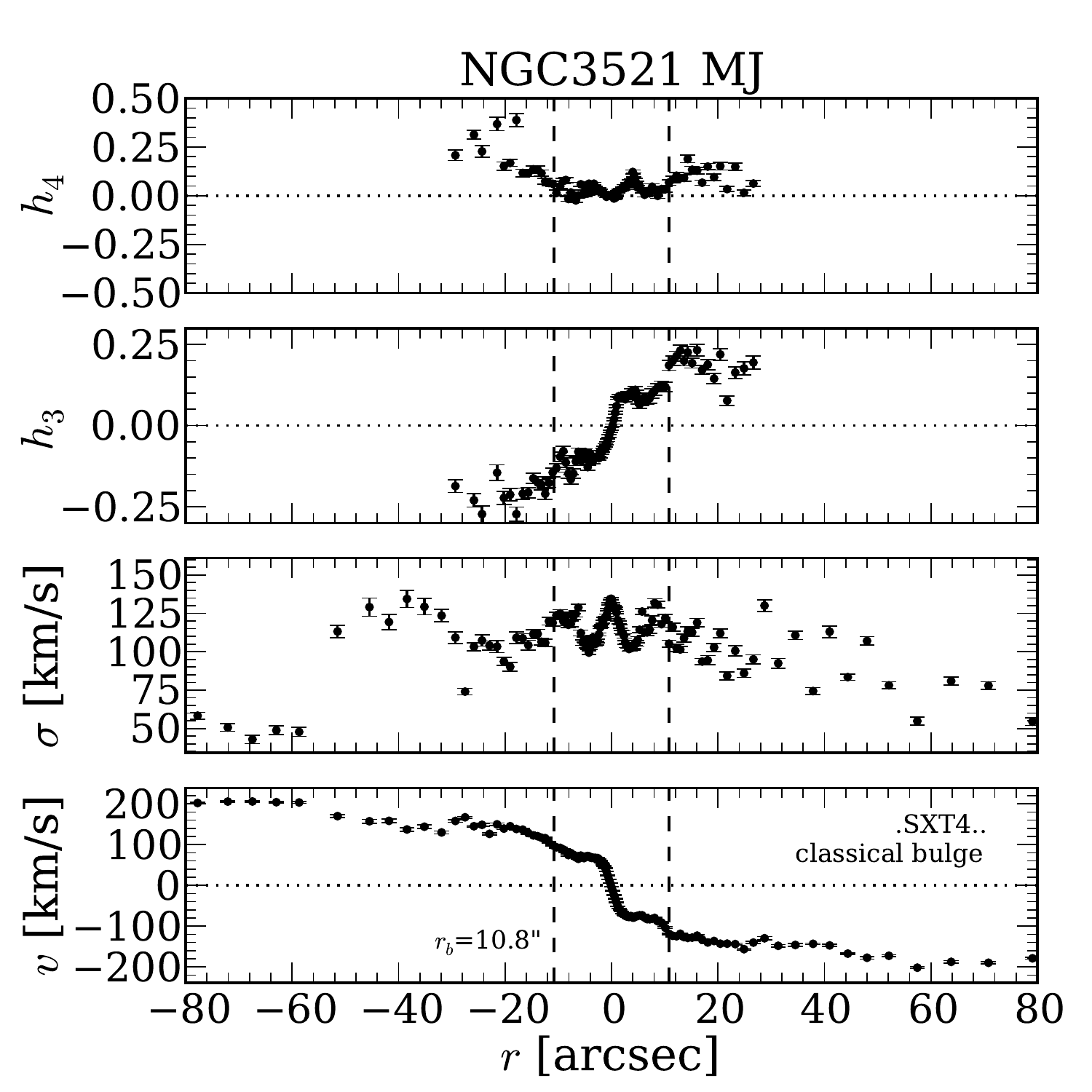} &
	\includegraphics[viewport=0 50 420 400,width=0.35\textwidth]{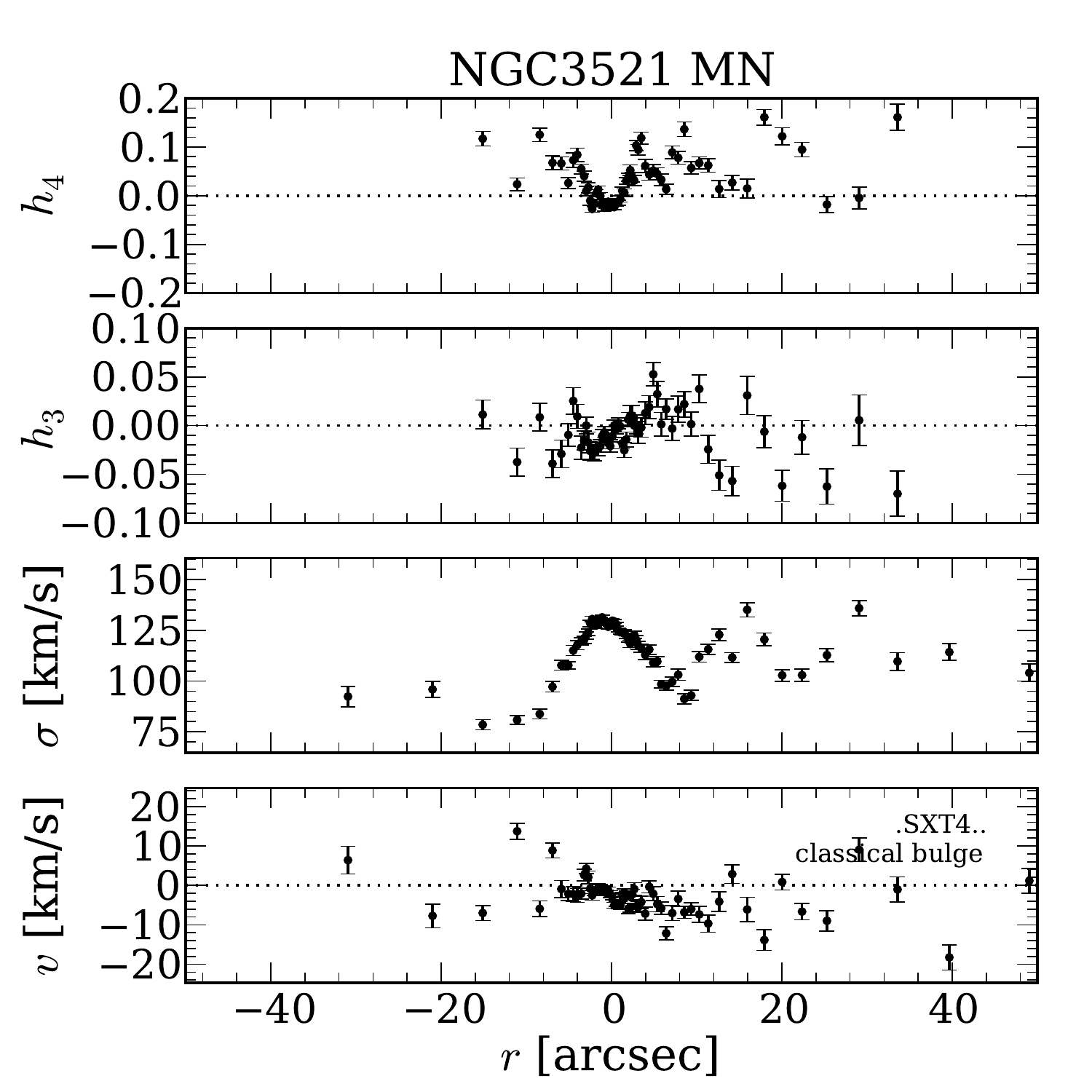}\\
        \end{tabular}
        \end{center}
        \caption{{\it continued --}\small Major and minor axis kinematic profiles for NGC\,3521,
				see also figures \ref{fig:n3521_losvds} and \ref{fig:kinDecomp}.
				}
\end{figure}
\setcounter{figure}{15}
\begin{figure}[H]
        \begin{center}
        \begin{tabular}{lll}
	\begin{minipage}[b]{0.185\textwidth}
	\includegraphics[viewport=0 55 390 400,width=\textwidth]{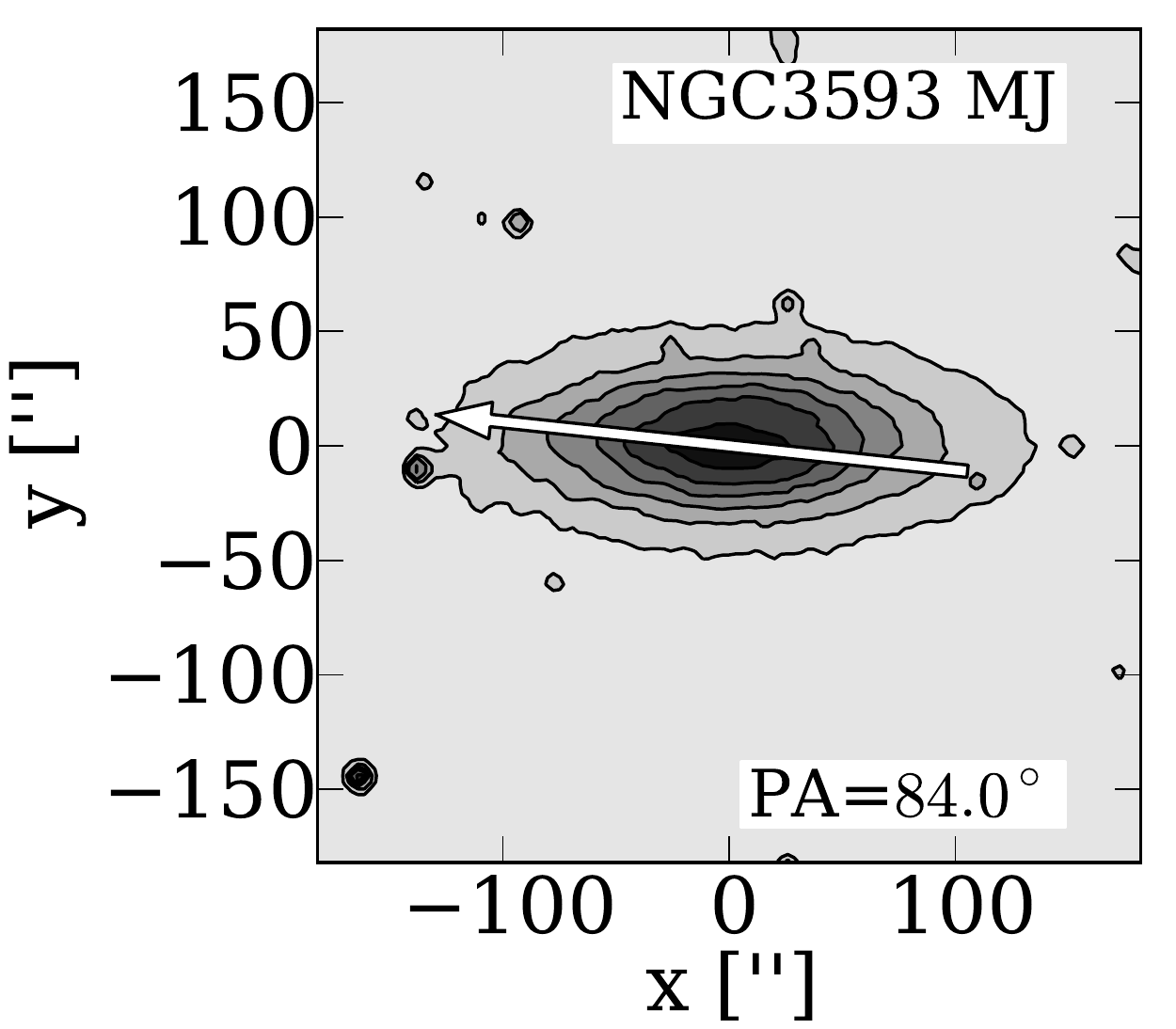}\\
        \includegraphics[viewport=0 55 390 400,width=\textwidth]{empty}
	\end{minipage} & 
	\includegraphics[viewport=0 50 420 400,width=0.35\textwidth]{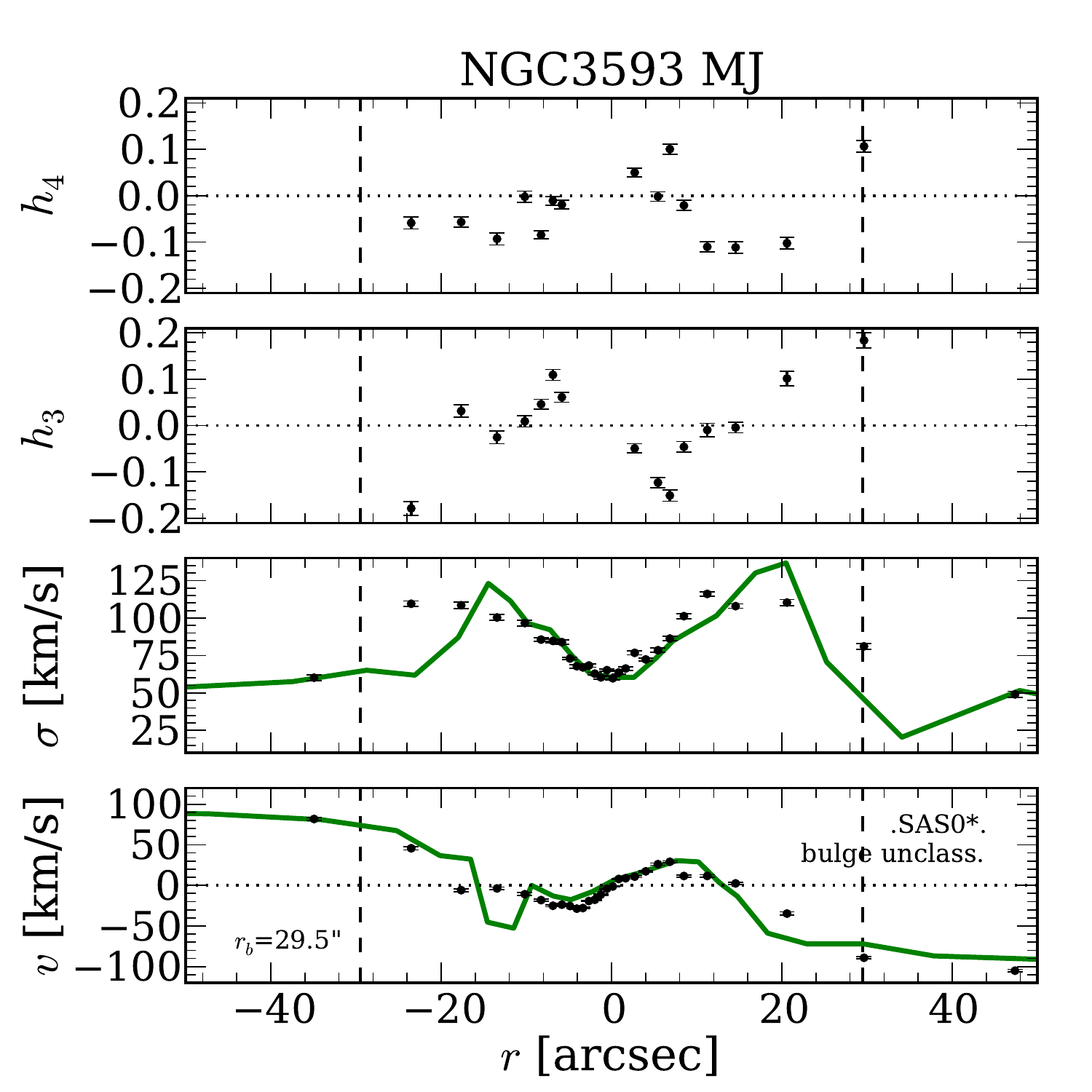}&
        \includegraphics[viewport=0 50 420 400,width=0.35\textwidth]{empty}
        \end{tabular}
        \end{center}
		\caption{{\it continued --}\small Major axis kinematic profile for
		NGC\,3593, we plot the results of \citet{Bertola1996} in green.
	}
\end{figure}
\setcounter{figure}{15}
\begin{figure}
        \begin{center}
        \begin{tabular}{lll}
	\begin{minipage}[b]{0.185\textwidth}
	\includegraphics[viewport=0 55 390 400,width=\textwidth]{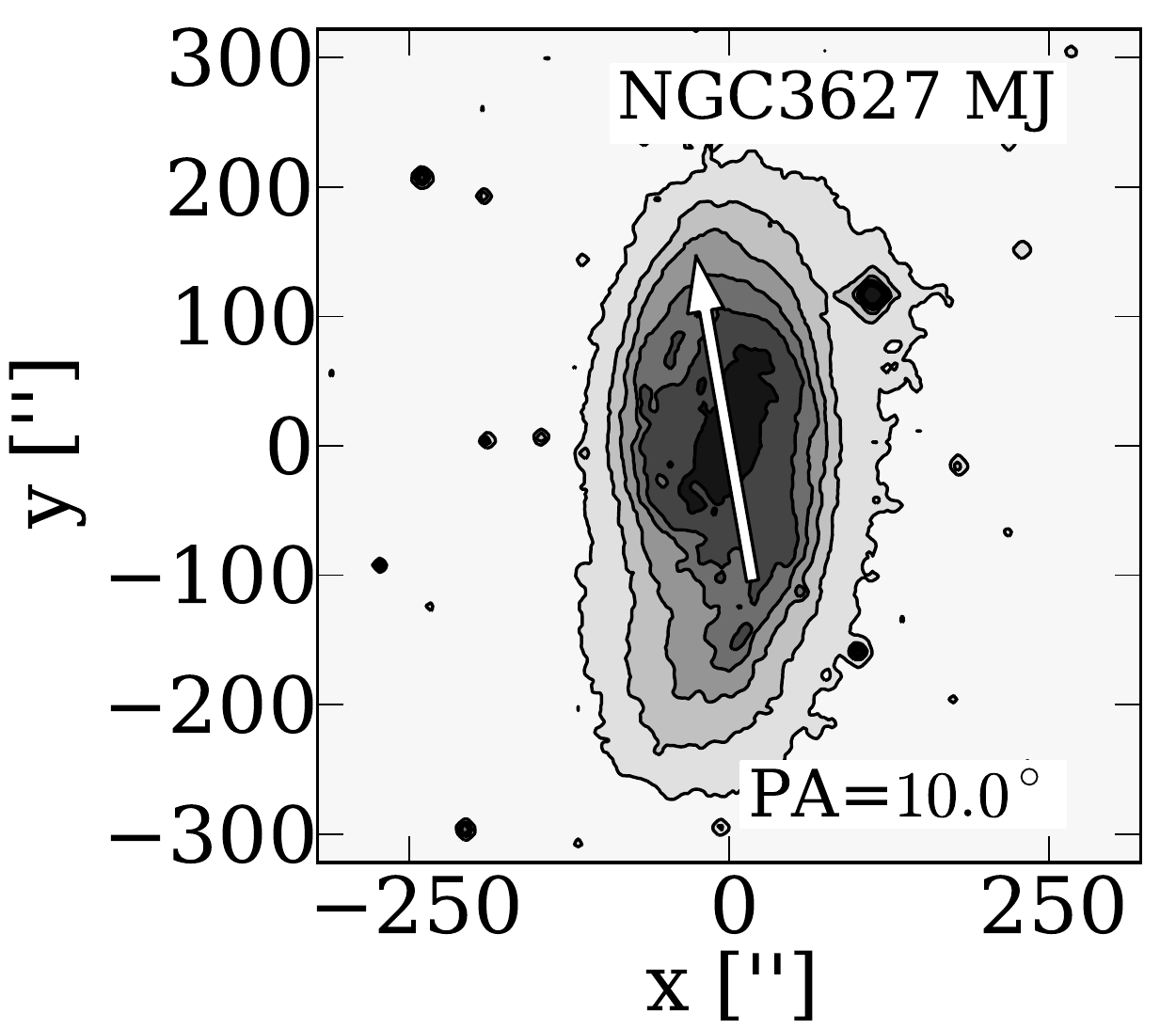}\\
	\includegraphics[viewport=0 55 390 400,width=\textwidth]{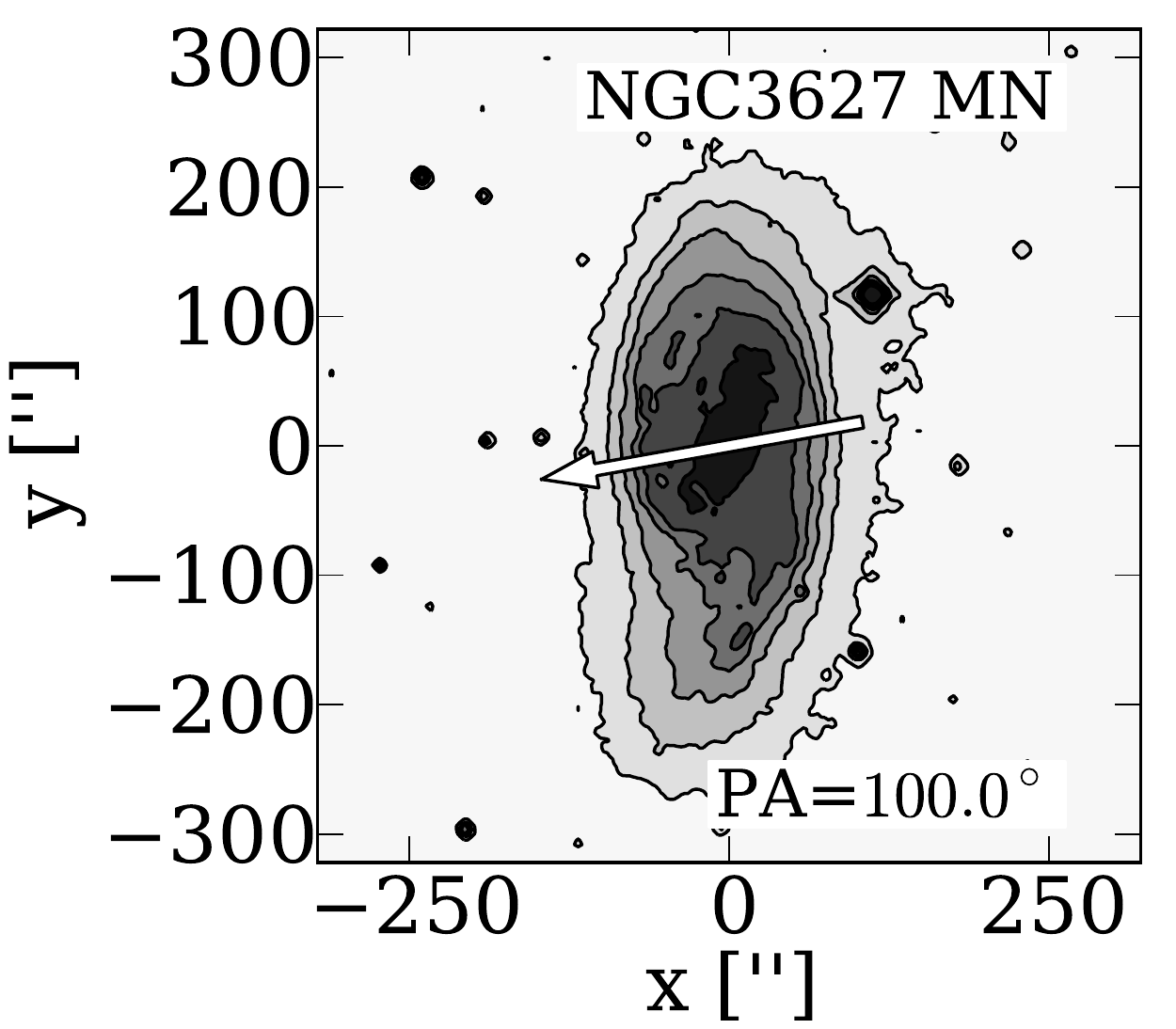}
	\end{minipage} & 
	\includegraphics[viewport=0 50 420 400,width=0.35\textwidth]{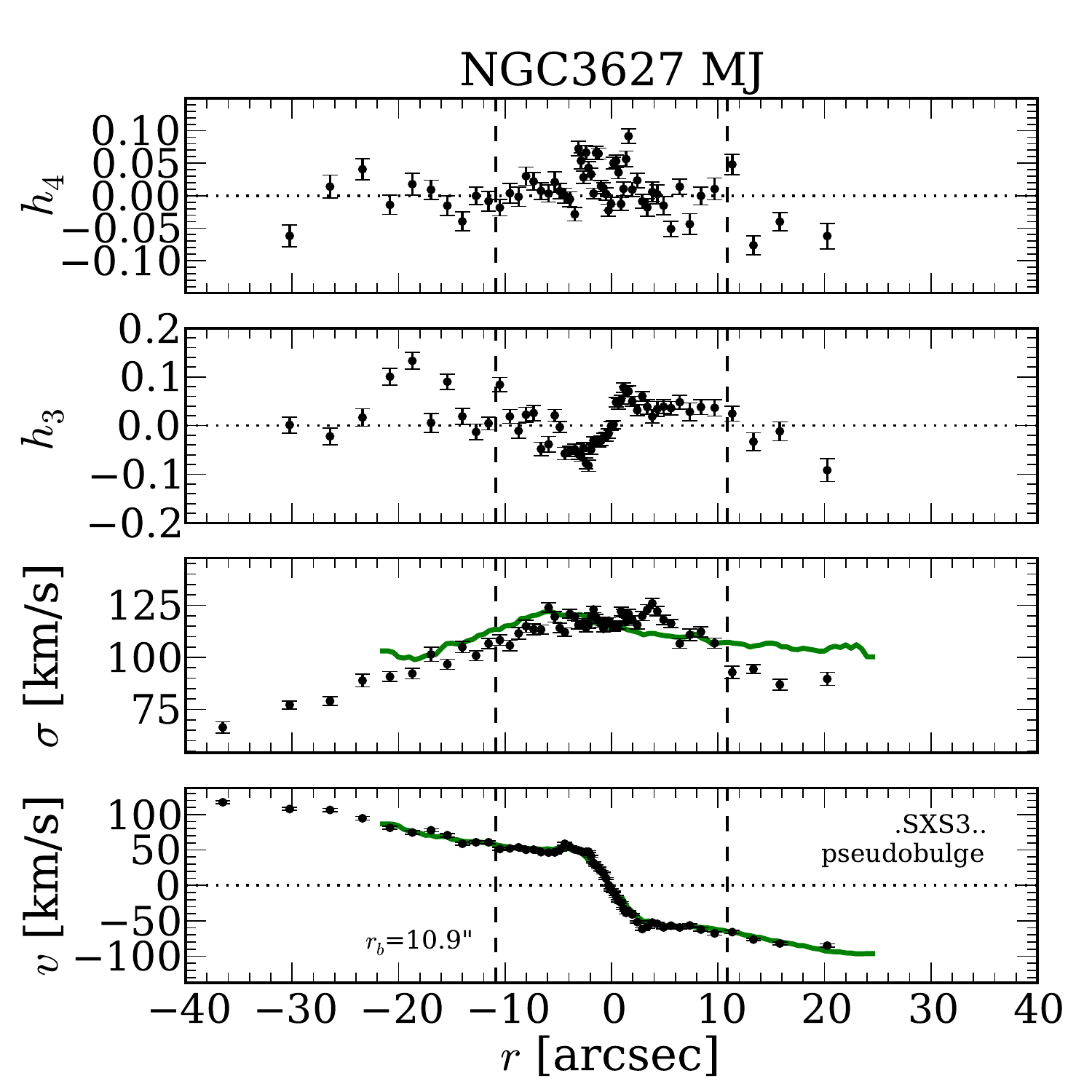} &
	\includegraphics[viewport=0 50 420 400,width=0.35\textwidth]{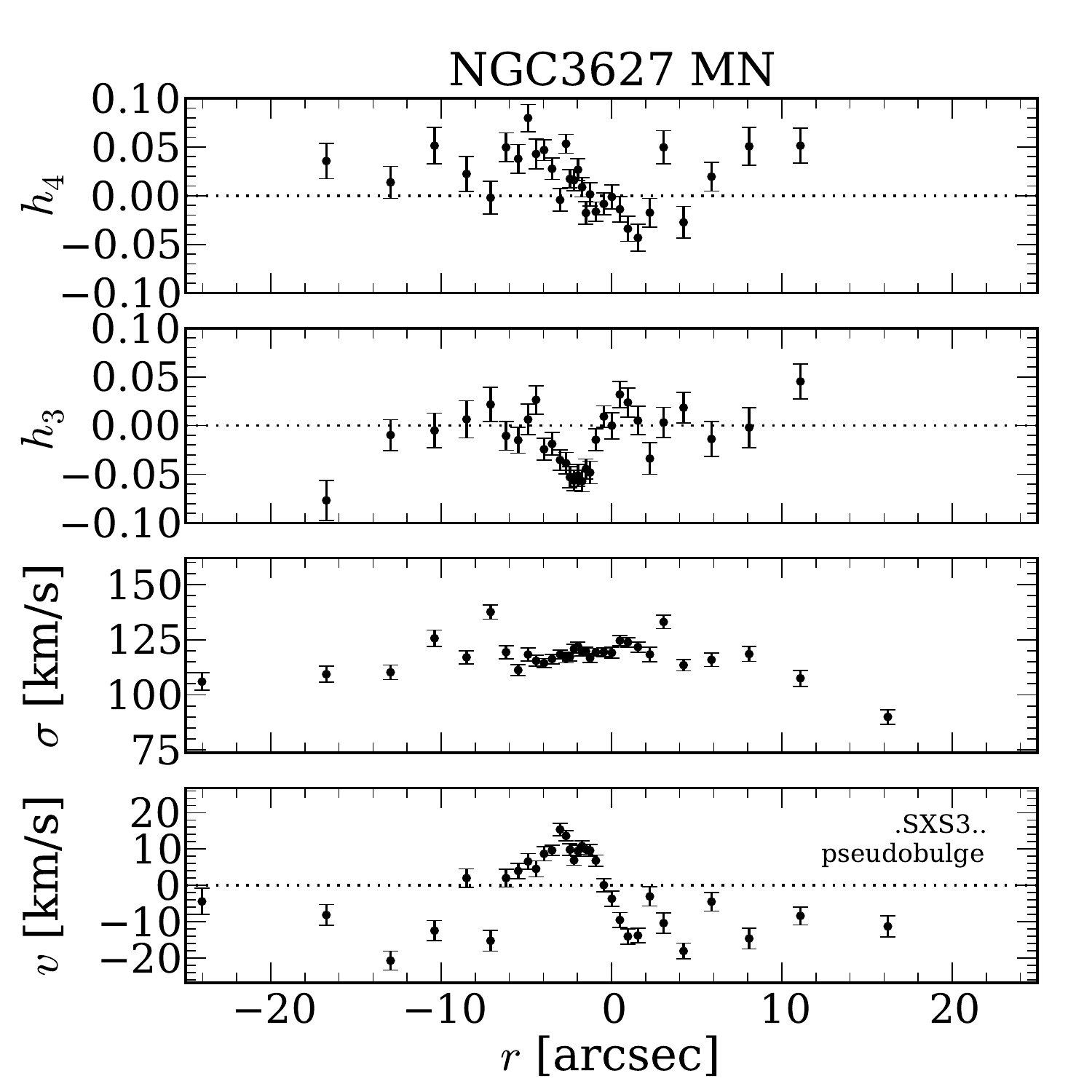}\\
        \end{tabular}
        \end{center}
        \caption{{\it continued --}\small Major and minor axis kinematic profiles for NGC\,3627.
	We plot data of \citet{Heraudeau1998} in green. Note: Their data were taken at a slit position angle of 
	173\Deg\ whereas our adopted value for the major axis position angle is 10\Deg.
	}
\end{figure}
\setcounter{figure}{15}
\begin{figure}
        \begin{center}
        \begin{tabular}{lll}
	\begin{minipage}[b]{0.185\textwidth}
	\includegraphics[viewport=0 55 390 400,width=\textwidth]{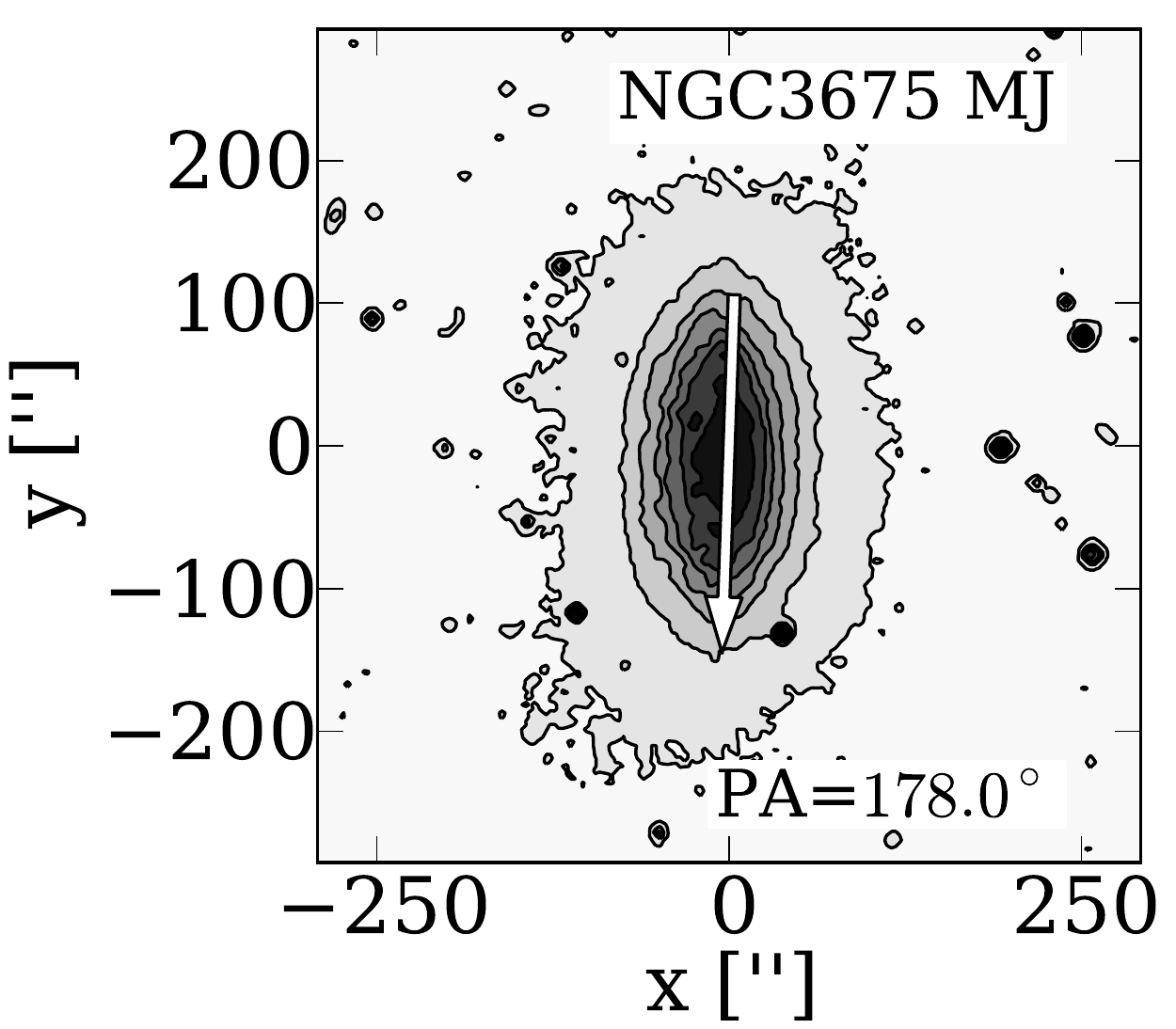}\\
        \includegraphics[viewport=0 55 390 400,width=\textwidth]{empty}
	\end{minipage} & 
	\includegraphics[viewport=0 50 420 400,width=0.35\textwidth]{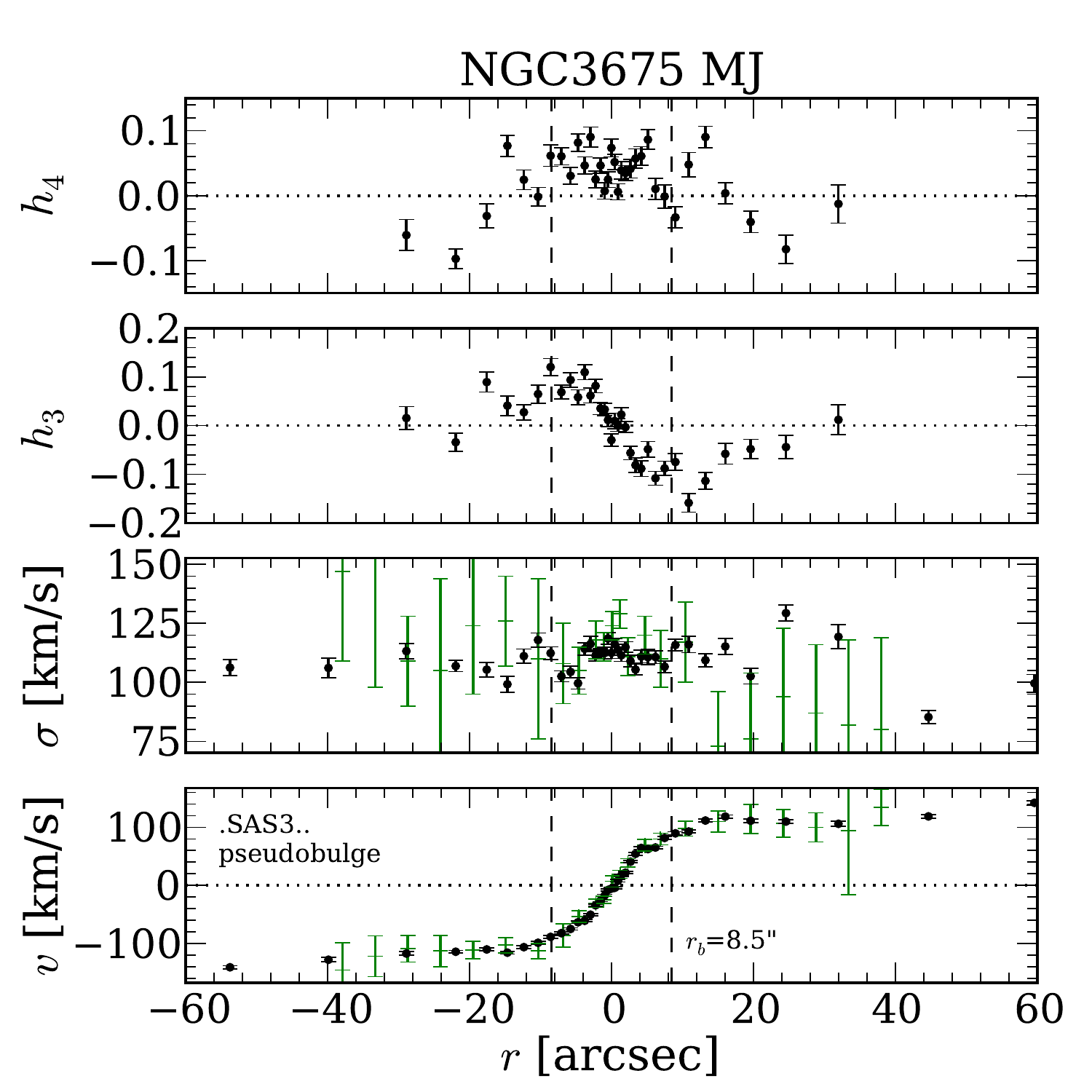}&
        \includegraphics[viewport=0 50 420 400,width=0.35\textwidth]{empty}
        \end{tabular}
        \end{center}
        \caption{{\it continued --}\small Major axis kinematic profile for NGC\,3675 we plot the results of \citet{Heraudeau1998} in green.
	}
\end{figure}
\setcounter{figure}{15}
\begin{figure}
        \begin{center}
        \begin{tabular}{lll}
	\begin{minipage}[b]{0.185\textwidth}
	\includegraphics[viewport=0 55 390 400,width=\textwidth]{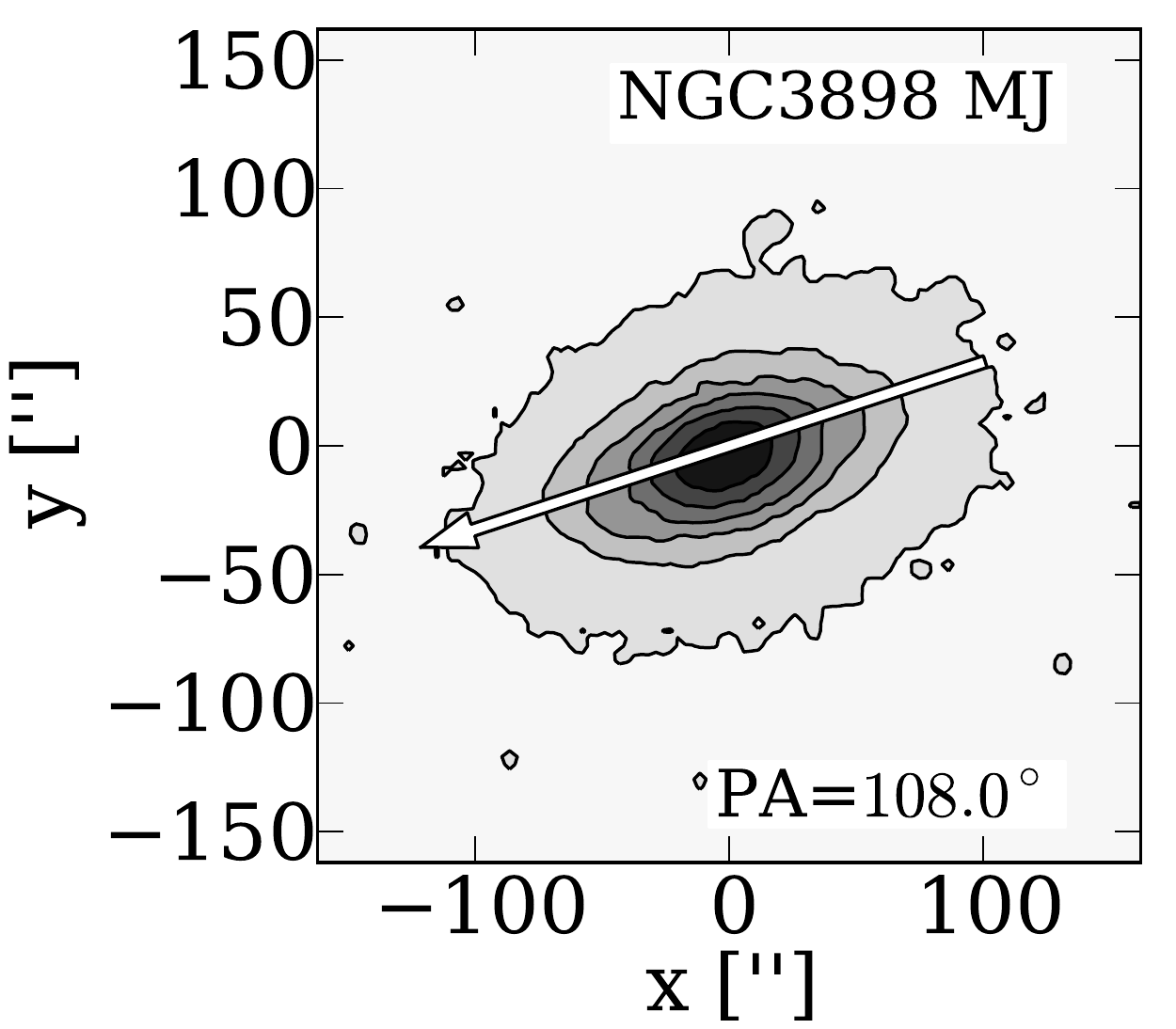}\\
        \includegraphics[viewport=0 55 390 400,width=\textwidth]{empty}
	\end{minipage} & 
	\includegraphics[viewport=0 50 420 400,width=0.35\textwidth]{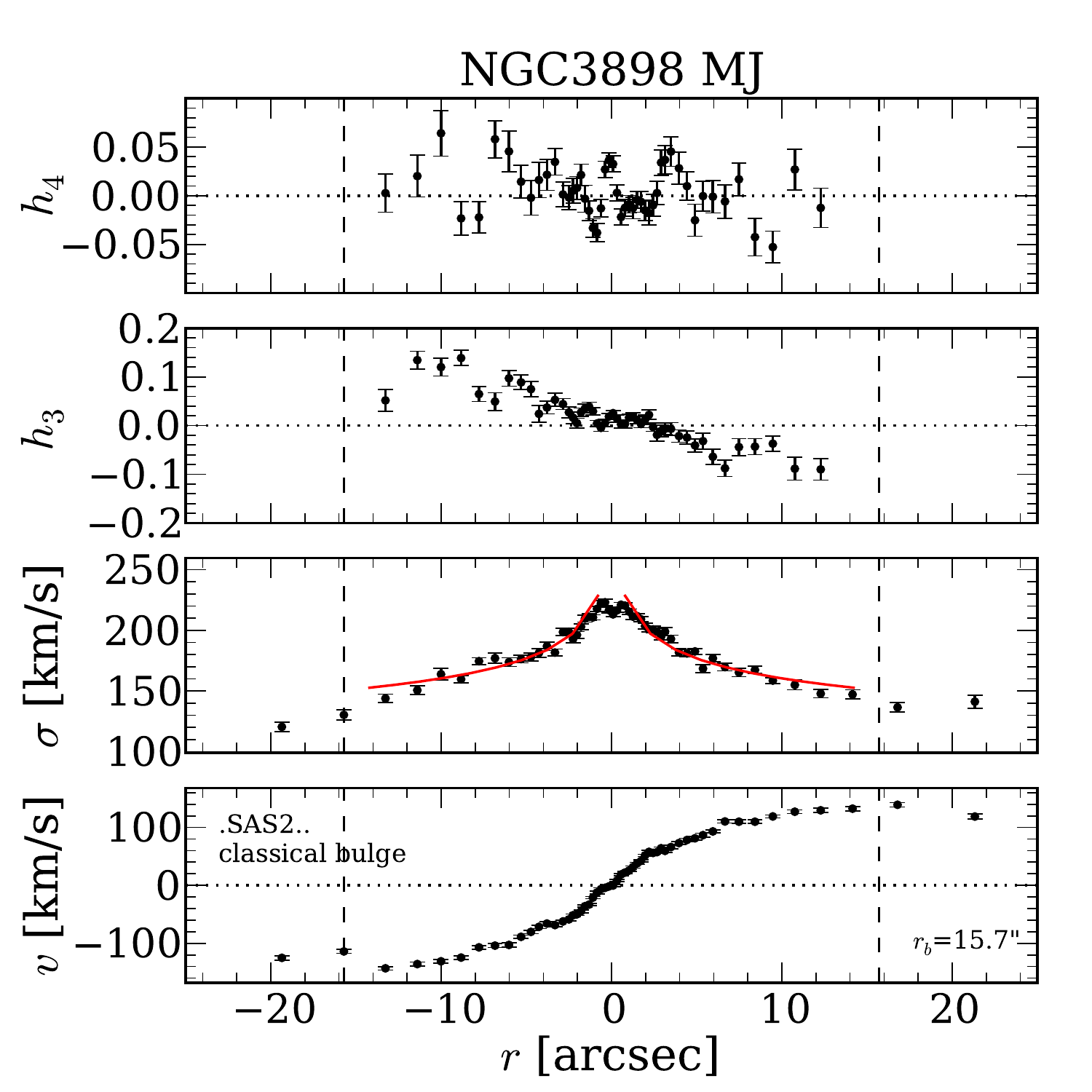}&
        \includegraphics[viewport=0 50 420 400,width=0.35\textwidth]{empty}
        \end{tabular}
        \end{center}
        \caption{{\it continued --}\small Major axis kinematic profiles for NGC\,3898, reproduced from Fig.~\ref{fig:twoprofiles}. 
	We plot the results of \citet{Pignatelli2001} in green.
	}
\end{figure}
\setcounter{figure}{15}
\begin{figure}
        \begin{center}
        \begin{tabular}{lll}
	\begin{minipage}[b]{0.185\textwidth}
	\includegraphics[viewport=0 55 390 400,width=\textwidth]{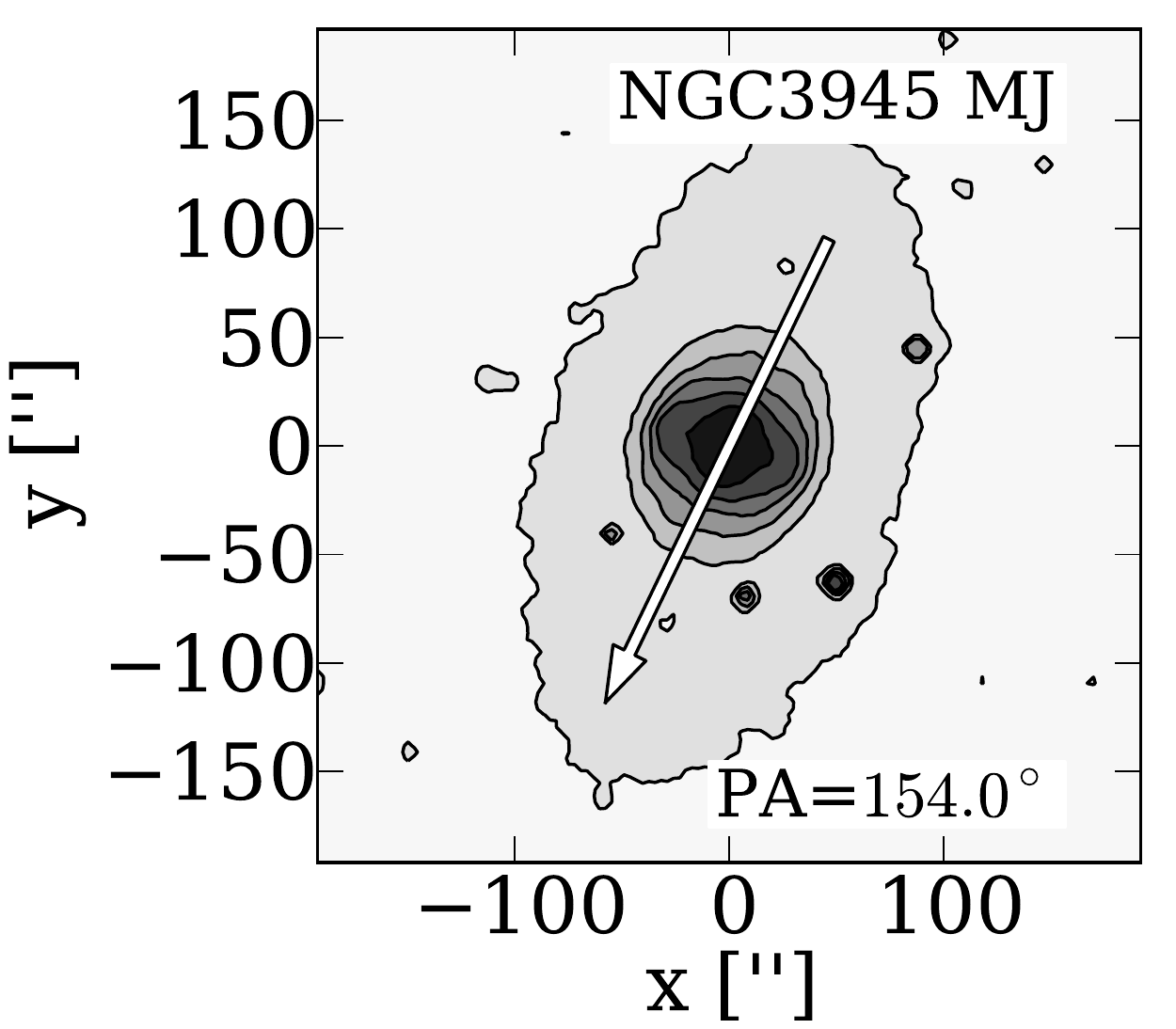}\\
	\includegraphics[viewport=0 55 390 400,width=\textwidth]{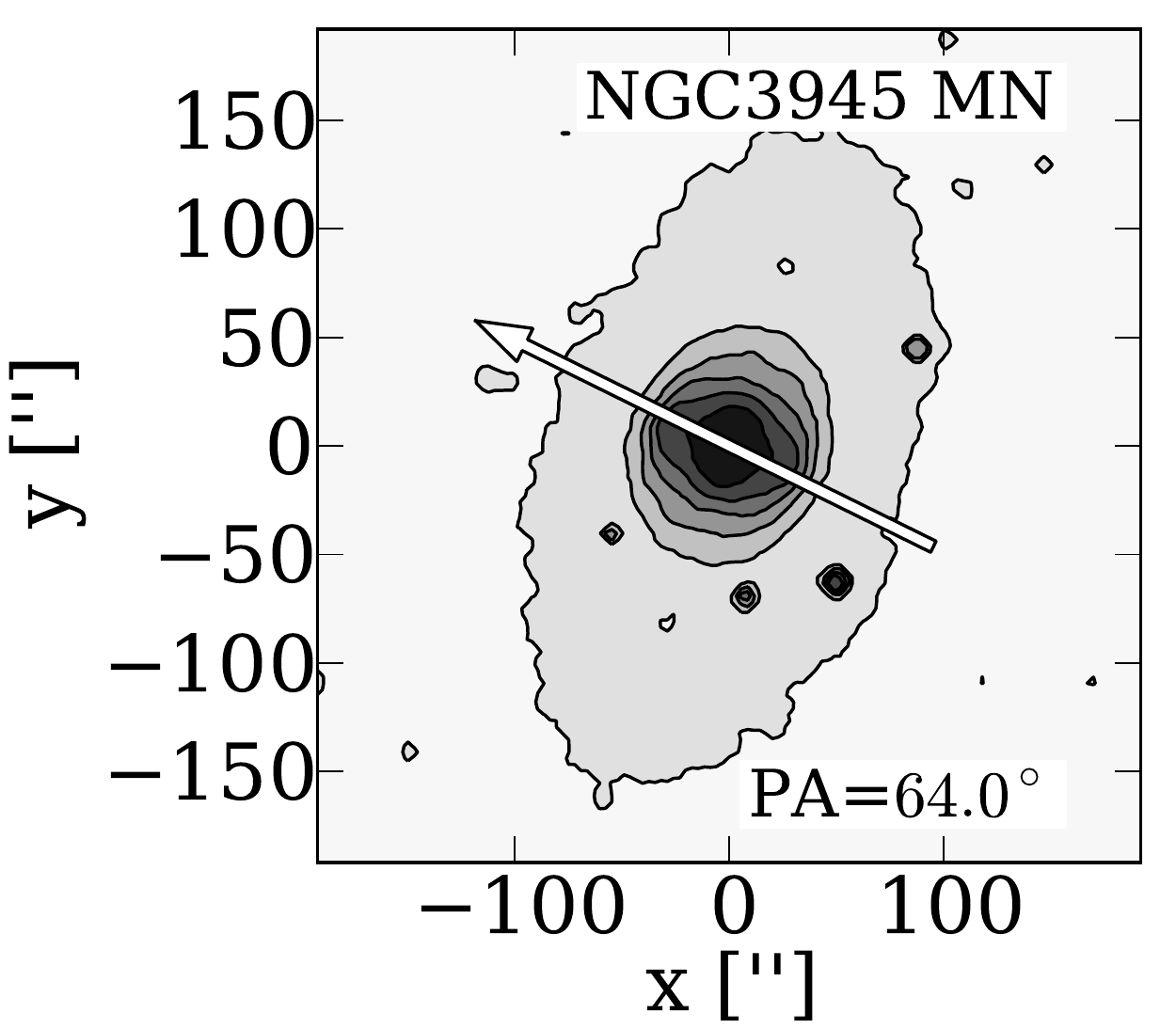}
	\end{minipage} & 
	\includegraphics[viewport=0 50 420 400,width=0.35\textwidth]{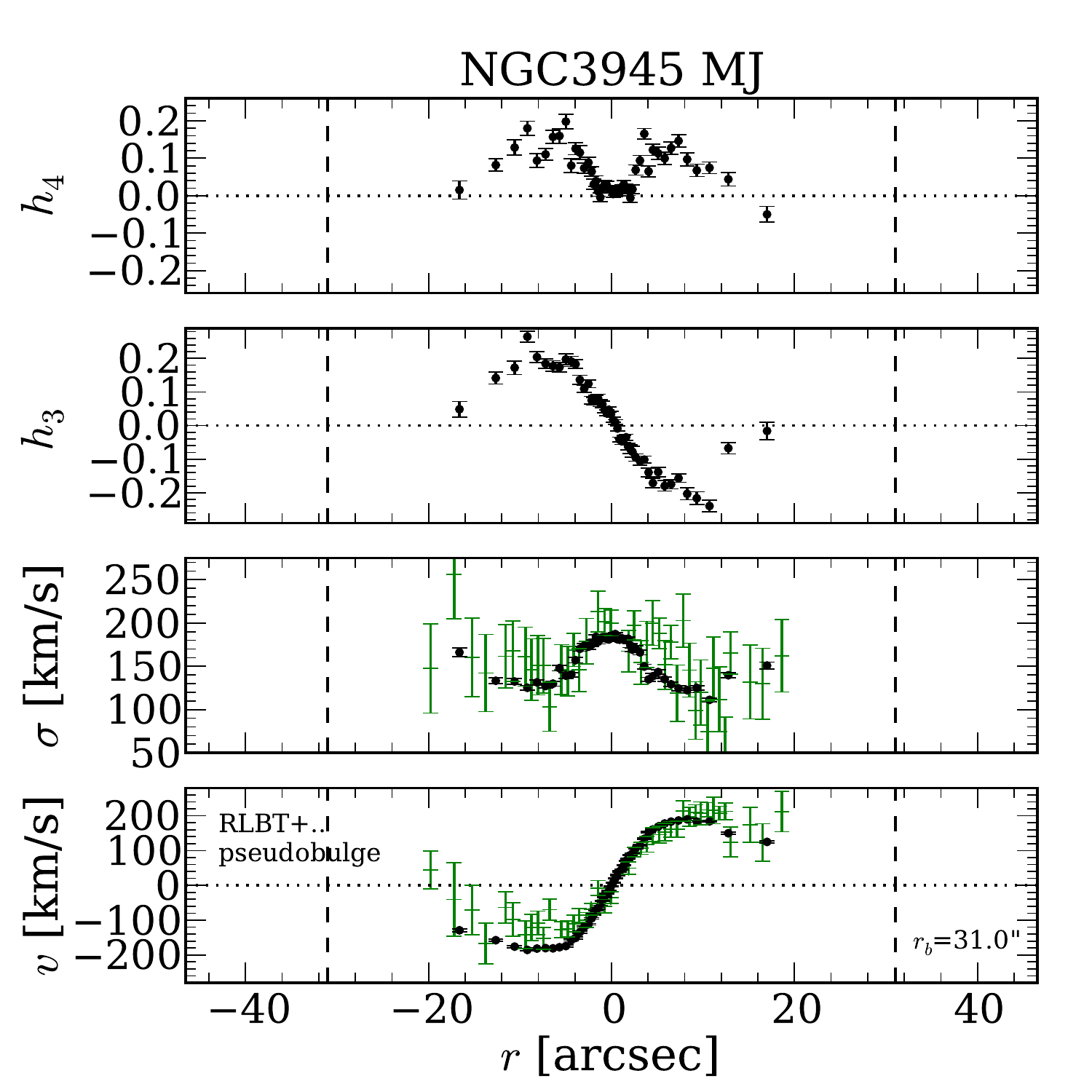} &
	\includegraphics[viewport=0 50 420 400,width=0.35\textwidth]{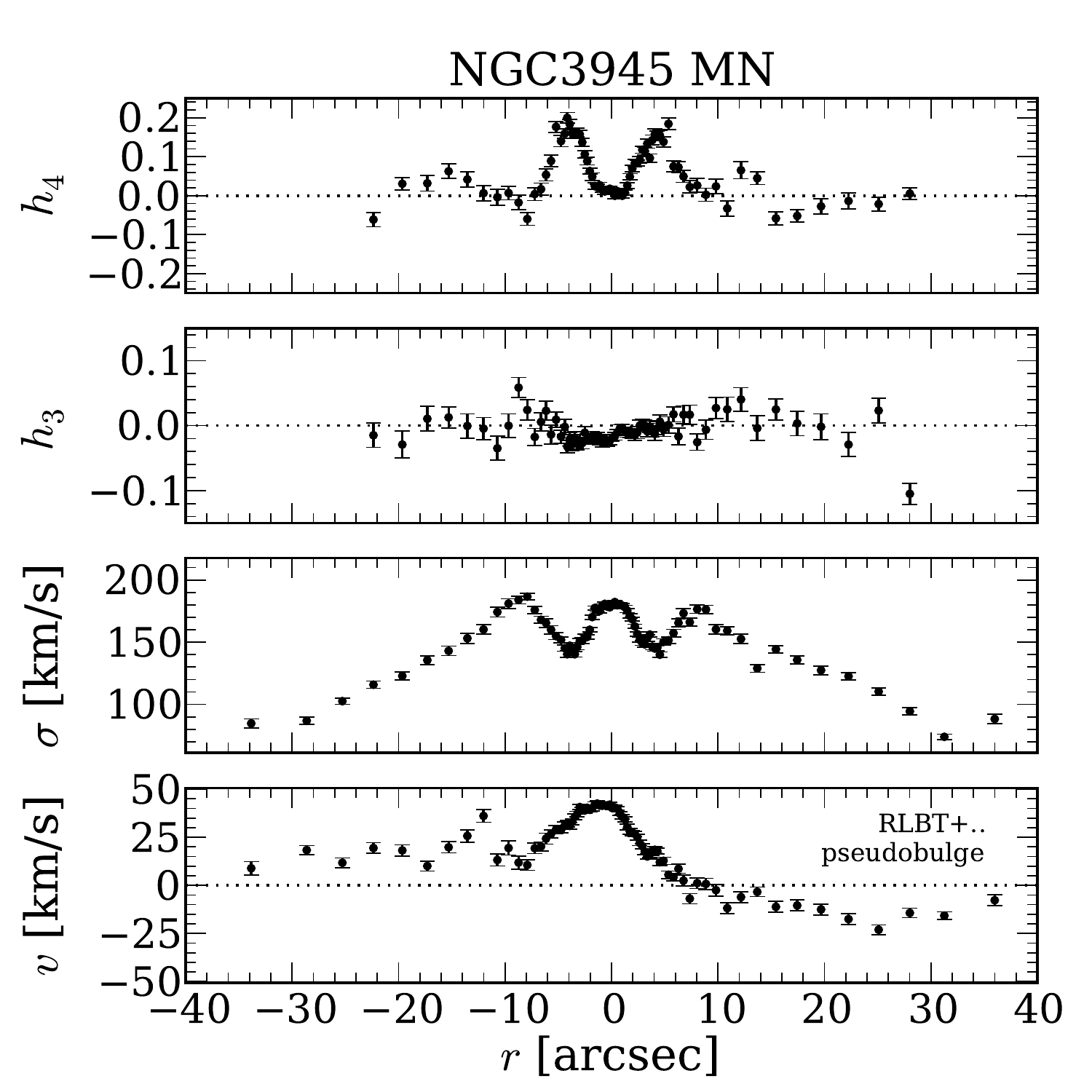}\\
        \end{tabular}
        \end{center}
        \caption{{\it continued --}\small Major and minor axis kinematic profiles for NGC\,3945,
	see also Fig.~\ref{fig:kinDecomp}.
	We plot data from \citet{Bertola1995} in green.
	}
\end{figure}
\clearpage
\setcounter{figure}{15}
\begin{figure}
        \begin{center}
        \begin{tabular}{lll}
	\begin{minipage}[b]{0.185\textwidth}
	\includegraphics[viewport=0 55 390 400,width=\textwidth]{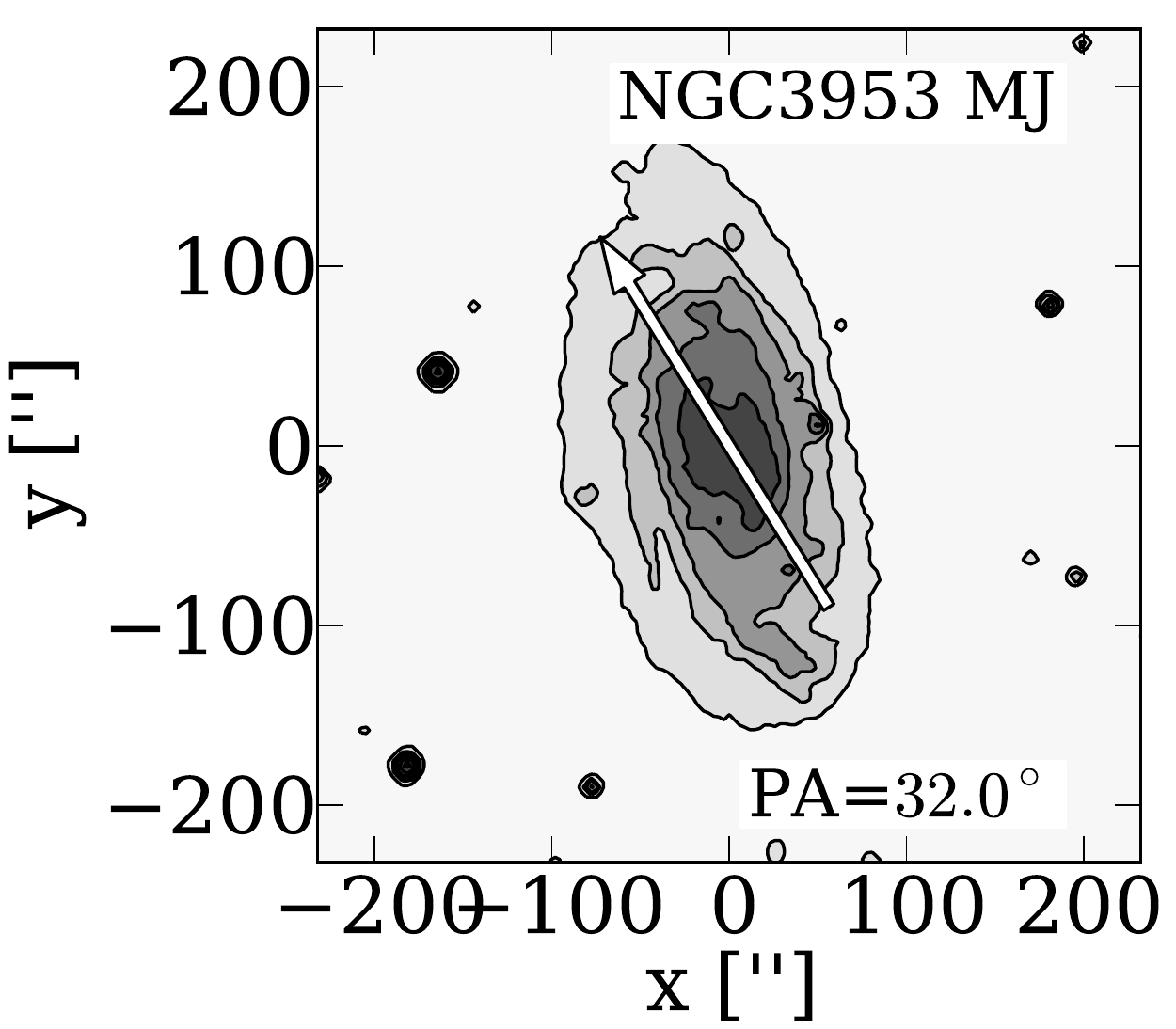}\\
        \includegraphics[viewport=0 55 390 400,width=\textwidth]{empty}
	\end{minipage} & 
	\includegraphics[viewport=0 50 420 400,width=0.35\textwidth]{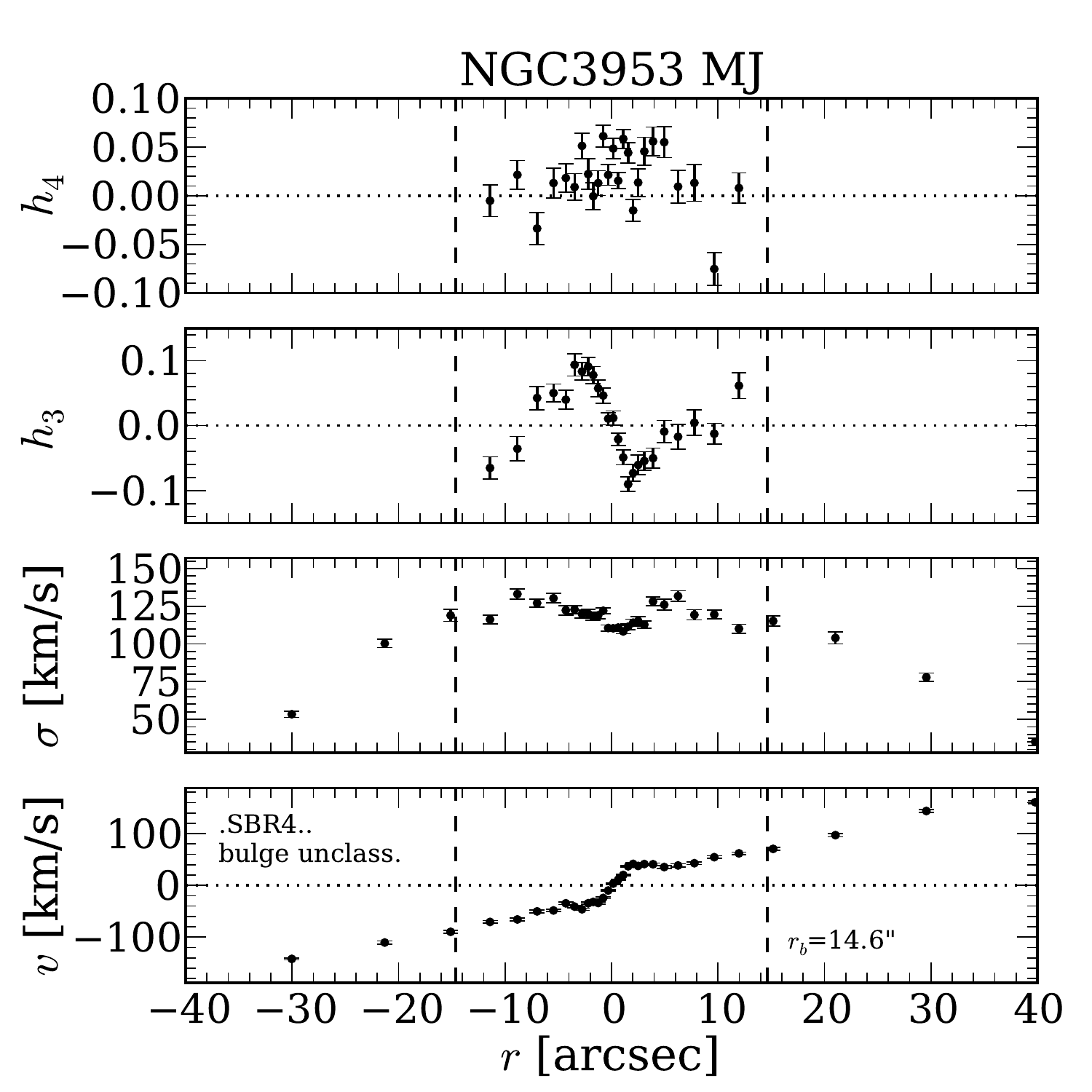}&
        \includegraphics[viewport=0 50 420 400,width=0.35\textwidth]{empty}
        \end{tabular}
        \end{center}
        \caption{{\it continued --}\small Major axis kinematic profile for NGC\,3953.
	}
\end{figure}
\setcounter{figure}{15}
\begin{figure}
        \begin{center}
        \begin{tabular}{lll}
	\begin{minipage}[b]{0.185\textwidth}
	\includegraphics[viewport=0 55 390 400,width=\textwidth]{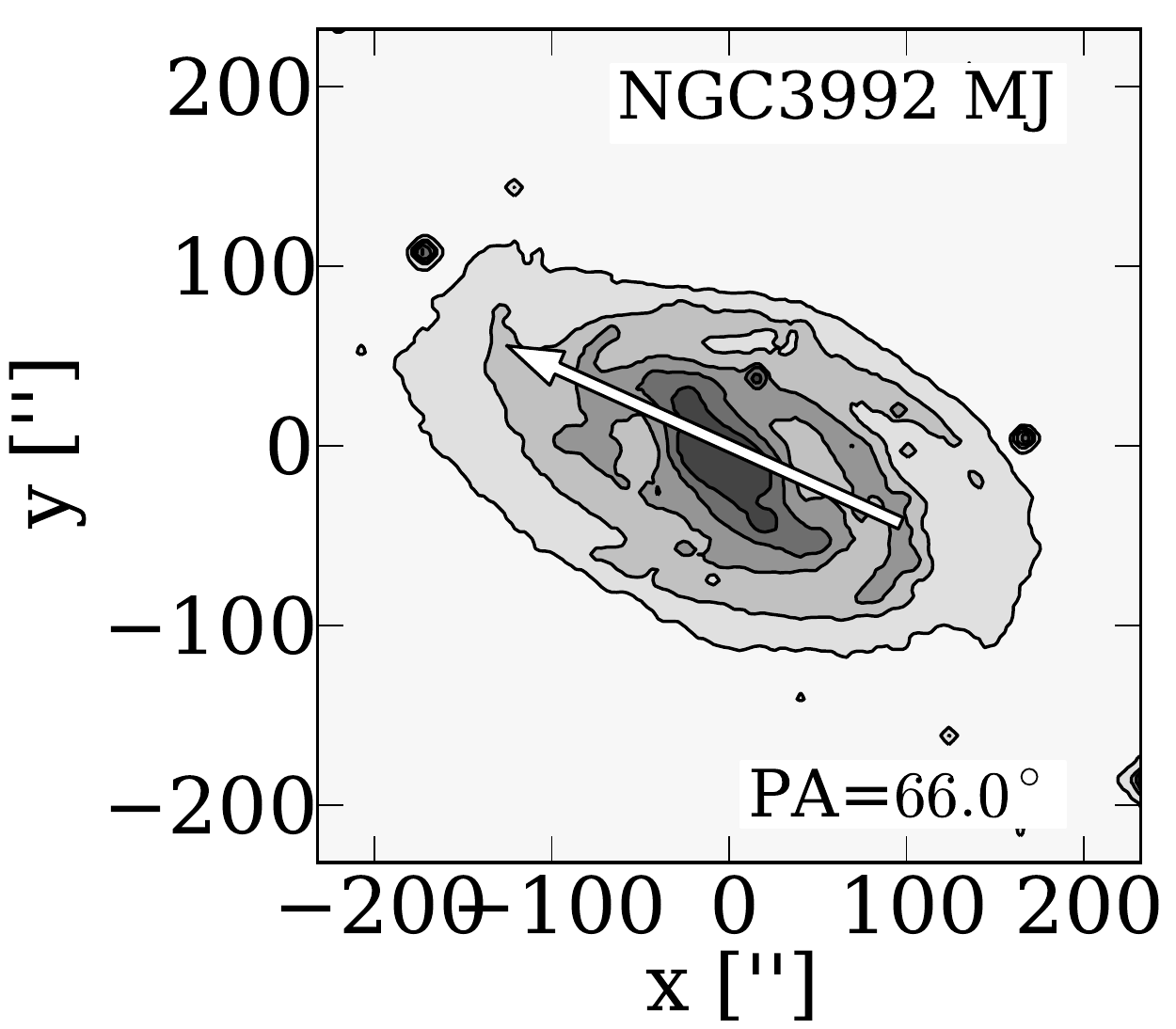}\\
        \includegraphics[viewport=0 55 390 400,width=\textwidth]{empty}
	\end{minipage} & 
	\includegraphics[viewport=0 50 420 400,width=0.35\textwidth]{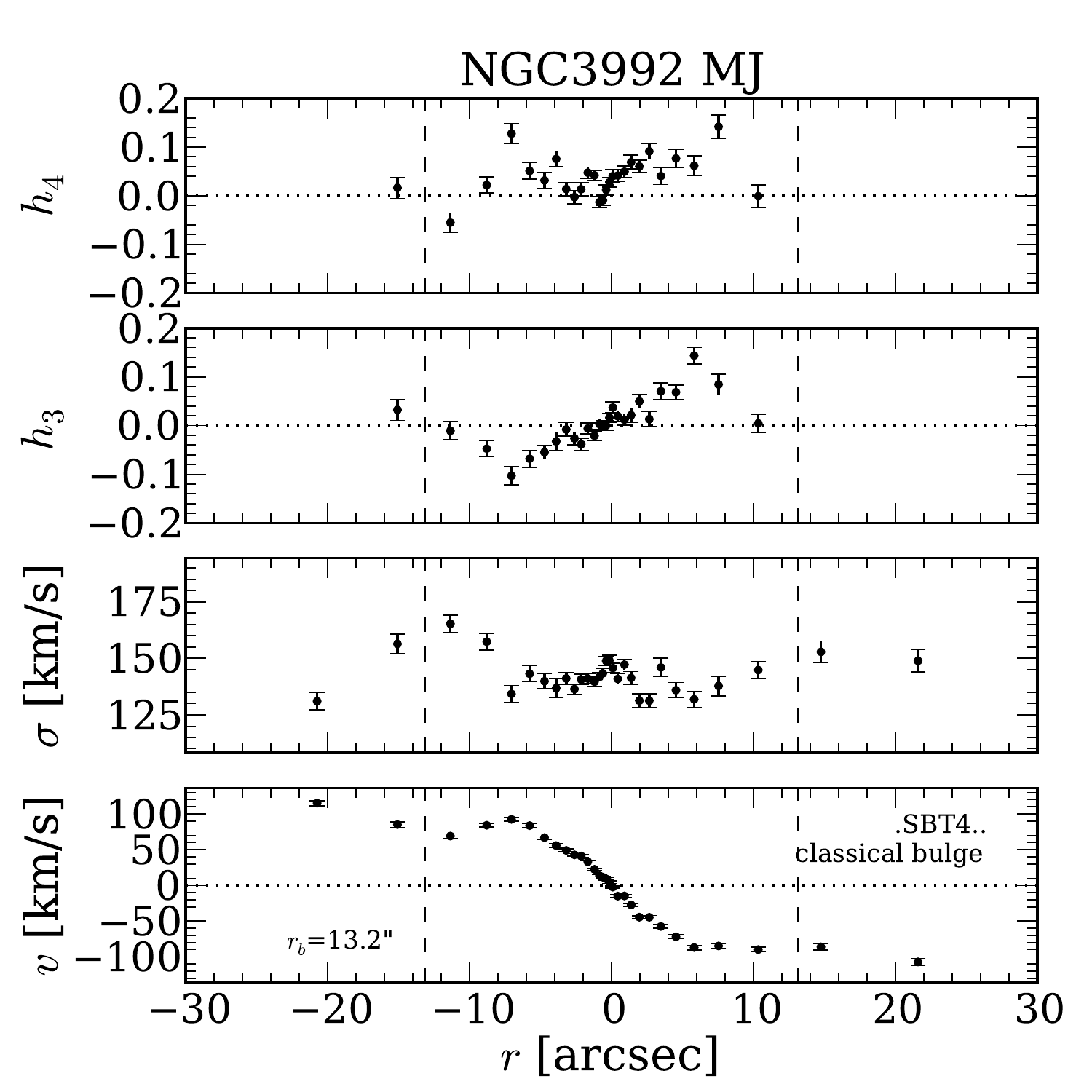}&
        \includegraphics[viewport=0 50 420 400,width=0.35\textwidth]{empty}
        \end{tabular}
        \end{center}
        \caption{{\it continued --}\small Major axis kinematic profile for NGC\,3992.
	}
\end{figure}
\setcounter{figure}{15}
\begin{figure}
        \begin{center}
        \begin{tabular}{lll}
	\begin{minipage}[b]{0.185\textwidth}
	\includegraphics[viewport=0 55 390 400,width=\textwidth]{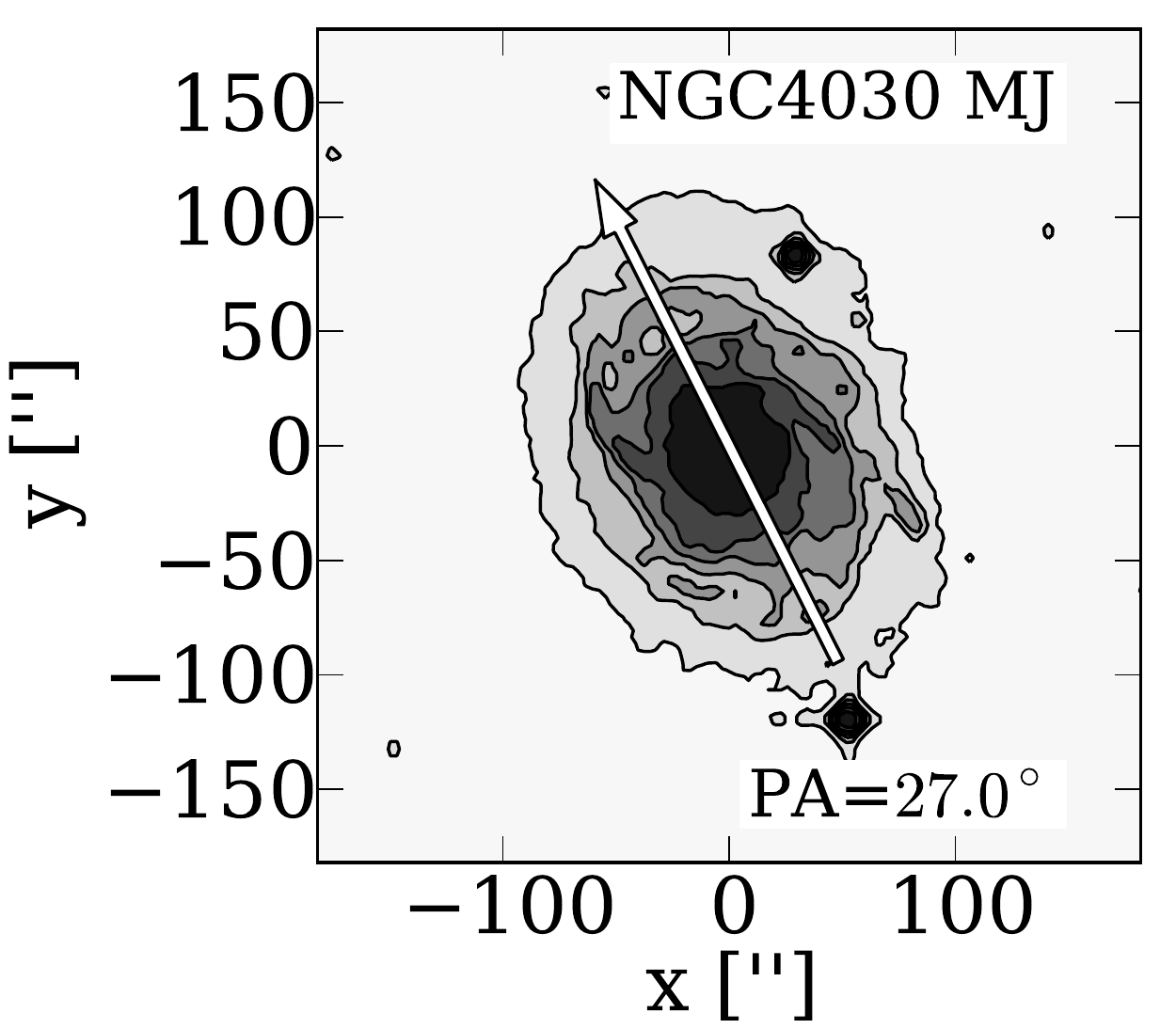}\\
        \includegraphics[viewport=0 55 390 400,width=\textwidth]{empty}
	\end{minipage} & 
	\includegraphics[viewport=0 50 420 400,width=0.35\textwidth]{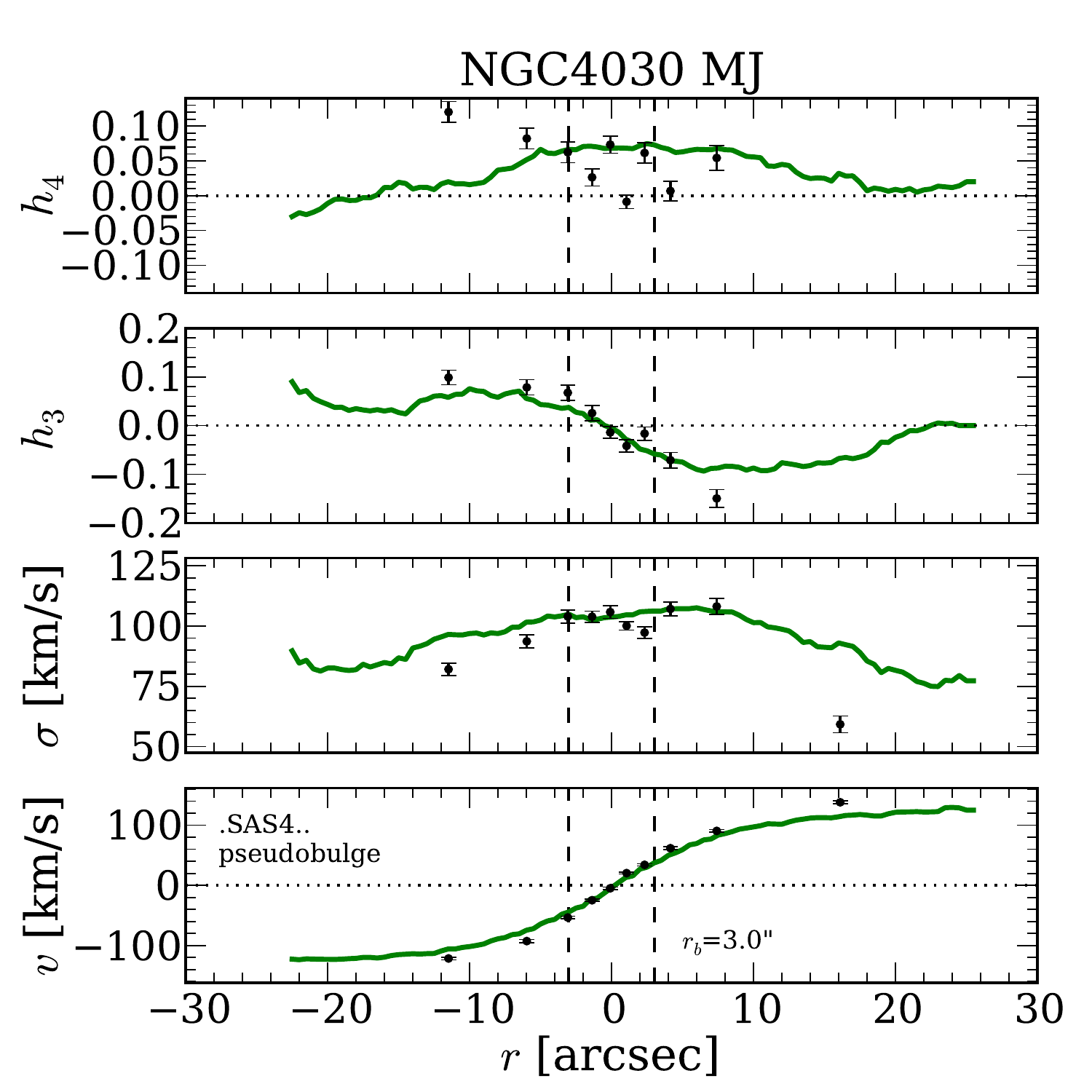}&
        \includegraphics[viewport=0 50 420 400,width=0.35\textwidth]{empty}
        \end{tabular}
        \end{center}
        \caption{{\it continued --}\small Major axis kinematic profiles for NGC\,4030. 
	We plot the SAURON results of \citet{Ganda2006} 
	} 
\end{figure}
\clearpage
\setcounter{figure}{15}
\begin{figure}
        \begin{center}
        \begin{tabular}{lll}
	\begin{minipage}[b]{0.185\textwidth}
	\includegraphics[viewport=0 55 390 400,width=\textwidth]{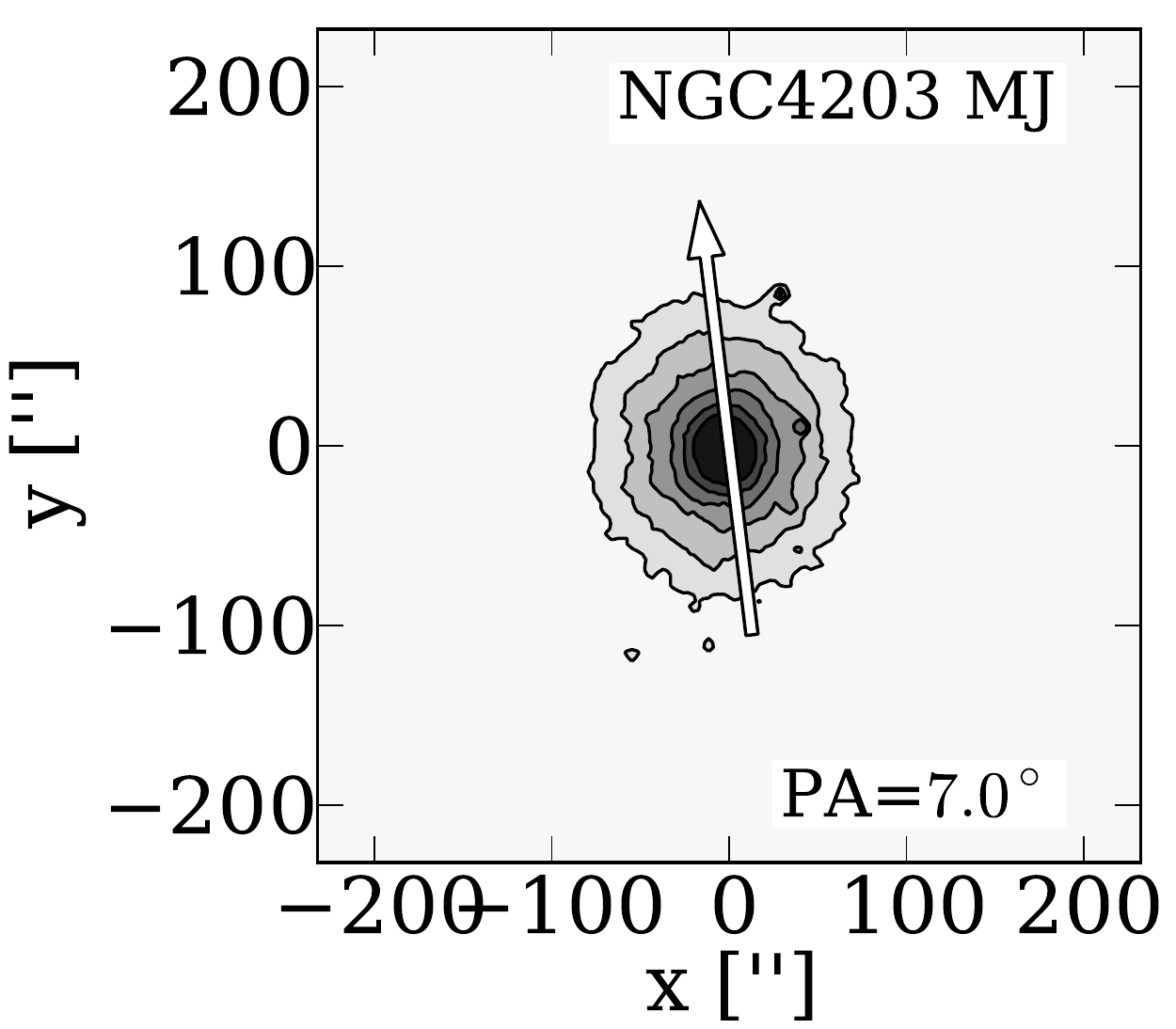}\\
        \includegraphics[viewport=0 55 390 400,width=\textwidth]{empty}
	\end{minipage} & 
	\includegraphics[viewport=0 50 420 400,width=0.35\textwidth]{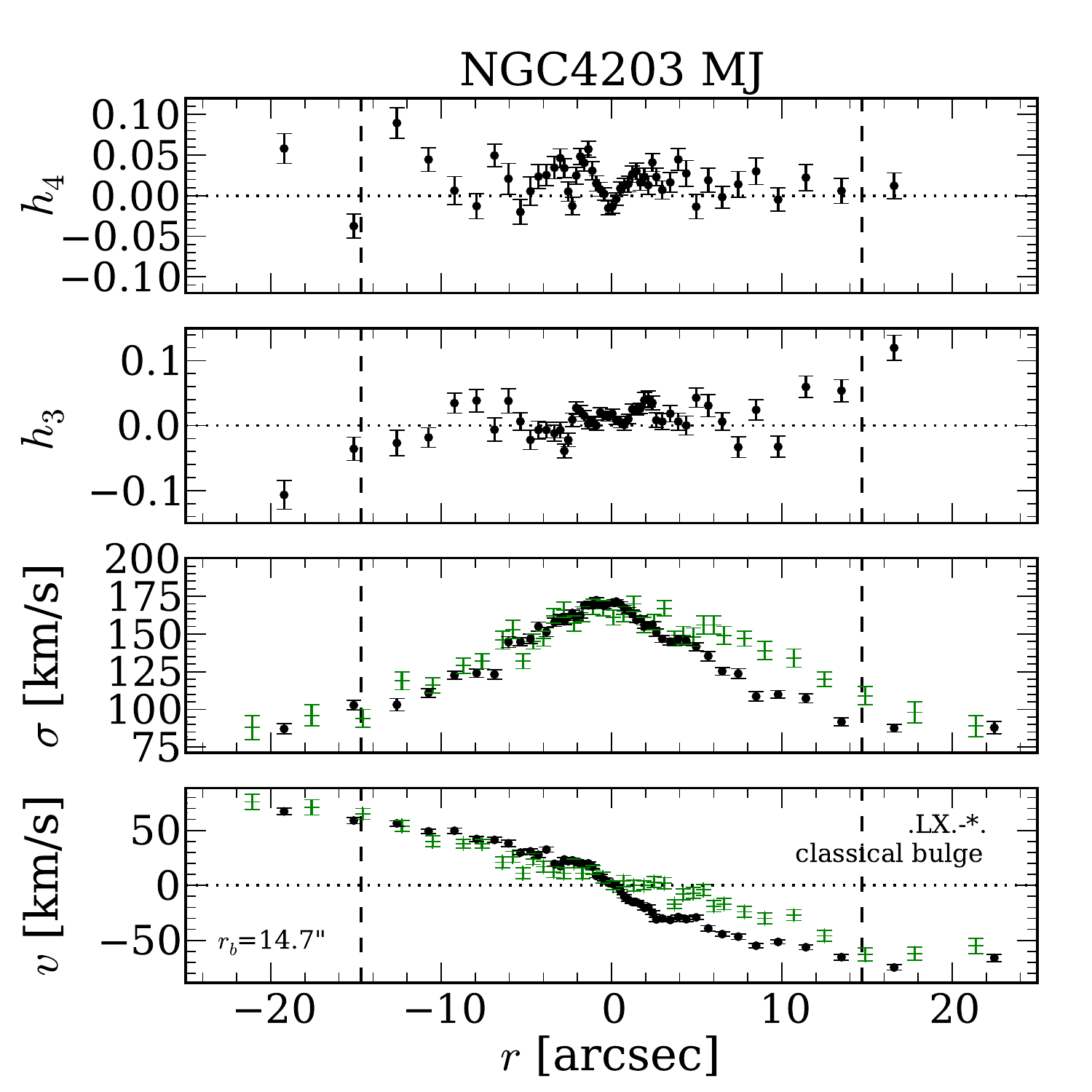} &
        \includegraphics[viewport=0 50 420 400,width=0.35\textwidth]{empty}
        \end{tabular}
        \end{center}
        \caption{{\it continued --}\small Major axis kinematic profiles for NGC\,4203. 
	We plot results of \citet{Simien2002} in green.
	The latter were taken at a position angle of 30\Deg\ which is quite different from our adopted value for the major axis
	position angle of 7\Deg. The difference in velocity and dispersion seen on the east side is probably explained by this. 
	}
\end{figure}
\setcounter{figure}{15}
\begin{figure}
        \begin{center}
        \begin{tabular}{lll}
	\begin{minipage}[b]{0.185\textwidth}
	\includegraphics[viewport=0 55 390 400,width=\textwidth]{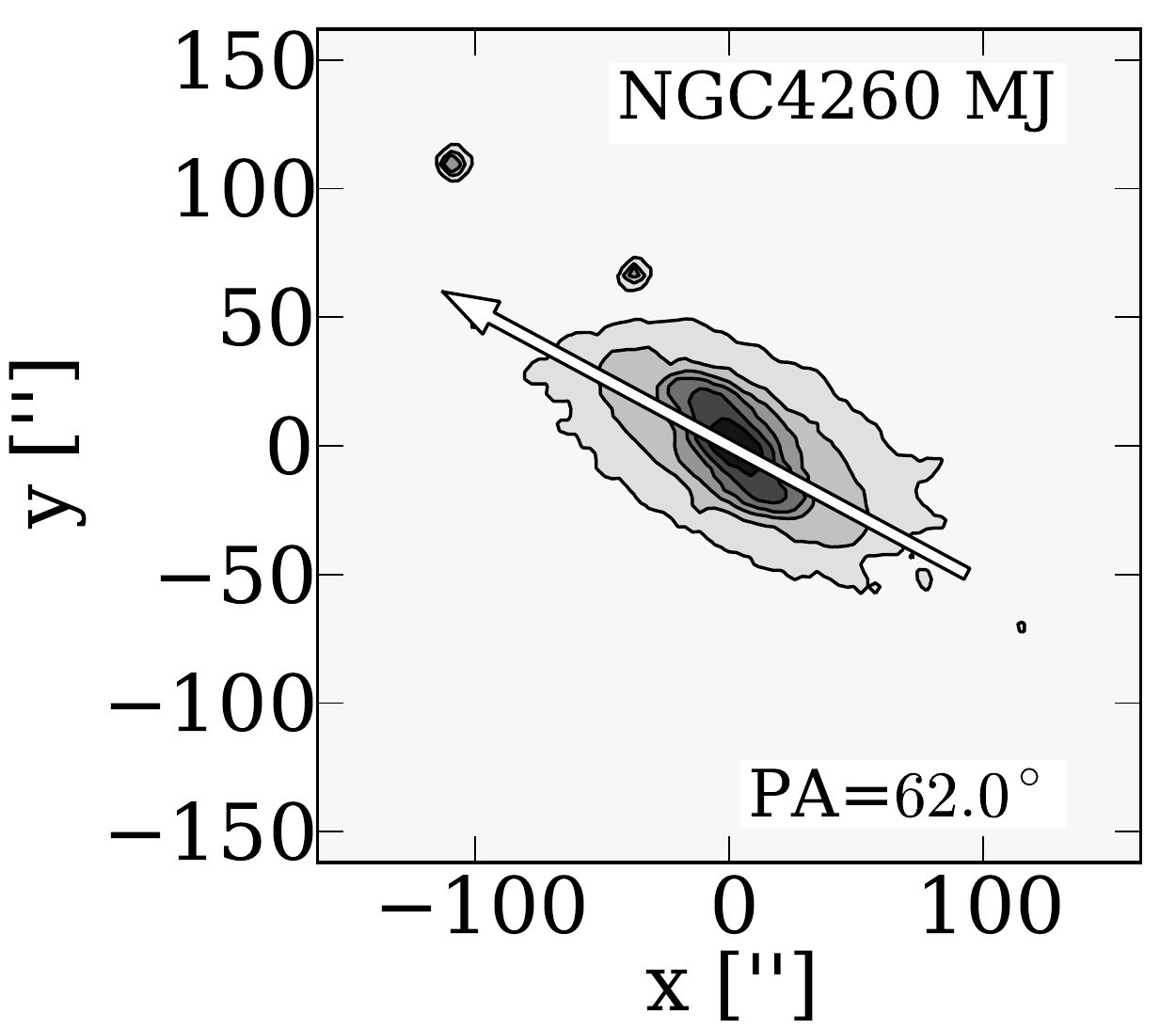}\\
        \includegraphics[viewport=0 55 390 400,width=\textwidth]{empty}
	\end{minipage} & 
	\includegraphics[viewport=0 50 420 400,width=0.35\textwidth]{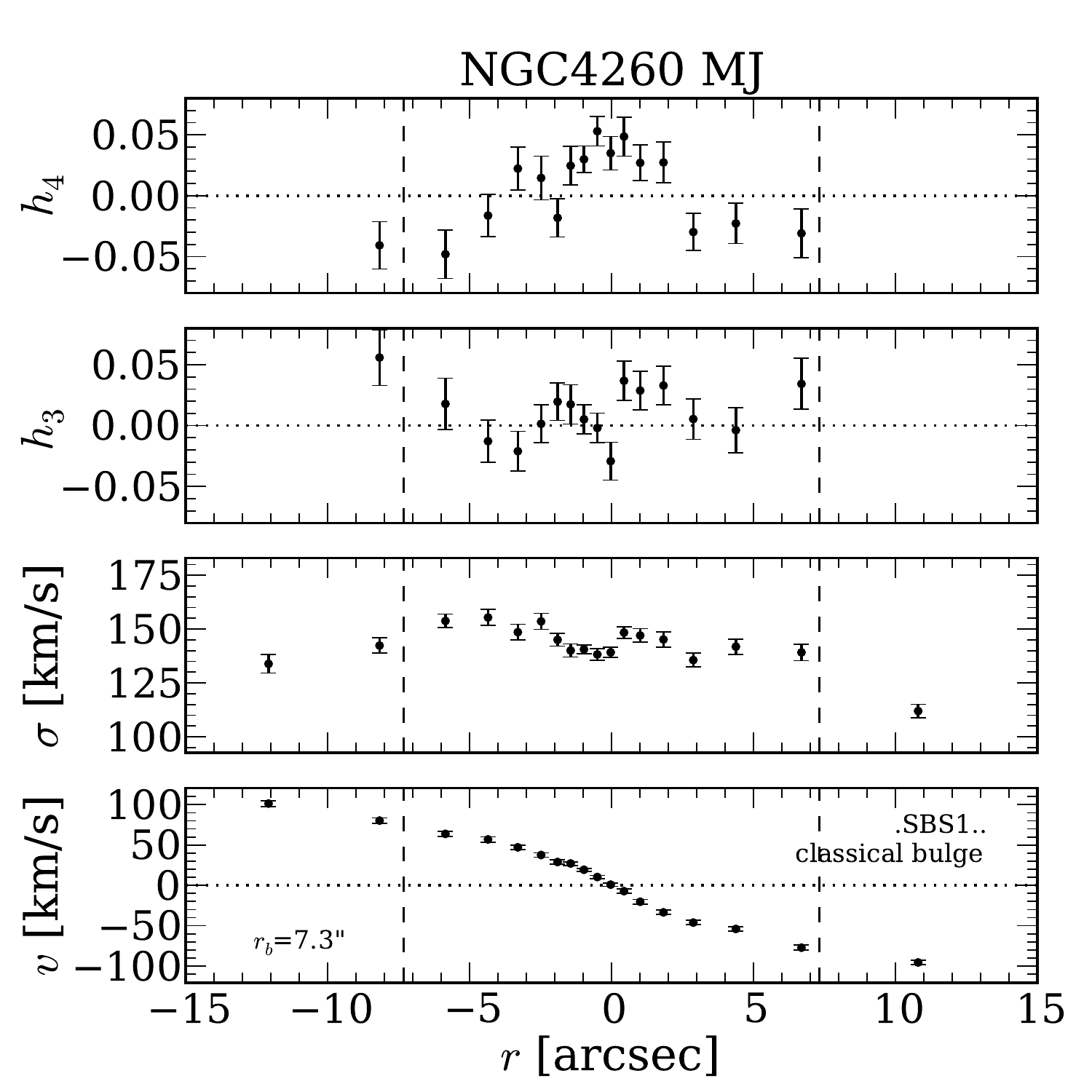}&
        \includegraphics[viewport=0 50 420 400,width=0.35\textwidth]{empty}
        \end{tabular}
        \end{center}
        \caption{{\it continued --}\small Major axis kinematic profile for NGC\,4260.} 
\end{figure}
\setcounter{figure}{15}
\begin{figure}
        \begin{center}
        \begin{tabular}{lll}
	\begin{minipage}[b]{0.185\textwidth}
	\includegraphics[viewport=0 55 390 400,width=\textwidth]{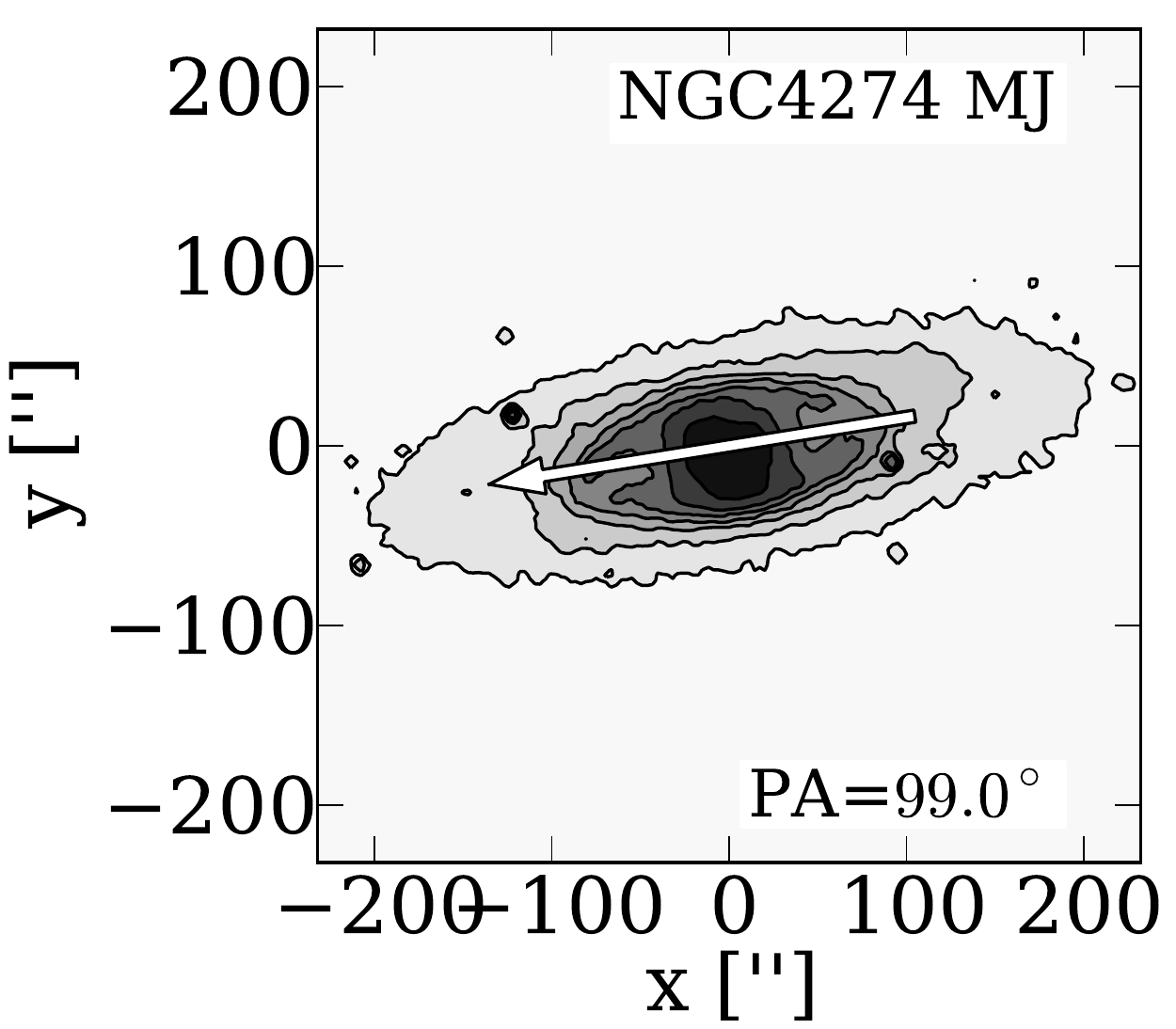}\\
        \includegraphics[viewport=0 55 390 400,width=\textwidth]{empty}
	\end{minipage} & 
	\includegraphics[viewport=0 50 420 400,width=0.35\textwidth]{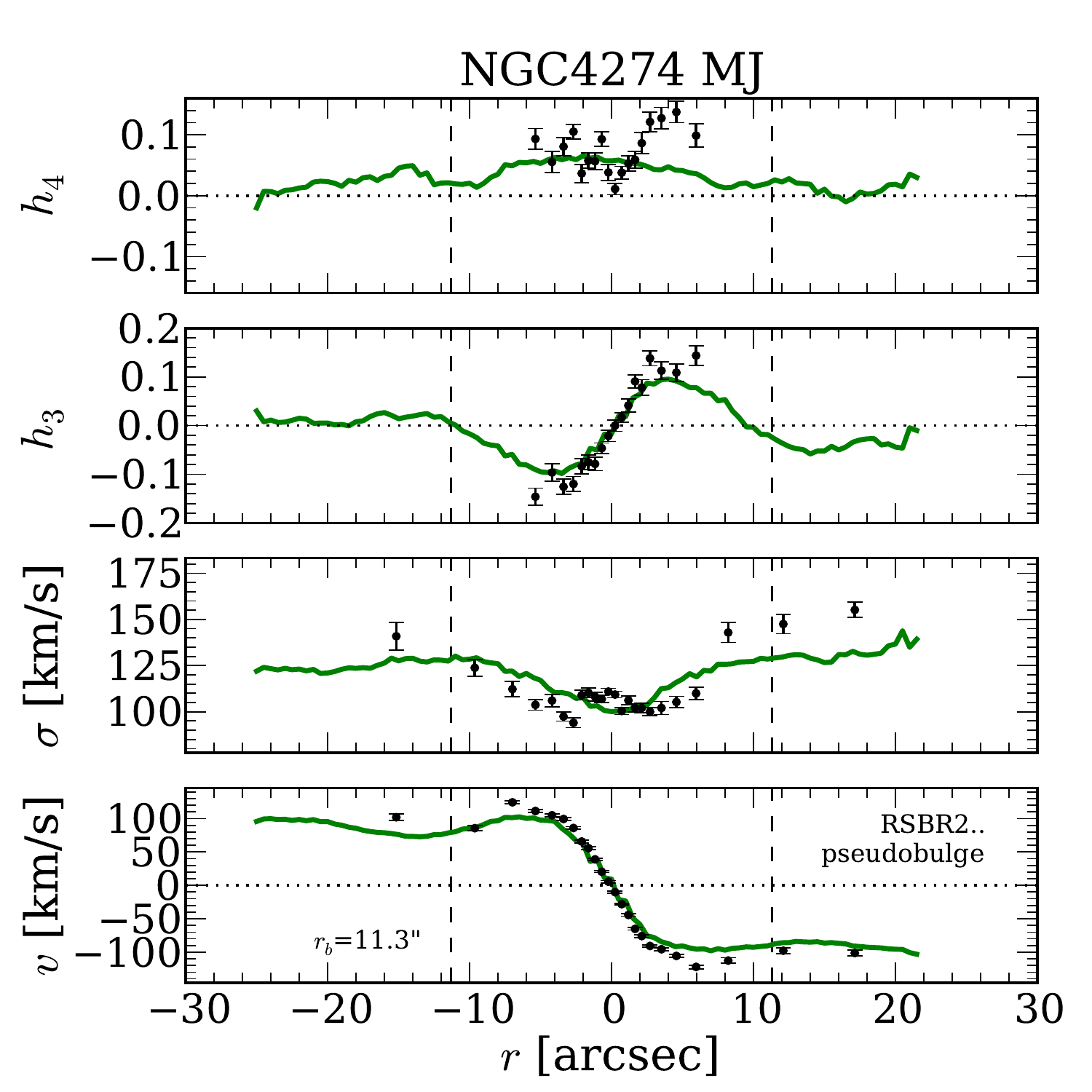}&
        \includegraphics[viewport=0 50 420 400,width=0.35\textwidth]{empty}
        \end{tabular}
        \end{center}
        \caption{{\it continued --}\small Major axis kinematic profile for NGC\,4274.
	We plot the SAURON results \citep{Falcon-Barroso2006} in green.} 
\end{figure}
\setcounter{figure}{15}
\begin{figure}
        \begin{center}
        \begin{tabular}{lll}
	\begin{minipage}[b]{0.185\textwidth}
	\includegraphics[viewport=0 55 390 400,width=\textwidth]{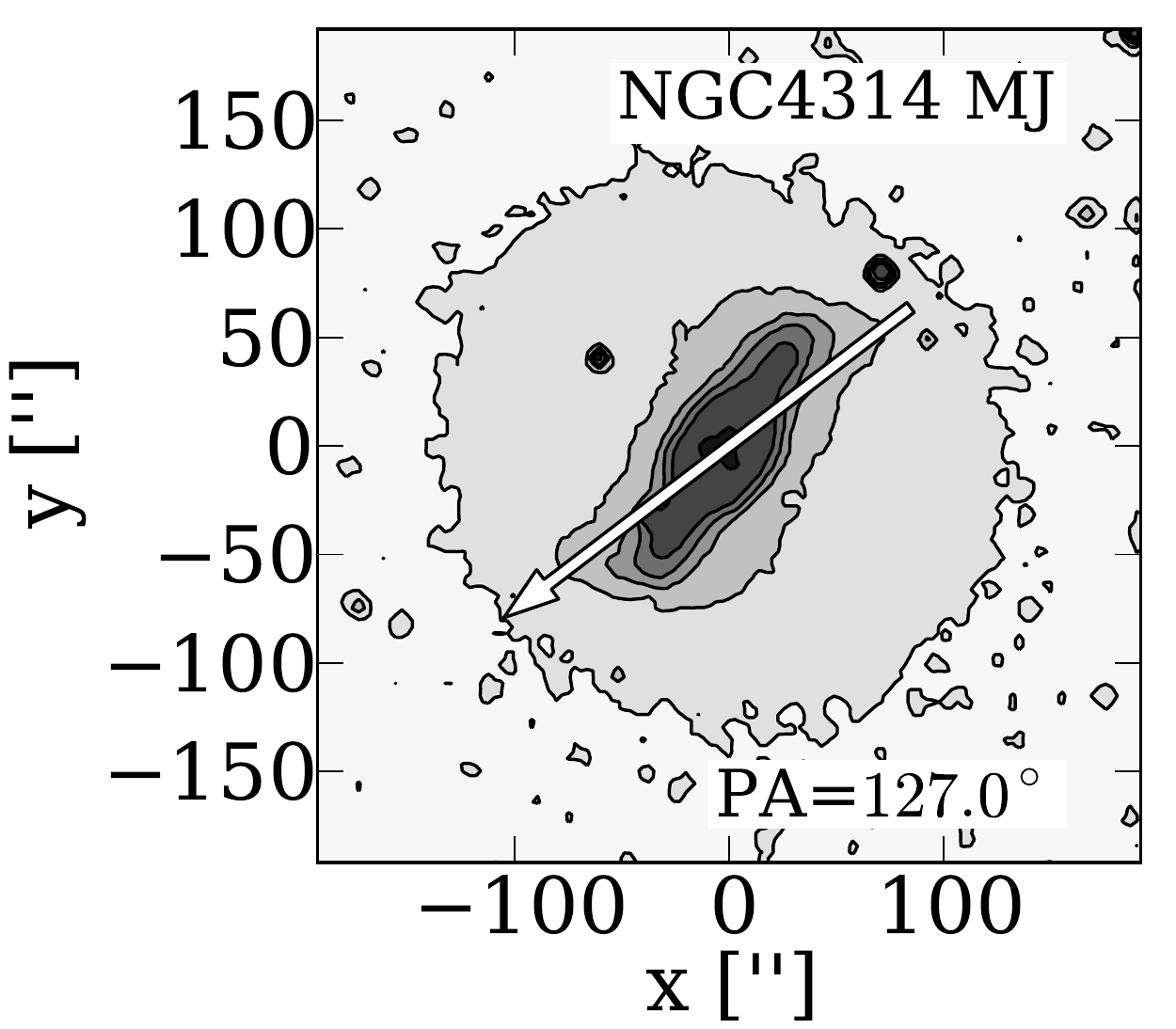}\\
        \includegraphics[viewport=0 55 390 400,width=\textwidth]{empty}
	\end{minipage} & 
	\includegraphics[viewport=0 50 420 400,width=0.35\textwidth]{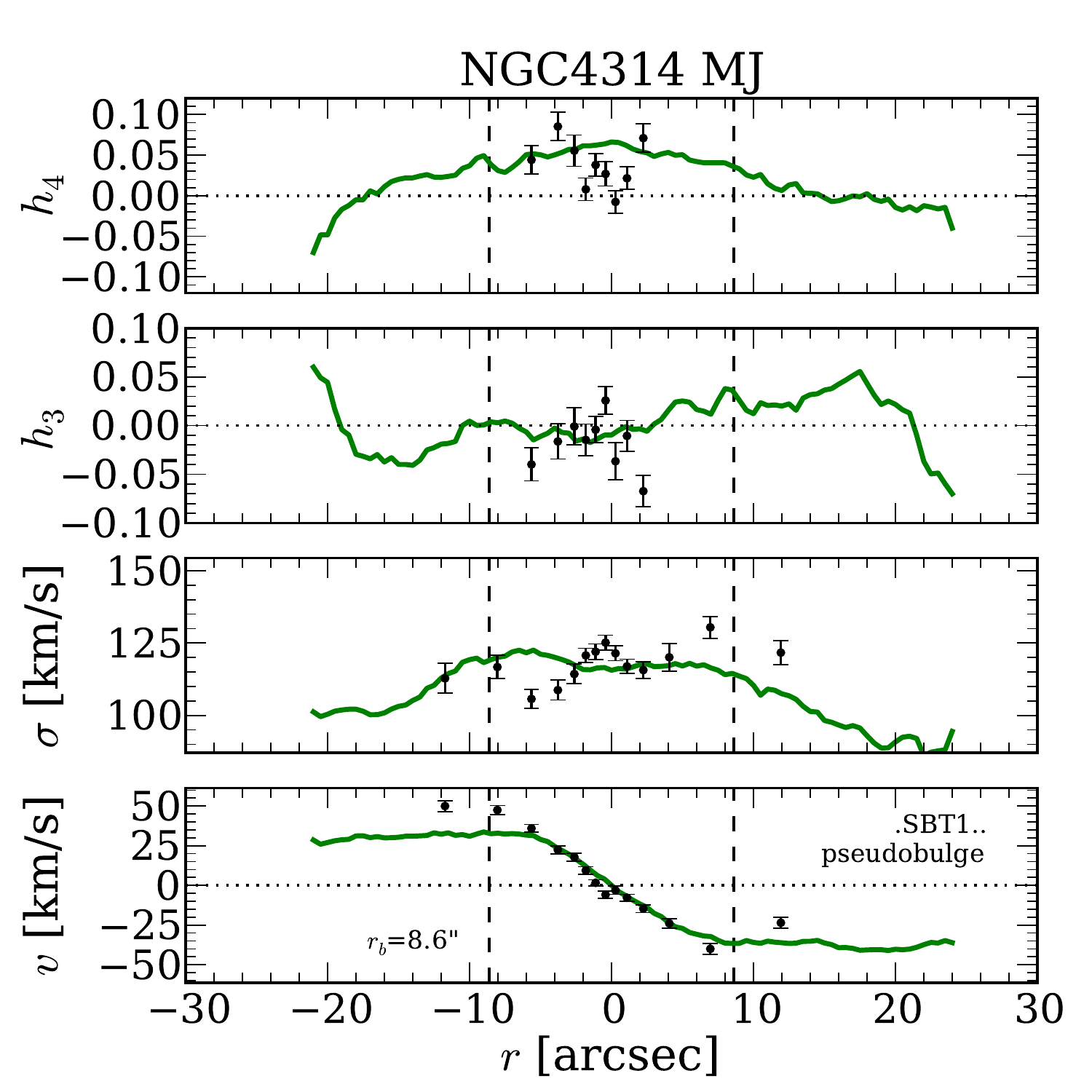} &
        \includegraphics[viewport=0 50 420 400,width=0.35\textwidth]{empty}
        \end{tabular}
        \end{center}
        \caption{{\it continued --}\small Major axis kinematic profile for NGC\,4314.
	We plot the SAURON results \citep{Falcon-Barroso2006} in green.
	}
\end{figure}
\setcounter{figure}{15}
%
%
\begin{figure}
        \begin{center}
        \begin{tabular}{lll}
	\begin{minipage}[b]{0.185\textwidth}
	\includegraphics[viewport=0 55 390 400,width=\textwidth]{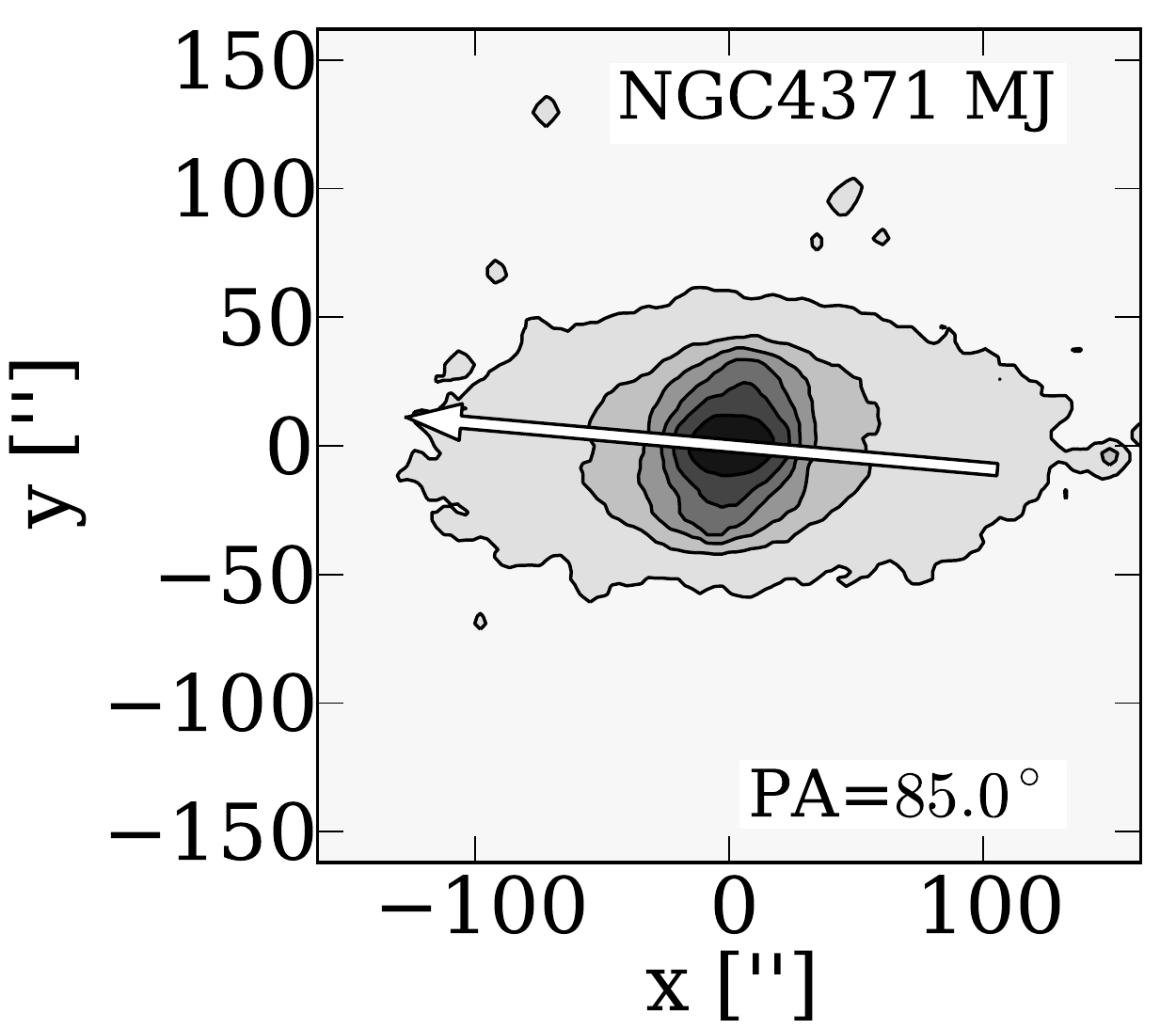}\\
	\includegraphics[viewport=0 55 390 400,width=\textwidth]{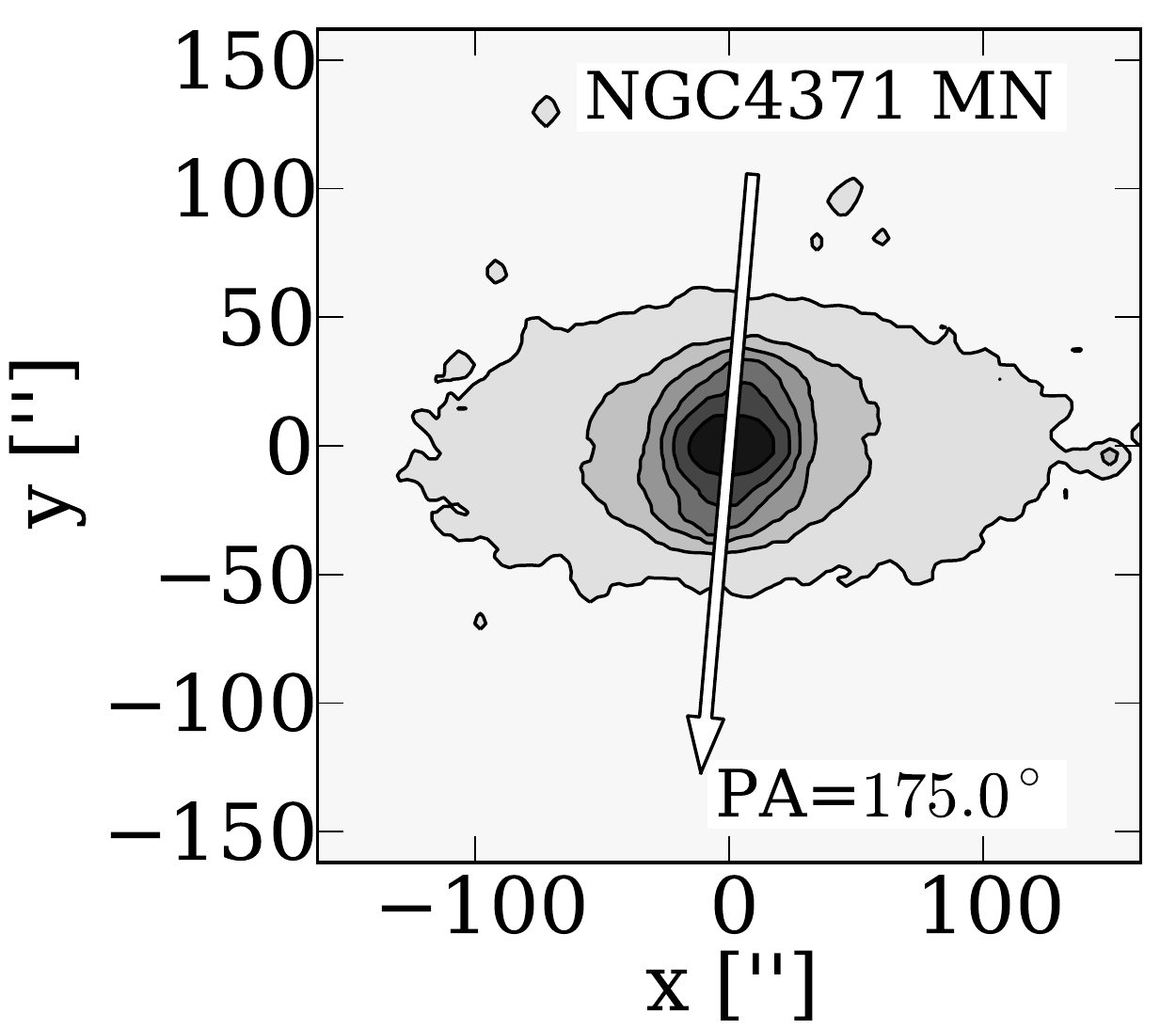}
	\end{minipage} & 
	\includegraphics[viewport=0 50 420 400,width=0.35\textwidth]{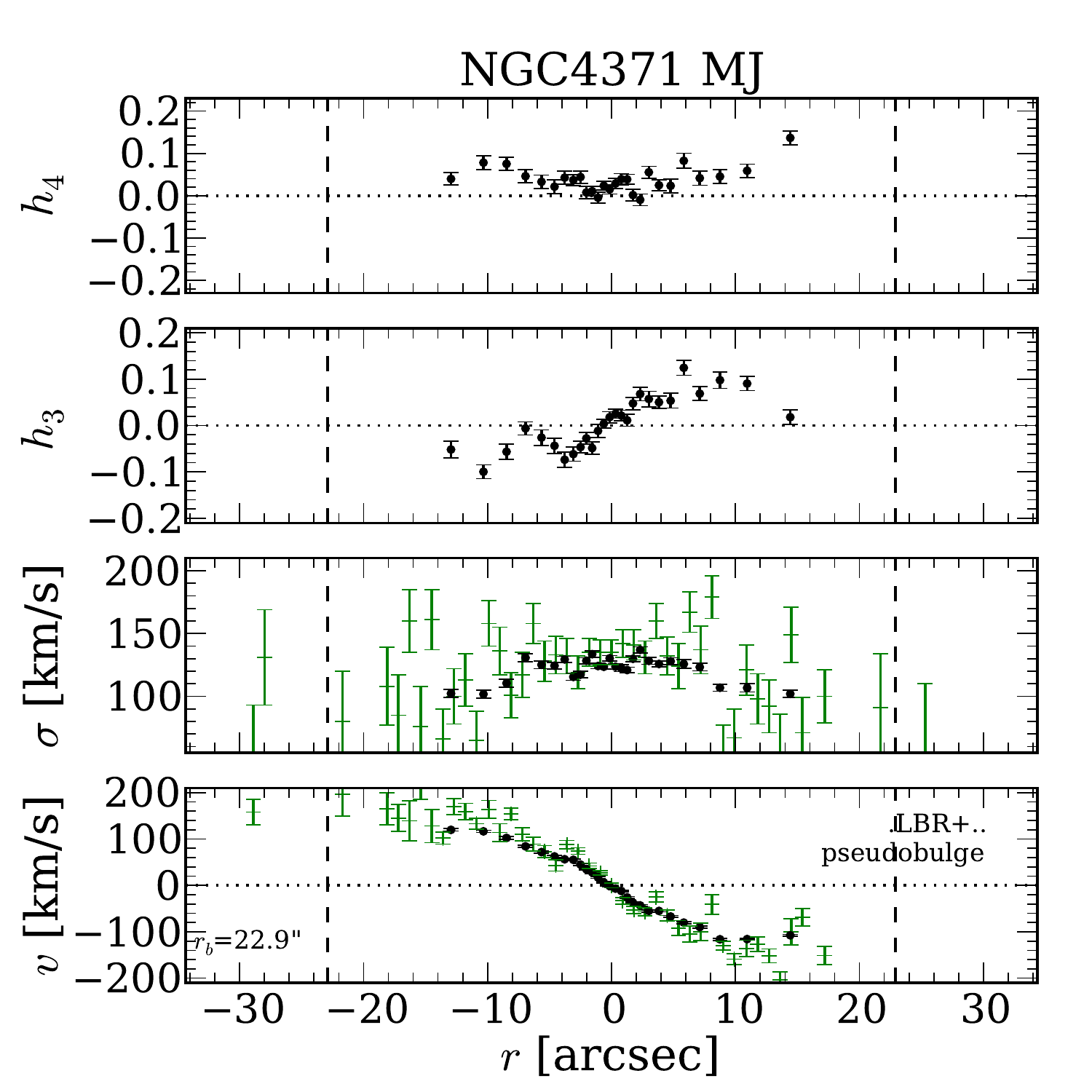} &
	\includegraphics[viewport=0 50 420 400,width=0.35\textwidth]{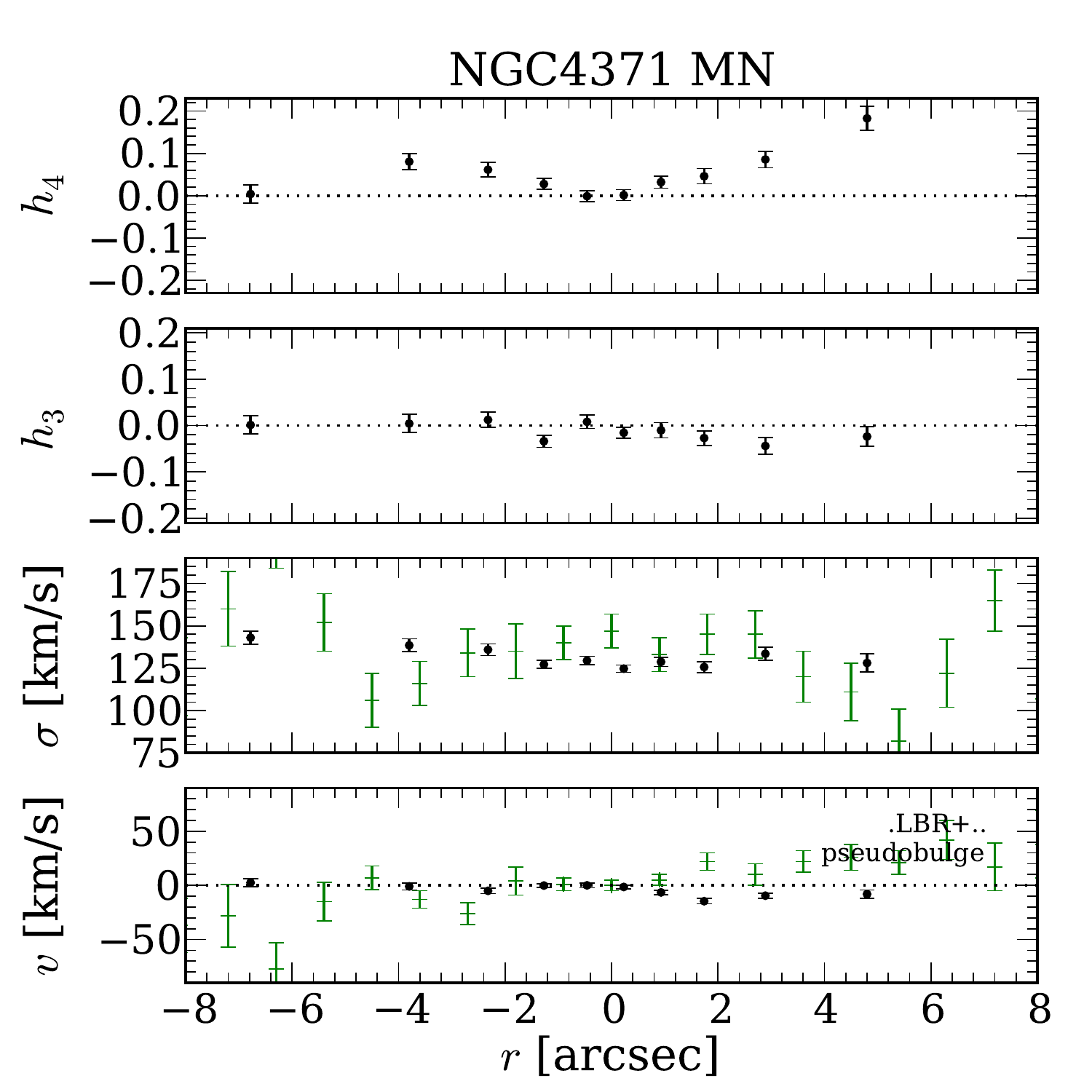}\\
        \end{tabular}
        \end{center}
        \caption{{\it continued --}\small Major and minor axis kinematic profiles for NGC\,4371. We 
	plot the PA=95\Deg\ data from \citet{Bettoni1997} on our major axis data
	and their PA=175\Deg\ on our minor axis data in green.
	}
\end{figure}
\setcounter{figure}{15}
\begin{figure}
        \begin{center}
        \begin{tabular}{lll}
	\begin{minipage}[b]{0.185\textwidth}
	\includegraphics[viewport=0 55 390 400,width=\textwidth]{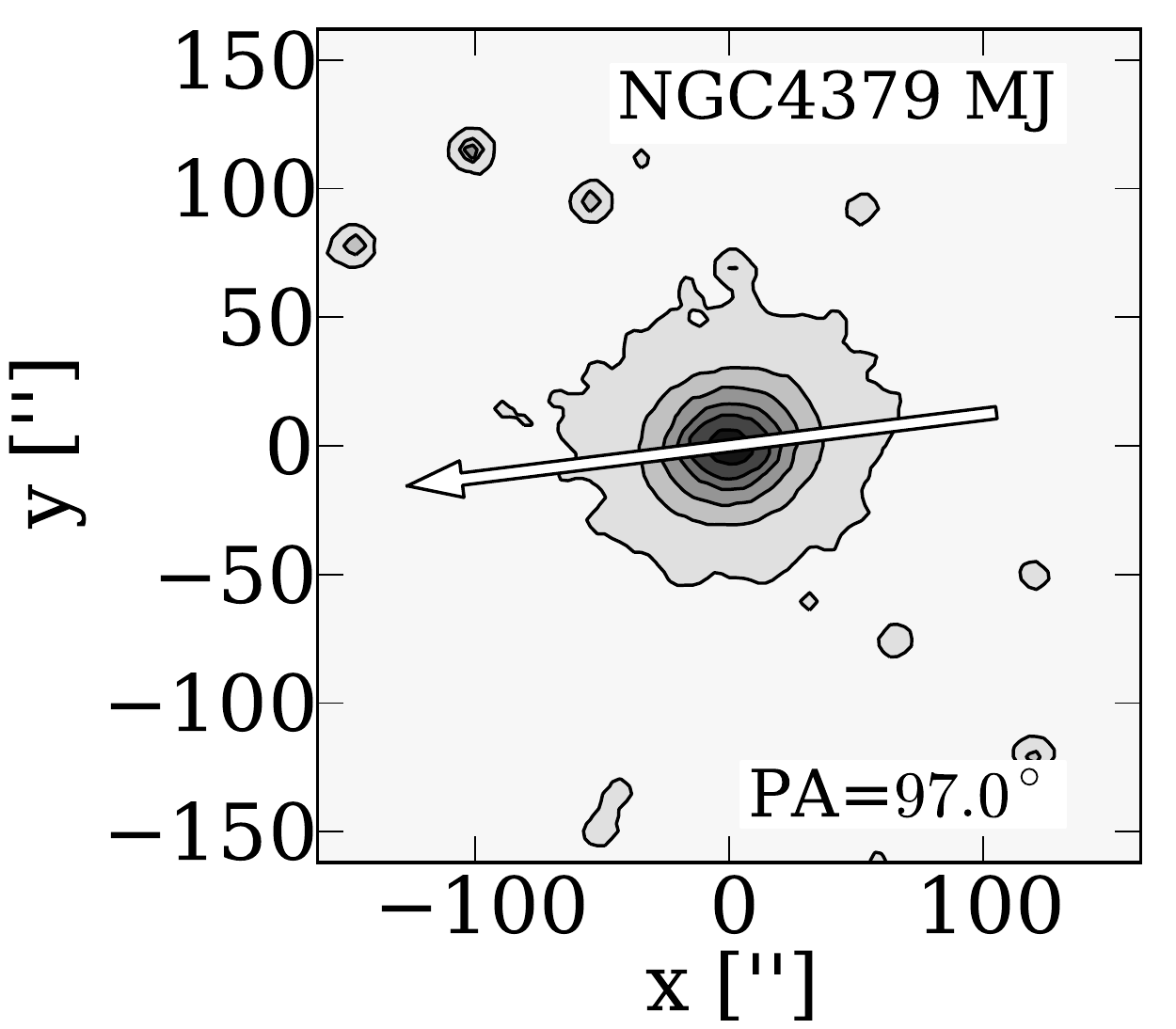}\\
        \includegraphics[viewport=0 55 390 400,width=\textwidth]{empty}
	\end{minipage} & 
	\includegraphics[viewport=0 50 420 400,width=0.35\textwidth]{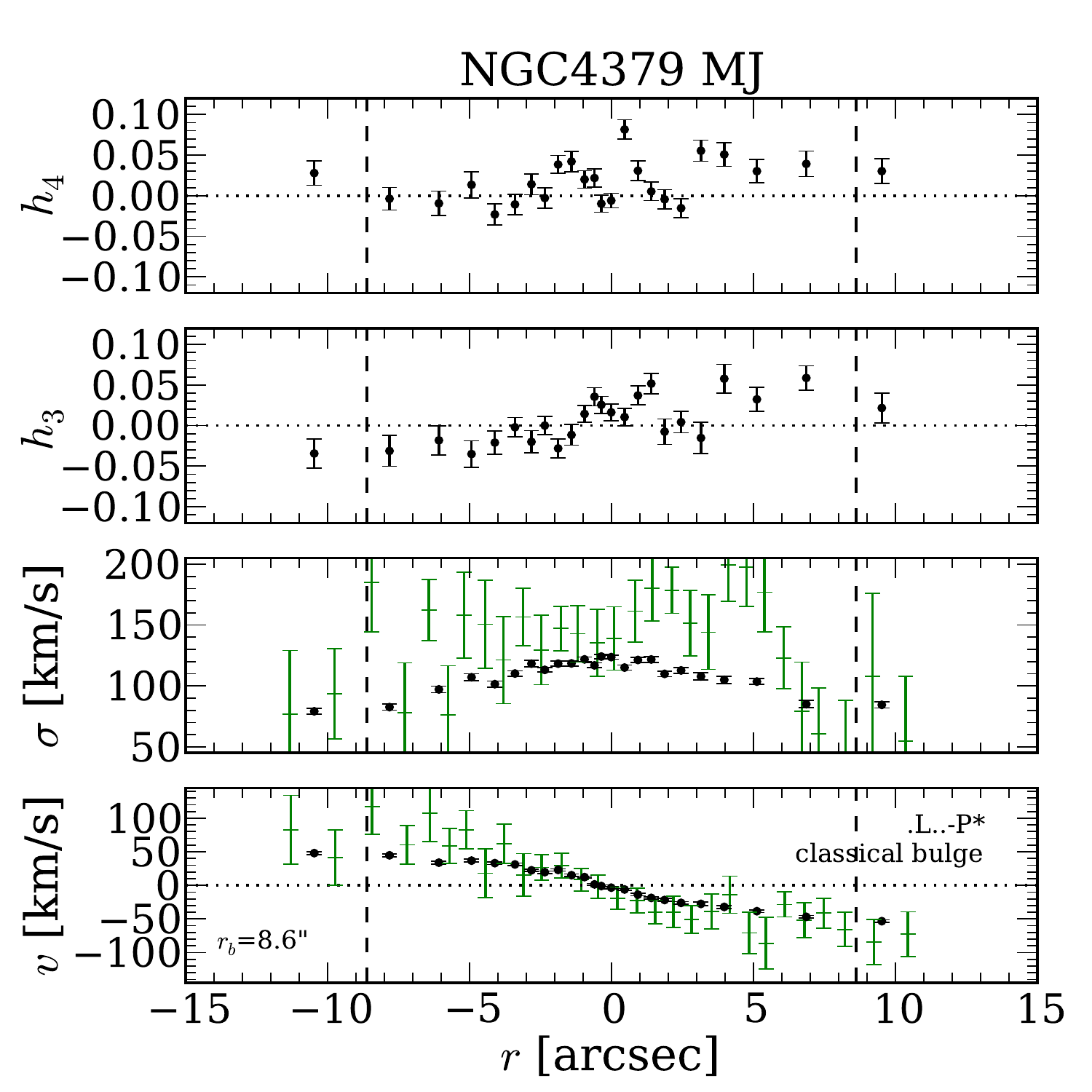}&
        \includegraphics[viewport=0 50 420 400,width=0.35\textwidth]{empty}
        \end{tabular}
        \end{center}
        \caption{{\it continued --}\small Major axis kinematic profile for NGC\,4379.
	We plot the results of \cite{Bertola1995} in green.
	}
\end{figure}
\setcounter{figure}{15}
\begin{figure}
        \begin{center}
        \begin{tabular}{lll}
	\begin{minipage}[b]{0.185\textwidth}
	\includegraphics[viewport=0 55 390 400,width=\textwidth]{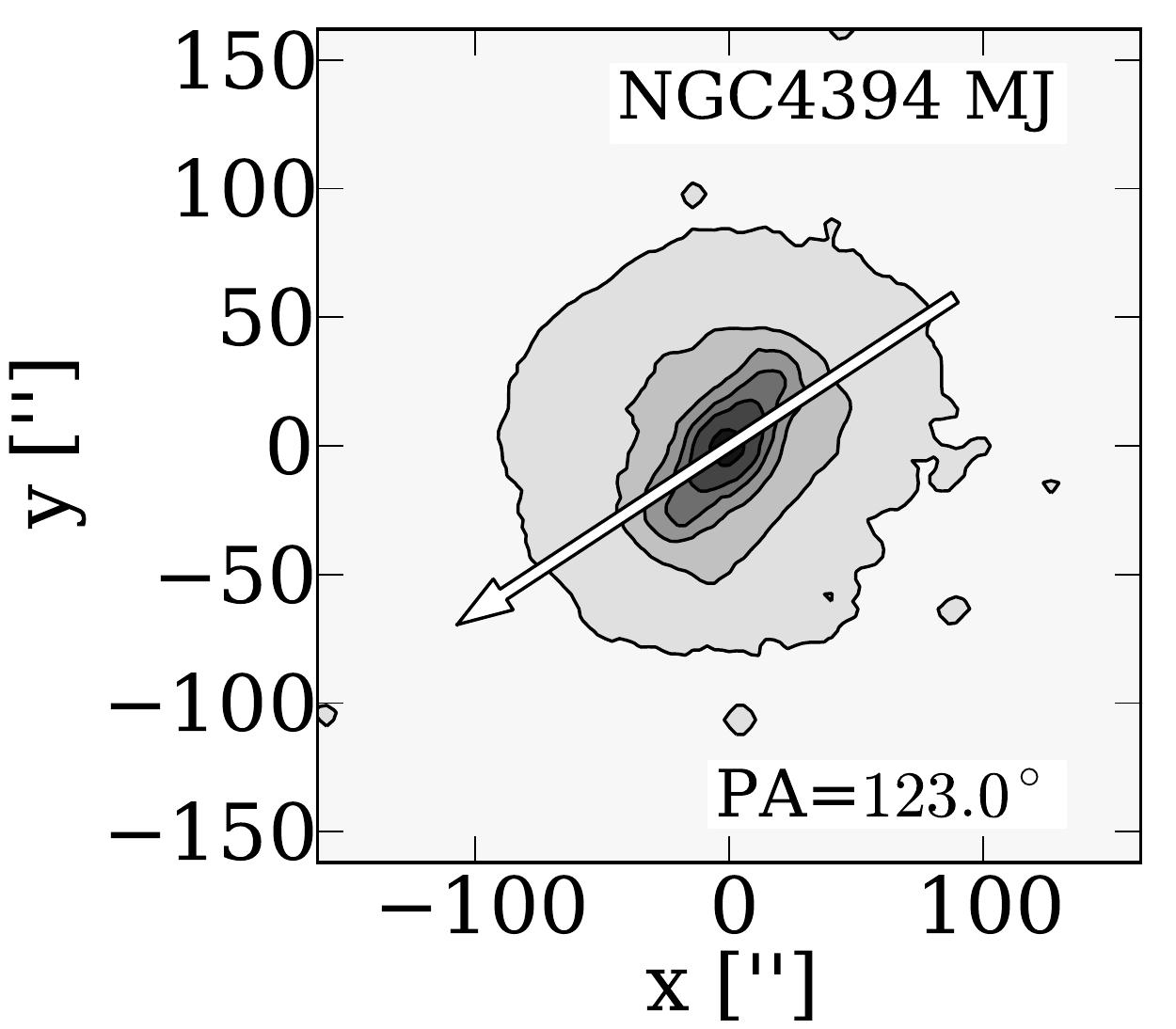} \\
        \includegraphics[viewport=0 55 390 400,width=\textwidth]{empty}
	\end{minipage} & 
	\includegraphics[viewport=0 50 420 400,width=0.35\textwidth]{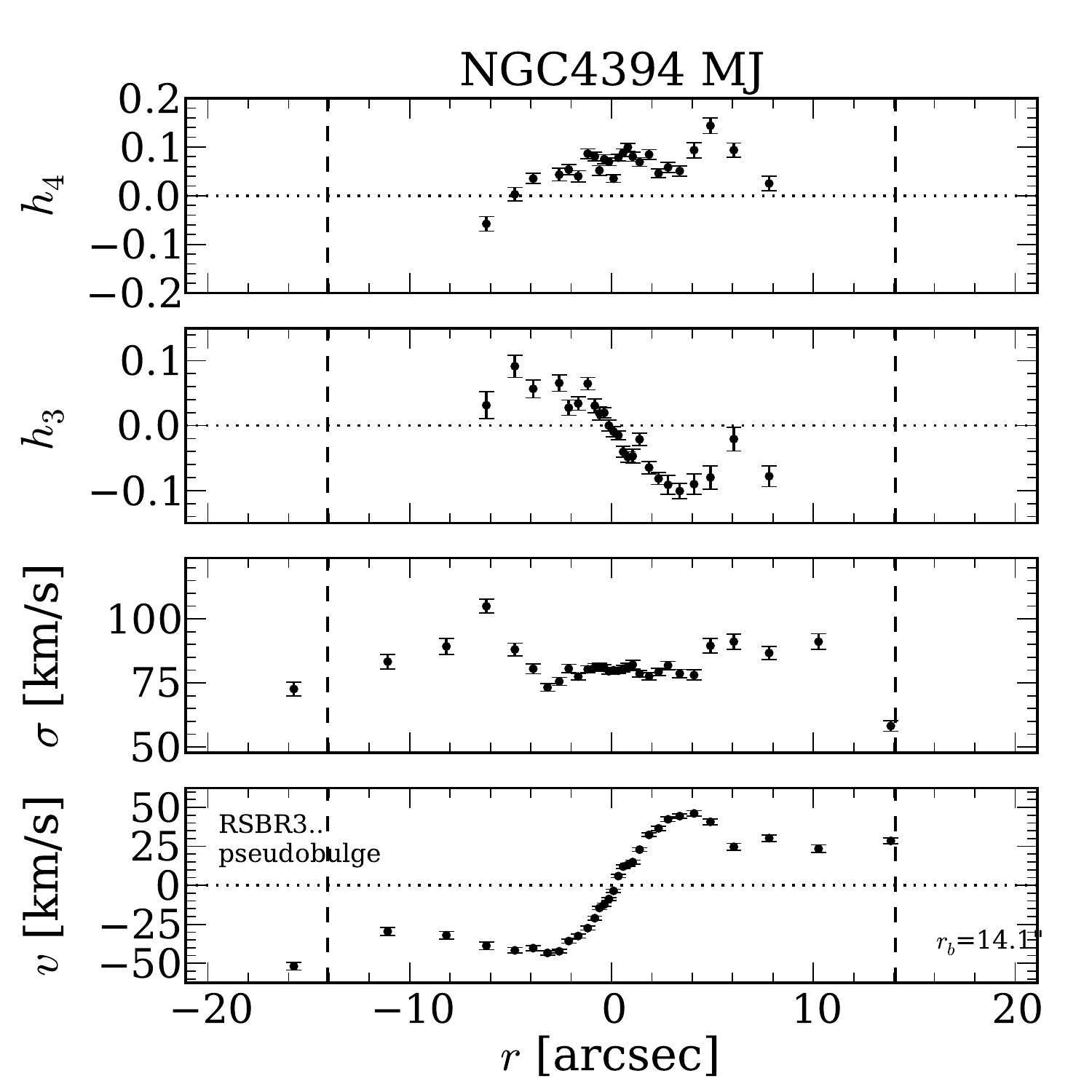} & 
        \includegraphics[viewport=0 50 420 400,width=0.35\textwidth]{empty}
        \end{tabular}
        \end{center}
        \caption{{\it continued --}\small Major axis kinematic profile for NGC\,4394.
	}
\end{figure}
\setcounter{figure}{15}
\begin{figure}
        \begin{center}
        \begin{tabular}{lll}
	\begin{minipage}[b]{0.185\textwidth}
	\includegraphics[viewport=0 55 390 400,width=\textwidth]{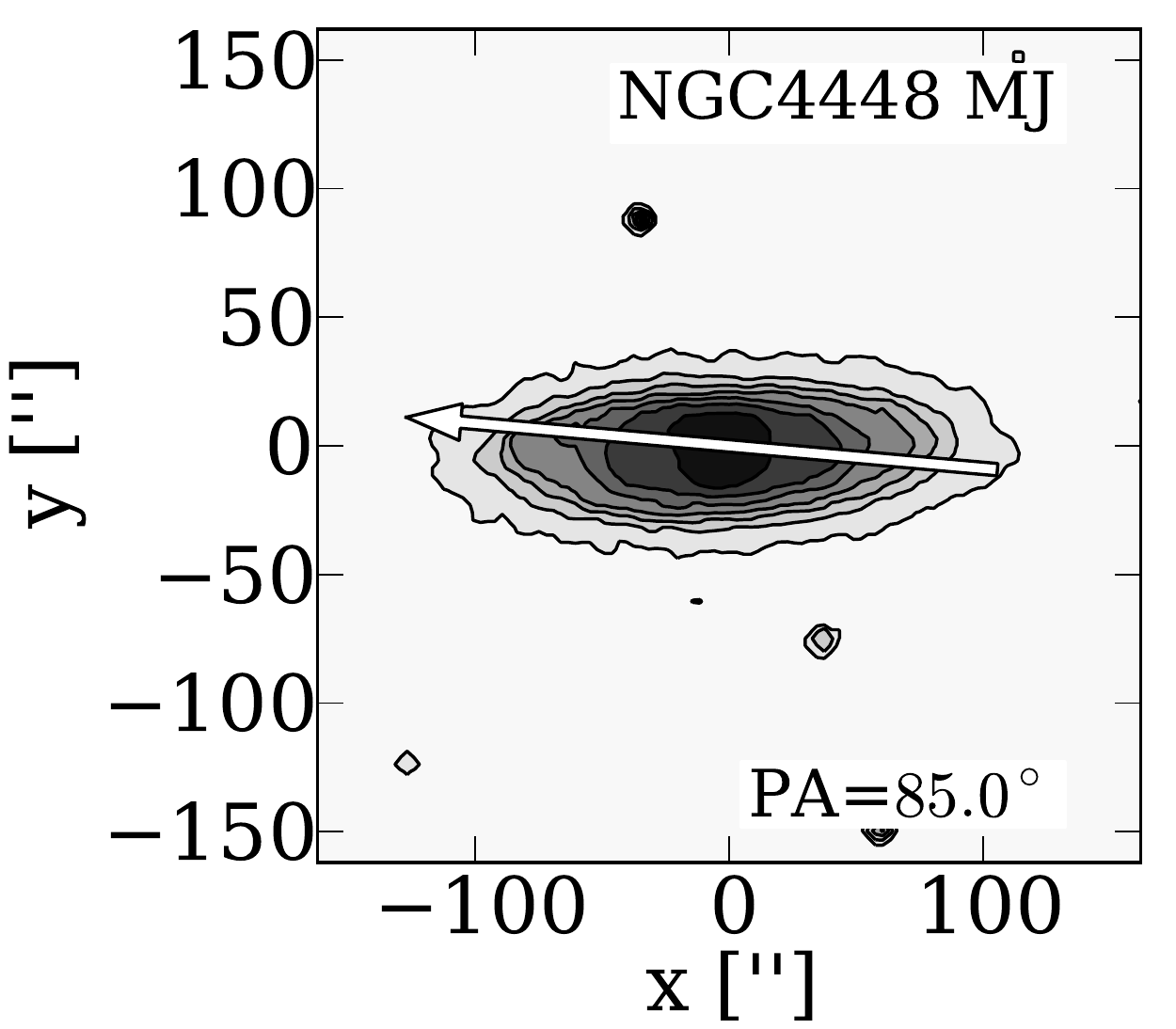}\\
        \includegraphics[viewport=0 55 390 400,width=\textwidth]{empty}
	\end{minipage} & 
	\includegraphics[viewport=0 50 420 400,width=0.35\textwidth]{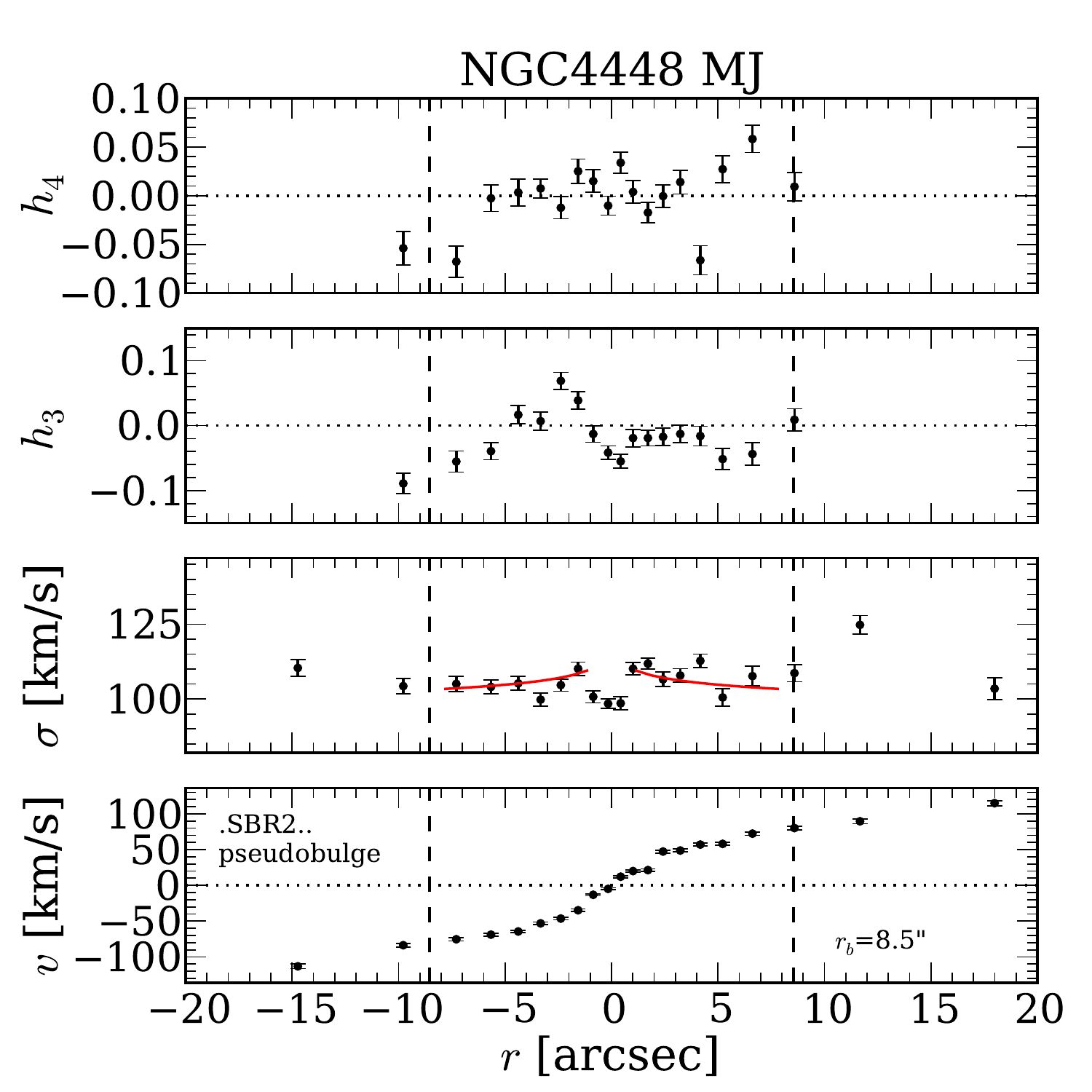} &
        \includegraphics[viewport=0 50 420 400,width=0.35\textwidth]{empty}
        \end{tabular}
        \end{center}
        \caption{{\it continued --}\small Major axis kinematic profile for 
	NGC\,4448, reproduced from Fig.~\ref{fig:twoprofiles}.
	}
\end{figure}
\clearpage
\setcounter{figure}{15}
\begin{figure}
        \begin{center}
        \begin{tabular}{lll}
	\begin{minipage}[b]{0.185\textwidth}
	\includegraphics[viewport=0 55 390 400,width=\textwidth]{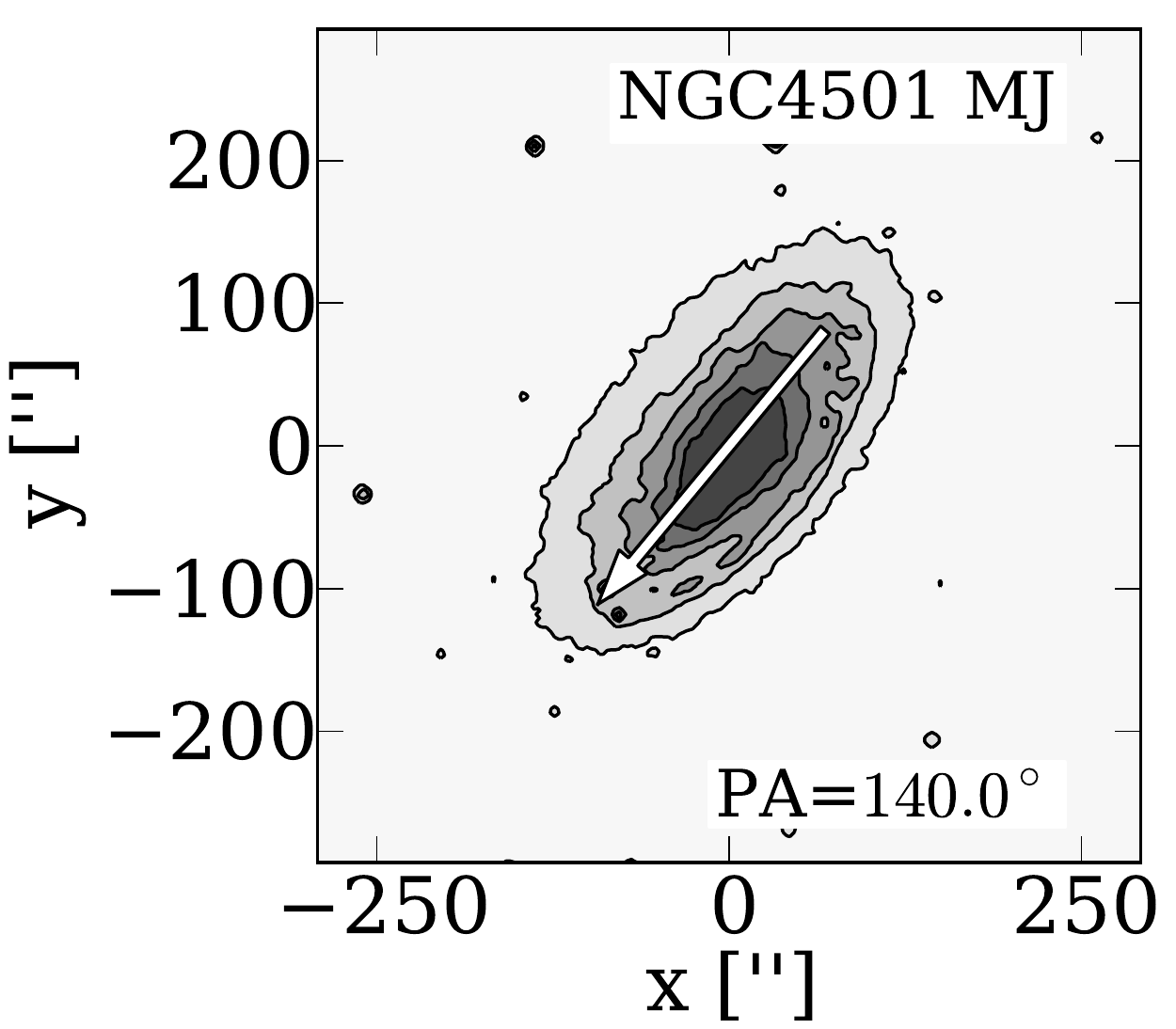}\\
	\includegraphics[viewport=0 55 390 400,width=\textwidth]{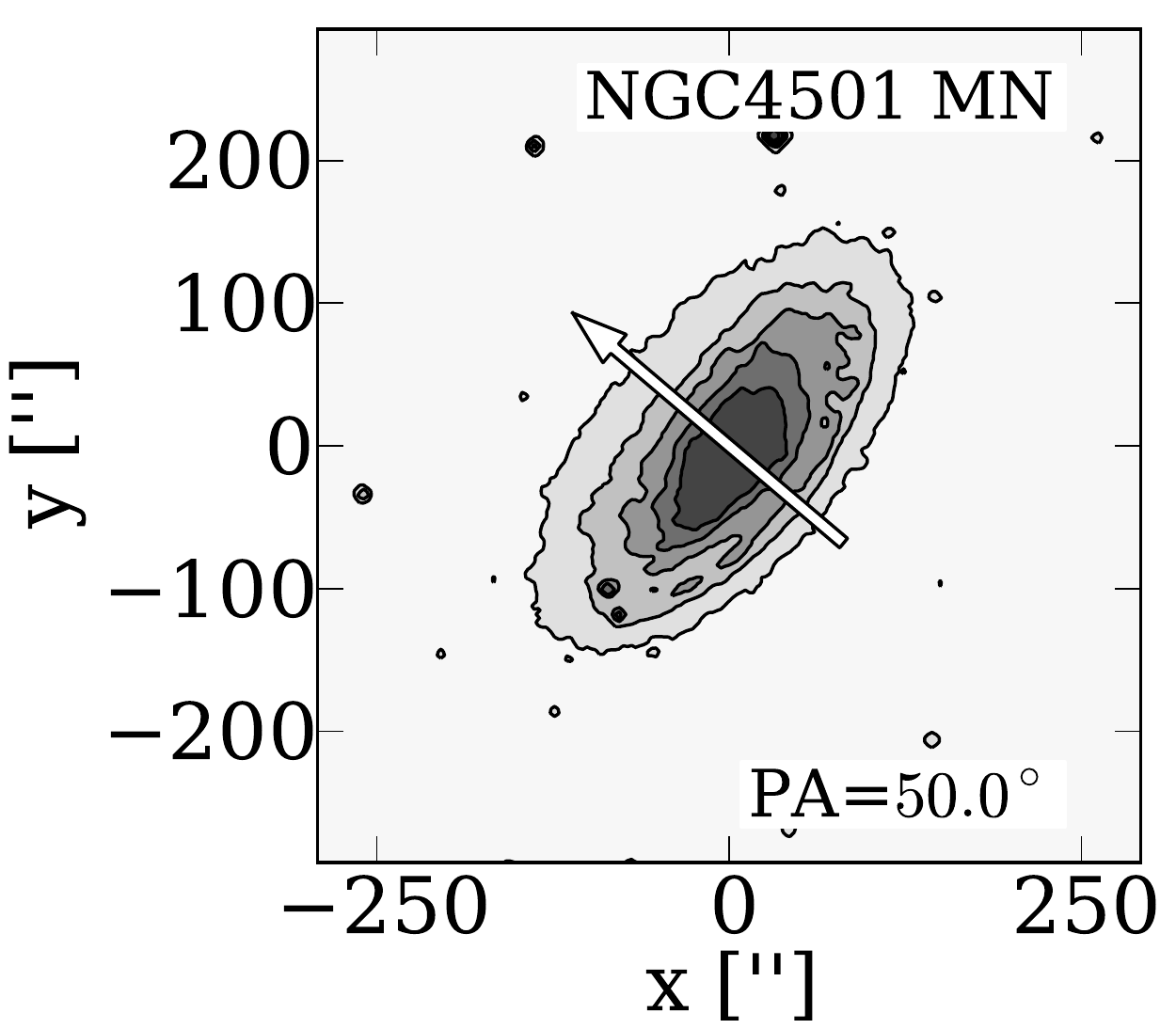}
	\end{minipage} & 
	\includegraphics[viewport=0 50 420 400,width=0.35\textwidth]{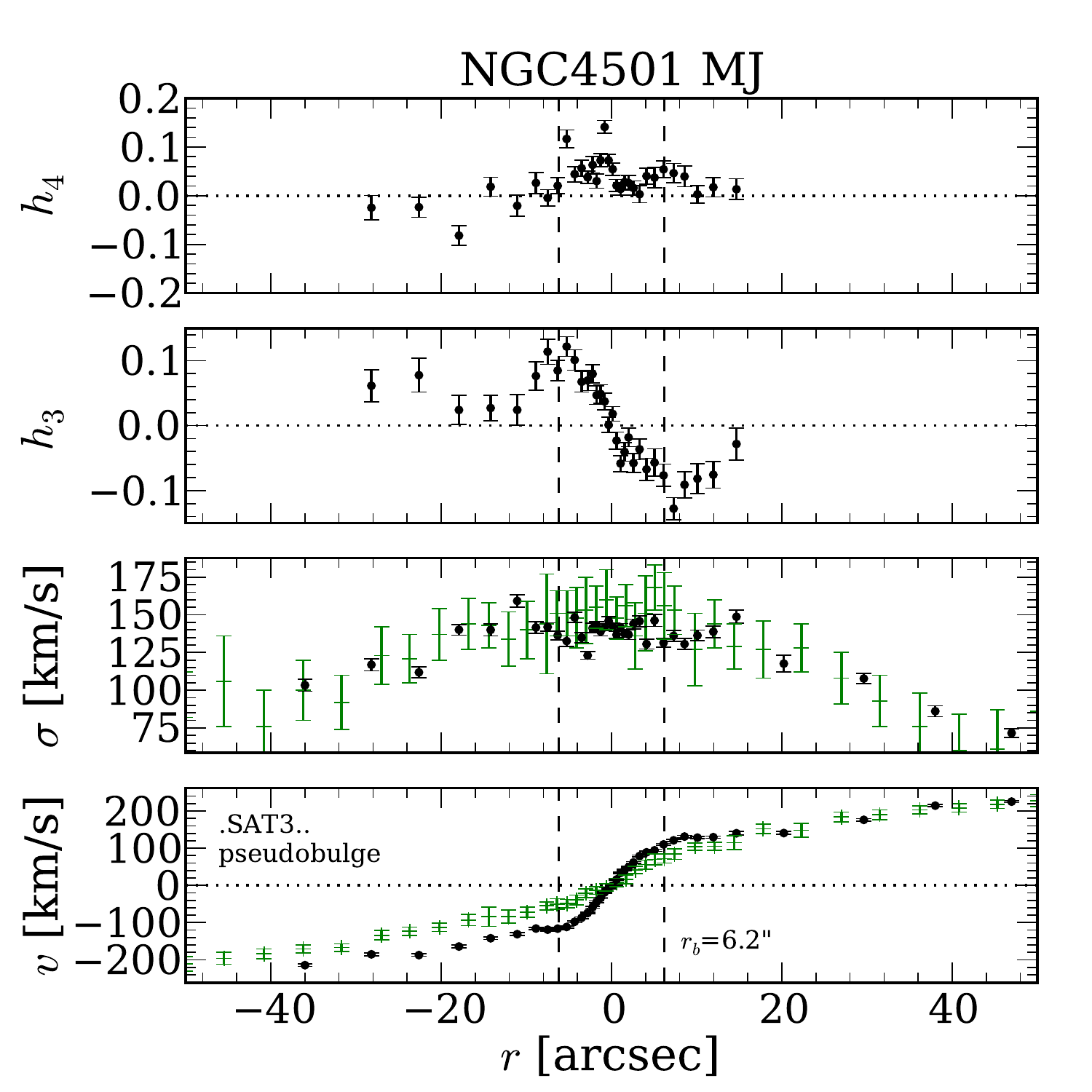} &
	\includegraphics[viewport=0 50 420 400,width=0.35\textwidth]{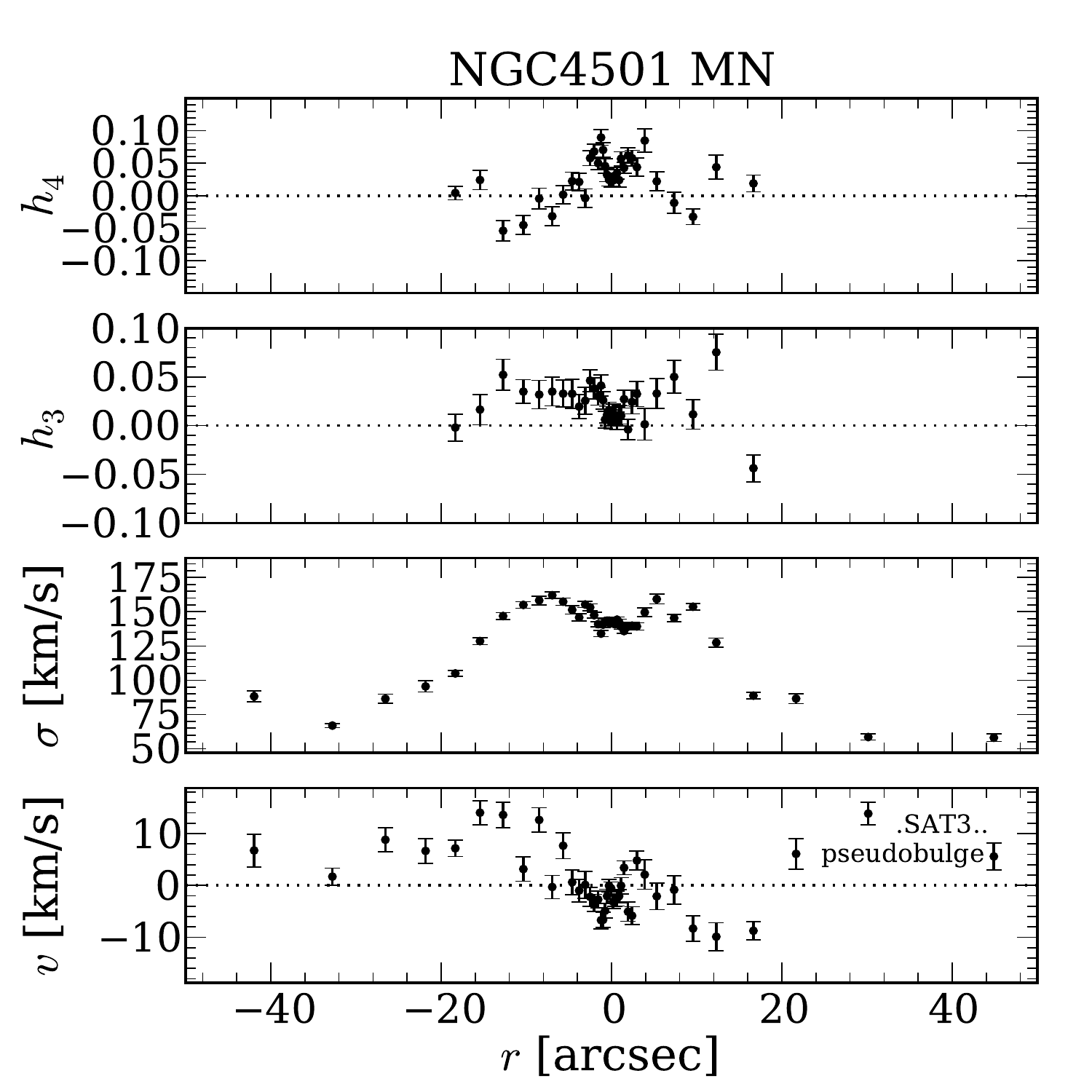}\\
        \end{tabular}
        \end{center}
        \caption{{\it continued --}\small Major and minor axis kinematic profiles for NGC\,4501.
	We plot data from \citet{Heraudeau1998} in green.
	}
\end{figure}
\setcounter{figure}{15}
\begin{figure}
        \begin{center}
        \begin{tabular}{lll}
	\begin{minipage}[b]{0.185\textwidth}
	\includegraphics[viewport=0 55 390 400,width=\textwidth]{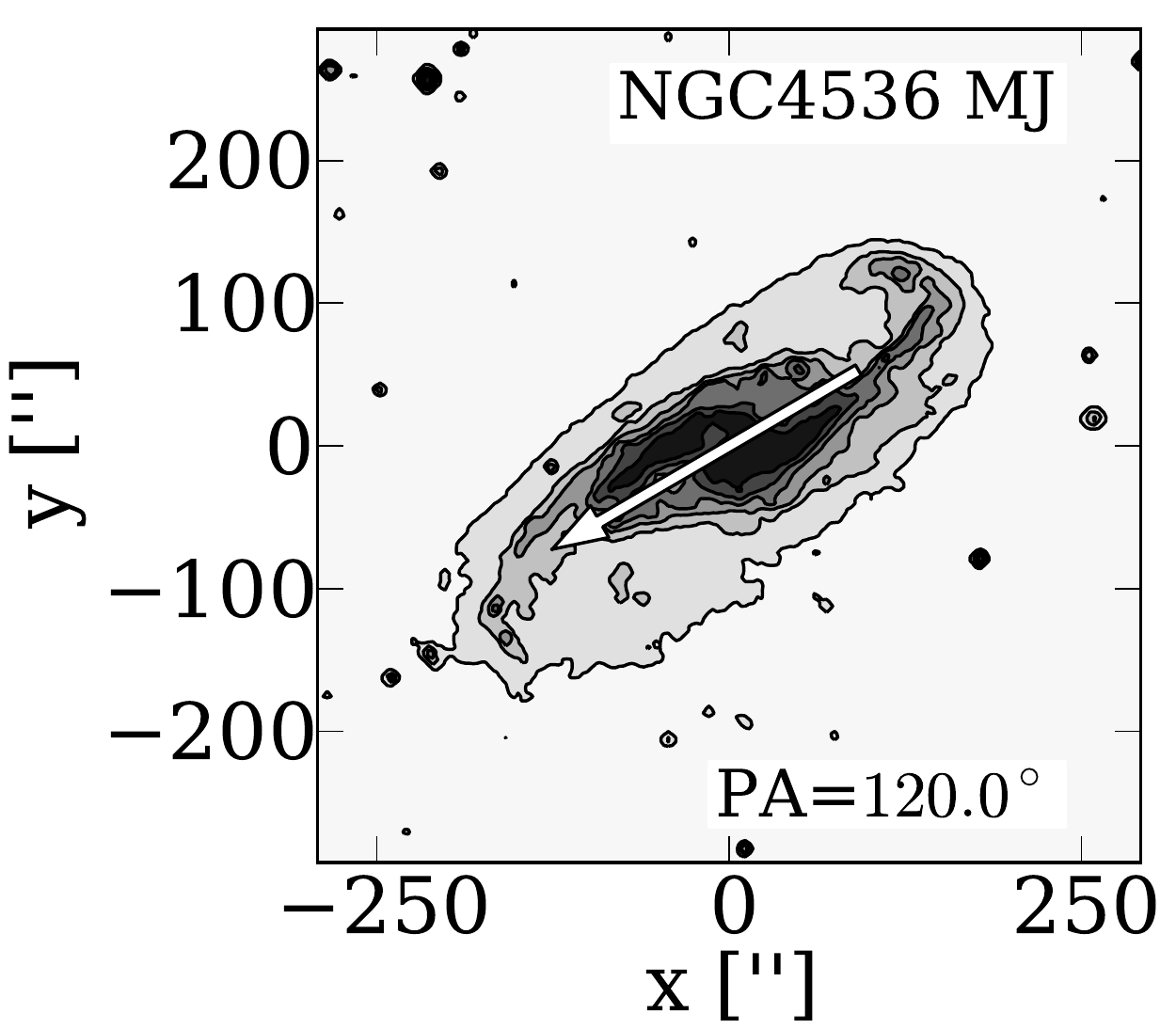}\\
	\includegraphics[viewport=0 55 390 400,width=\textwidth]{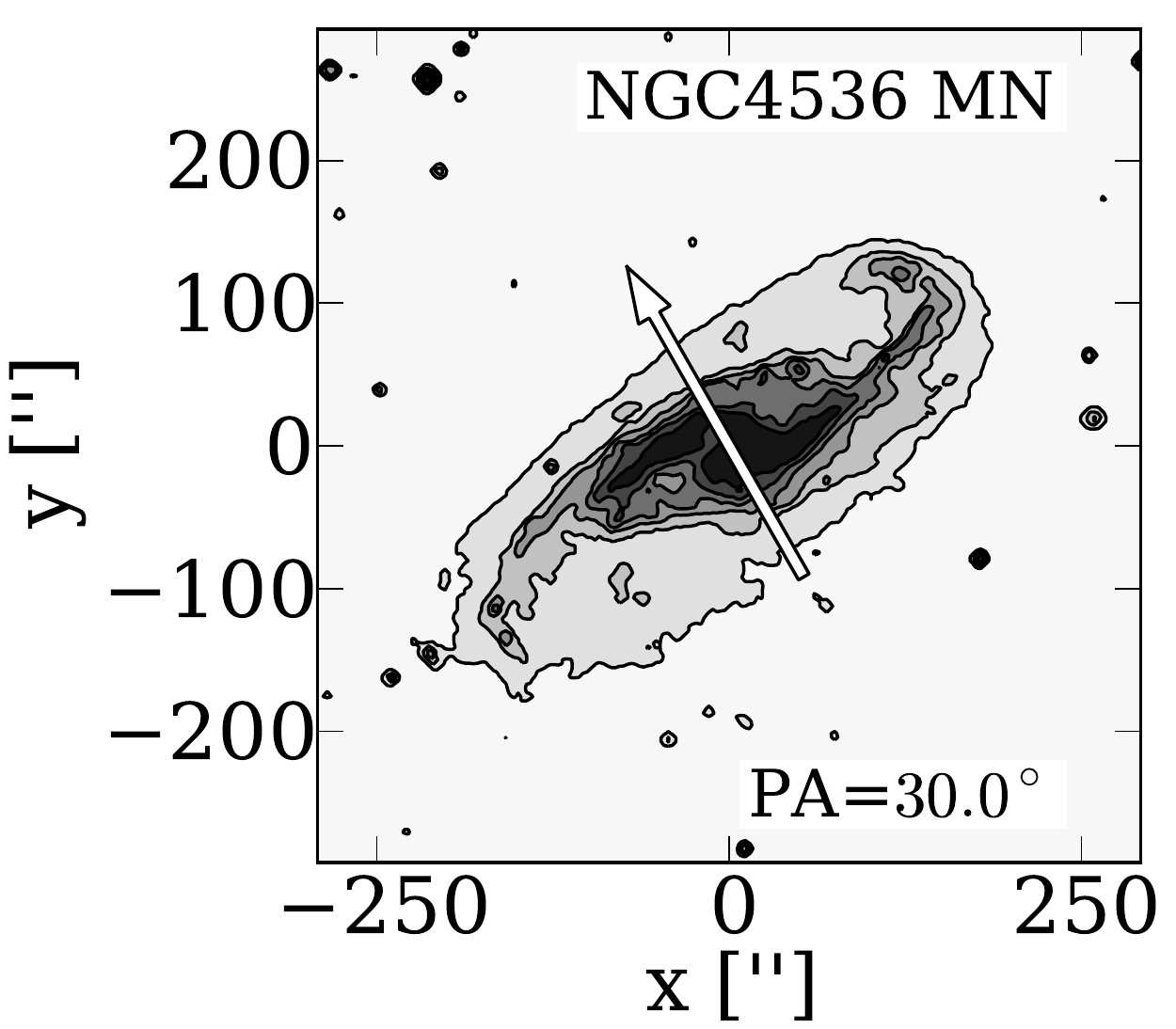}
	\end{minipage} & 
	\includegraphics[viewport=0 50 420 400,width=0.35\textwidth]{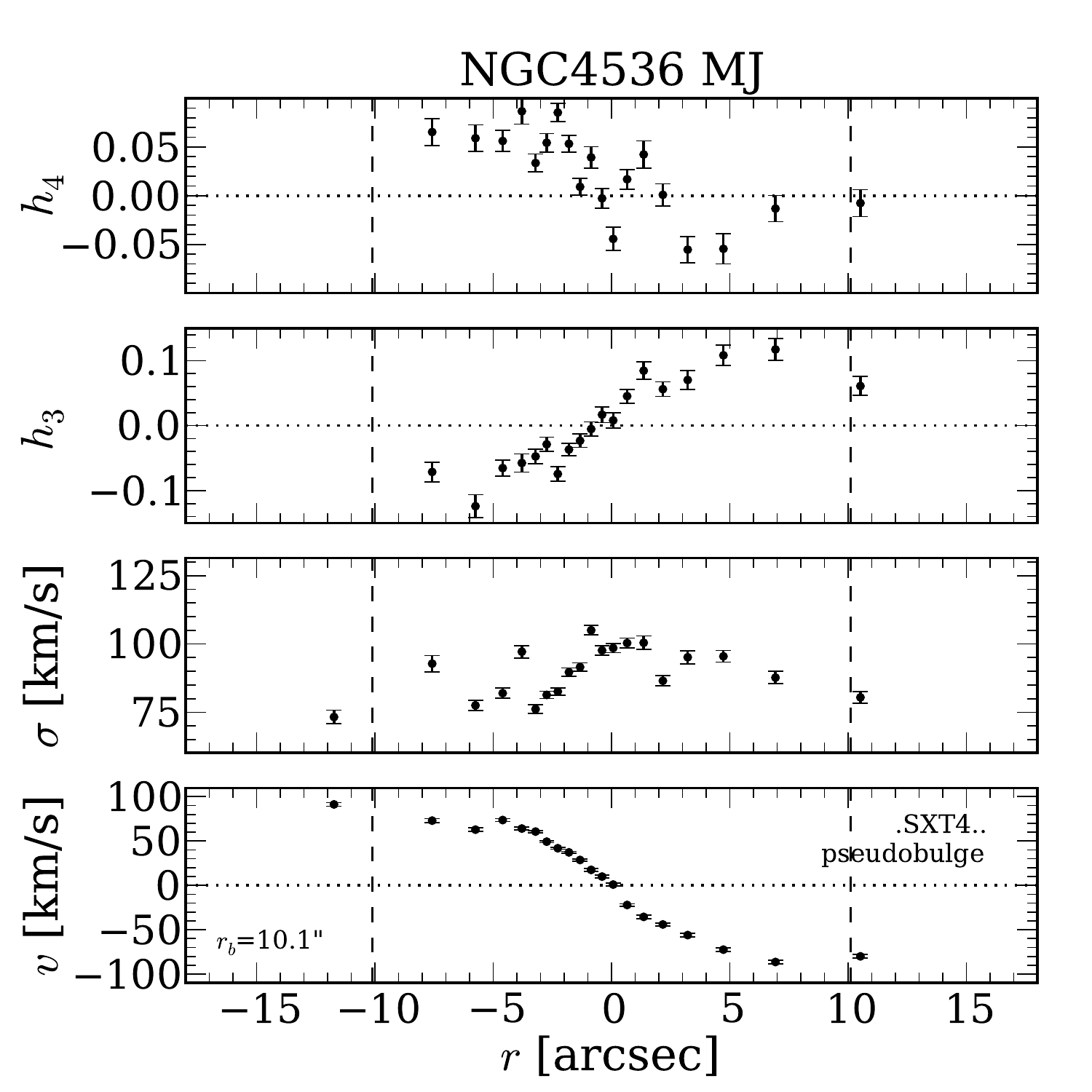} &
	\includegraphics[viewport=0 50 420 400,width=0.35\textwidth]{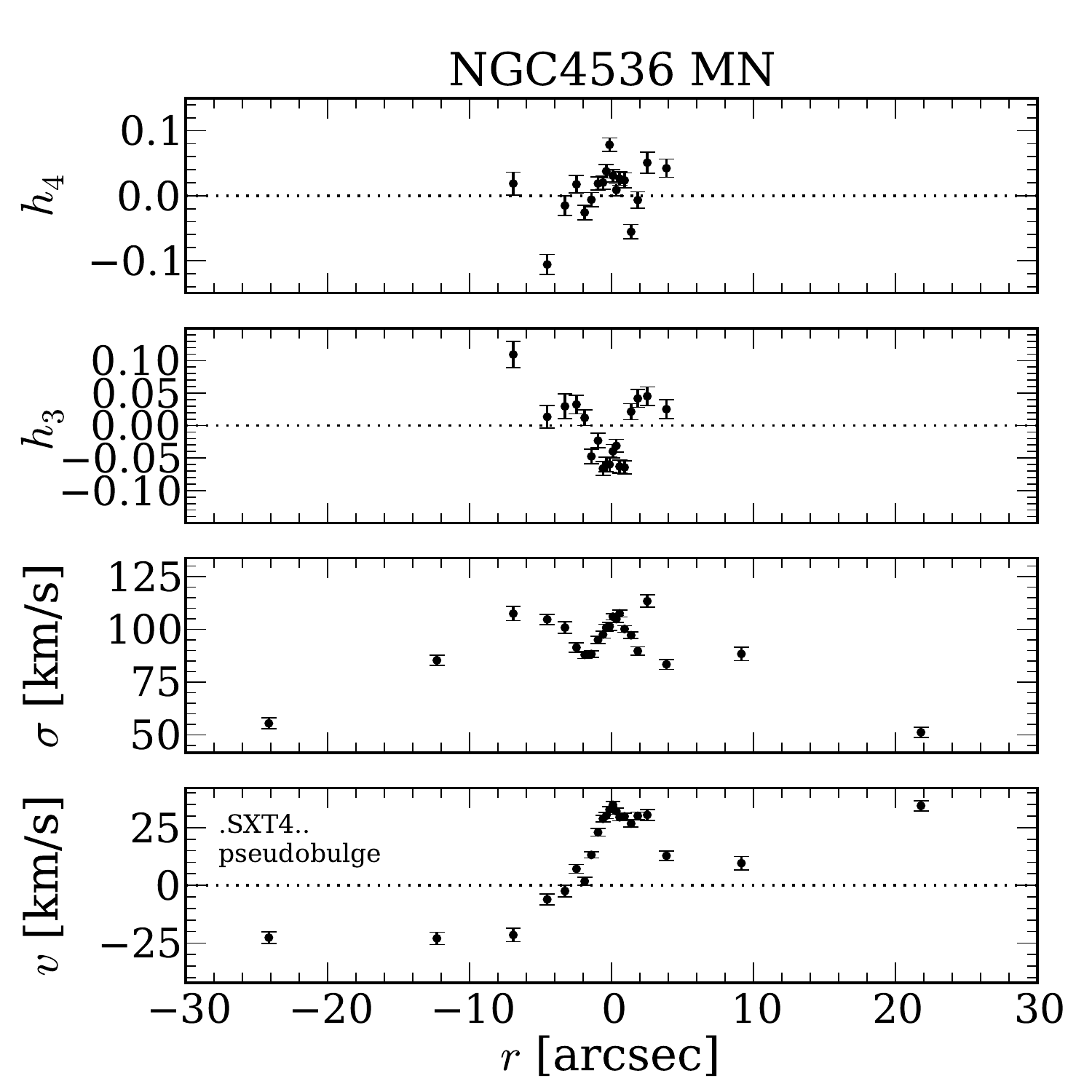}\\
        \end{tabular}
        \end{center}
        \caption{{\it continued --}\small Major and minor axis kinematic profiles for NGC\,4536.}
\end{figure}
\clearpage
\setcounter{figure}{15}
\begin{figure}
        \begin{center}
        \begin{tabular}{lll}
	\begin{minipage}[b]{0.185\textwidth}
	\includegraphics[viewport=0 55 390 400,width=\textwidth]{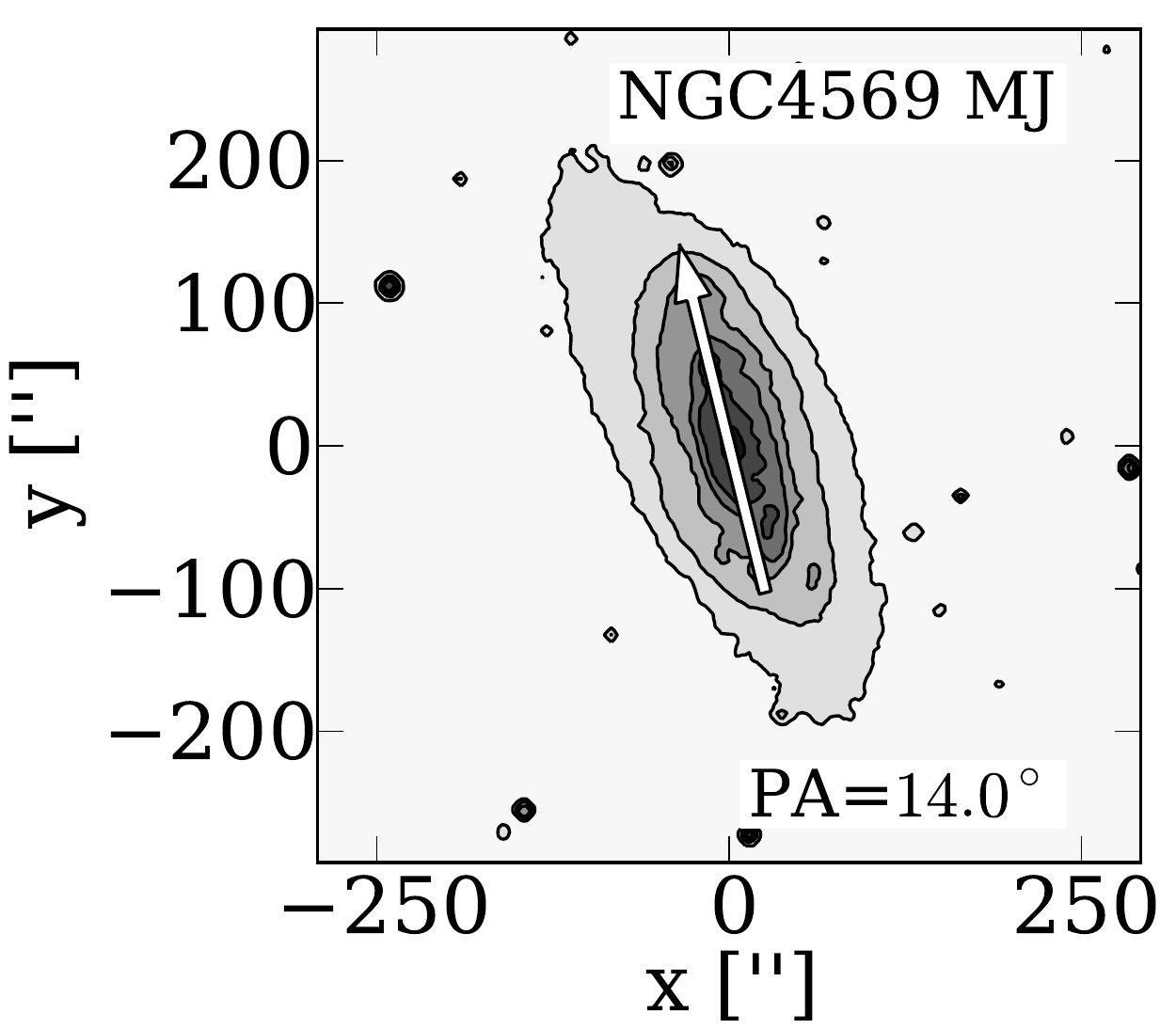}\\
	\includegraphics[viewport=0 55 390 400,width=\textwidth]{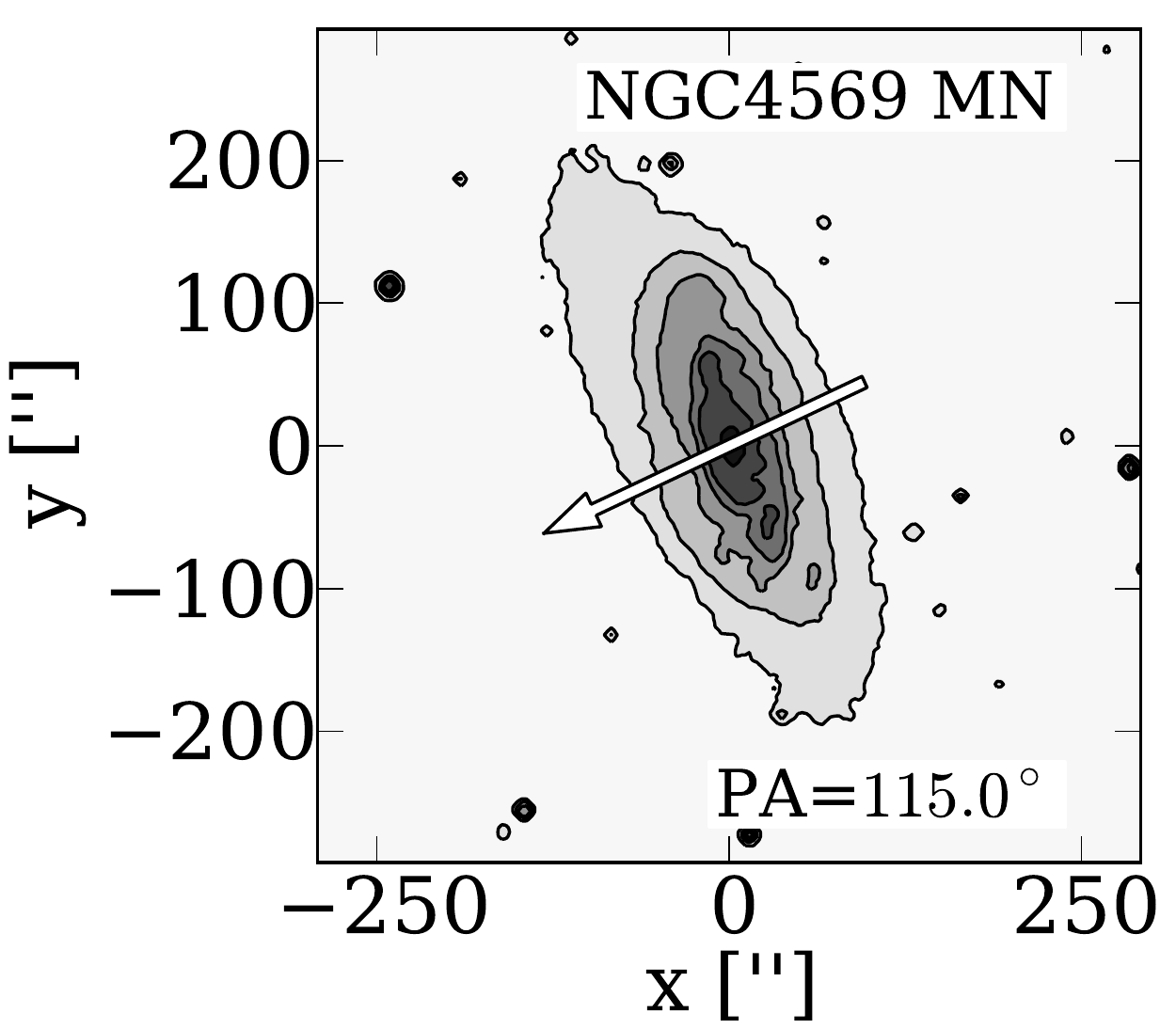}
	\end{minipage} & 
	\includegraphics[viewport=0 50 420 400,width=0.35\textwidth]{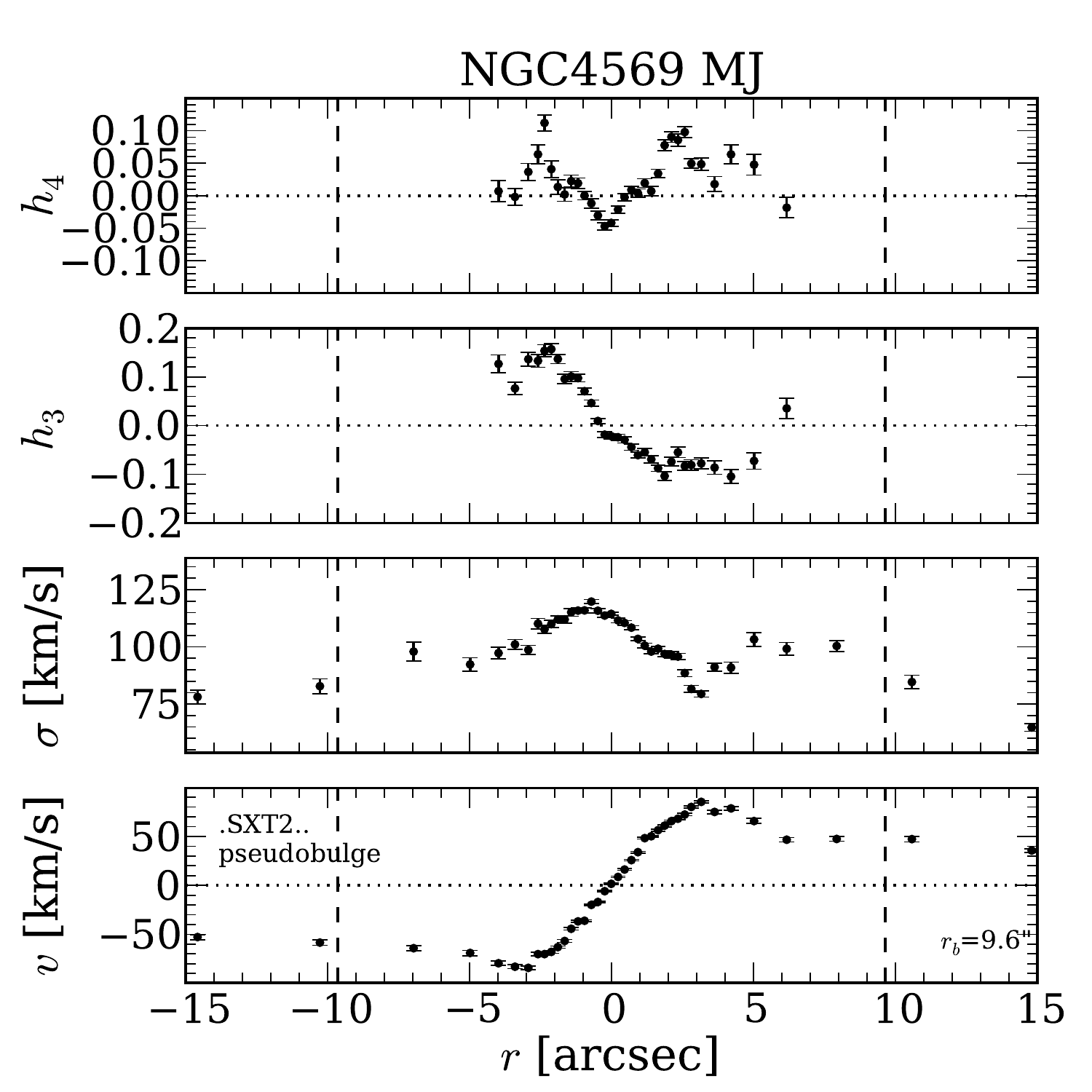} &
	\includegraphics[viewport=0 50 420 400,width=0.35\textwidth]{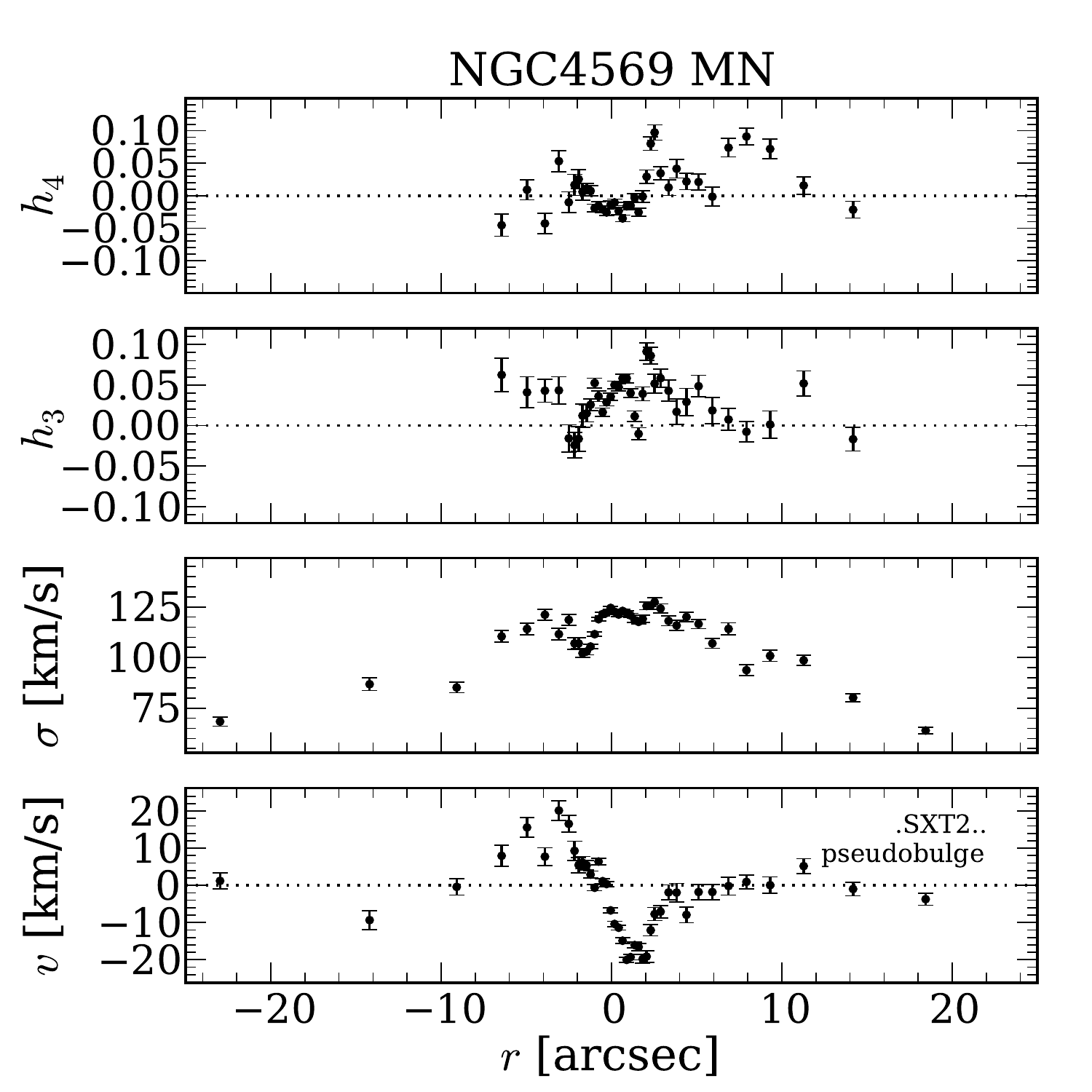}\\
        \end{tabular}
        \end{center}
        \caption{{\it continued --}\small Major and minor axis kinematic profiles for NGC\,4569.}
\end{figure}
\setcounter{figure}{15}
\begin{figure}
        \begin{center}
        \begin{tabular}{lll}
	\begin{minipage}[b]{0.185\textwidth}
	\includegraphics[viewport=0 55 390 400,width=\textwidth]{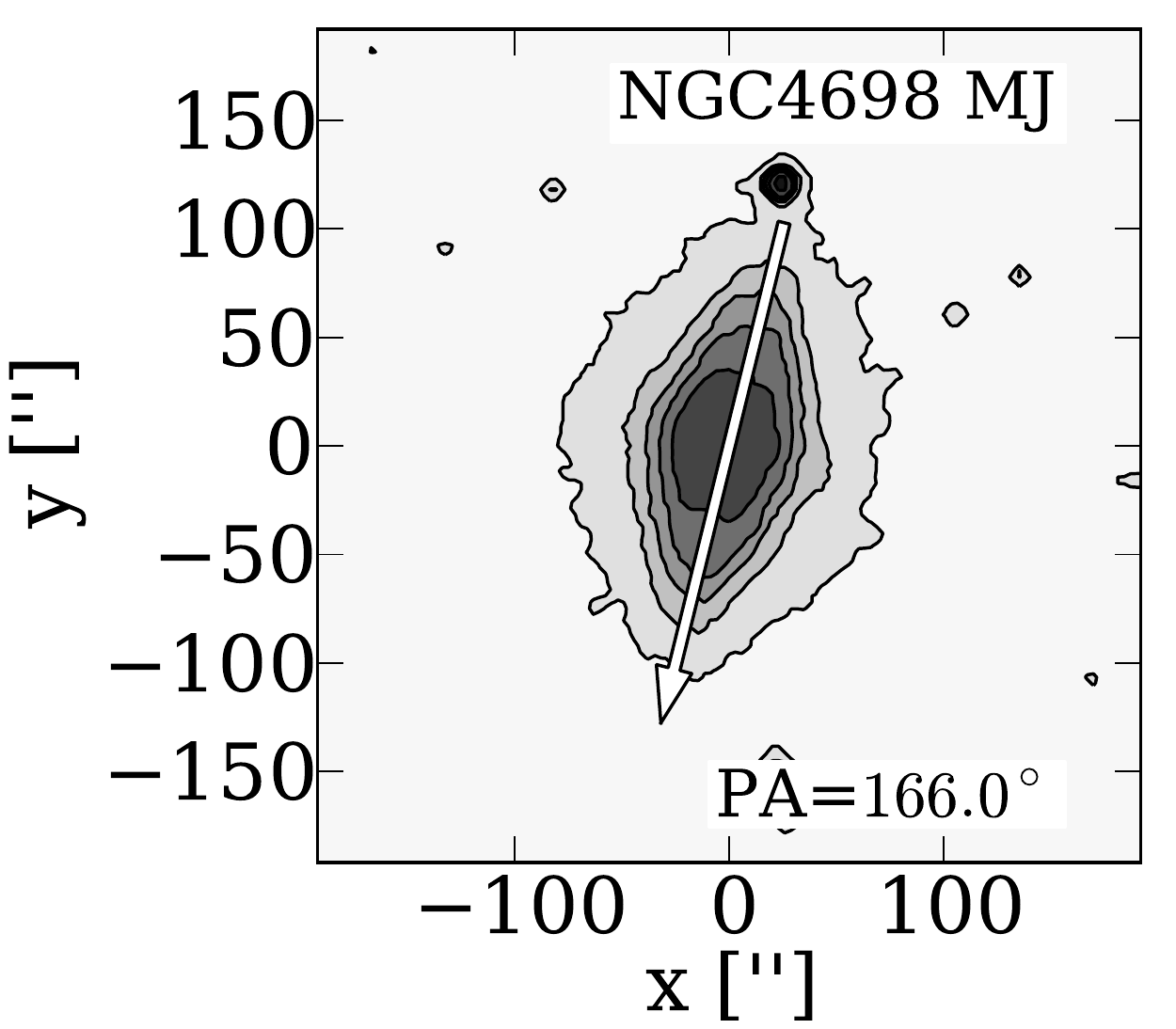} \\
        \includegraphics[viewport=0 55 390 400,width=\textwidth]{empty}
	\end{minipage} & 
	\includegraphics[viewport=0 50 420 400,width=0.35\textwidth]{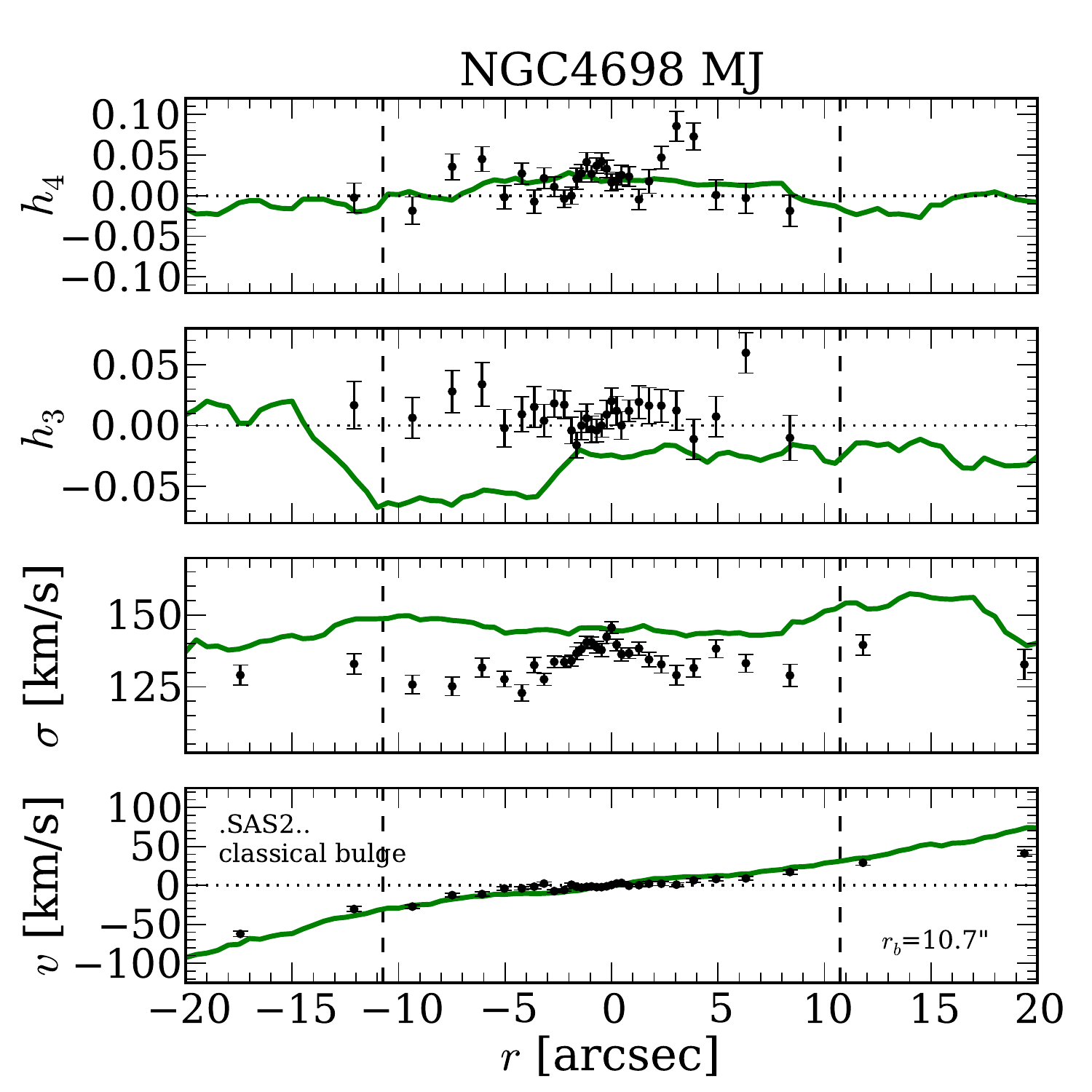} &
        \includegraphics[viewport=0 50 420 400,width=0.35\textwidth]{empty}
        \end{tabular}
        \end{center}
        \caption{{\it continued --}\small Major axis kinematic profile for NGC\,4698. We 
	plot the SAURON results \citep{Falcon-Barroso2006} in green.} 
\end{figure}
\clearpage
\setcounter{figure}{15}
\begin{figure}
        \begin{center}
        \begin{tabular}{lll}
	\begin{minipage}[b]{0.185\textwidth}
	\includegraphics[viewport=0 55 390 400,width=\textwidth]{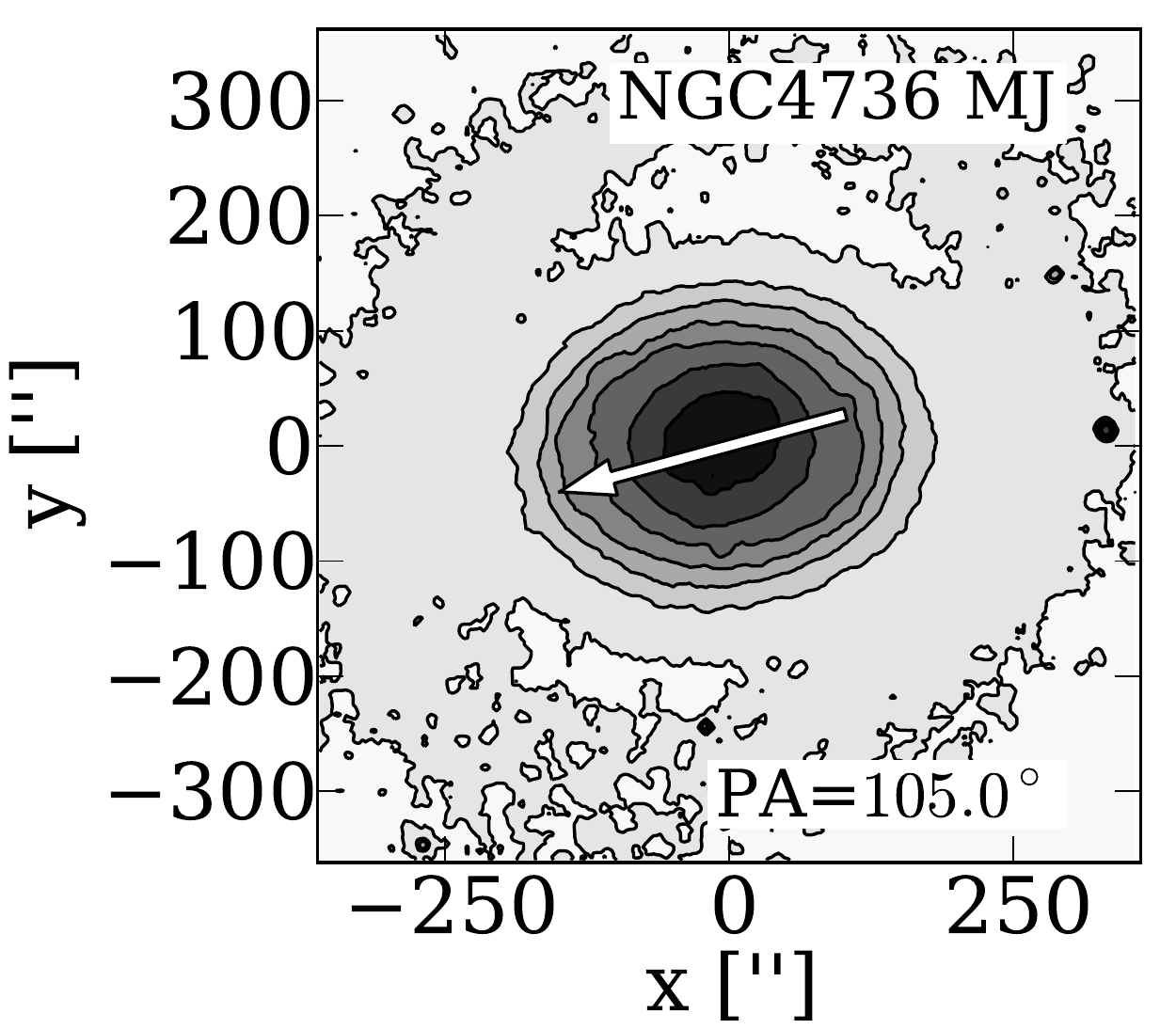}\\
	\includegraphics[viewport=0 55 390 400,width=\textwidth]{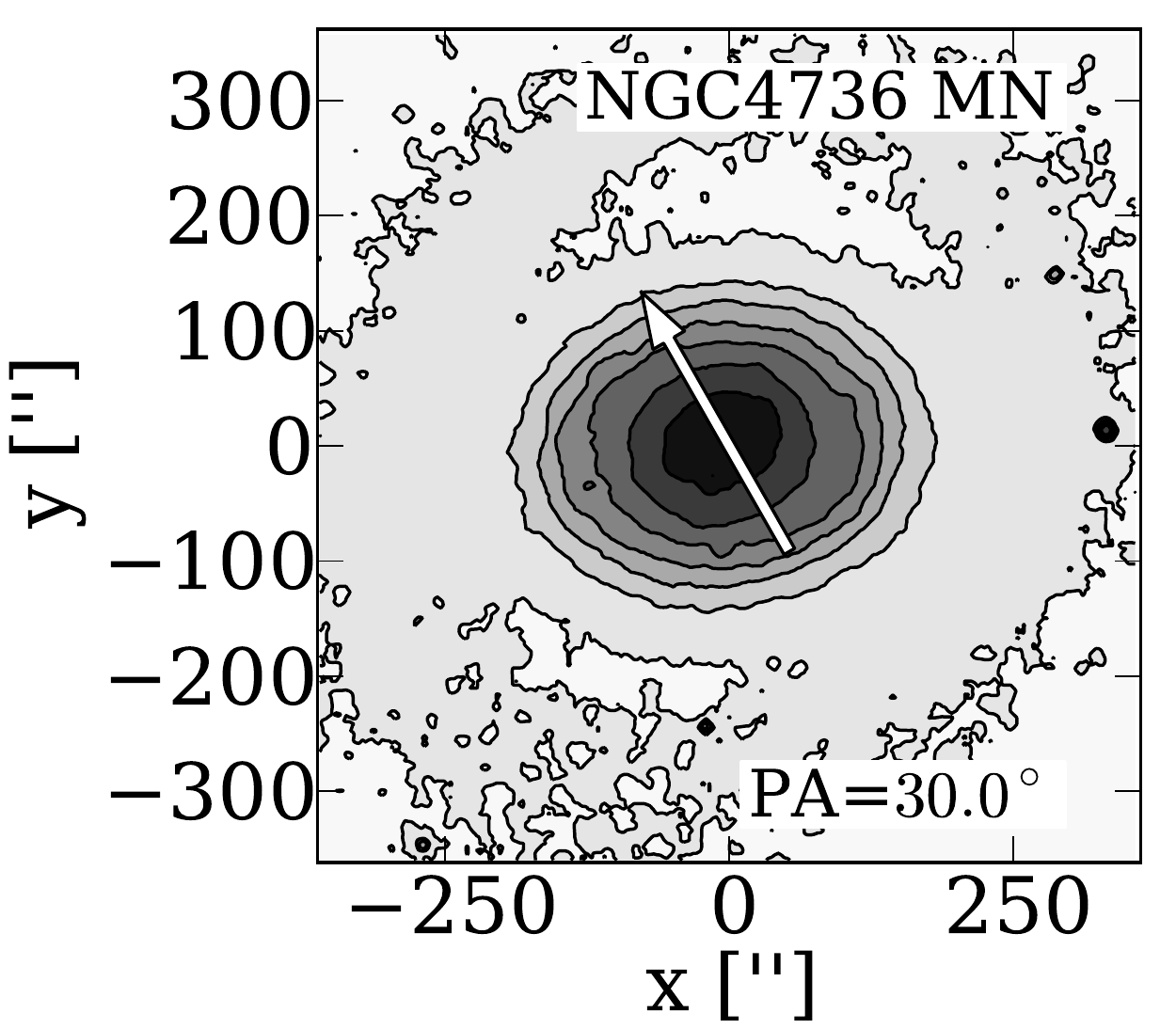}
	\end{minipage} & 
	\includegraphics[viewport=0 50 420 400,width=0.35\textwidth]{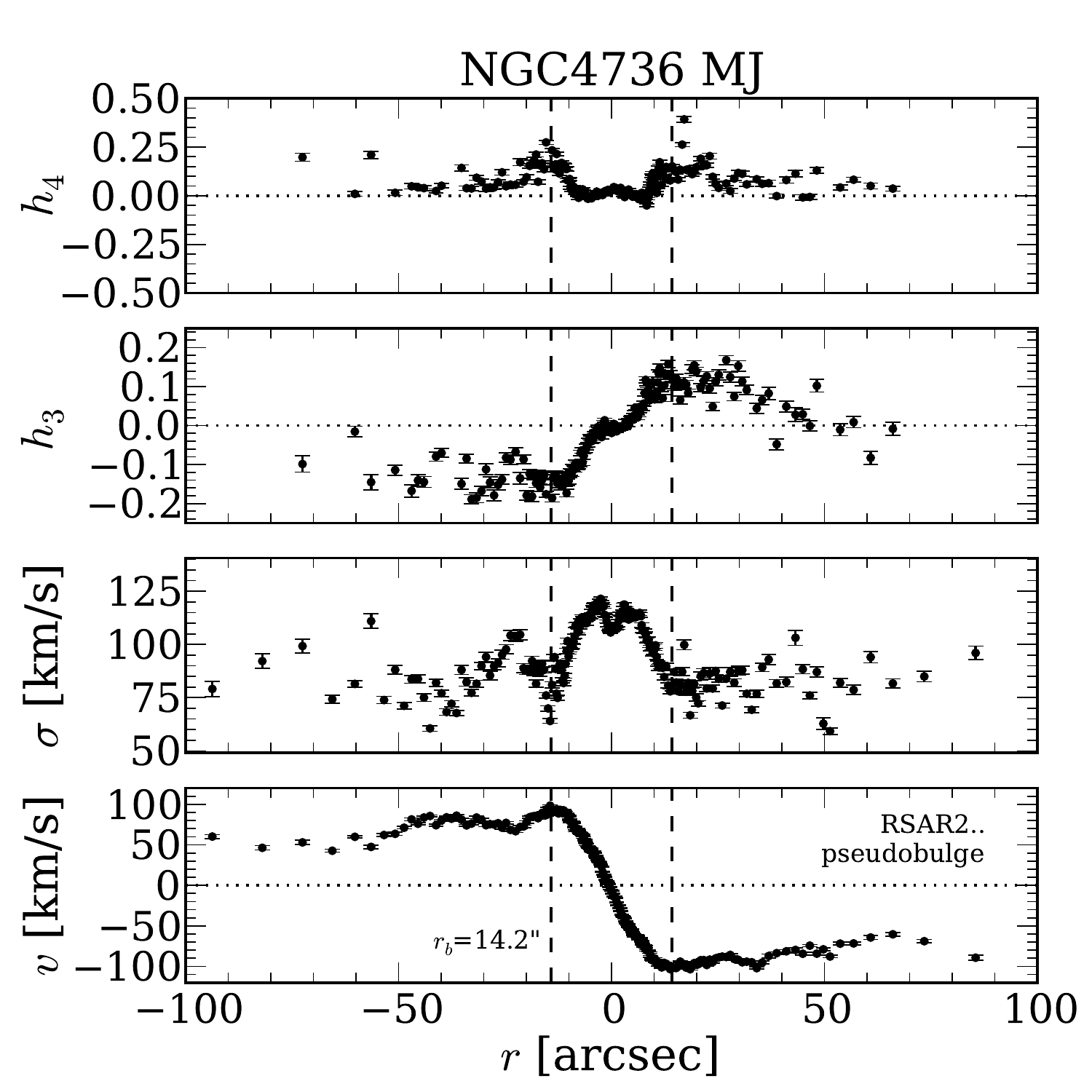} &
	\includegraphics[viewport=0 50 420 400,width=0.35\textwidth]{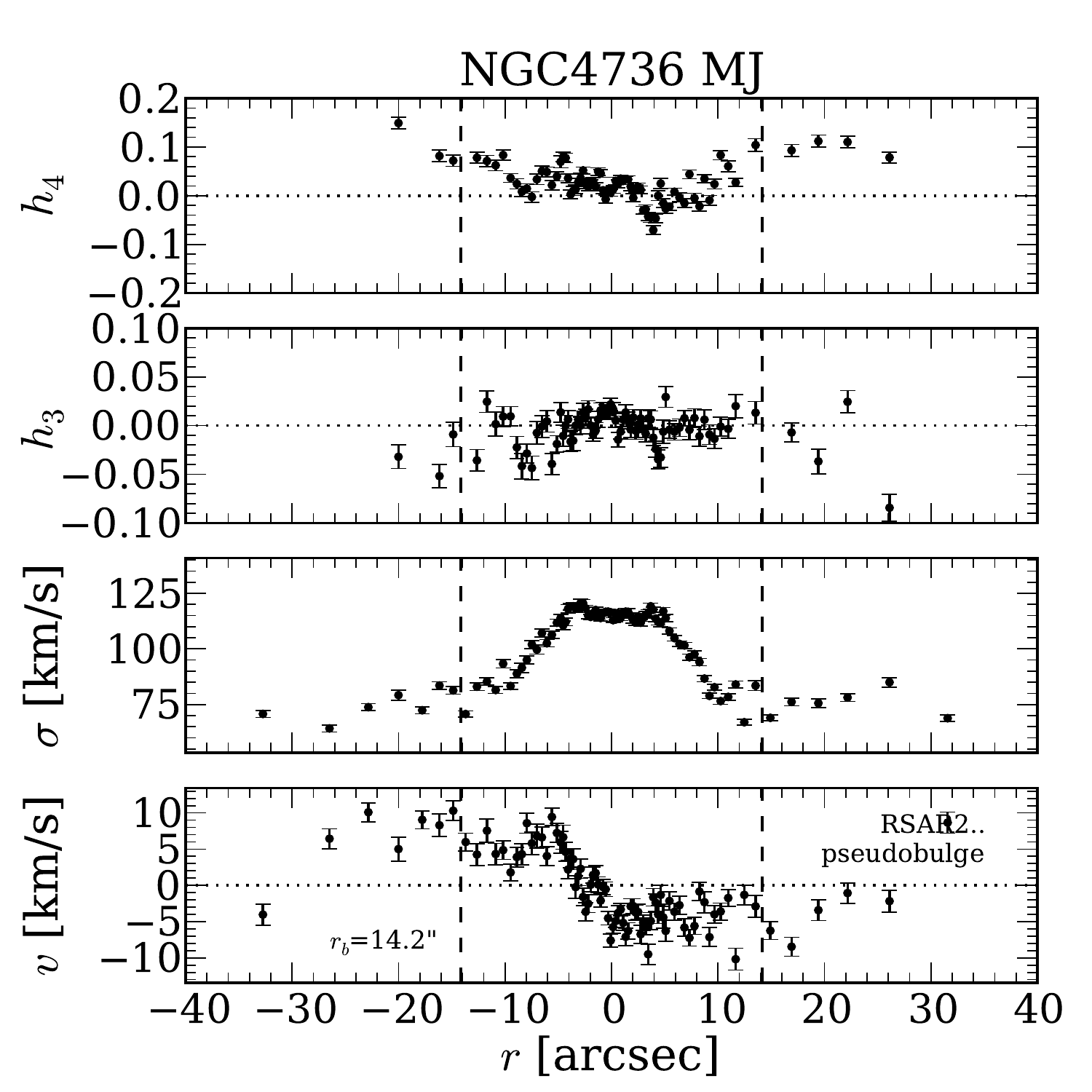}\\
        \end{tabular}
        \end{center}
        \caption{{\it continued --}\small Major and minor axis kinematic profiles for NGC\,4736,
				see also Fig.~\ref{fig:kinDecomp}.
				}
\end{figure}
\setcounter{figure}{15}
\begin{figure}
        \begin{center}
        \begin{tabular}{lll}
	\begin{minipage}[b]{0.185\textwidth}
	\includegraphics[viewport=0 55 390 400,width=\textwidth]{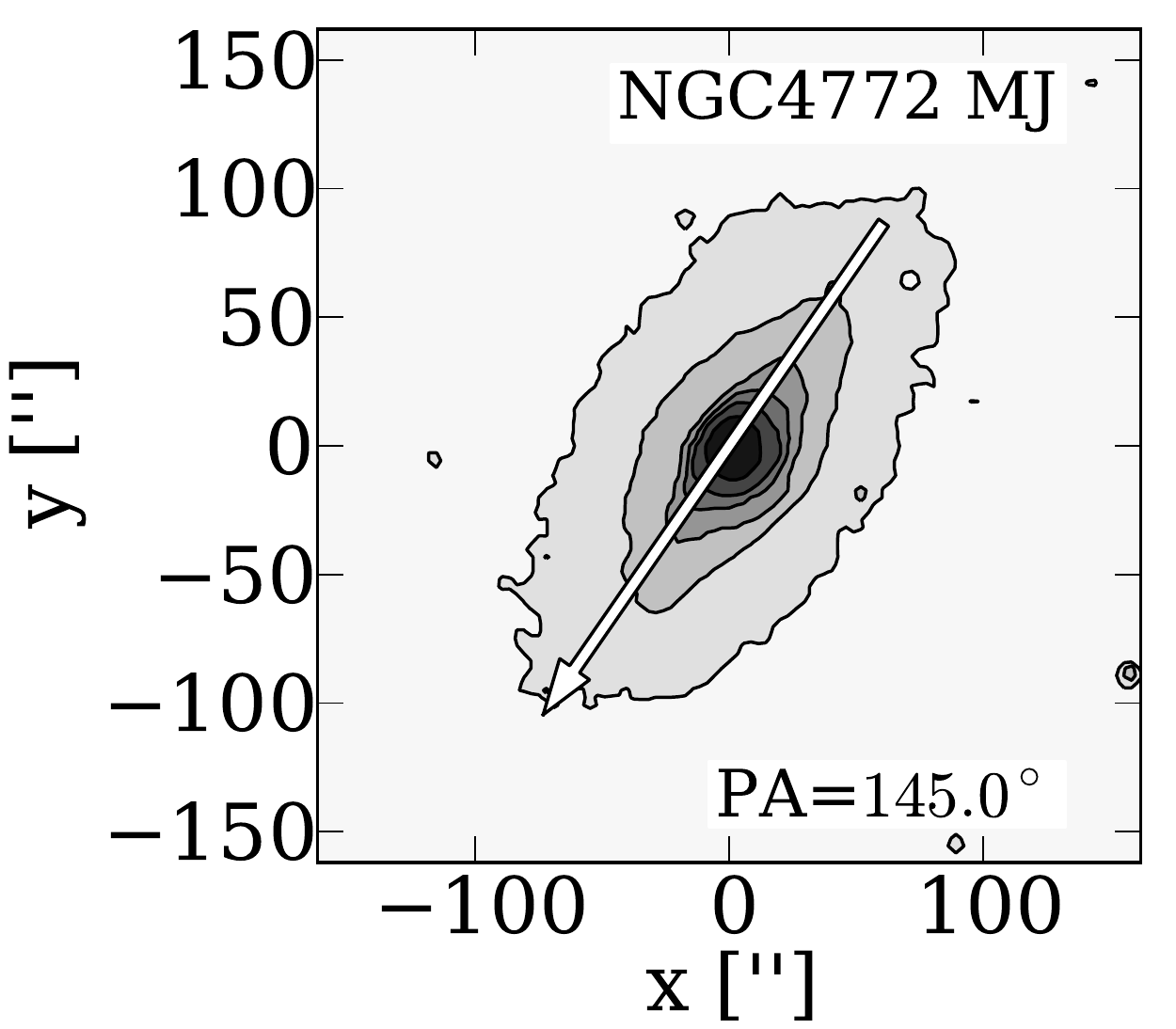} \\
        \includegraphics[viewport=0 55 390 400,width=\textwidth]{empty}
	\end{minipage} & 
	\includegraphics[viewport=0 50 420 400,width=0.35\textwidth]{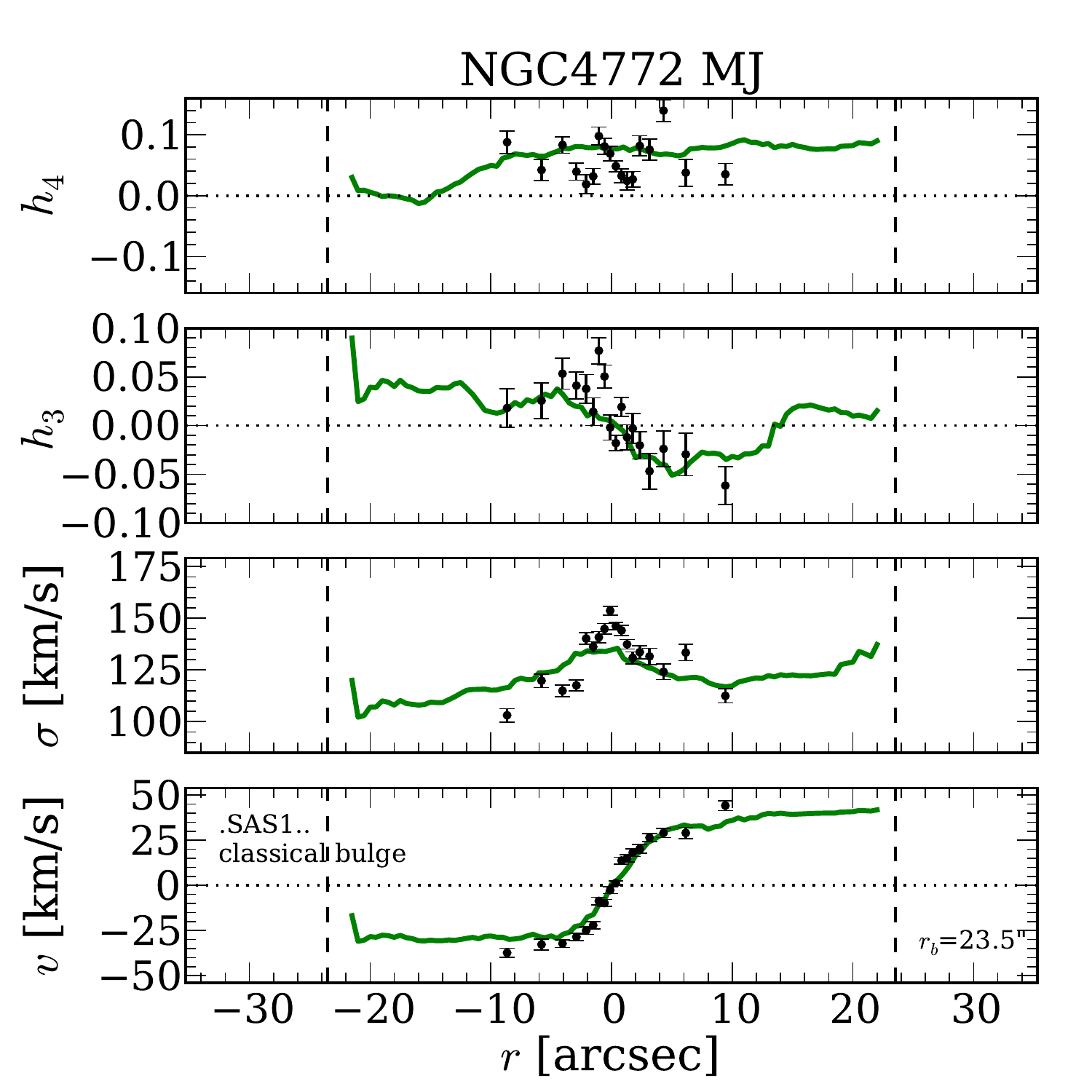}& 
        \includegraphics[viewport=0 50 420 400,width=0.35\textwidth]{empty}
        \end{tabular}
        \end{center}
        \caption{{\it continued --}\small Major axis kinematic profile for NGC\,4772.
	We plot the SAURON results \citep{Falcon-Barroso2006} in green.} 
\end{figure}
\clearpage
\setcounter{figure}{15}
\begin{figure}
        \begin{center}
        \begin{tabular}{lll}
	\begin{minipage}[b]{0.185\textwidth}
	\includegraphics[viewport=0 55 390 400,width=\textwidth]{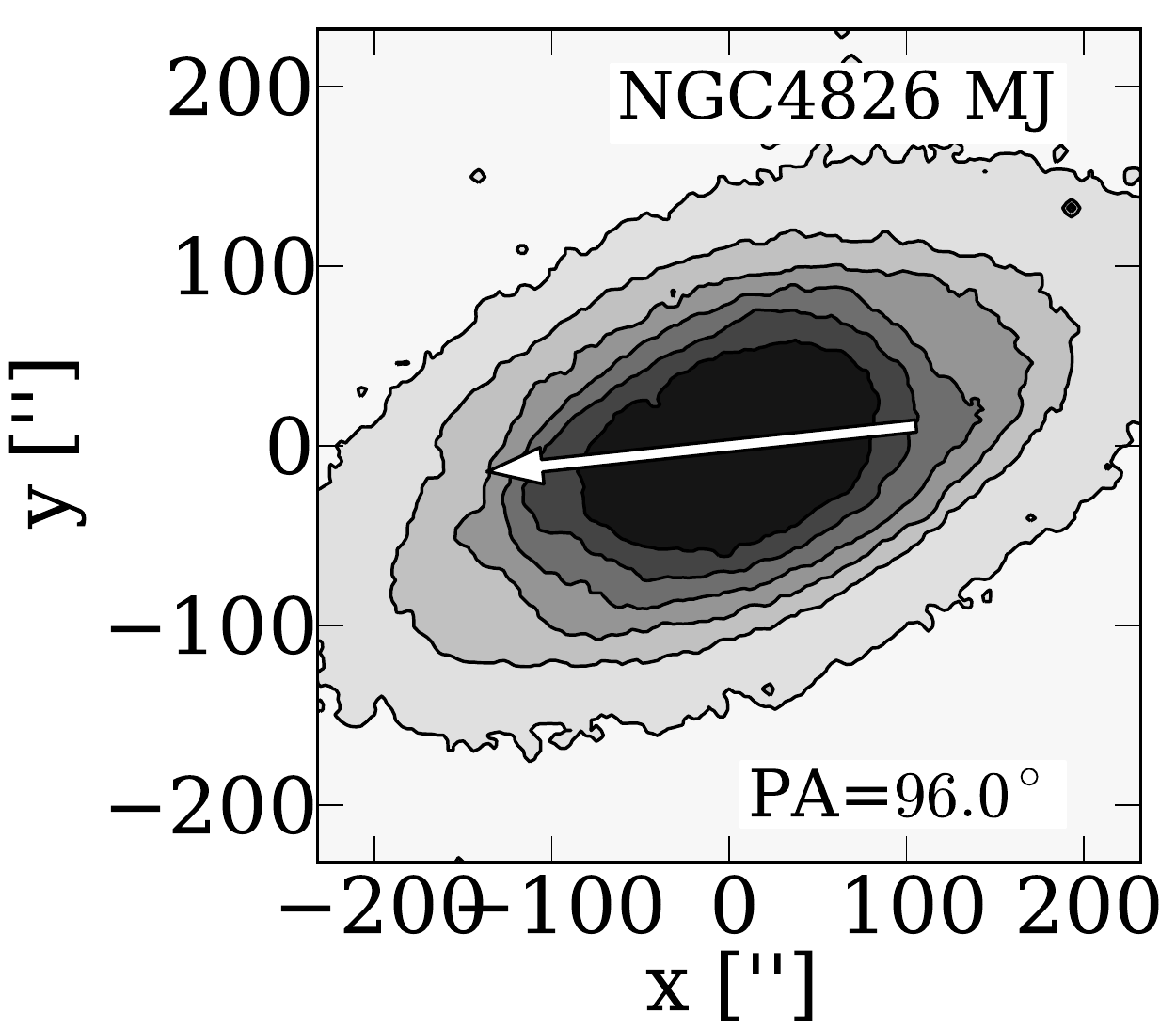}\\
	\includegraphics[viewport=0 55 390 400,width=\textwidth]{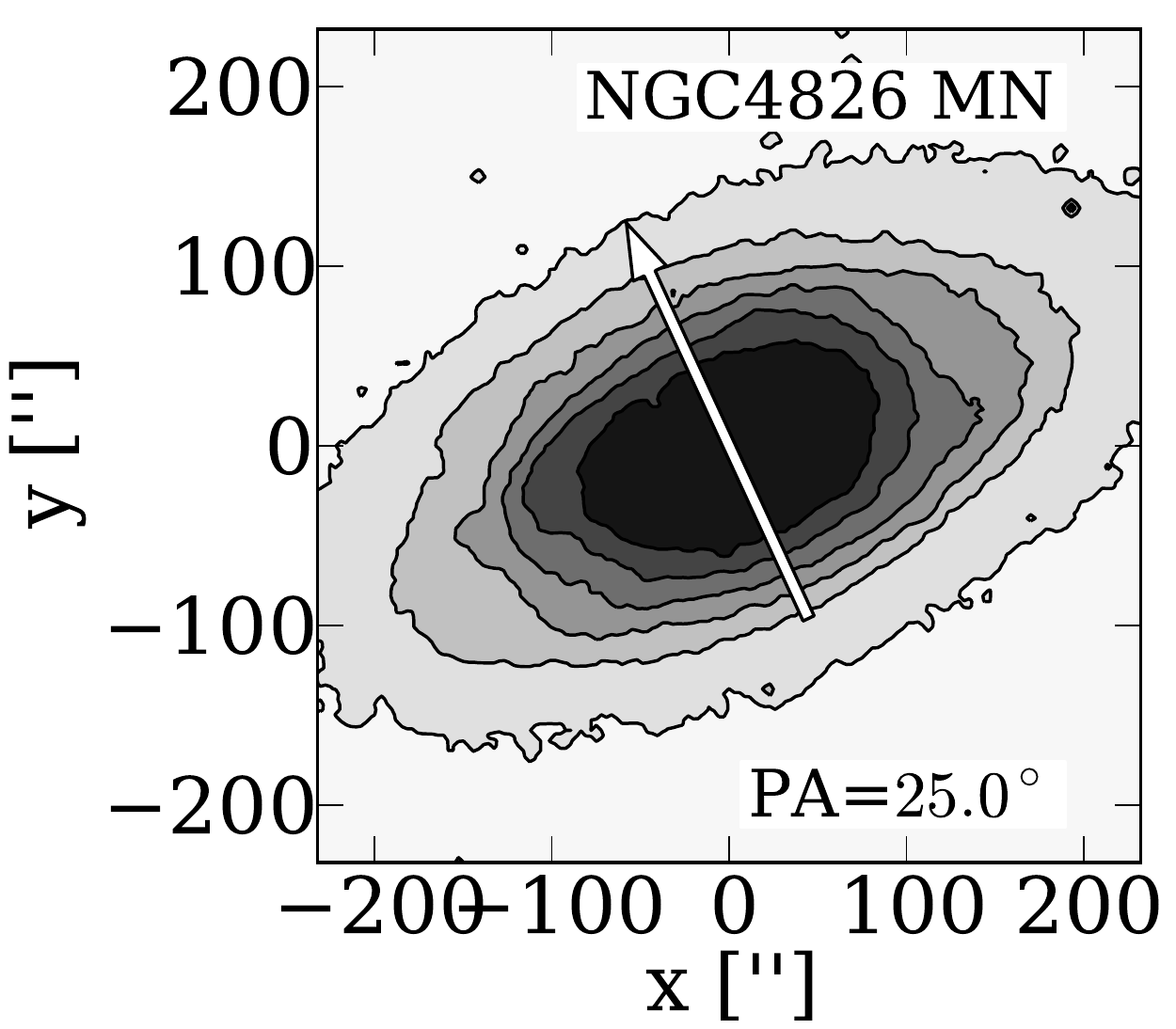}
	\end{minipage} & 
	\includegraphics[viewport=0 50 420 400,width=0.35\textwidth]{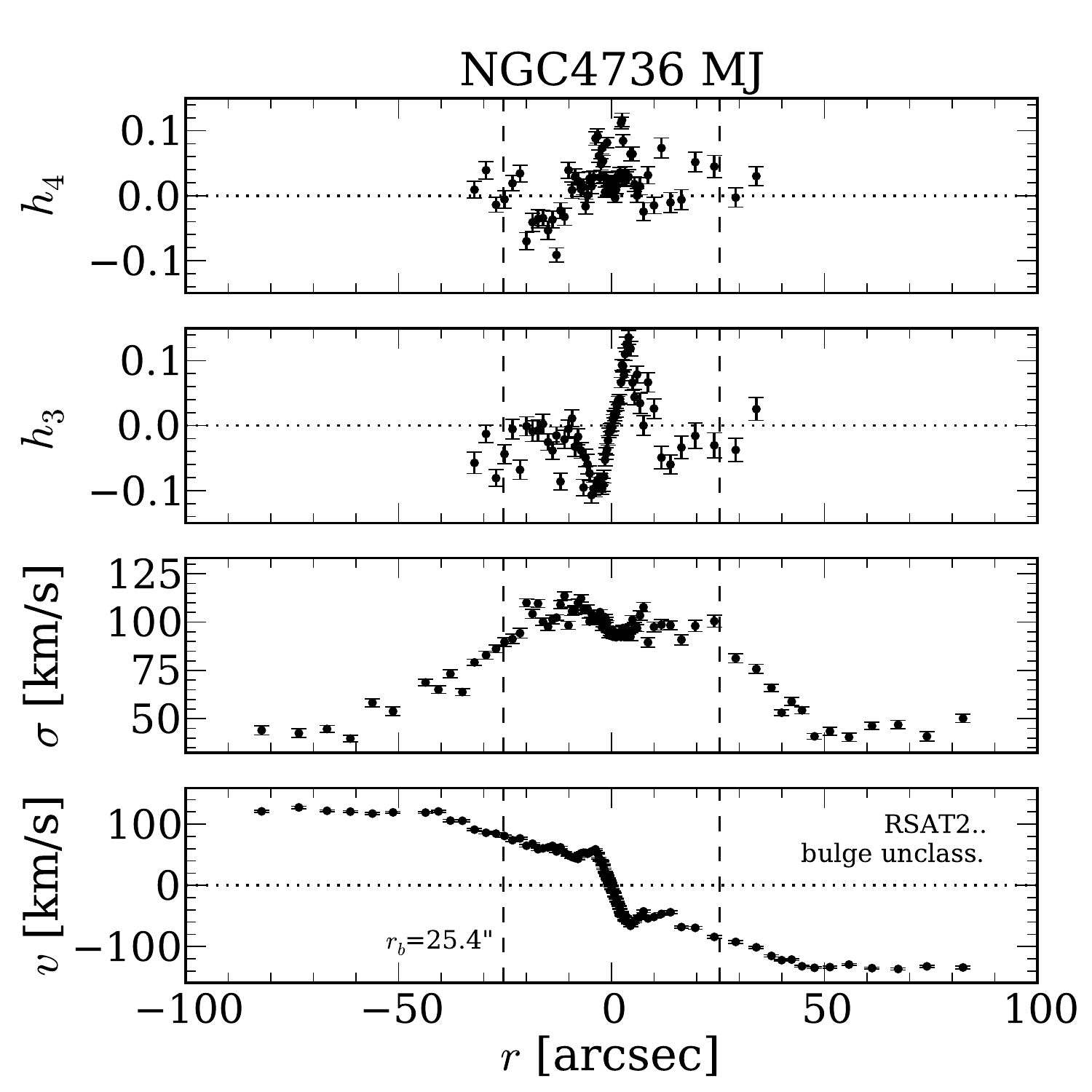} &
	\includegraphics[viewport=0 50 420 400,width=0.35\textwidth]{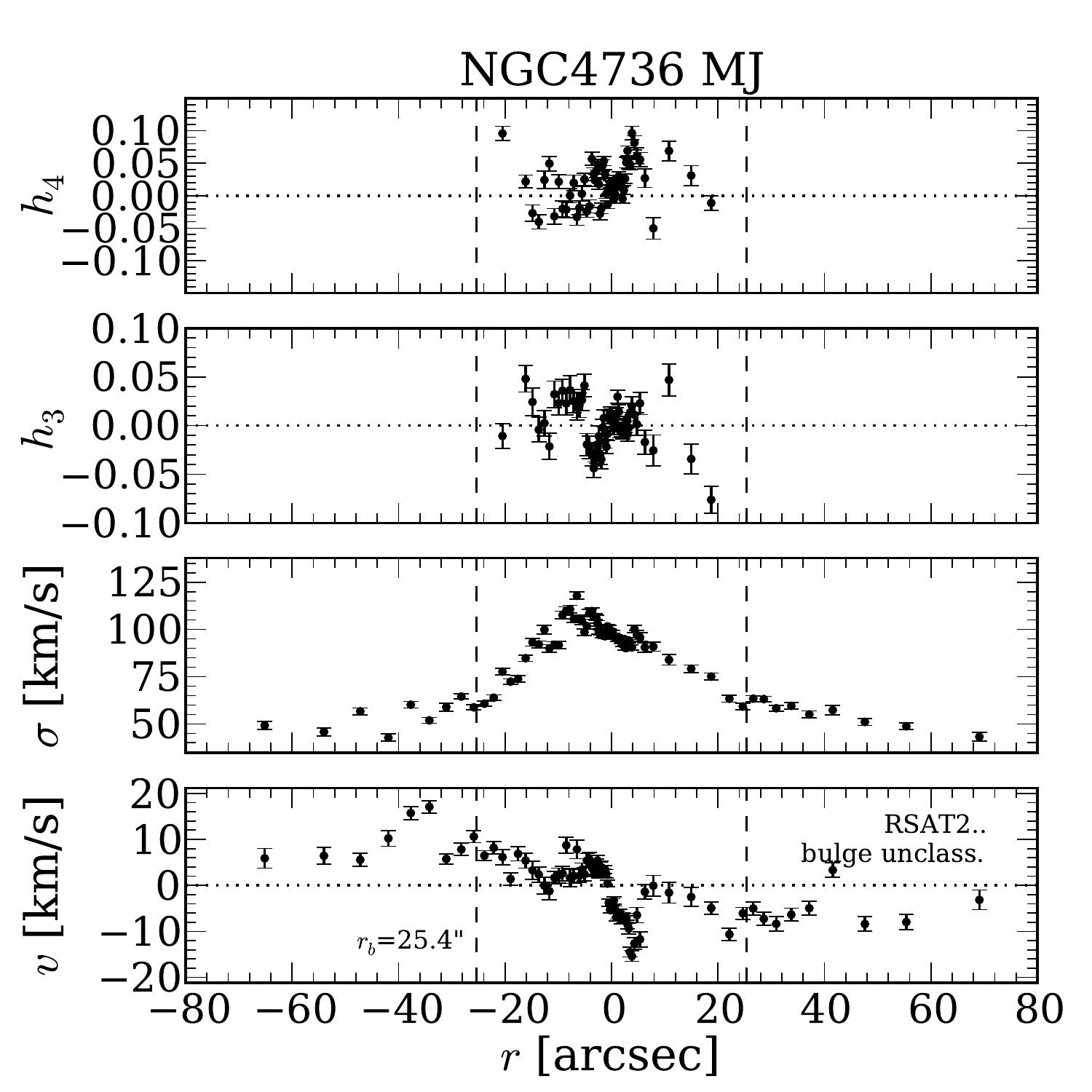}\\
        \end{tabular}
        \end{center}
        \caption{{\it continued --}\small Major and minor axis kinematic profiles for NGC\,4826.
	We plot results of \citet{Heraudeau1998} in green. Note: Their data were observed at a slit position angle of
	115\Deg\ whereas our adopted value for the major axis position angle is 96\Deg.
	}
\end{figure}
\setcounter{figure}{15}
\begin{figure}
        \begin{center}
        \begin{tabular}{lll}
	\begin{minipage}[b]{0.185\textwidth}
	\includegraphics[viewport=0 55 390 400,width=\textwidth]{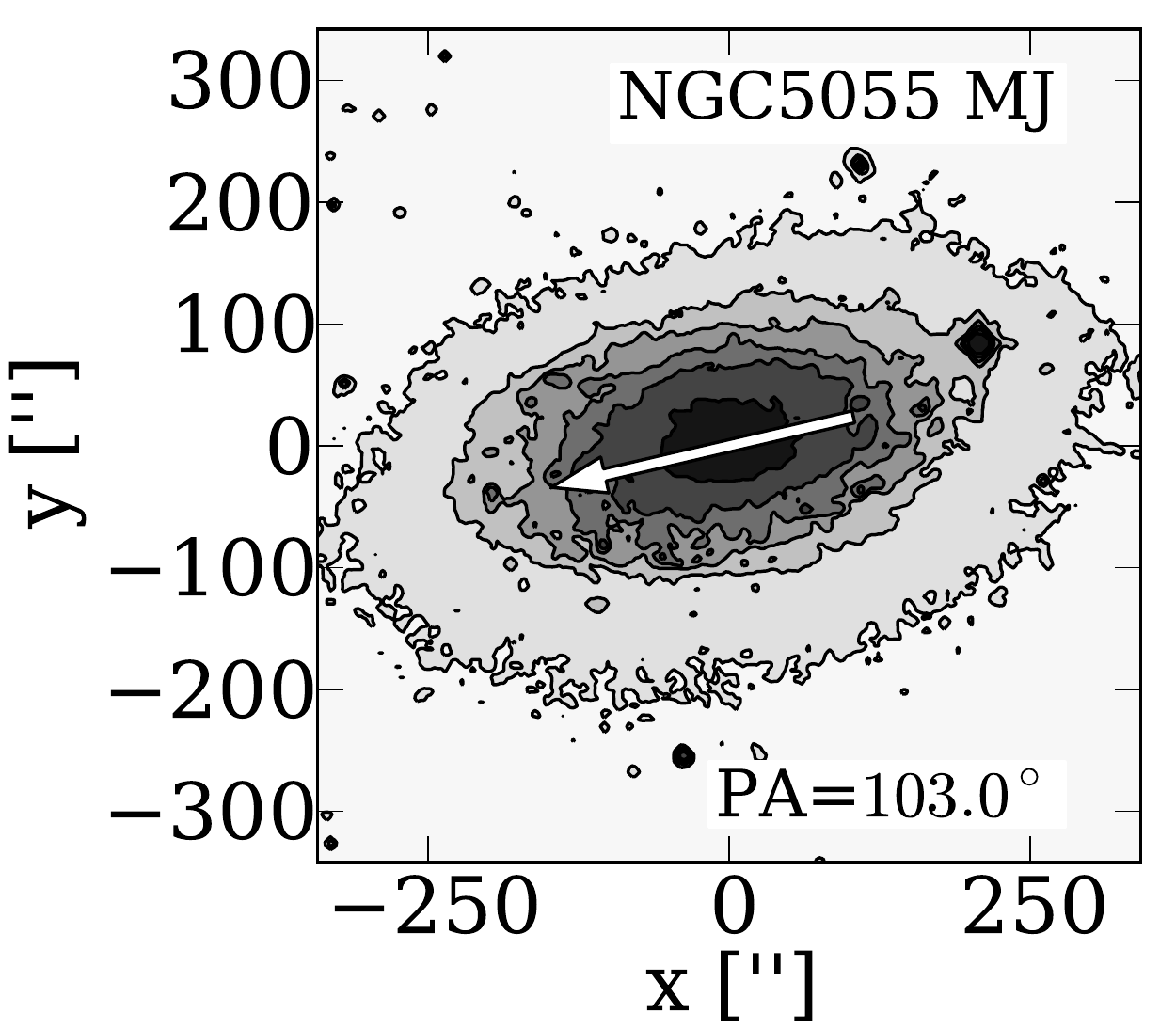}\\
	\includegraphics[viewport=0 55 390 400,width=\textwidth]{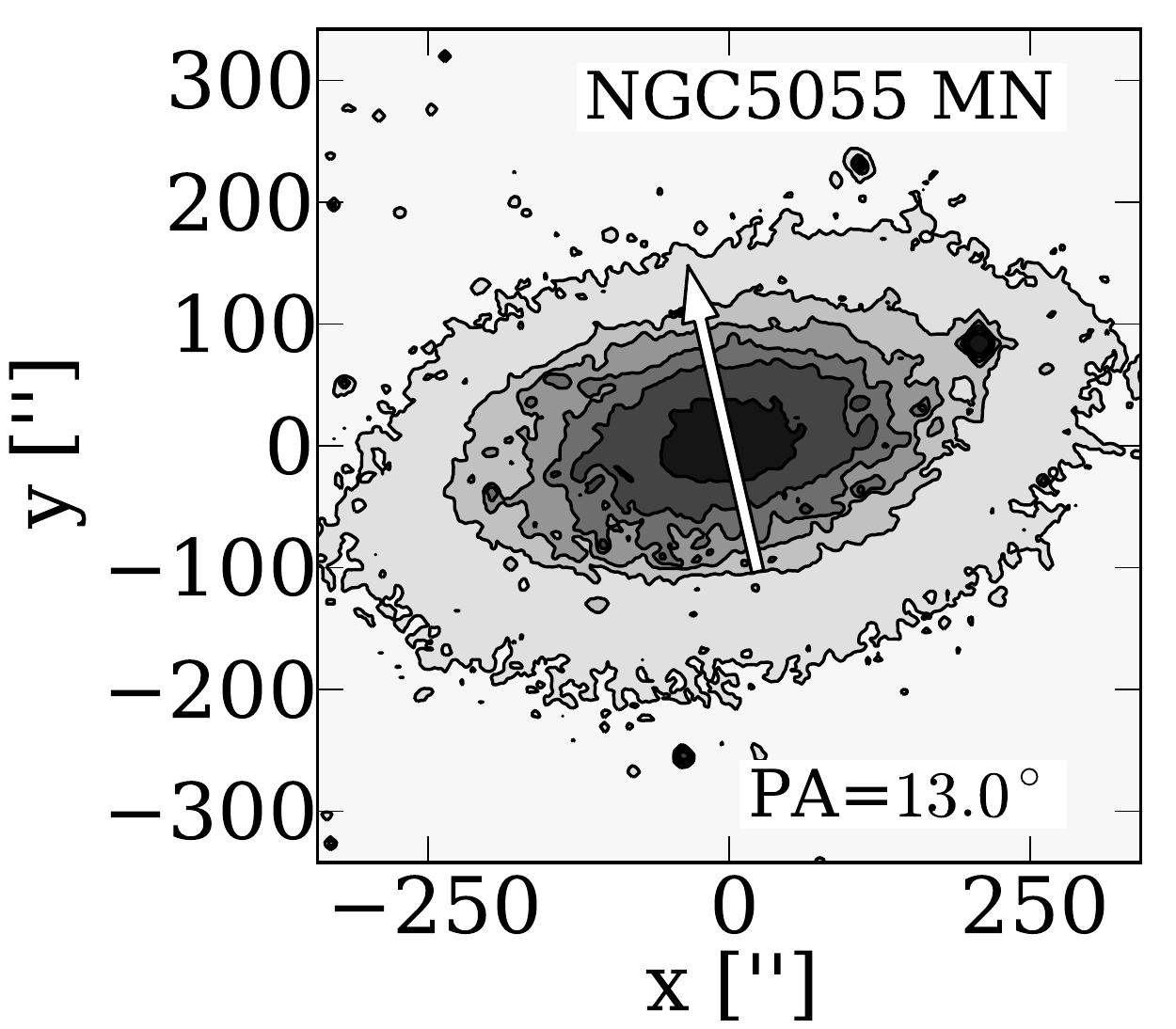}
	\end{minipage} & 
	\includegraphics[viewport=0 50 420 400,width=0.35\textwidth]{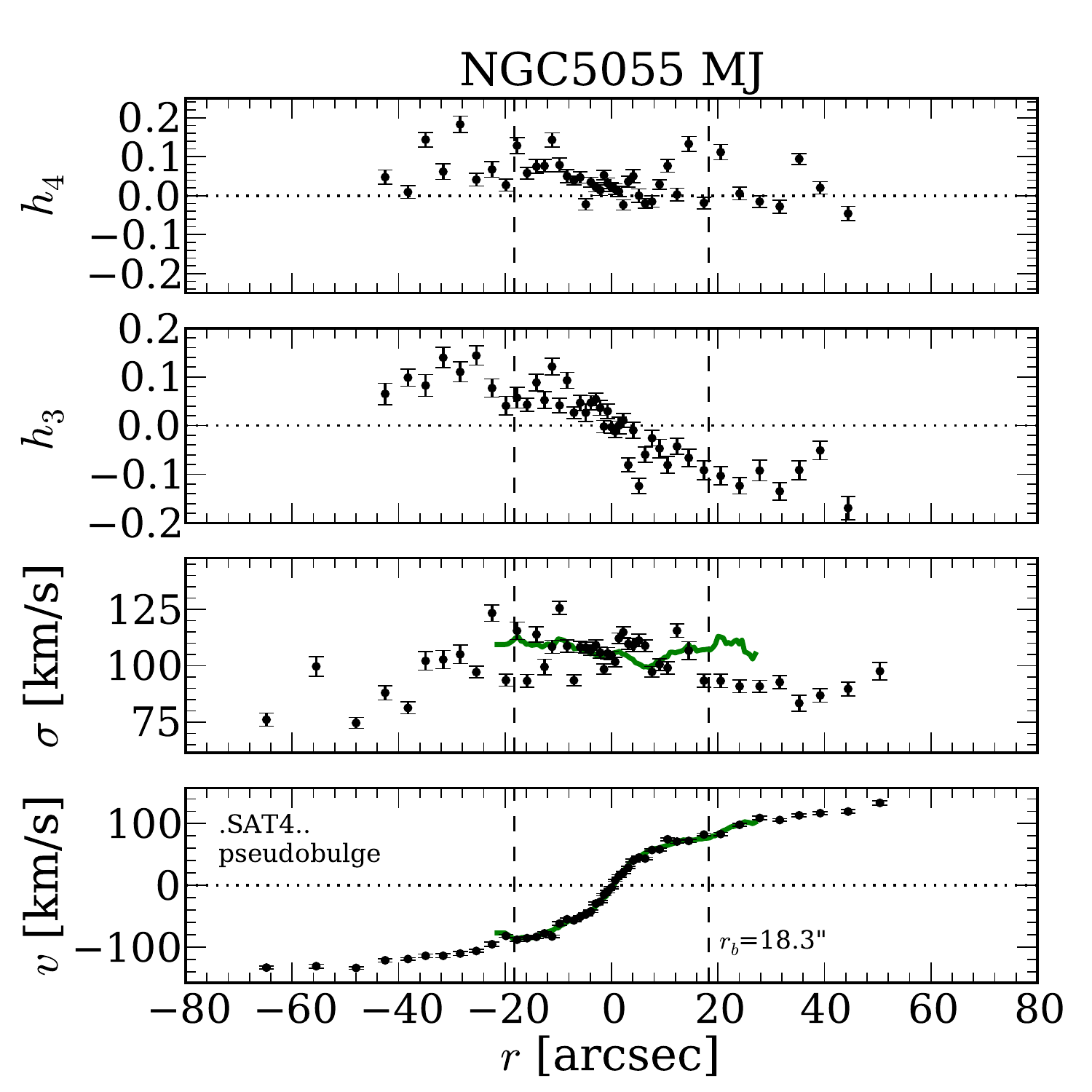} &
	\includegraphics[viewport=0 50 420 400,width=0.35\textwidth]{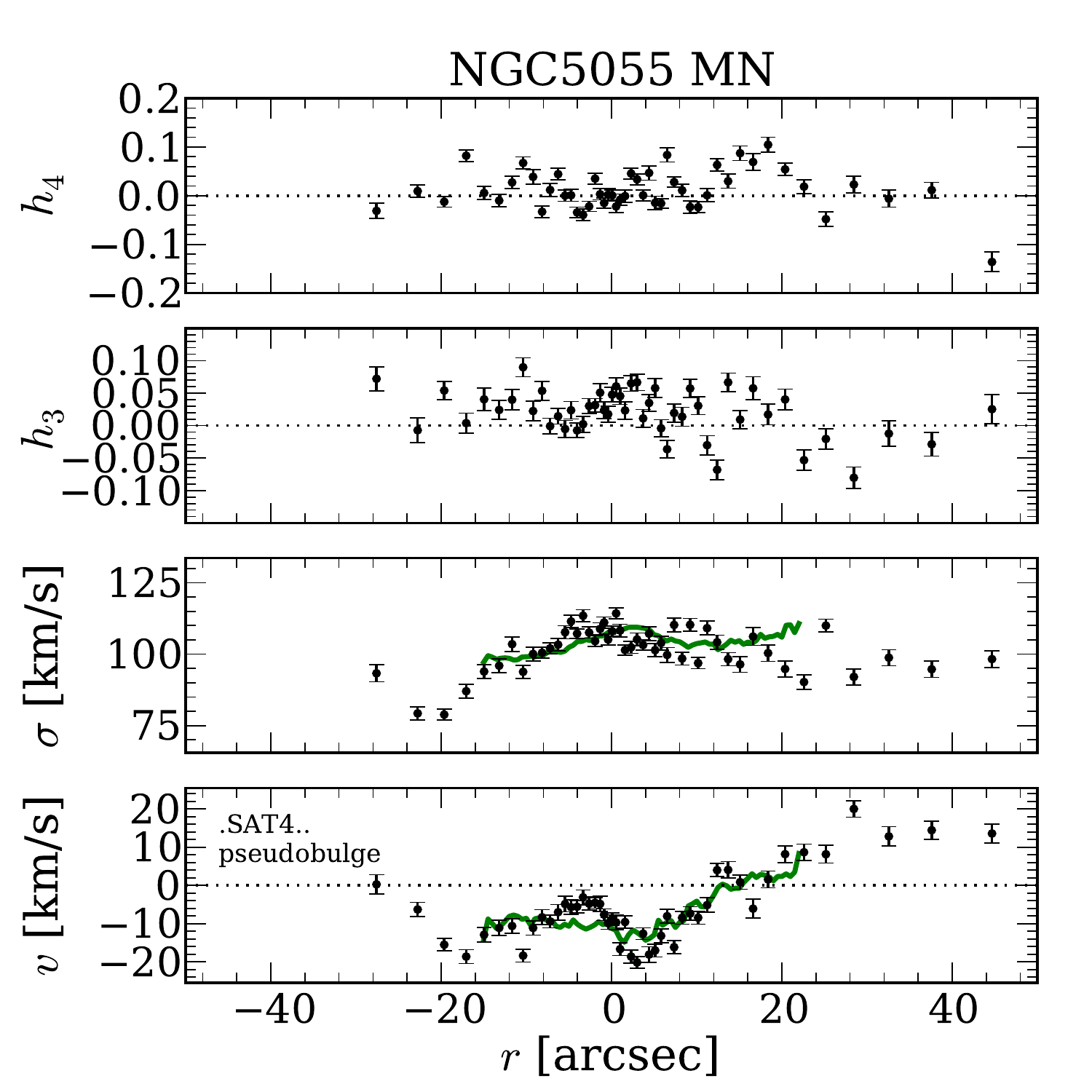}\\
        \end{tabular}
        \end{center}
        \caption{{\it continued --}\small Major and minor axis kinematic profiles for NGC\,5055.
	We plot SAURON results from \citet{Dumas2007} in green.
	}
\end{figure}
\setcounter{figure}{15}
\begin{figure}
        \begin{center}
        \begin{tabular}{lll}
	\begin{minipage}[b]{0.185\textwidth}
	\includegraphics[viewport=0 55 390 400,width=\textwidth]{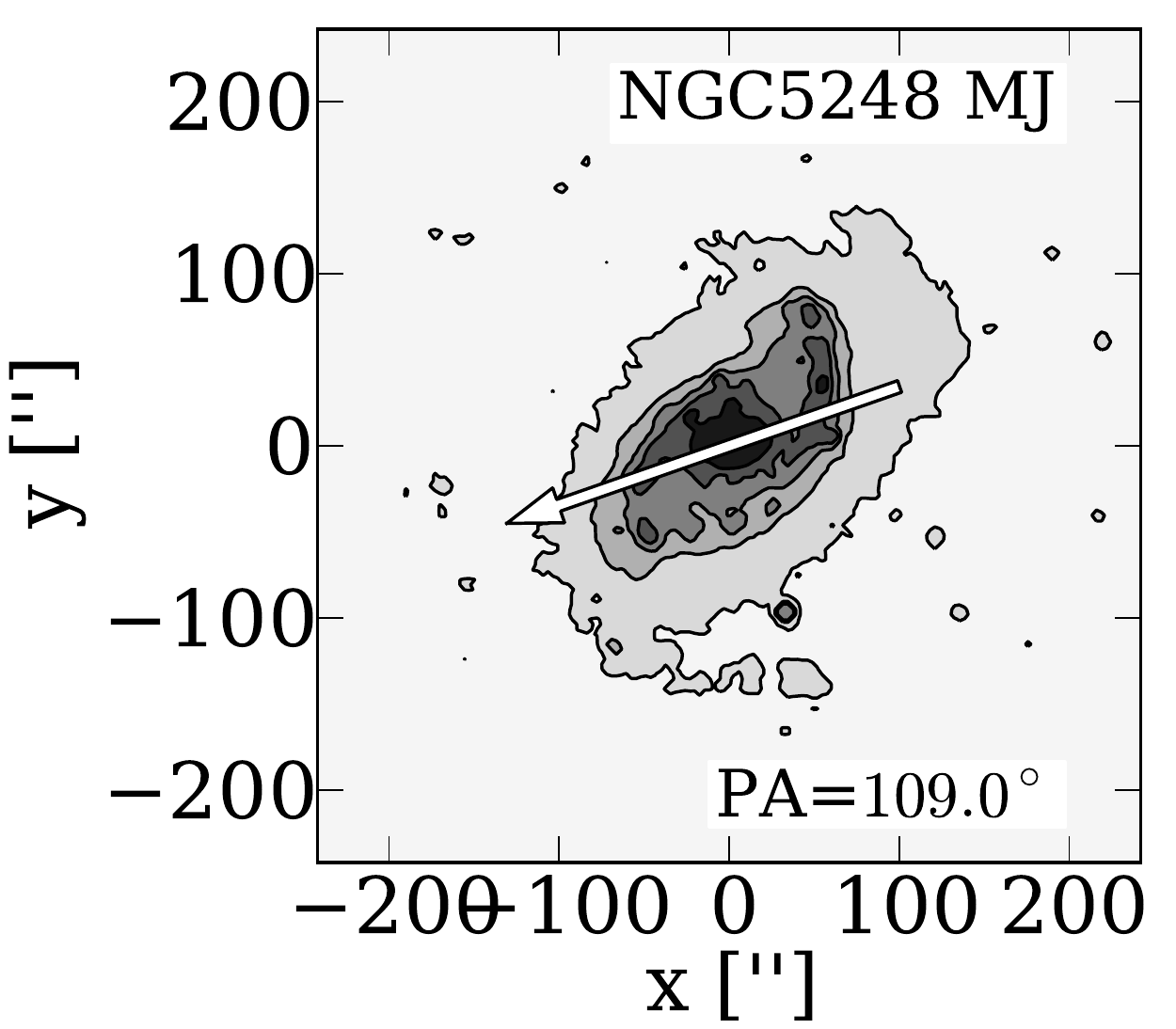} \\
        \includegraphics[viewport=0 55 390 400,width=\textwidth]{empty}
	\end{minipage} & 
	\includegraphics[viewport=0 50 420 400,width=0.35\textwidth]{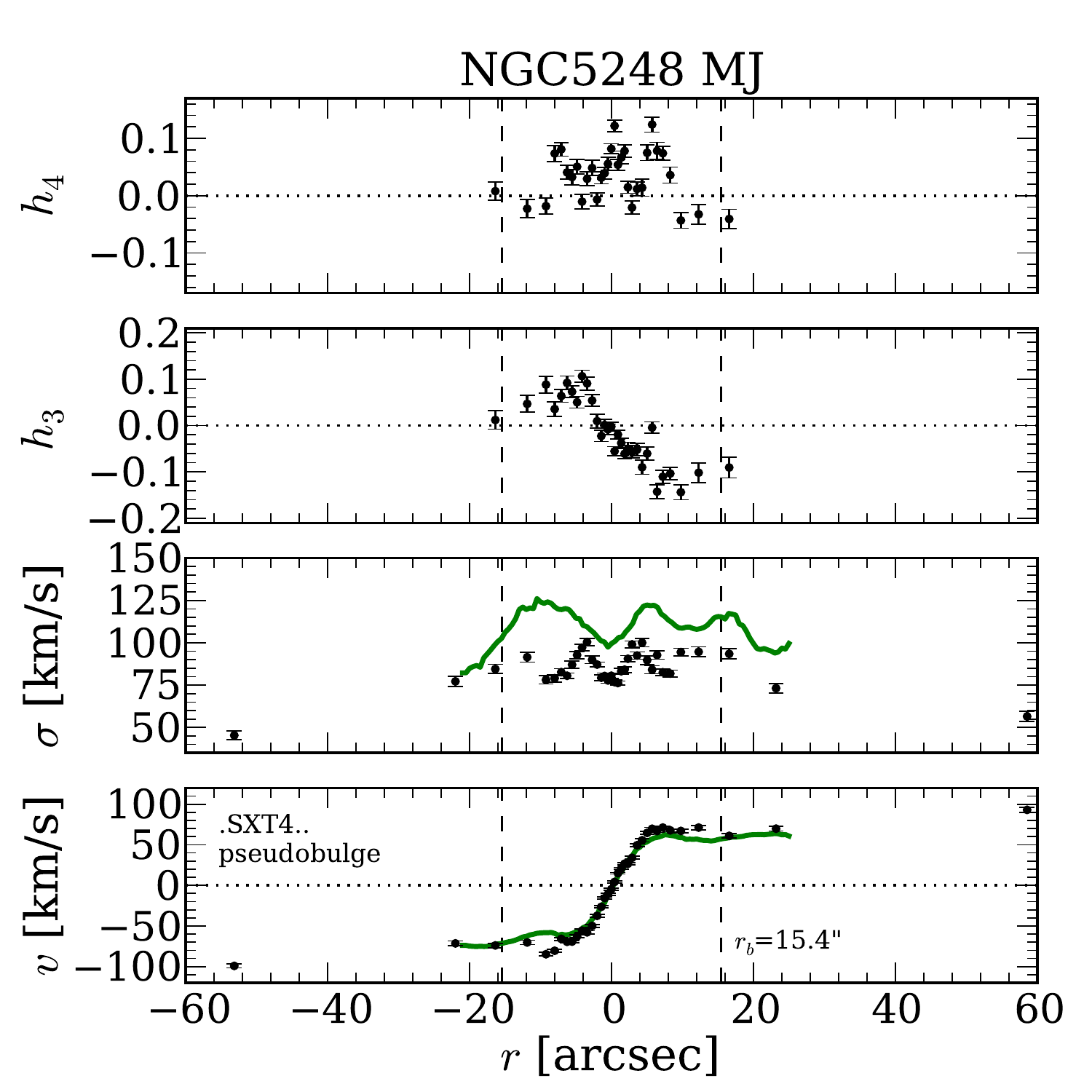}
        \includegraphics[viewport=0 50 420 400,width=0.35\textwidth]{empty}
        \end{tabular}
        \end{center}
        \caption{{\it continued --}\small Major axis kinematic profile for NGC\,5248.
		We plot results from \citet{Dumas2007}.} 
\end{figure}
\setcounter{figure}{15}
\begin{figure}
        \begin{center}
        \begin{tabular}{lll}
	\begin{minipage}[b]{0.185\textwidth}
	\includegraphics[viewport=0 55 390 400,width=\textwidth]{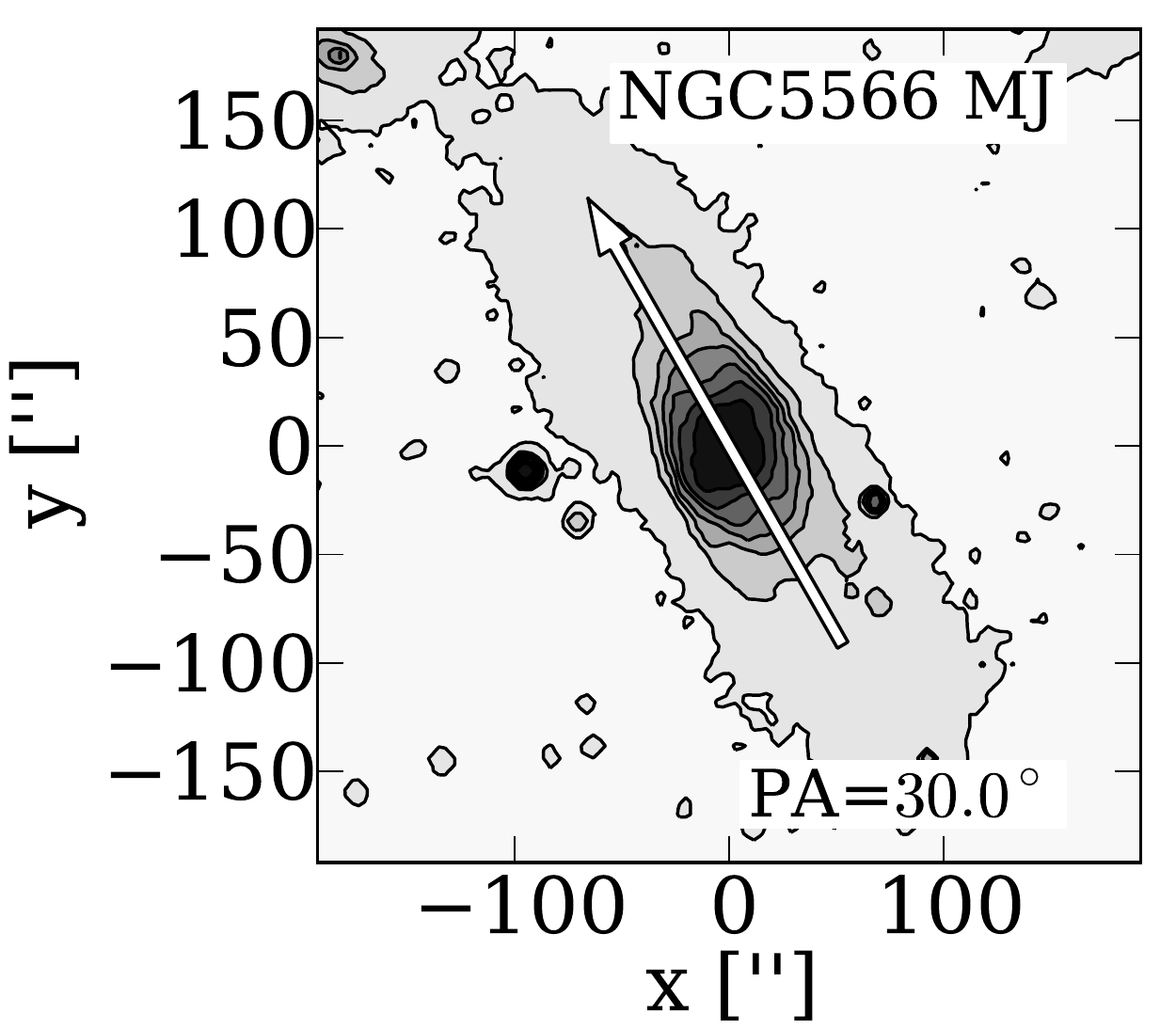}\\
	\includegraphics[viewport=0 55 390 400,width=\textwidth]{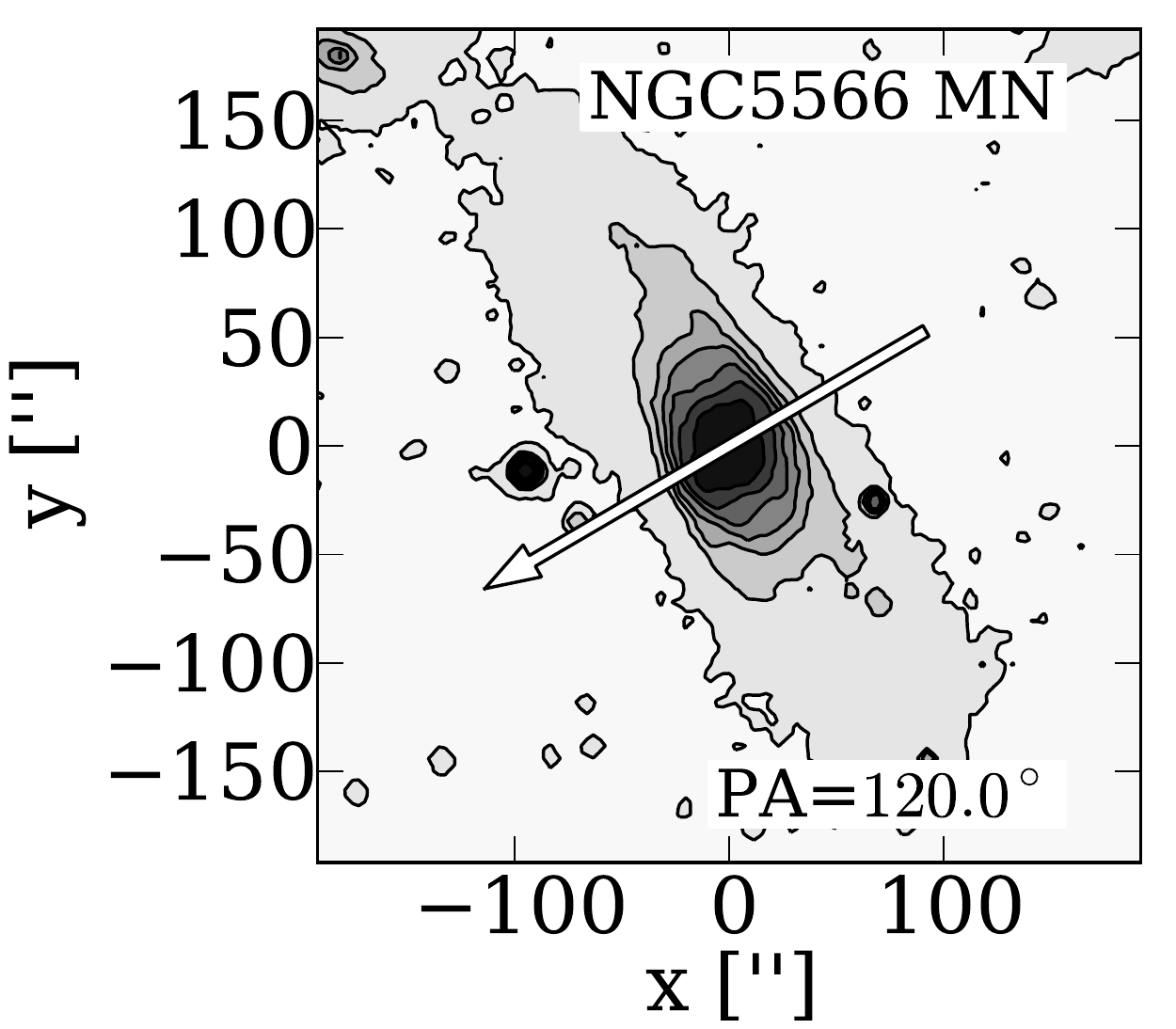}
	\end{minipage} & 
	\includegraphics[viewport=0 50 420 400,width=0.35\textwidth]{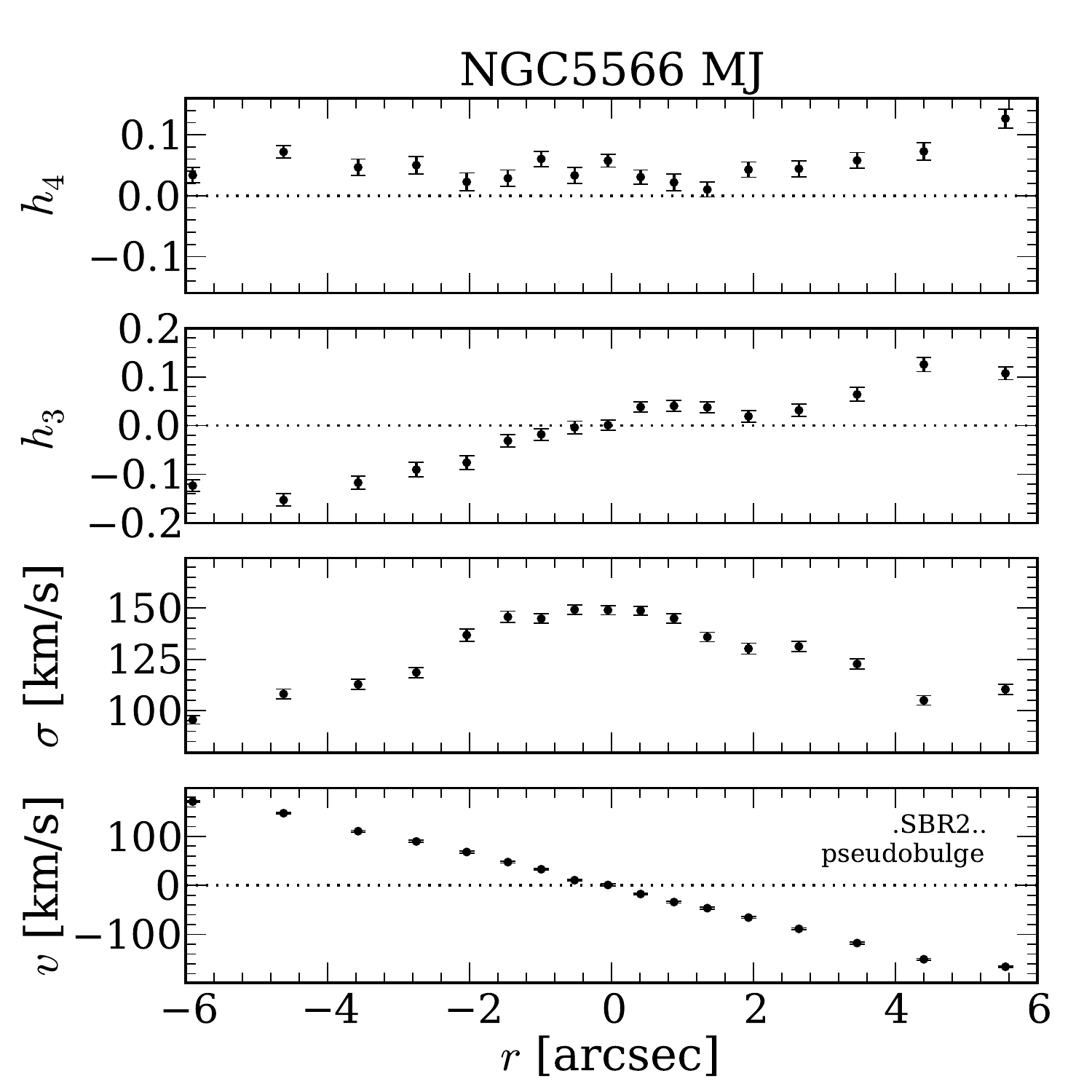} &
	\includegraphics[viewport=0 50 420 400,width=0.35\textwidth]{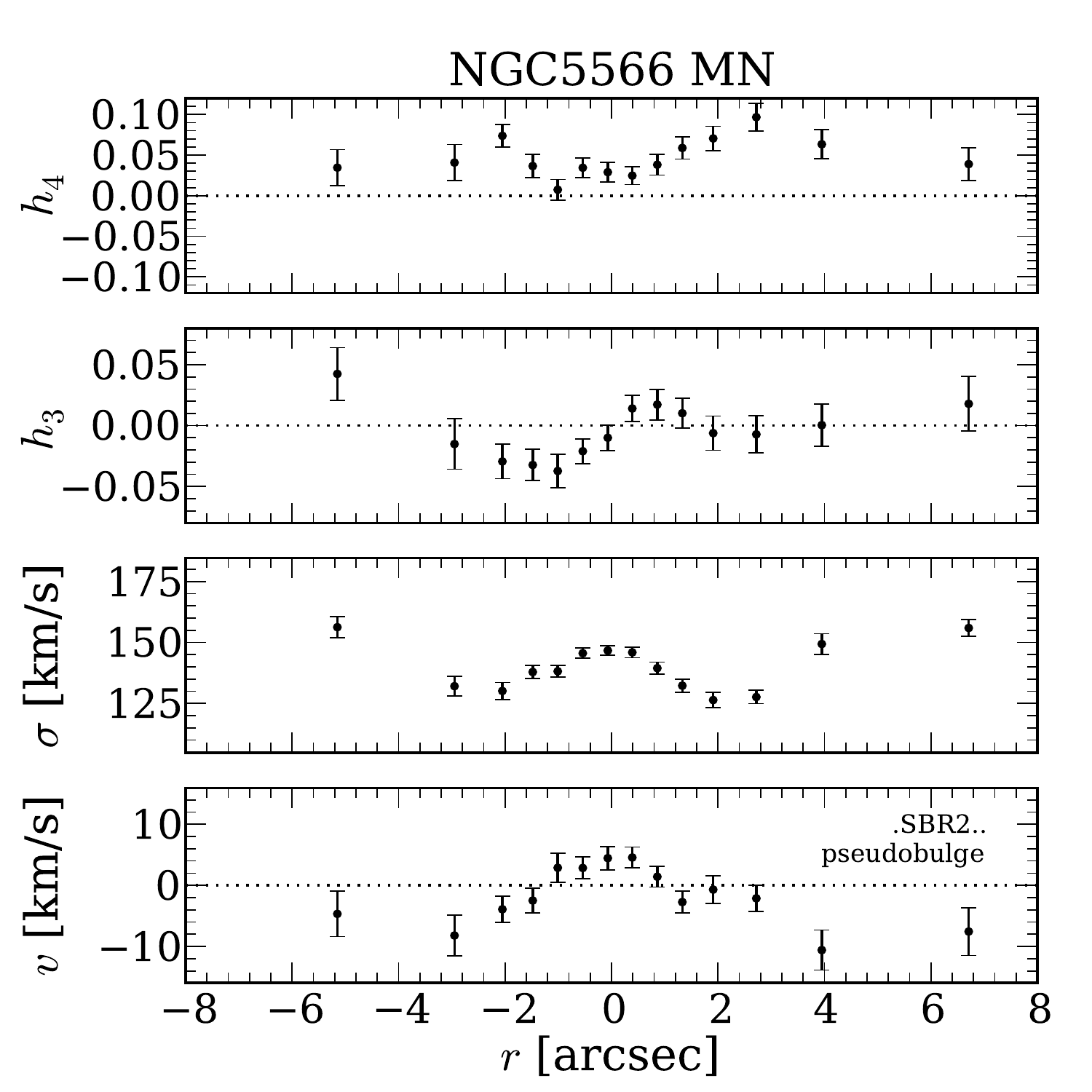}\\
        \end{tabular}
        \end{center}
		\caption{{\it continued --} \small Major and minor axis kinematic
		profiles for NGC\,5566.  Note: We do not derive a decomposition for this galaxy
		(see discussion in Appendix~\ref{sec:individuals}), hence there is no bulge
		radius indicator.}
\end{figure}
\clearpage
\setcounter{figure}{15}
\begin{figure}
        \begin{center}
        \begin{tabular}{lll}
	\begin{minipage}[b]{0.185\textwidth}
	\includegraphics[viewport=0 55 390 400,width=\textwidth]{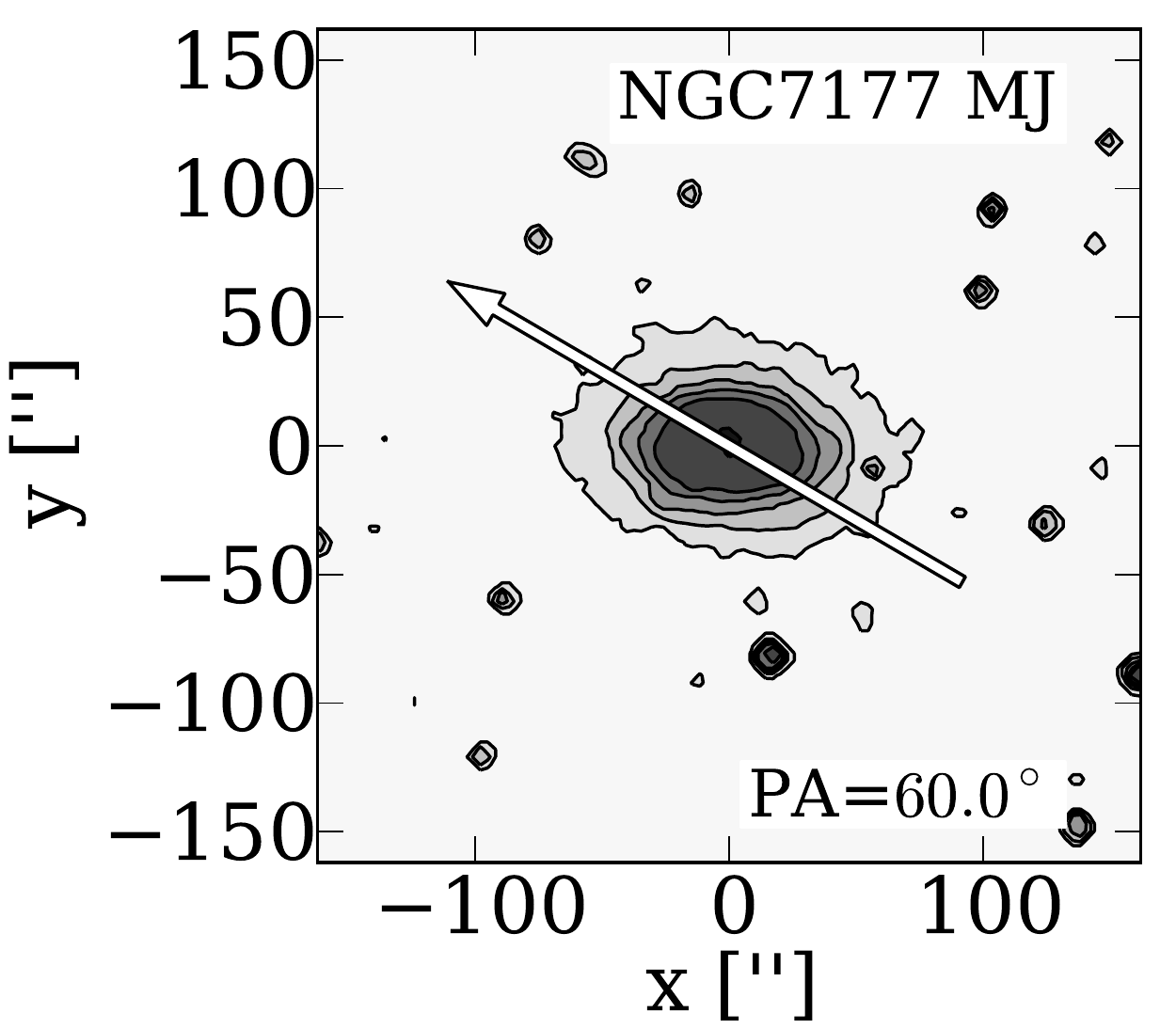}\\
	\includegraphics[viewport=0 55 390 400,width=\textwidth]{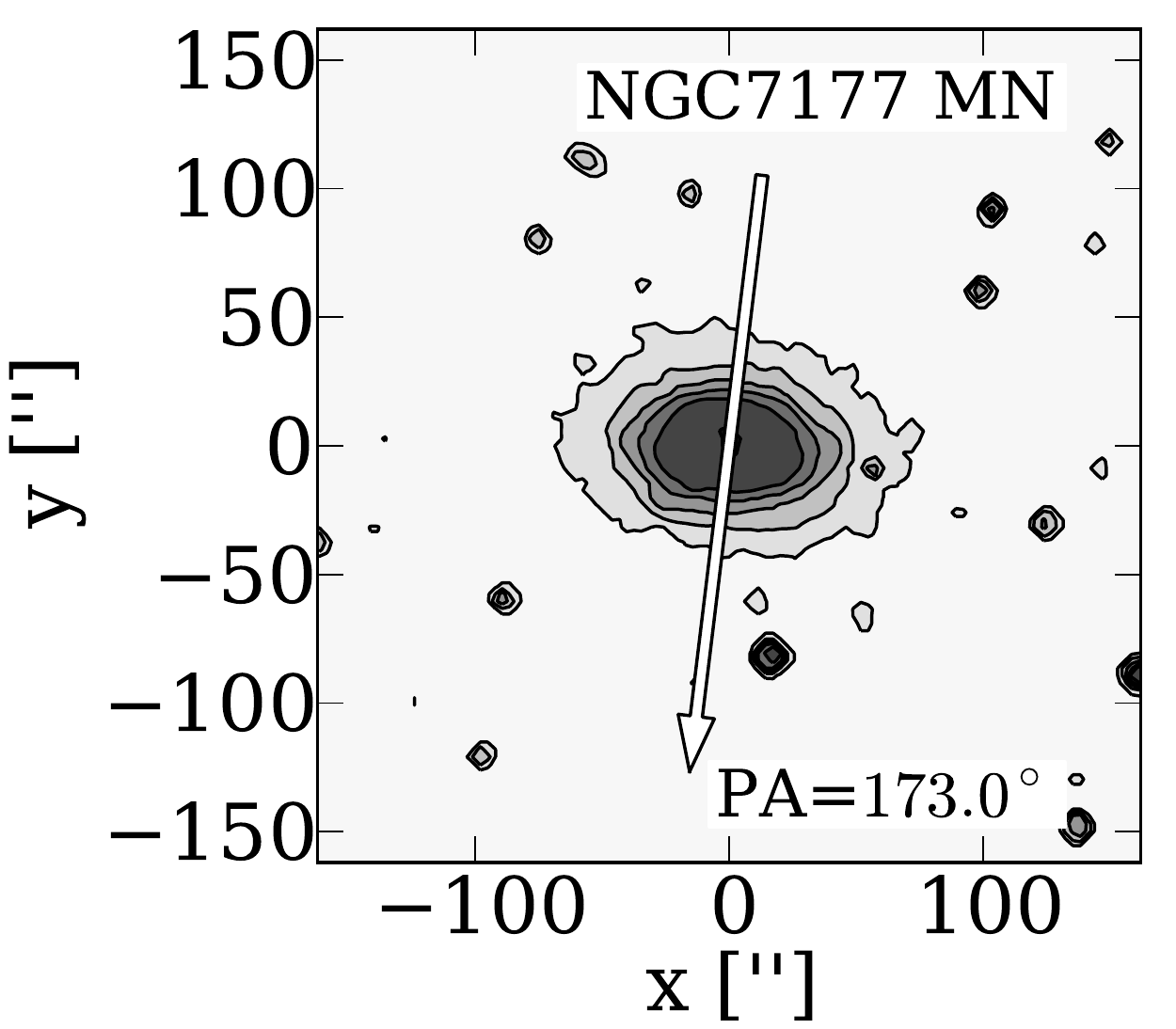}
	\end{minipage} & 
	\includegraphics[viewport=0 50 420 400,width=0.35\textwidth]{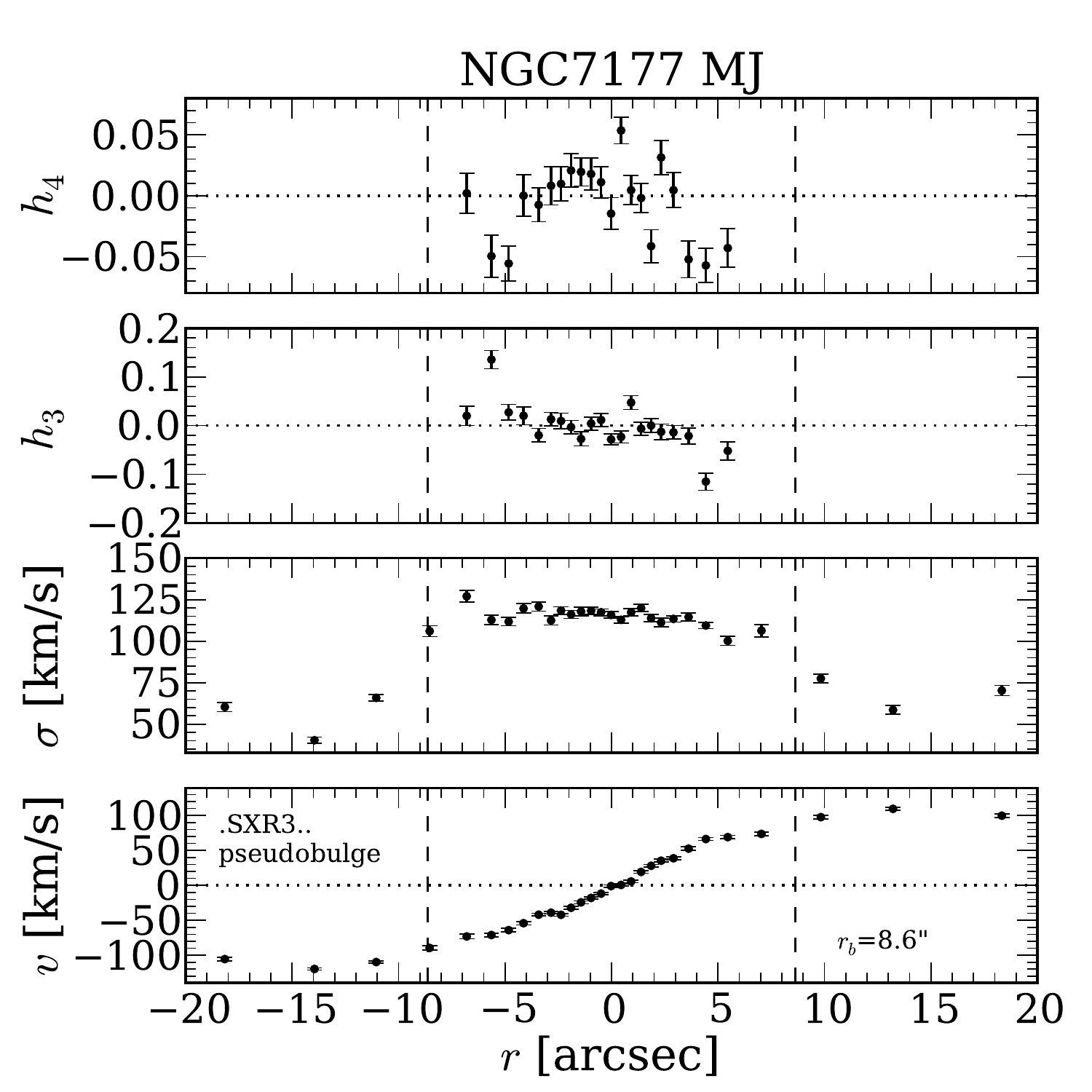} &
	\includegraphics[viewport=0 50 420 400,width=0.35\textwidth]{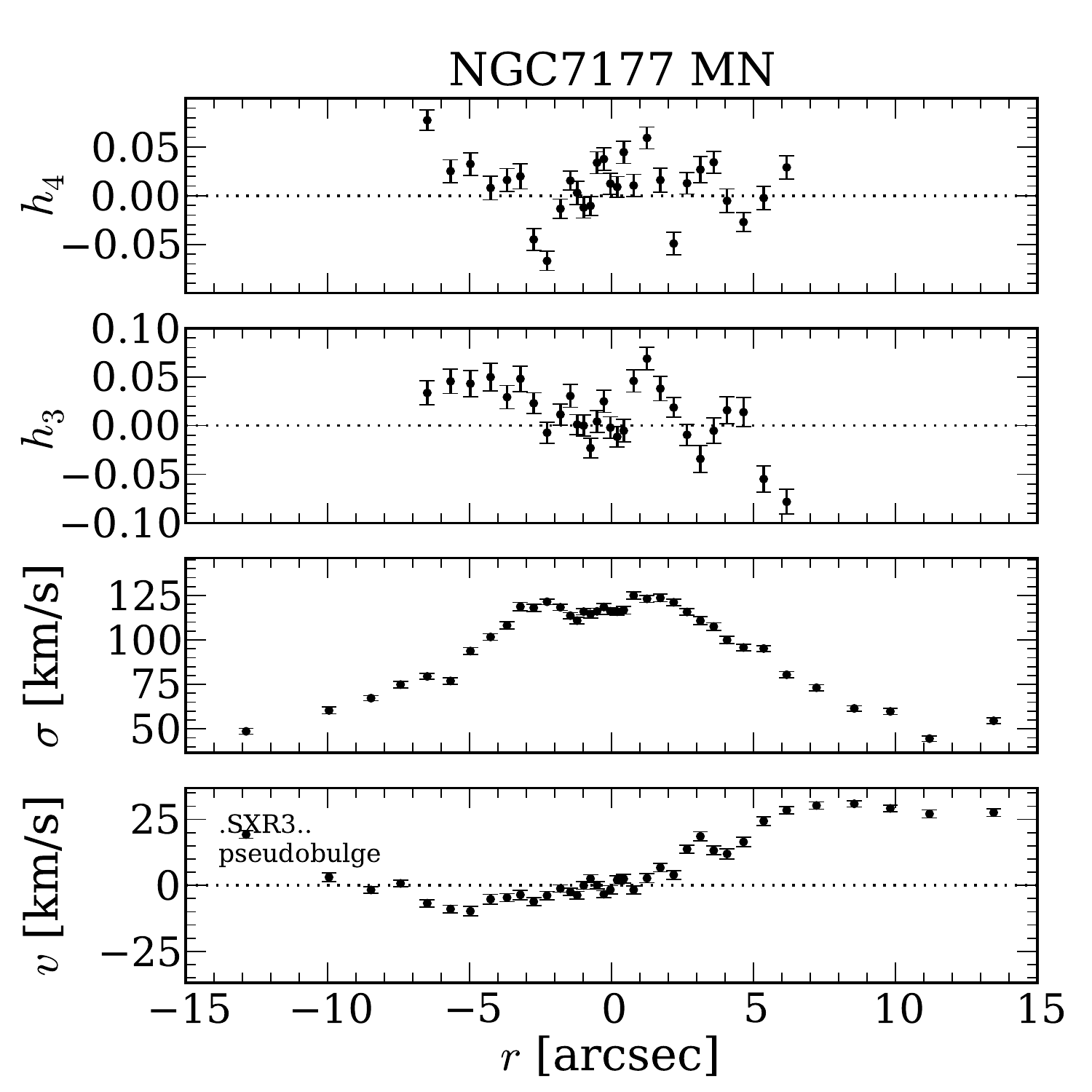}\\
        \end{tabular}
        \end{center}
        \caption{{\it continued --}\small Major and minor axis kinematic profiles for NGC\,7177.}
\end{figure}
\setcounter{figure}{15}
\begin{figure}
        \begin{center}
        \begin{tabular}{lll}
	\begin{minipage}[b]{0.185\textwidth}
	\includegraphics[viewport=0 55 390 400,width=\textwidth]{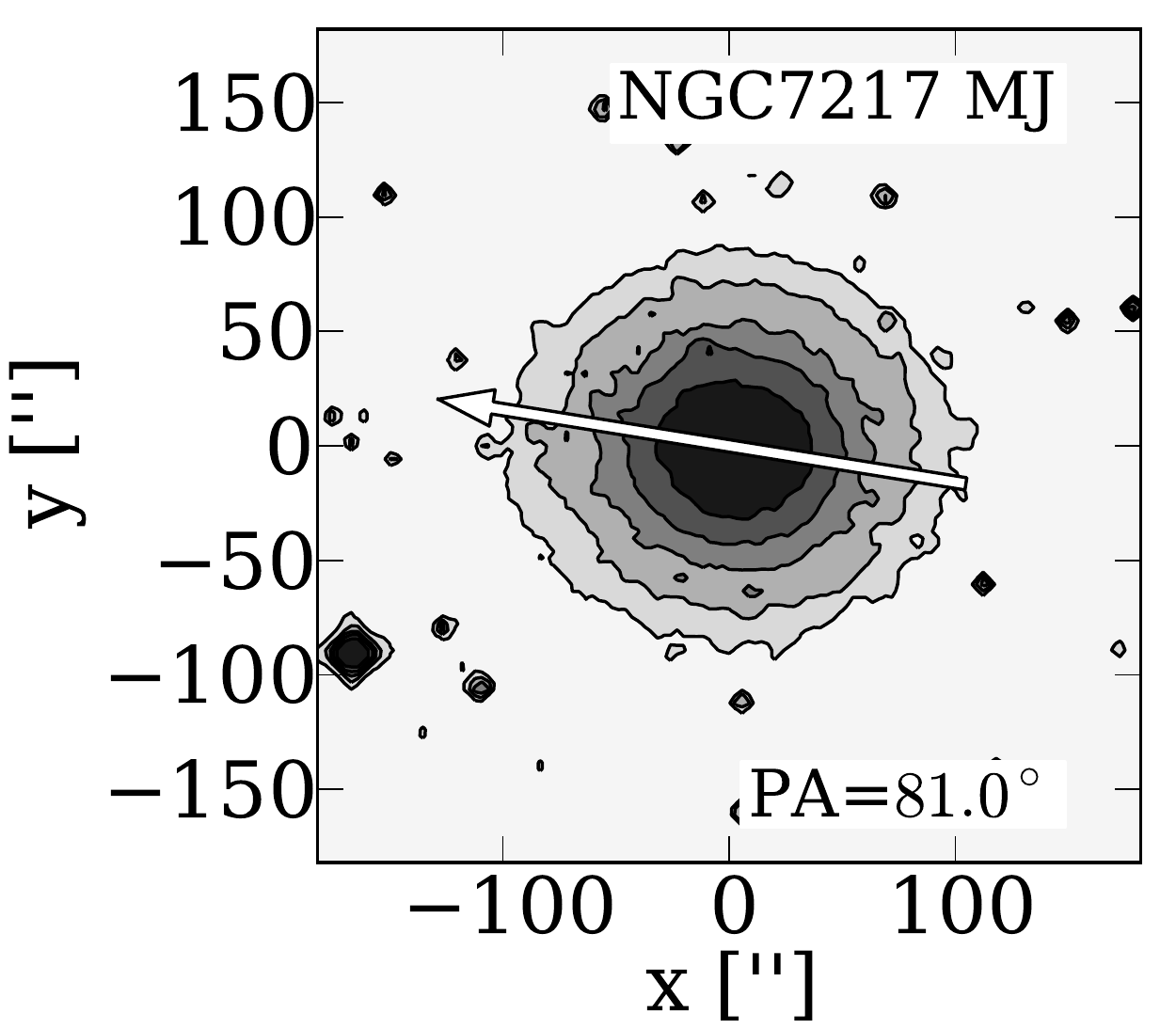}\\
	\includegraphics[viewport=0 55 390 400,width=\textwidth]{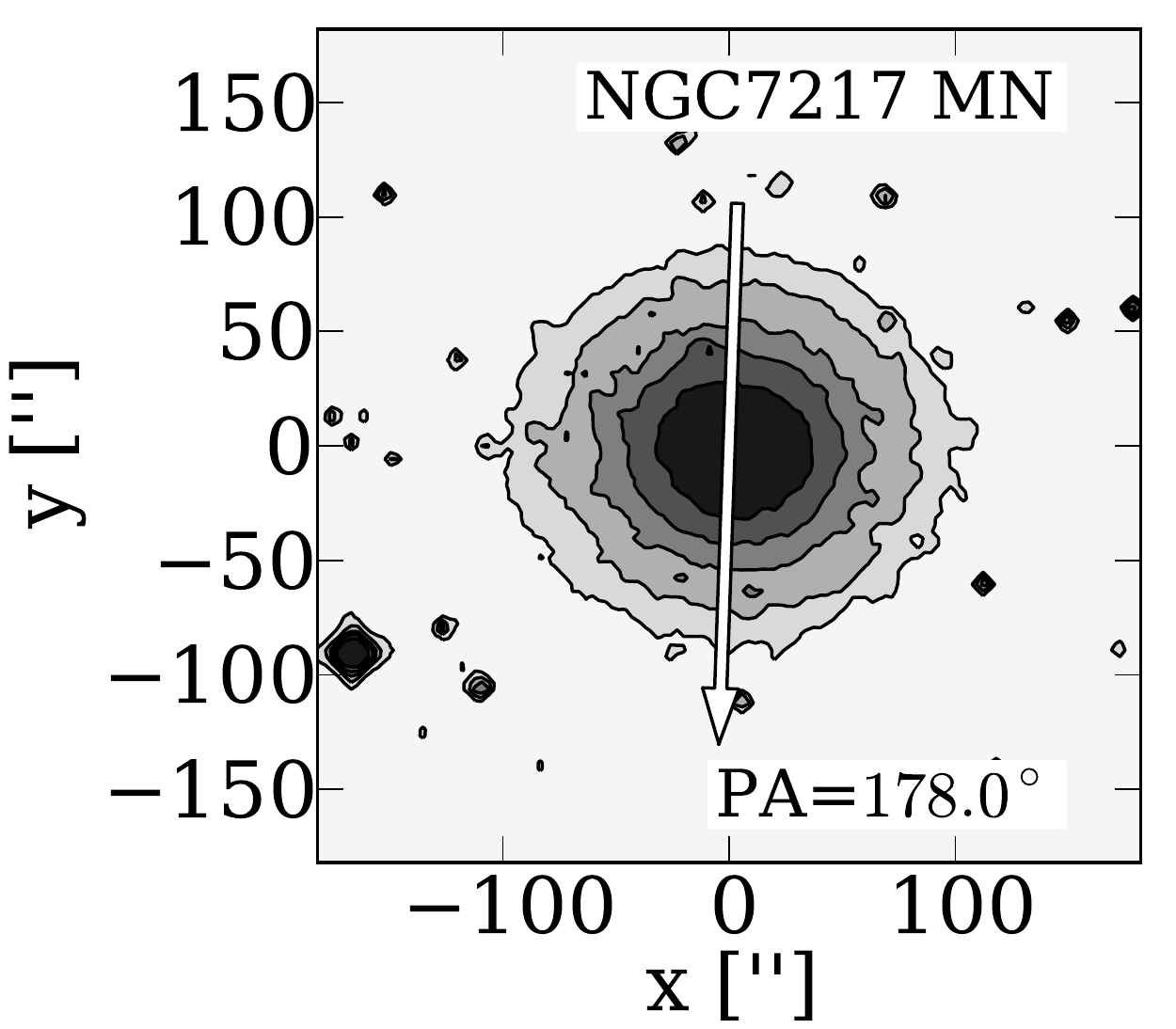}
	\end{minipage} & 
	\includegraphics[viewport=0 50 420 400,width=0.35\textwidth]{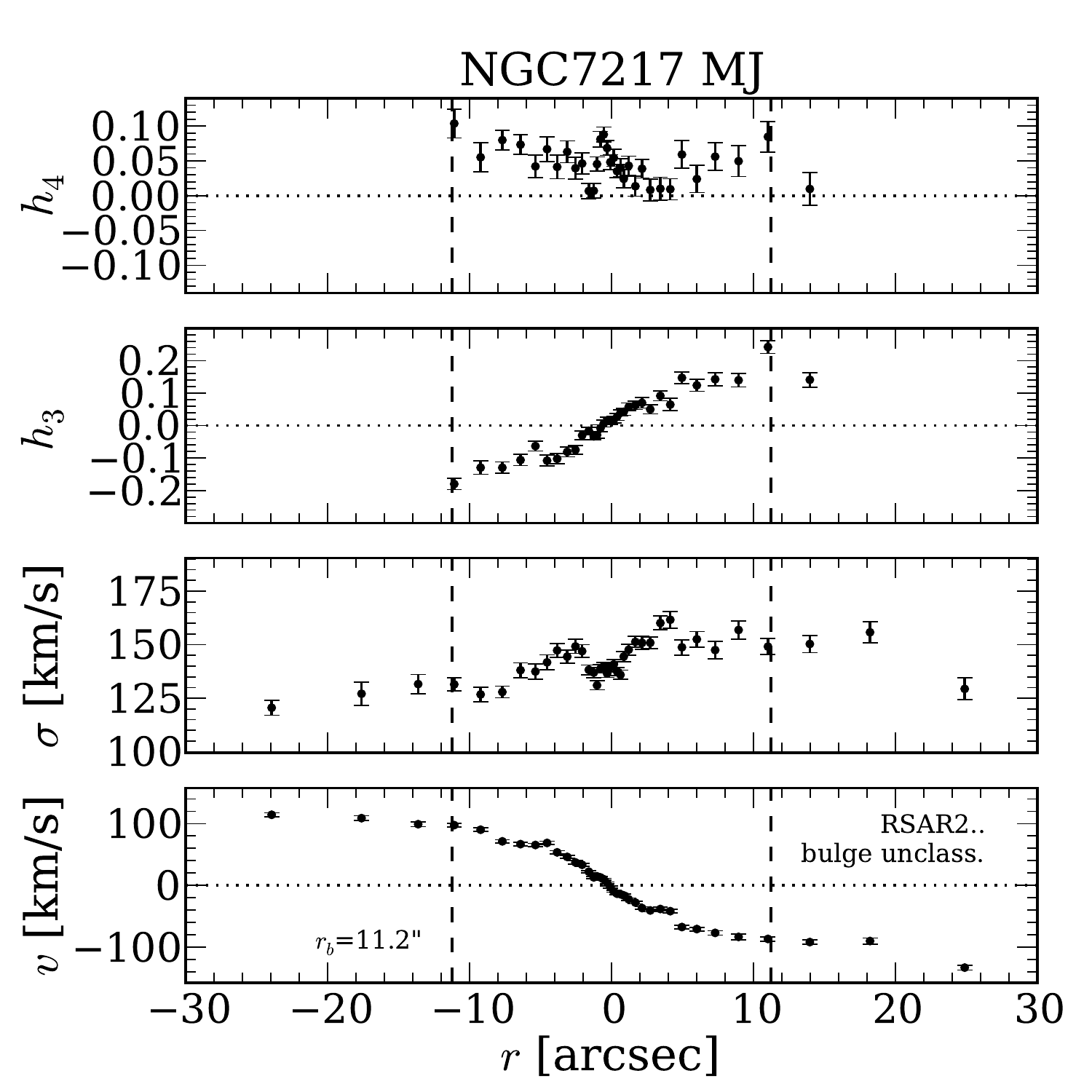} &
	\includegraphics[viewport=0 50 420 400,width=0.35\textwidth]{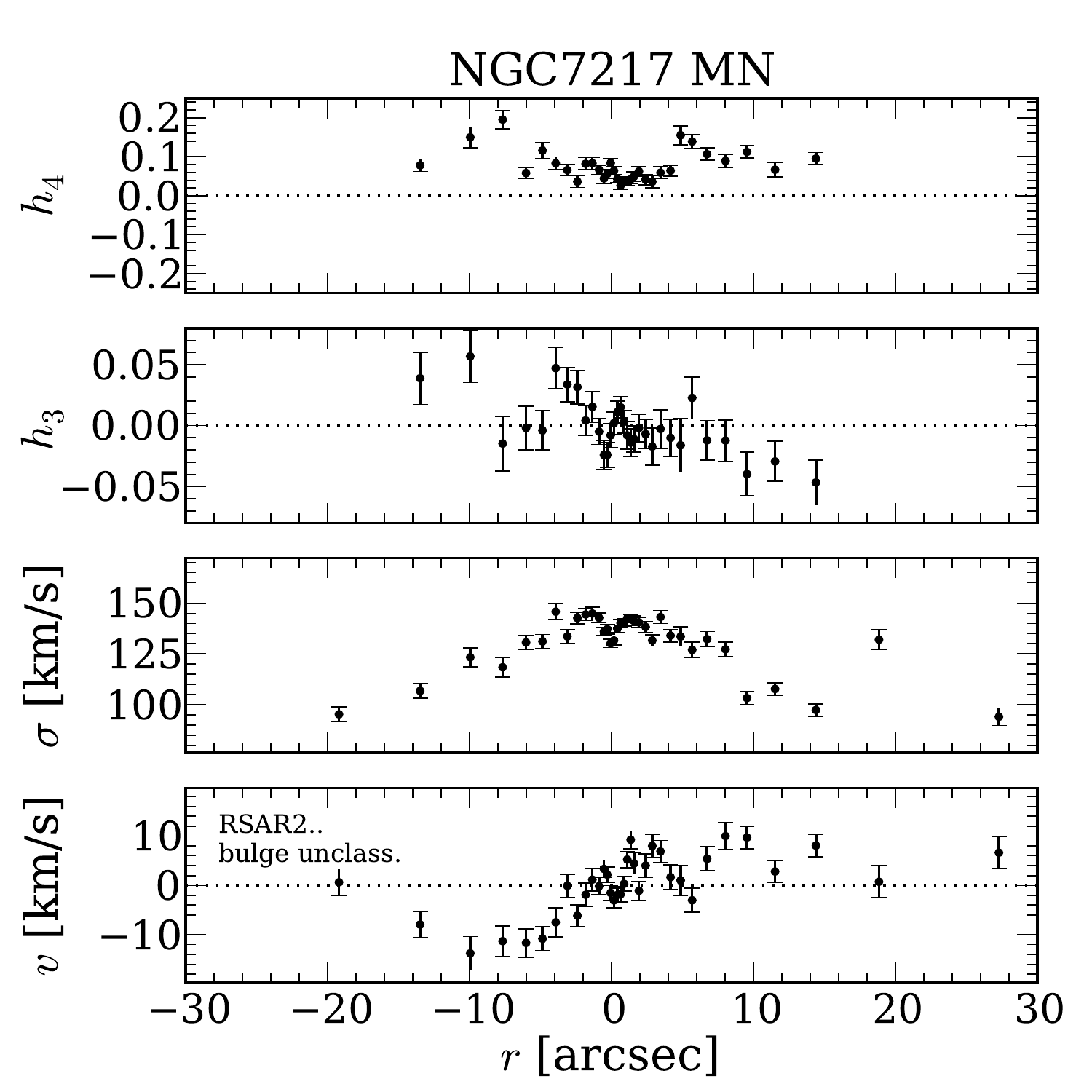}\\
        \end{tabular}
        \end{center}
        \caption{{\it continued --}\small Major and minor axis kinematic profiles for NGC\,7217,
	see also Fig.~\ref{fig:kinDecomp}.
	}
\end{figure}
\clearpage
\setcounter{figure}{15}
\begin{figure}
        \begin{center}
        \begin{tabular}{lll}
	\begin{minipage}[b]{0.185\textwidth}
	\includegraphics[viewport=0 55 390 400,width=\textwidth]{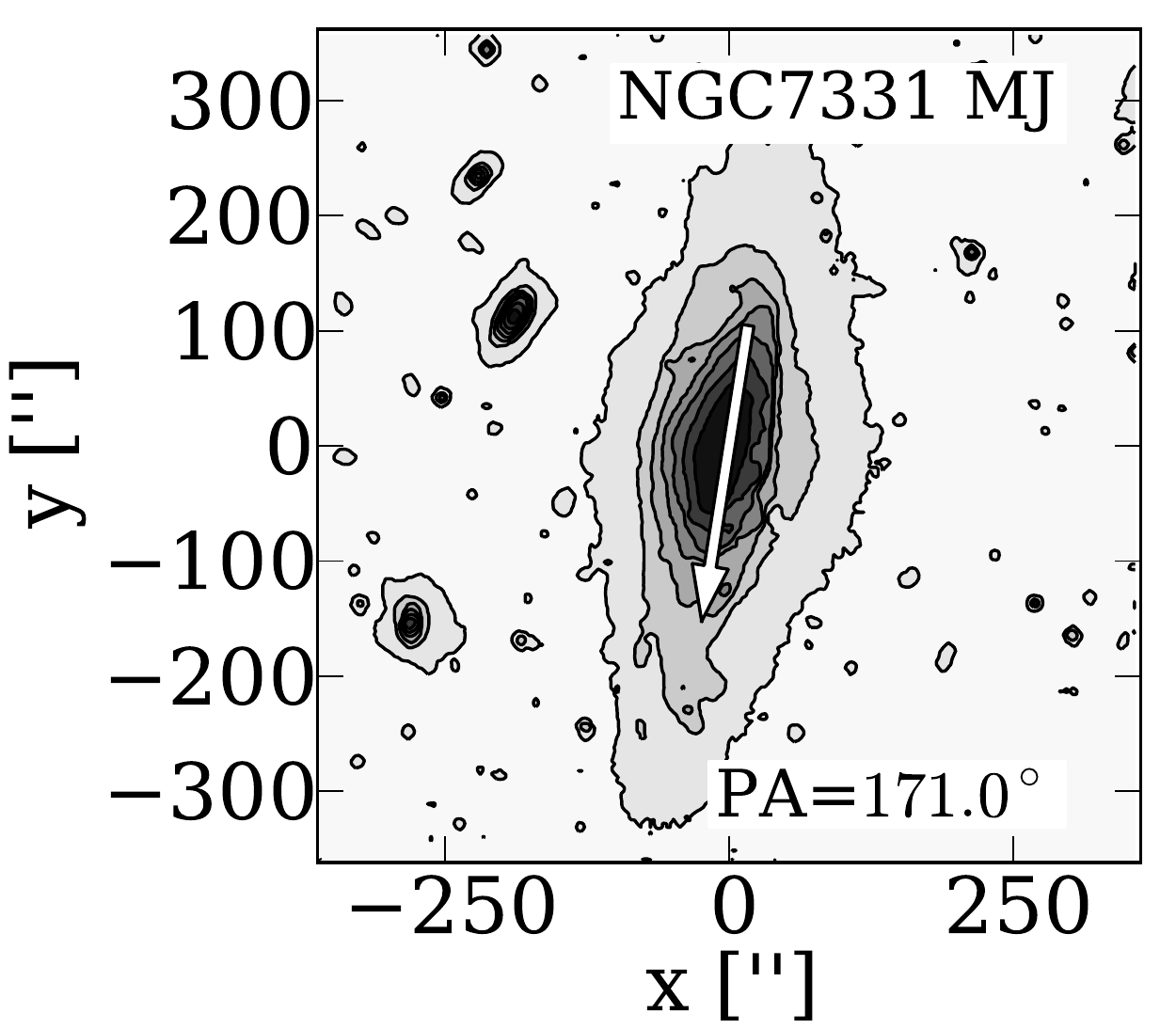} \\ 
        \includegraphics[viewport=0 55 390 400,width=\textwidth]{empty}
	\end{minipage} & 
	\includegraphics[viewport=0 50 420 400,width=0.35\textwidth]{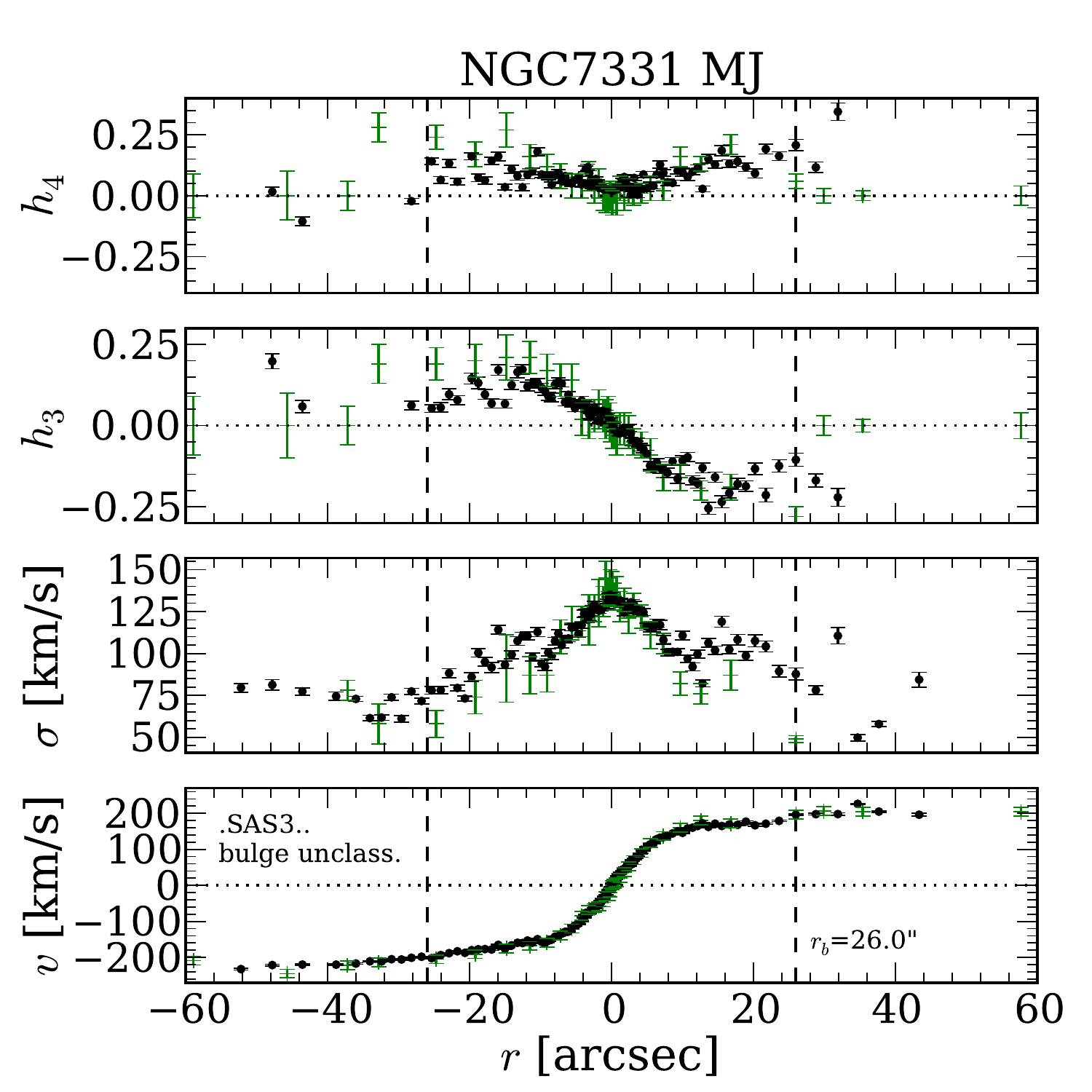} & 
        \includegraphics[viewport=0 50 420 400,width=0.35\textwidth]{empty}
        \end{tabular}
        \end{center}
        \caption{{\it continued --}\small Major axis kinematic profile for NGC\,7331,
	see also Fig.~\ref{fig:kinDecomp}.
	We plot results   
	from \citet{Vega-Beltran2001} for NGC\,7331 in green.} 
\end{figure}
\setcounter{figure}{15}
\begin{figure}
        \begin{center}
        \begin{tabular}{lll}
	\begin{minipage}[b]{0.185\textwidth}
	\includegraphics[viewport=0 55 390 400,width=\textwidth]{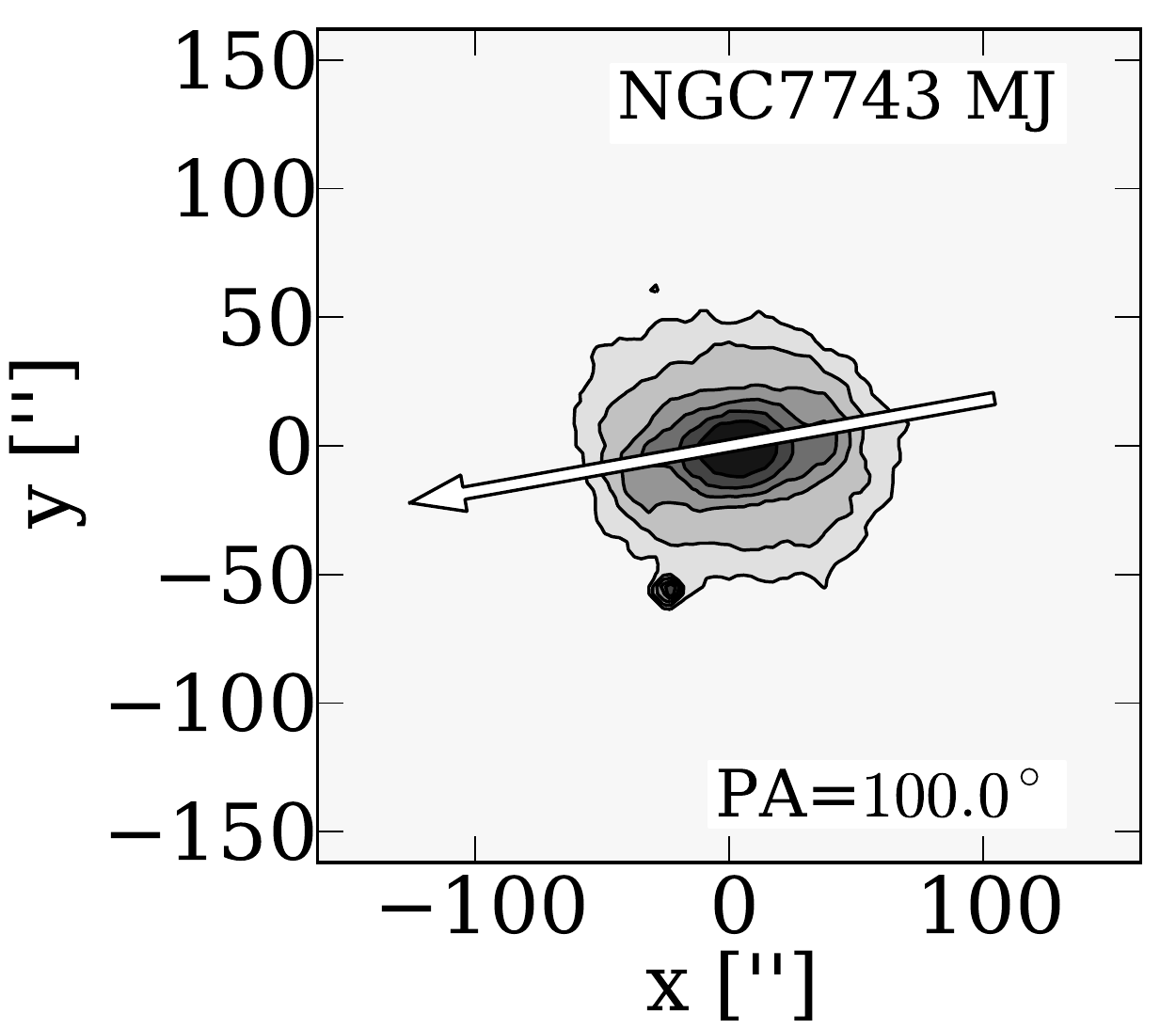}\\
	\includegraphics[viewport=0 55 390 400,width=\textwidth]{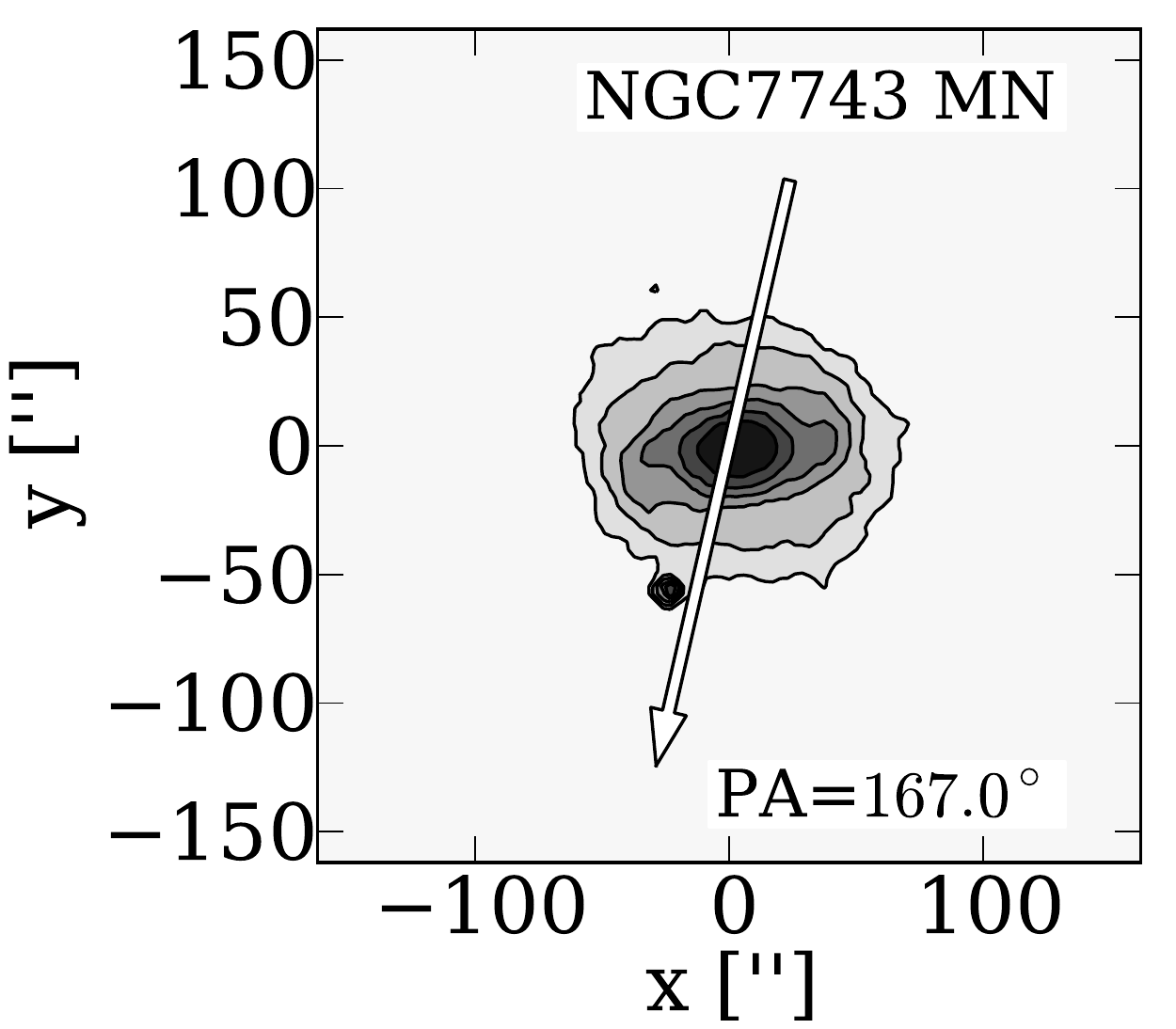}
	\end{minipage} & 
	\includegraphics[viewport=0 50 420 400,width=0.35\textwidth]{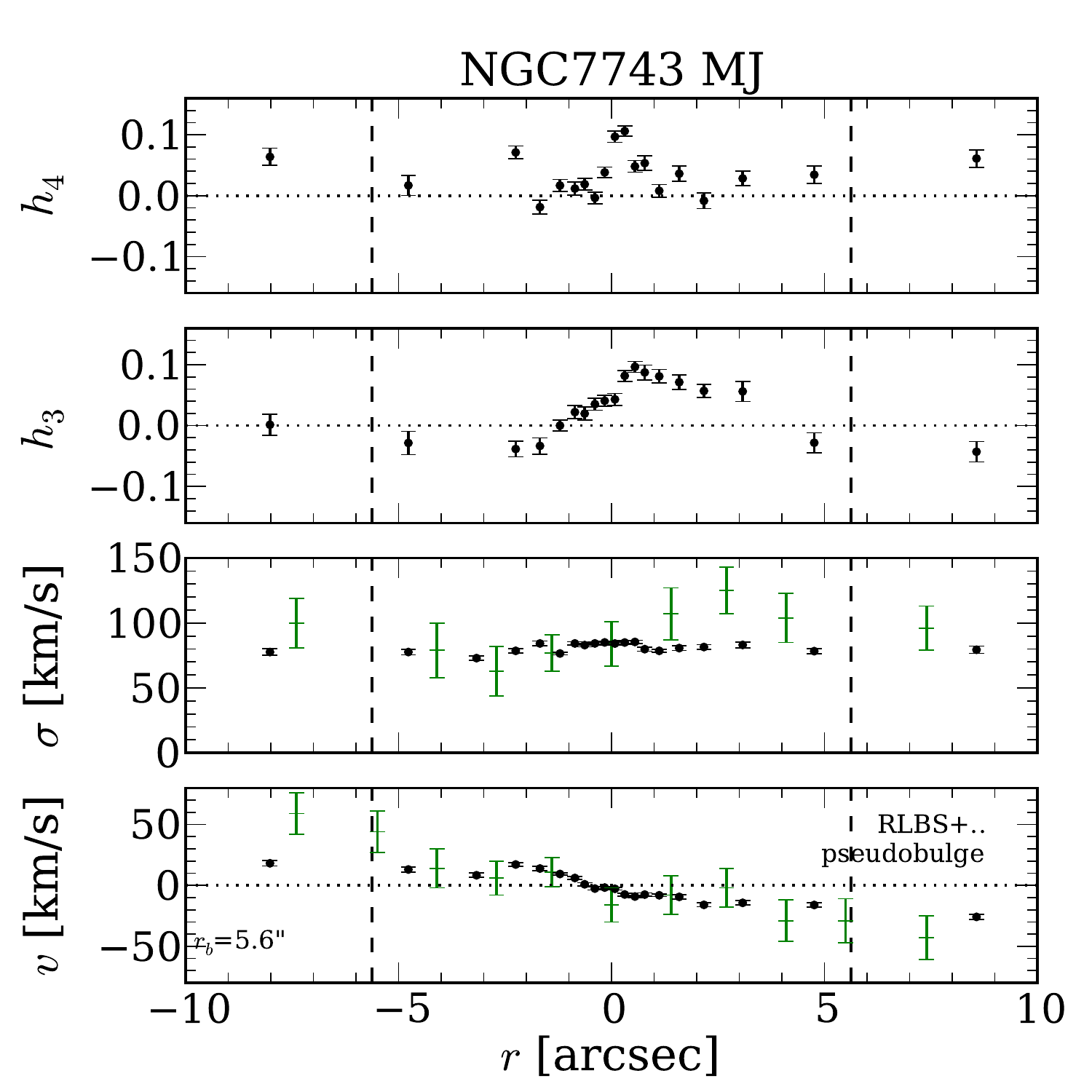} &
	\includegraphics[viewport=0 50 420 400,width=0.35\textwidth]{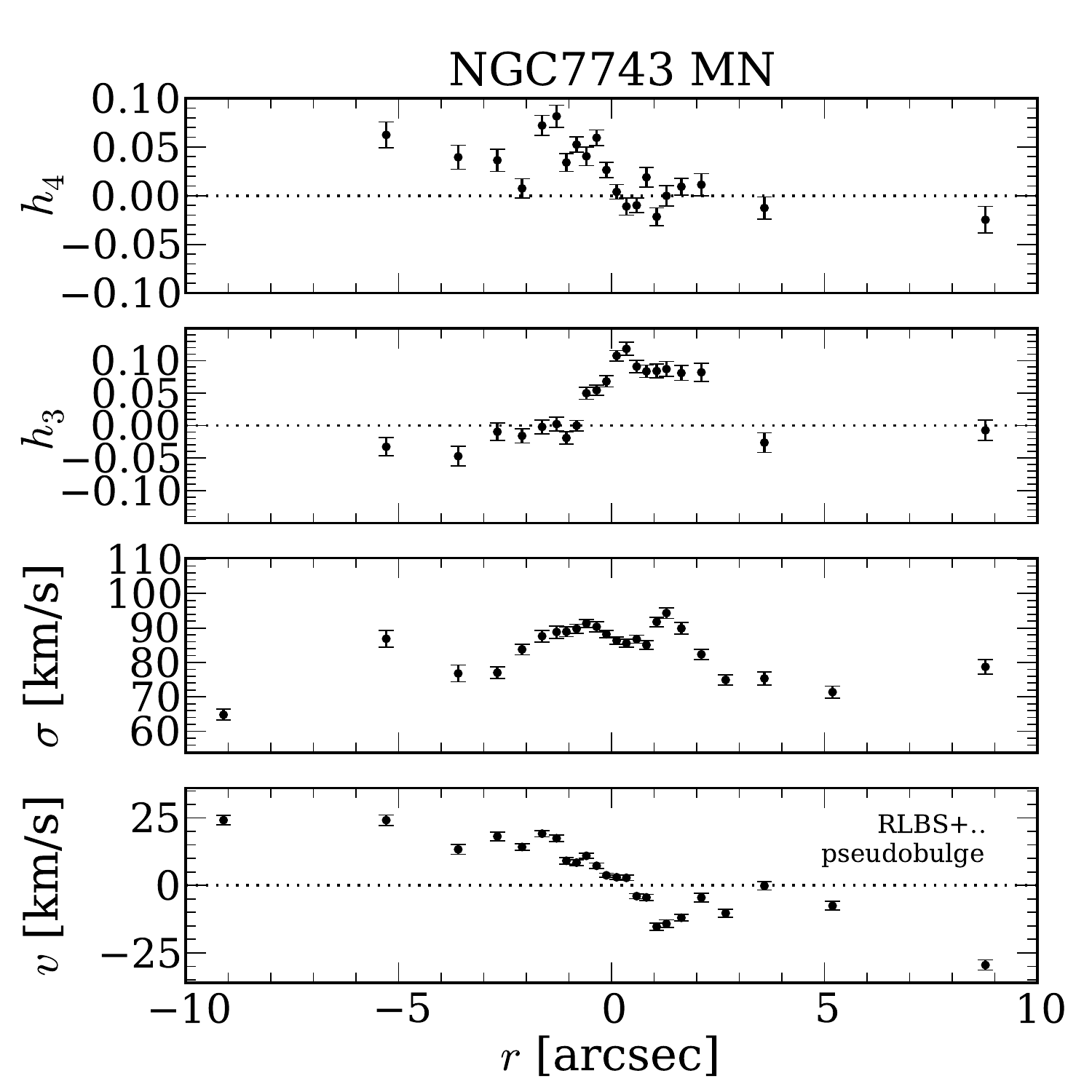}\\
        \end{tabular}
        \end{center}
        \caption{{\it continued --}\small Major and minor axis kinematic profiles for NGC\,7743.
	We plot results from \citet{Kormendy1982b} in green.}
\end{figure}
%
\clearpage
\section{Notes on individual galaxies}
\label{sec:individuals}
\subsection{Classical bulges}
\paragraph{NGC\,1023}
.LBT-.. --- Clean classical morphology in HST F555W. The rotation curve steps
rapidly from $\approx$ +60~kms$^{-1}$ to $\approx$ -60~kms$^{-1}$ in the central two
arcseconds and rises then gradually to a value of $\approx$ 200~kms$^{-1}$ at
50~arcseconds. The $h_3$ profile shows an equally fast change within the inner
$\approx 2$~arcseconds from -0.03 to 0.03 with opposite sign. Like
\citet{Emsellem2004} we see that the $v$/$h_3$ anti-correlation turns into a
correlation outwards of $\approx$10~arcseconds. The velocity dispersion profile
rises all the way to the centre. Outside of 50~arcseconds the velocity
dispersion profile flattens out at about 100~kms$^{-1}$ which coincides with a
flattening in the rotation curve. This is significantly beyond the bulge
radius of $\approx$~19~arcseconds. The minor axis rotation is mostly
close to zero at larger radii but becomes negative inwards of 4~arcseconds
($\approx$ -25~kms$^{-1}$ at the centre). The acquisition image does not show an
obvious offset of the minor axis slit but we note that due to the rapid rise of
major axis rotation in the central arcseconds already a small offset of
($\approx$ 0.5~arcsecond) to the west suffices to explain the observed
behaviour.  The $h_3$ moments become positive in the same radial range, which
is expected if the velocity offset is due to actual rotation. The minor axis
$h_4$ moments show a double peak at $\pm~7$~arcseconds with maximum values of
$h_4 \approx 0.04$.  The continuously centrally rising velocity dispersion of
the major axis is reproduced on the minor axis.
%
\paragraph{NGC\,2775}
.SAR2.. --- The HST~F606W image shows a very clear classical morphology, the
F450W image shows very little amounts of dust in the central region.  We see a
depression in the velocity dispersion profile inwards of 5~arcseconds in the
major axis profile as well as in the minor axis profile which coincides with a
steeper part in the rotation curve. \citet{Eskridge2002} describe a large,
slightly elliptical bulge which contains a bright nuclear point source.  The
Spitzer MIPS 24~\um~image shows a resolved nuclear source of emission which
clearly stands out from a region of low emission which again in size roughly
corresponds to the bulge radius. This may hint at a cold nuclear component with
active star formation which dominates the kinematics. We exclude the data
inwards of 5~arcseconds from our analysis but note that this choice does
significantly affect the position of NGC\,2775 in the plane of S\'ersic index
versus velocity dispersion slope.
\paragraph{NGC\,2841}
.SAR3*. --- The HST~F438 image shows a weak nuclear dust spiral that is
misaligned with the outer disc. The larger scale bulge morphology is smooth and
shows little sign of dust and no spiral pattern. The $h_3$ profile is
anti-correlated with the rotation curve inwards of 4~arcseconds but then
changes sign and becomes correlated with the velocity until about
20~arcseconds.  The velocity dispersion profile is centrally rising and may
show a little shelf inwards of $\approx$~5~arcseconds. The minor axis rotation
curve shows an offset within the bulge radius.
\paragraph{NGC\,2859}
RLBR+.. --- Prominent outer ring galaxy. This galaxy has no close-to $V$-band HST
image available. The bulge morphology in HST~F814W and in the acquisition
images is generally smooth and classical with few weak dust lanes. Our HET long
slit data do not cover the full bulge region ($r_b = 27.6$~arcsec). Still the
rotation curve starts to flatten out at our outermost data point at $r \approx
8$~arcseconds.  Within this region the velocity dispersion profile rises
centrally from about 125~kms$^{-1}$ at $\pm 6$~arcseconds to 175~kms$^{-1}$
in the centre. The $h_3$ moments are clearly anti-correlated with velocity. The
$h_4$ moments show indication of the double peak signature at $r \approx \pm 5$~
arcseconds. The minor axis kinematic data only reach out to 6~arcseconds with
little rotation along the minor axis (less than 25~kms$^{-1}$). The coverage of
the dispersion profile in insufficient to judge whether the central rise seen
along the major axis is reproduced. The minor axis $h_3$ and $h_4$ moments are
somewhat noisy but do not exceed values of 0.05 and show no significant trends.

%
\paragraph{NGC\,2880}
.LB.-.. --- The bulge morphology is classical in HST~F555W. \citet{Erwin04}
finds indication of a weak inner disk but acknowledges that this is the weakest
case in his sample. The velocity dispersion profile rises centrally with a weak
non-symmetric shelf-like structure inwards of $\approx$~4~arcseconds.  The $h_3$
moments are anti-correlated with velocity within the bulge.
\paragraph{NGC\,3031}
.SAS2.. --- M81, interacting with the M81 group. Prominent central emission features
connected to a liner type activity prevent us from deriving the central
kinematics ($r < \pm 2$ arcseconds) reliably. This galaxy exhibits an interesting
shape of its velocity dispersion profile.  The profile first rises gradually
until $\approx$~150~kms$^{-1}$ at radius of about 25~arcseconds. It then drops quickly
to a minimum of about 130~kms$^{-1}$ at 18~arcseconds and rises then again to
$\approx$~160~kms$^{-1}$ at the centre. The drop around 18~arcseconds is accompanied
by a rapid change of slope of the rotational velocity which stays relatively
flat outwards of this radius and a strengthening of the $h_3$ moments. Also the
otherwise vanishing $h_4$ moments rise to positive values ($\approx$~0.1) at
$r\approx$~15~arcseconds where they form the most prominent double peak feature
of our sample. The minor axis profile shows similar local minima in the
dispersion profile that are accompanied by local maxima in the $h_4$ moments at
radii of about 9~arcseconds. Given the inclination of $59 \Deg$ this points to
a flattened structure within the bulge.
\paragraph{NGC\,3245}
.LAR0*. --- The HST~F547M image reveals a nuclear dust disk while the larger scale bulge
morphology is classical. This galaxy potentially hosts a lens \citep{RC1}. The
rotational velocity exhibits a shelf reaching from about 3~arcseconds out to
the bulge radius at 15~arcseconds. The velocity dispersion profile rises
steadily towards the centre. The $h_3$ moments are anti-correlated with
velocity inwards of 5~arcseconds. At 5~arcseconds they reach a maximum and
then drop to zero towards the bulge radius. The $h_4$ 
are compatible with zero at the bulge radius ($r_b = 8.5$~arcseconds) 
but with decreasing radius
briefly drop to about -.05 at $\pm 5$~arcseconds. Finally inside of
2.~arcseconds they become compatible with zero.
\paragraph{NGC\,3521}
.SXT4.. --- The latest Hubble type with classical bulge in our sample.  In
HST~F606W the classical bulge morphology stands in strong contrast to the
strong outer disk spiral structure with a sudden transition of those two
morphologies at r~$\approx$~10~arcseconds.  The rotation curve forms a shelf at
about 3~arcseconds but rises again slightly towards larger radii and starts to
flatten out at 16~arcseconds. The velocity dispersion reaches a local maximum
of $\approx$~115~kms$^{-1}$ at the bulge radius ($r_b = 10.8$~arcseconds) but
then strongly drops with decreasing radius and reaches a local minimum of
$\approx$~100~kms$^{-1}$ at 4~arcseconds. Further inwards the dispersion is
centrally peaked.  Outside the bulge region the dispersion shows another local
maximum around 40~arcseconds which also corresponds to a slight secondary shelf
on the rotational velocity.  The $h_3$ moments are anti-correlated with
velocity and form a shelf at about the same radii where the inner shelf in
rotational velocity is seen. Outside the bulge $h_3$ becomes rather strong (up
to 0.22) which is accompanied by strong, positive $h_4$ moments of up to 0.15.
These large moments are a consequence of the double-peak structure of the
LOSVDs which has been previously reported by \citet{Zeilinger2001} caused by a
secondary kinematic component (see \S~\ref{sec:kindecomp}). The minor axis
velocity dispersion profile does not show the strong local minima that the
major axis profile shows but is similarly disturbed outside the bulge.
\paragraph{NGC\,3898}
.SAS2.. --- Outer disk dust spiral transitions into weak dust lanes over a
smooth and classical bulge morphology in HST F606W. The rotation curve rises
slowly out to r~$\approx$~10~arcseconds and flattens at about the
bulge radius of $\approx$~15.7~arcseconds. The velocity dispersion profile is
strongly centrally peaked and rises smoothly from $\approx$~140~kms$^{-1}$ to
$\approx$~220~kms$^{-1}$ in the centre. The $h_3$ moments are anti-correlated
with velocity and reach local maximum at r~$\approx$~10~arcseconds, the same
radius where the flattening of the rotation curve sets in.
\paragraph{NGC\,3992}
.SBT4.. --- This galaxy is prominently barred. While the bulge morphology generally
appears classical it exhibits a few randomly distributed dust lanes in the only
available optical HST image in F547M.  With increasing radius the velocity
profile reaches a first plateau at r~$\approx$~7~arcseconds but outside the
bulge radius of $13.2$~arcseconds it starts rising again slowly.  The velocity
dispersion profile is significantly depressed within the bulge region.  The
$h_3$ moments are strongly anti-correlated with velocity inwards of
7~arcseconds but become weaker further out. The $h_4$ moments show double peak
feature around 8~arcseconds.
\paragraph{NGC\,4203}
.LX.-*. --- This galaxy represents a borderline case in the morphological bulge
classification. The bulge has a few dust lanes in HST~F555W superimposed on a
generally smooth morphology.  \cite{Fisher2008} classify the bulge as
classical.  The major axis velocity profile flattens out at about the bulge
radius of 14.7~arcseconds. The velocity dispersion profile shows a prominent
rise from about 105~kms$^{-1}$ at the bulge radius to 175~kms$^{-1}$ in the
centre. The $h_3$ moments are mostly compatible with zero in the bulge region.
The $h_4$ moments are noisy and show no significant trend within the covered
region.
\paragraph{NGC\,4260}
.SBS1.. --- Strongly barred. The bulge morphology is generally smooth.
\citet{Martini2003} acknowledge the presence of dust structures in the central
region but claim that they do not imply a sense of rotation. The bulge is
classified as a classical bulge by \citet{Fisher2008}.  The velocity dispersion
profile is somewhat irregular and S-shaped within the bulge region.  The $h_3$
moments are small and mostly compatible with zero, The $h_4$ moments show a
gradual drop from a central value of zero to about $-0.06$ at about
10~arcseconds.
\paragraph{NGC\,4379}
.L..-P* --- Exhibits a very smooth and featureless, classical central
morphology in HST~F555W.  The bulge radius is 8.6~arcseconds. The inset of the
flattening of the rotational velocity occurs already at 8-10~arcseconds.  The
velocity dispersion profile is centrally peaked, rising from
$\approx$~90~kms$^{-1}$ at 10~arcseconds to $\approx$~120~kms$^{-1}$ in the
centre. The $h_3$ moments are generally small (within the errors mostly
compatible with zero) but an overall trend points to anti-correlation with
velocity. The $h_4$ values are are noisy but mostly compatible with zero.
\paragraph{NGC\,4698}
.SAS2.. --- HST~F606W shows some weak dust lanes in the central region which
however do not imply any sense of rotation. It is classified as classical bulge by
\citet{Fisher2010}. 
\citet{Falcon-Barroso2006} find that the stellar velocity
field displays rotation perpendicular to the major axis within the central
$\approx \pm 5$~arcseconds.  
The major axis rotational velocity is very slowly
rising indicative of counter rotation. The velocity dispersion profile is
mostly flat within the bulge region ($r_b = 10.7$~arcseconds) with a weak
central peak. In the small radial range that our data cover the $h_3$ and $h_4$
moments are mostly compatible with zero.
\paragraph{NGC\,4772}
.SAS1.. --- The bulge morphology is smooth and featureless in HST~F606W, the
bulge is consequently classified as classical by \citet{Fisher2008}.  
The [O\,\textsc{iii}] map of \citet{Falcon-Barroso2006} shows almost counter
rotation with respect to the stars in the central $\approx \pm 5$~arcseconds
(see also \citealp{Haynes2000}).
Our data
virtually only cover the bulge region. The velocity dispersion profile is
generally noisy with values between 100~kms$^{-1}$ and 150~kms$^{-1}$ but
features a clear central peak. The $h_3$ moments are anti-correlated with
velocity and reach values of up to about $\pm 0.07$. $h_4$ moments scatter
around values of 0.05 at all radii.
\subsection{Pseudobulges}
\paragraph{NGC\,2681}
PSXT0.. --- Possibly a triple barred system \citep{Erwin99}. In HST~F555W a
dust spiral is seen, which extends all the way into the centre. The rotational
velocity curve shows a shelf between $\approx$~2~and $\approx$~12~arcseconds. The
outer radius of the shelf region coincides with the bulge radius ($r_b =$
13.2~arcseconds). Inwards of 2~arcseconds the rotation curve drops quickly to
zero. With decreasing radius, the velocity dispersion rises from about
50~kms$^{-1}$ to a values of $\approx$~75~kms$^{-1}$ at the bulge radius.
Towards smaller radii it first stays relatively constant but then shows a step
at about 5~arcseconds and rises again inwards of 2~arcseconds. The $h_3$
moments are anti-correlated with velocity in the region of the fast velocity
rise but become correlated in the velocity shelf region. The $h_4$ moments show
a double peak feature in the radial range of 2-4~arcseconds.
\paragraph{NGC\,2964}
.SXR4*. --- High contrast dust spiral in HST~F606W. 
The bulge is small ($r_b =
3.1$~arcseconds) and we do not sufficiently resolve it to include this galaxy
in any of our structural plots. Here we publish the kinematic profiles.  The
velocity dispersion rises from values of about 40~kms$^{-1}$ in the disk to $\approx$
106~kms$^{-1}$ at the bulge radius and exhibits a depression within the bulge. $h_3$
and $h_4$ moments scatter strongly within values of $\pm 0.1$ which is possibly
a consequence of the dust.
%
\paragraph{NGC\,3166}
.SXT0.. --- The HST~F547M image shows strong dust features and a dust spiral
which extends all the way into the centre. \citet{Laurikainen04} describe this
galaxy as strongly barred.  The velocity dispersion profile shows a strong
depression at $\approx$~4.5~arcseconds which is accompanied by relatively strong
$h_3$ moments of $\pm 0.17$ and positive $h_4$ moments and a local maximum in
the rotation curve. The minor axis shows small but significant rotation within
the bulge region ($\approx\pm$~10~kms$^{-1}$) while the $h_3$ moments are
largely compatible with zero. The depression of the velocity dispersion is also
seen on the minor axis whilst not as strong.
\paragraph{NGC\,3351}
.SBR3.. --- The bulge hosts a clear spiral structure and a nuclear ring
\citep{Fisher2010} and shows signs of active star formation. 
SAURON data \citep{Dumas2007} show a drop in the gas velocity dispersion 
derived from H$\beta$ and lowered [O\,\textsc{iii}]/H$\beta$ 
ratios in the ring indicative of star formation.
The velocity
dispersion profile shows a depression within the bulge region --- most strongly
at r~$\approx$~5~arcseconds down to a value of 70~kms$^{-1}$ --- and rises again
centrally to $\approx$~90~kms$^{-1}$ which is still below the values of about
100~kms$^{-1}$ seen just inside of the bulge radius ($r_b = 12.9$~arcseconds).
The $h_3$ moments are clearly anti-correlated with velocity. The minor axis
profile shows significant rotation ($v_{max } \approx $50~kms$^{-1}$) indicative
of a slit misalignment, in fact there is a 25\Deg difference between our major
axis slit position of 165\Deg and the Hyperleda published value of 9.9\Deg.
%
\paragraph{NGC\,3368}
.SXT2.. --- Complex morphology with a number of stellar components \citep{Erwin04,
Nowak2010}. This galaxy is possibly double-barred \citep{Jungwiert97}.  The
bulge hosts a strong nuclear spiral and an inner disk \citep{Erwin04} and is
classified as pseudobulge by \cite{Fisher2010}.
The bulge radius is 20.4~arcseconds. The complex morphology is
reflected in the kinematic structure. The rotational velocity reaches a local
maximum at about 7~arcseconds.  This is accompanied by local minimum in the
velocity dispersion profile that has been rising inwards until
$\approx$~13~arcseconds.  Inwards of 7~arcseconds the dispersion rises again, but
asymmetrically about the centre, see also \citet{Nowak2010}. The local maximum
in velocity and local minimum in dispersion coincide with
strengthened $h_3$ moments. $h_3$ is anti-correlated with velocity inwards of
15~arcseconds but correlated outside. The $h_4$ moments are close to zero at
13~arcseconds but become positive and reach a local maximum at about the
same radii where the local maxima in velocity are observed and the drops in
velocity dispersion and strengthening of $h_3$ moments occur. The minor axis
profile shows similar depressions in velocity dispersion. The minor axis
velocity profile shows a central peak of about 30~kms$^{-1}$. Visual inspection of
the HET pre-acquisition images reveals that a minor offset of the slit position
($\approx 1$~arcsec) to the west is responsible for the peak.
\paragraph{NGC\,3384}
.LBS-*. --- Contains a nuclear bar \citep{Fisher2010} and a rapidly rotation disk described by
\citet{Busarello96,Fisher1997} and \citet{Emsellem2004} already find strong $h_3$
moments in anti-correlation with velocity. The velocity dispersion profile
changes slope at 10~arcseconds and becomes more shallow towards smaller radii,
but then exhibits a pronounced peak inwards of
3~arcseconds that is accompanied by a dip (as seen by \citet{Fisher1997, Emsellem2004}
as well) in the $h_4$ moments that become positive just outside of this region.
%
\paragraph{NGC\,3627}
.SXS3.. --- Prominently barred galaxy with wide, open arms, interacting with the
Leo group. High contrast dust lanes that extend to the very centre are seen in
HST F606W and let \citet{Fisher2008} classify this as a pseudobulge.
After a fast rise the rotational velocity
forms a shelf between 3~arcseconds and the bulge radius of about 11~arcseconds.
Towards larger radii the velocity rises again. The velocity dispersion rises
inwards, starting already far outside the bulge radius.  At
$\approx$~4~arcseconds it flattens out and stays essentially constant.  The $h_3$
moments are anti-correlated with velocity inside of 9~arcseconds but change
sign at larger radii and become correlated with velocity. While the minor axis
rotation is compatible with zero at larger radii it exhibits significant
rotation inwards of about 7~arcseconds that is also seen in anti-correlated
$h_3$ moments.
\paragraph{NGC\,3675}
.SAS3.. --- The HST~F606W image clearly shows a high contrast flocculent spiral that extends all
the way into the centre. The velocity dispersion reaches a maximum of about 110~kms$^{-1}$
at $\approx$~40~arcseconds and stays rather constant inside of this radius. The $h_3$
moments are anti-correlated with velocity within the bulge and reach maximum
values of about 0.1 at the bulge radius of 8.5~arcseconds.
\paragraph{NGC\,3945}
RLBT+.. --- Double barred \citep{Kormendy79, Kormendy1982b, Wozniak95, Erwin99,
Erwin03, Erwin04} galaxy with prominent outer ring. Exhibits complicated
kinematic structure. The rotational velocity has local minima around
18~arcseconds, rises then towards smaller radii and reaches a local maximum
around 8~arcseconds before it falls off to the centre.  The dispersion profile
has very strong local minima --- drops from $\approx$~150~kms$^{-1}$ to
$\approx$~110~kms$^{-1}$ --- at r~$\approx$~8~arcseconds but then rises again
towards the centre. This galaxy shows exceptionally strong $h_3$ moments of up
to $\approx$~0.25 which are anti-correlated with velocity in the inner region
but become positively correlated at about 20~arcseconds. The $h_4$ moments are
similarly strong (up to 0.2) with significant central depression. The LOSVDs do
show significant low velocity shoulders at radii between $2'' < \|r\| < 10''$
which is indicative of a kinematic distinct component (see Sec.
\ref{sec:kindecomp}).  The minor axis profile also shows a local depression in
velocity dispersion but at r~$\approx$~3.5~arcseconds, the central dip in $h_4$
is seen as well.  We measure slight rotation ($i\approx\pm$~25~kms$^{-1}$)
along the observed position angle of $i = 64 \Deg$ and a significant central
offset of the velocity ($\approx$~40~kms$^{-1}$). The major axis velocity
profile shows that a slit position offset of one arcsecond is sufficient to
explain the central velocity peak. A visual inspection of the pre-acquisition
image confirmed that such an offset was indeed present (about 0.6 arcseconds).
\paragraph{NGC\,4030}
.SAS4.. --- The flocculent spiral structure --- easily seen in HST F606W ---
extends all the way to the centre with star-forming knots in the inner disk
\citep{Eskridge2002}. \cite{Fisher2008} classify it as pseudobulge.  The major
axis rotational velocity reaches $\approx$100~kms$^{-1}$ at about $\pm
10$~arcseconds. The velocity dispersion stays moderately flat within the bulge
region and drops off outside
, this was also observed by \citep{Ganda2006}
. The $h_3$ moments are clearly anti-correlated
with velocity and reach a values of up to 0.1 at a radius of about
7~arcseconds. $h_4$ moments are compatible with zero at all radii that are
covered by our data. The bulge is small ($r_b = 3.0$~arcseconds) and we do not
sufficiently resolve it to include this galaxy in any of our structural plots.

\paragraph{NGC\,4274}
RSBR2.. --- 
Double barred galaxy \citep{Shaw1995,Erwin04}.
It is classified as pseudobulge with a S\'ersic index of $1.60~\pm~0.35$ by
\citep{Fisher2010}. The bulge hosts a prominent nuclear spiral including strong
dust lanes and a nuclear ring. 
The ring is seen as a fast-rotating, low-dispersion component in SAURON
velocity and dispersion maps \citep{Falcon-Barroso2006}. It also appears in
their ionized gas maps through increased H$\beta$ emission and lowered
[O\,\textsc{iii}]/H$\beta$ ratios which indicates star-formation.
The major axis rotational velocity
rises quickly with increasing radius and starts flattening out at about
4~arcseconds, well within the bulge radius of 11.3~arcseconds. The velocity
dispersion profile shows a strong depression inside the bulge with values of
about 100~kms$^{-1}$. Outside the bulge the dispersion rises to values exceeding
130~kms$^{-1}$. The $h_3$ moments are well anti-correlated with velocity in the radial range
that is covered by our data. They reach values of up to $\pm 0.14$ at about the
same radius where the rotational velocity starts to flatten out. This is
accompanied by peaks in the $h_4$ moments with values of up to 0.14.
\paragraph{NGC\,4314}
.SBT1.. --- Strongly barred galaxy with prominent nuclear ring. 
The ring appears in SAURON ionized gas maps \citep{Falcon-Barroso2006} through
lowered velocity dispersions, increased H$\beta$ and lowered
[O\,\textsc{iii}]/H$\beta$ which indicates star-formation.
The bulge is consequently classified as pseudobulge by \citet{Fisher2008}. The
rotational velocity starts to flatten at the bulge radius of 8.6 ~arcseconds.
The velocity dispersion profile is asymmetric but relatively flat, it varies
between values of 105~kms$^{-1}$ and 130~kms$^{-1}$. The $S/N$ only allows us
to derive $h_3$ and $h_4$ moments inside a radius of 6~arcseconds. The $h_3$
moments are mostly compatible with zero, the $h_4$ moments rise from zero in
the centre to values around 0.05 at 4~arcseconds.
\paragraph{NGC\,4371}
.LBR+.. --- Strongly barred galaxy. \citet{Erwin99} find a bright stellar ring
that is notable by adjusting the contrast of the e.g. the HST~F606W image
carefully or through unsharp masking.  While free of obvious dust or spiral
structures the ring with a radius of about 5~arcseconds falls within the bulge
radius which lets \citet{Fisher2010} classify this as a pseudobulge.  The major
axis rotational velocity starts to flatten out at a radius of 8~arcseconds ---
well inside the bulge radius or 22.9~arcseconds --- with a weak shelf around
3-8~arcseconds.  The velocity dispersion rises from about 105~kms$^{-1}$ at the
bulge radius to about 130~kms$^{-1}$ at 7~arcseconds. Inside of 7~arcseconds
the velocity dispersion stays relatively constant. The $h_3$ moments are
somewhat asymmetric but generally anti-correlated with velocity within the
bulge. The $h_4$ moments are compatible with zero in the centre but rise
gradually to values of about 0.05 at 10~arcseconds. The minor axis velocity
dispersion stays mostly constant within the covered radial range. The minor
axis $h_3$ moments remain compatible with zero while the $h_4$ moments rise
from zero at the centre to values of about 0.05 at $\pm 9$~arcseconds.
\paragraph{NGC\,4394}
RSBR3.. --- Strongly barred galaxy with a face-on spiral in the central
r~$\approx 5 $~arcseconds.  \citet{Fisher2008} classify it as a pseudobulge. The
major axis velocity profile rises quickly to about 50~kms$^{-1}$ at
3~arcseconds. The velocity dispersion exhibits two prominent maxima around $\pm
7$~arcseconds. The maxima reach values of about 105~kms$^{-1}$ but drop
quickly to about 80~kms$^{-1}$ at 4~arcseconds and stay relatively constant
towards smaller radii from there on. The $h_3$ moments are anti-correlated with
velocity inside the bulge. The $h_4$ values are asymmetric and somewhat large
within the bulge with values of up to 0.07 but we note that with dispersion
values around 80~kms$^{-1}$ we reach the limit of our ability to resolve those
values properly.
\paragraph{NGC\,4448}
.SBR2.. --- \citet{Fisher2010} note a mild spiral structure that extends into
the centre and classify it as a pseudobulge. The major axis
rotational velocity starts to flatten out at about 5~arcseconds --- just inside
of the bulge radius of 8.5~arcseconds. Outside of 5~arcseconds the velocity
profile exhibits a weak shelf.  The velocity dispersion is mostly constant at a
value of $\approx$~115~kms$^{-1}$ at all radii covered by our data. The $h_3$ moments
are somewhat asymmetric, close to zero at positive radii but vary strongly at
negative radii. The $h_4$ moments scatter about zero in the bulge region with a
few outliers at -0.05.
\paragraph{NGC\,4501}
.SAT3.. --- A nuclear spiral extends all the way to the centre in HST~F606W.
\citet{Fisher2010} classify the bulge as pseudobulge. 
The major axis velocity profile rises quickly from the
centre to the bulge radius of 6.2~arcseconds but then flattens out at the bulge
radius and forms a shelf out to about 20~arcseconds.  The velocity dispersion
rises from 75~kms$^{-1}$ in the disk to about 150~kms$^{-1}$ at 15~arcseconds.  The $h_3$
moments are strongly anti-correlated with velocity inside the bulge and reach
values of $\pm$~0.1 at the bulge radius. $h_4$ moments are mostly compatible
with zero at all radii. The minor axis velocity dispersion profile starts
rising with decreasing radius inwards of 30~arcseconds and reaches a maximum of
$\approx$~155~kms$^{-1}$ at $r = 10$~arcseconds. The minor axis velocity dispersion
exhibits a central depression of about 15~kms$^{-1}$. The minor axis velocity, $h_3$
and $h_4$ moments are mostly compatible with zero.
\paragraph{NGC\,4536}
.SXT4.. --- A strong dust spiral extends into the very centre.
\citet{Fisher2008} classify the bulge as pseudobulge. 
The bulge radius is 10.1~arcseconds. The major axis
rotational velocity flattens well inside bulge radius at around 2~arcseconds.
The velocity dispersion profile is slightly asymmetric but mostly constant
within the bulge region. The $h_3$ moments are anti-correlated with velocity
inside the bulge region and reach values of about 0.1. The $h_4$ profile is
asymmetric. The minor axis velocities show slight asymmetric rotation
($\approx$~25~kms$^{-1}$). The velocity dispersion profile rises centrally but
stays relatively flat within $\pm$~10~arcseconds. The minor axis $h_3$ moments
are somewhat noisy but seem to show a central depression.
\paragraph{NGC\,4569}
.SXT2.. --- A nuclear spiral extends all the way to the centre. The bulge is
classified as pseudobulge by \citet{Fisher2010}.  The major axis rotational
velocity rises with increasing radius to a local maximum of about 80~kms$^{-1}$
at $\pm$~3~arcseconds. The bulge radius is $r_b = 9.6$~arcseconds. The velocity
then drops to about 50~kms$^{-1}$ at the maximum radius covered by our data.
The velocity dispersion rises with decreasing radius to about 100~kms$^{-1}$ at
a radius of 6~arcseconds and then drops and reaches a local minimum at around
3~arcseconds, roughly coinciding with the locations of the local maxima in the
velocity profile. The $h_3$ moments are anti-correlated with velocity inside
the bulge. Their absolute values reach up to 0.15. The $h_4$ moments show a
strong double peak feature at about 3~arcseconds and fall off to zero at the
bulge radius. The minor axis profile shows rotation in the bulge region. While
somewhat asymmetric, the minor axis dispersion profile does not show the same
complicated structure of the major axis profile. 
The $h_3$ moments on the minor axis are mostly compatible with zero. 
The $h_3$ moments on the minor axis are mostly slightly positive with a mean value 
of 0.03. 
The $h_4$ moments are generally noisy but
the double peak feature of the major axis is reproduced.
\paragraph{NGC\,4736}
RSAR2.. --- Hosts a nuclear bar \citep{Sakamoto1999} and prominent nuclear
spiral which extends all the way into the centre in HST~F555W.  The bulge is
classified as pseudobulge by \citet{Fisher2010}.
The obtained kinematic data extend well into the disk. The
rotational velocity flattens out abruptly at about the bulge radius of
14.2~arcseconds and shows a shallow negative gradient out to about
70~arcseconds where our data points start to become sparse. The velocity
dispersion rises abruptly from about 75~kms$^{-1}$ to 115~kms$^{-1}$ at about the bulge
radius. Well within the disk at radii larger than 50~arcseconds we see again a
gradual increase of velocity dispersion. Inside of 2.5~arcseconds the velocity
dispersion exhibits a central drop. The $h_3$ moments are anti-correlated with velocity
but show s-shape around the centre. They reach exceptionally large values of
$\pm 0.2$ at the bulge radius. The $h_4$ moments are compatible with zero in the
inner bulge but reach pronounced local maxima of values as large as 0.25 at
about the bulge radius. They fall off to zero at $r\approx 35$~arcseconds.
These strong higher moments are a consequence of the multi-component structure
of the LOSVDs at the respective radii (see Sec. \ref{sec:kindecomp}). The
minor axis profile reflects the rich structure seen in the major axis profile.
The velocity dispersion rises significantly inwards of 10~arcseconds. The $h_3$
moments are mostly compatible with zero at all radii, $h_4$ moments are zero
inside of 10~arcseconds but rise to about 0.1 at 20~arcseconds.
\paragraph{NGC\,5055}
.SAT4.. --- The HST~F606W image shows that the outer disk flocculent spiral
extends into the very centre. Consequently this galaxy is classified as
pseudobulge by \citet{Fisher2010}. The velocity dispersion profile remains flat
inside the bulge radius of 18.3~arcseconds. The $h_3$ moments are
anti-correlated with velocity inside the bulge and reach absolute values of up
to 0.1 at the bulge radius. The $h_4$ moments are compatible with zero inwards
of 10~arcseconds, they become noisy further out but show a weak tendency
towards more positive values towards the bulge radius.  The minor axis
velocities appear somewhat irregular but small ($< 20$~kms$^{-1}$). The minor
axis $h_3$ moments are noisy but mostly scatter close to zero. Again the $h_4$
moments are mostly compatible with zero inwards 10~arcseconds.
but show a weak increase further out but only on the east side.
%
\paragraph{NGC\,5248}
.SXT4.. --- Has a prominent nuclear spiral clearly visible in HST~F814W.
SAURON maps show the presence of a nuclear ring in H$\beta$ and
[O\,\textsc{iii}] emission. A lowered [O\,\textsc{iii}]/H$\beta$ shows that the
ring is star-forming.
The bulge was classified as pseudobulge by
\citet{Fisher2010}.  The rotational velocity starts to flatten at about
6~arcseconds --- well inside the bulge radius of 15.4~arcseconds. The velocity
profile shows a shelf between $\approx$~10~arcseconds and
$\approx$~40~arcseconds.  The velocity dispersion profile is mostly flat at
about 80~kms$^{-1}$ with two small peaks at $r \approx \pm 4$~arcseconds. The
$h_3$ moments are anti-correlated with velocity and reach values of up to $\pm
0.1$. The $h_4$ moments scatter around values of 0.05.
\paragraph{NGC\,5566}
.SBR2.. --- Shows a nuclear spiral in HST F606W. The surface brightness profile does not
resemble a traditional bulge plus disk structure. We do not include this galaxy
in any of the structural plots and publish only the kinematic profile here.
The rotational velocity starts to flatten at 6~arcseconds. The velocity
dispersion profile rises towards the centre and peaks at a value of about
150~kms$^{-1}$. The $h_3$ moments are anti-correlated with velocity and reach values
of up to 0.15. The $h_4$ moments are small in the central arcseconds but rise
to values of about 0.1 at $\approx \pm 5$~arcseconds.
\paragraph{NGC\,7177}
.SXR3.. --- A nuclear bar extends out to about $r = 10$~arcseconds. The bulge
is classified as pseudobulge by \citet{Fisher2010}. The flattening of the
major axis rotational velocity coincides with bulge radius of $r_b =
8.6$~arcseconds. The velocity dispersion rises from about 50~kms$^{-1}$ in the
disk to values of $\approx$ 115~kms$^{-1}$ inside the bulge but remains
relatively flat inside the bulge radius. The $h_3$ moments are weakly
anti-correlated with velocity but remain small. The $h_4$ moments drop to
values of -0.05 at $\approx \pm 5$~arcseconds. The minor axis shows an
asymmetric velocity profile with values of up to $\pm 25$~kms$^{-1}$. The minor
axis velocity dispersion again rises from about 50~kms$^{-1}$ in the disk to
values of about 115~kms$^{-1}$ at 2.5~arcseconds and remains flat inside. The
$h_3$ moments are anti-correlated with velocity. The $h_4$ moments are noisy
and scatter around zero.
\paragraph{NGC\,7743}
RLBS+.. --- The central region exhibits some weak dust lanes overlaid on a
generally \citep{Martini2003} smooth light distribution. The bulge is
classified as classical bulge by \citet{Fisher2008}. The amplitude of the
rotation is small at about $\pm 25$~kms$^{-1}$ due to the low inclination. We
find rotation of similar value along the minor axis slit due to a misplacement
of the slit ($166 \Deg$ rather than $10 \Deg$). The velocity dispersion is flat
for both position angles and takes values of about 80~kms$^{-1}$. Both, major
axis and minor axis $h_3$ moments are anti-correlated with velocity and
become compatible with zero at the bulge radius ($r_b = 5.6$~arcseconds). The
$h_4$ moments are very noisy, probably due to the low velocity dispersion of
this object.
\subsection{Bulges without classification}
\paragraph{NGC\,2460}
.SAS1.. --- Mixed type morphology in HST~F606W with a weak asymmetric dust
structure in the bulge region that is overlaid on an otherwise smooth light
distribution. We label this bulge as unclassified.  The decomposition gave a
value of $3.5 \pm 0.32$ for the S\'ersic index and 6.6~arcseconds for the bulge
radius.  Within this region the velocity dispersion profile is flat and $h_3$
and $h_4$ moments scatter around zero.
\paragraph{NGC\,3593}
.SAS0*. --- Peculiar bulge structure with prominent spiral visible even in
NIC~F160W. The bulge is classified a pseudobulge by \citet{Fisher2010}.  We
label it as {\it non-classified} because its high inclination inhibits an
unperturbed view into the bulge region. This is the only example in our sample
where counter rotation is seen in the velocity profile as an actual change of
the sign of the velocity with respect to the systemic velocity (this was found
also by \citealp{Bertola1996}). The counter rotation within the bulge radius is
reflected in the anti-correlated $h_3$ moments. The velocity dispersion drops
dramatically from $\approx$~115~kms$^{-1}$ at the bulge radius to
$\approx$~60~kms$^{-1}$ in the centre.
\paragraph{NGC\,3953}
.SBR4.. --- \cite{Fisher2010} classify this bulge as pseudobulge but
acknowledge that there is no optical HST image available.  Here we label it as
{\it non-classified}.  The rotational velocity profile first reaches a weak
local maximum at r~$\approx$~2.5~arcseconds before it starts rising again
outside of 8~arcseconds. The disk velocity dispersion rises centrally from
values of $\approx 50$~kms$^{-1}$ at $\pm$~30~arcseconds to $\approx
130$~kms$^{-1}$ at $\approx 8$~arcseconds. Inside of a radius of 6~arcseconds it
then falls toward a central value of $\approx 110$~kms$^{-1}$. The fast central
increase of velocity is accompanied by strong anti-correlated $h_3$ moments
with values of up to $\pm$~0.1 at $r \approx 6$~arcseconds, they become
correlated with velocity outside of 8~arcseconds.  The $h_4$ moments are
generally noisy in the bulge region and scatter between zero and 0.05.
\paragraph{NGC\,4826}
RSAT2.. --- Also named the {\it black eye} galaxy. An extreme dust spiral in
the central 50~arcseconds stands in strong contrast to a virtually dust free
outer disk. The central dust content leads to a classification as pseudobulge
in \citet{Fisher2008}. The major axis kinematic profile is rich in structure.
The rotational velocity rises quickly from the centre to a value of $\approx
50$~kms$^{-1}$ at $r = 4$~arcseconds. It then forms a shallow trough around
$8$~arcseconds and then rises again --- more slowly --- out to 50~arcseconds
where it finally flattens out. 
The velocity dispersion in the disk is
$\approx$~45~kms$^{-1}$, it shows a distinct central increase inwards of
50~arcseconds. 
The dispersion reaches values of up to 110~kms$^{-1}$ inside of
the bulge radius ($r_b = 25.4$~arcseconds). From there on it stays relatively
constant with decreasing radius except for a mild depression down to
90~kms$^{-1}$ in the central few arcseconds. The $h_3$ moments are strongly
anti-correlated with velocity for $r < 9$~arcseconds and reach absolute values
of up to 0.15. The $h_4$ moments show two peaks at about $\pm 3$~arcseconds. We
find weak rotation along the minor axis ($\approx \pm 10$~kms$^{-1}$). The
central increase in velocity dispersion is also seen along the minor axis, the
increase sets in at a radius of about 25~arcseconds. 
This is much closer to the centre than in the case of the major axis and points
to a flattened structure.  For the photometry we obtain a mean bulge
ellipticity of 0.23.  The radial difference of the dispersion increase along
the major and the minor axis would point to an ellipticity of about 0.45.  
The centre of the velocity dispersion profile is asymmetric which may be a
consequence of the strong dust. The minor axis $h_3$ and $h_4$ moments are
compatible with zero.  The fact that the final flattening of the rotational
velocity and the inset of the dispersion increase at 50~arcseconds falls
together with the sudden appearance of the strong dust structure is intriguing
and lets one suspect that the actual bulge radius fall closer to
50~arcseconds. The much smaller bulge radius from the decomposition may be a
consequence of the strong dust content in the central regions (see also
discussion in \S~\ref{sec:struct_vs_kin}).
\paragraph{NGC\,7217}
RSAR2.. --- \cite{Fisher2010} point out a sudden break in morphology at a
radius of about 8~arcseconds where the outer spiral transitions onto a
relatively smooth morphology with little dust. They consequently label this
bulge as classical. However we determine a bulge radius of 11.2~arcseconds
where there is already a pronounced spiral pattern visible. We label this
galaxy as {\it non-classified}.  The rotational velocity profile starts to
flatten out at about the bulge radius of. The major axis dispersion profile is
asymmetric with higher values on the east side of the centre. Within the
central $\pm$~2~arcseconds the velocity dispersion shows a mild depression of
about 20~kms$^{-1}$. The $h_3$ moments are well anti-correlated with velocity.
The $h_4$ moments scatter around values of 0.05. We observe mild rotation on
the minor axis (of the order of $\pm 10$~kms$^{-1}$) indicative of a slight
slit misalignment. The minor axis velocity dispersion is symmetric and rises
towards the centre from values of about 100~kms$^{-1}$ at radii of
$\pm$~20~arcseconds to 145~kms$^{-1}$ at 2~arcseconds. Within the central
arcseconds the mild depression which is seen on the major axis is reproduced on
the minor axis. The minor axis $h_3$ moments mostly scatter around zero while
$h_4$ moments fall closer to 0.05 with a few relatively large outliers at radii
around 6-10~arcseconds. \citet{Merrifield1994} found the 20\%-30\% of the light
is captured in a counterrotating component. We confirm this and present a
kinematic decomposition in \S\ref{sec:kindecomp}.
\paragraph{NGC\,7331}
.SAS3.. --- The HST~F555W image shows several dust lanes in the bulge region. However these
do not imply any sense of rotation and seem to be overlaid on a generally
smooth light distribution. \citet{Fisher2008} admit that the high inclination
leaves the classification as classical bulge questionable. Here we label it as
{\it non-classified}.  The rotational velocity profile is already flattened at
the bulge radius of $r_b = 26$~arcseconds. The velocity dispersion rises from
about 75~kms$^{-1}$ in the disk to 125~kms$^{-1}$ in the centre. The dispersion profile has
two steps or shoulders at $\approx \pm 20$~arcseconds. The $h_3$ moments are
generally anti-correlated with velocity and reach a local maximum of about $\pm 0.15$
at $r = 15$~arcseconds. Also the $h_4$ moments reach local maxima with values of up to 0.15
in the same radial range.These large moments are a consequence of the
double-peak structure of the LOSVDs caused by a counterrotating kinematic component
discovered by \citet{Prada1996} (see \S~\ref{sec:kindecomp}).

\end{document}